\pdfoutput=1
\documentclass[useAMS,usenatbib]{mn2e}
 \usepackage[pdftex]{graphicx}

\usepackage{txfonts}
\title[A near IR imaging survey of high and intermediate-mass young 
stellar outflow candidates]
{A near IR imaging survey of intermediate and high-mass young stellar
outflow candidates}

\author[W.~P. Varricatt,  C.~J. Davis, S. Ramsay \& S.~P. Todd]
{Watson P. Varricatt$^{1}$\thanks{E-mail:
w.varricatt@jach.hawaii.edu},  Christopher J.
Davis$^{1}$, Suzanne Ramsay$^{2}$ \& Stephen P. Todd$^{3}$ \\
$^{1}$Joint Astronomy Centre, 660 N. Aohoku Place, Hilo, HI-96720, USA,
$^{2}$European Southern Observatory, Karl-Schwarzschild-Strasse 2, \\
D-87548 Garching bei Muenchen, Germany,
$^{3}$UK Astronomy Technology Centre, ROE, Edinburgh, UK}

\begin{document}

\date{Accepted 2009 December 11. Received 2009 December 11; in original form 2008 November 17}

\pagerange{\pageref{firstpage}--\pageref{lastpage}} \pubyear{2010}

\maketitle

\label{firstpage}

\begin{abstract}

We have carried out a near-infrared imaging survey of luminous
young stellar outflow candidates using the United Kingdom Infrared
Telescope.  Observations were obtained in the broad band $K$ (2.2\,$\mu$m)
and through narrow band filters at the wavelengths of 
H$_{2}$ ${\it{v}}=$1--0 S(1) (2.1218\,$\mu$m) and Br$\gamma$ (2.166\,$\mu$m) 
lines.  Fifty regions were imaged with a field of view of 
2.2$\times$2.2\,arcmin$^{2}$. Several young embedded clusters are unveiled 
in our near-infrared images. 76\% of the objects exhibit H$_2$ emission and 
50\% or more of the objects exhibit aligned H$_2$ emission features 
suggesting collimated outflows, many of which are new detections. These 
observations suggest that disk accretion is probably the leading mechanism 
in the formation of stars, at least up to late O spectral types.  The 
young stellar objects responsible for many of these outflows are 
positively identified in our images based on their locations with 
respect to the outflow lobes, 2MASS colours and association with MSX, 
IRAS, millimetre and radio sources.  The close association of molecular 
outflows detected in CO with the H$_2$ emission features produced by 
shock excitation by jets from the young stellar objects suggests that 
the outflows from these objects are jet-driven.  Towards strong radio 
emitting sources, H$_{2}$ jets were either not detected or were weak 
when detected, implying that most of the accretion happens in the 
pre-UCH{\sc{ii}} phase;  accretion and outflows are probably weak 
when the YSO has advanced to its  UCH{\sc ii} stage.

\end{abstract}



\section{Introduction}

The formation of low and intermediate-mass stars (M $\leq$ 8\,M$_\odot$) 
is explained through gravitational collapse and subsequent accretion 
of their parent molecular clouds \citep{ps93}.  Observational evidence 
accumulated over the last two decades supports this hypothesis. During 
the process of accretion, they drive well collimated bipolar outflows, 
which point to the locations of the young stars and can be used as 
signposts of star formation.  Do massive stars form through a mechanism
similar to that of their lower mass counterparts?  

While the scenario of low-mass star formation is reasonably well 
understood, not much is known about high-mass (M $>$ 8 M$_{\odot}$;
O and early B spectral types) star formation. When compared to those 
from their lower mass counterparts,  the outflows from High Mass Young 
Stellar Objects (HMYSOs) are much more energetic, massive, faster 
(100s of km~s$^{-1}$; \citealt{churchwell99}) and have shorter time 
scales of evolution.  There are far fewer HMYSOs than low mass YSOs 
and most are at kpc distances.  Additionally, since these objects are 
mostly confined to the galactic plane and are located in giant 
molecular clouds extending over several parsecs or tens of parsecs, 
they are subjected to large amounts of extinction and reddening.  Thus, 
most of them appear fainter at optical and near-IR wavelengths,  
even though they are intrinsically much 
more luminous than their low-mass counterparts.  It is known that 
massive stars form in clusters.  Hence, to understand the formation 
of massive stars and their relation to their environments, we need 
to observe them at wavelengths at which we can see the emission from 
them, i.e. at IR and longer wavelengths, at high spatial resolution.

It has been proposed that the model of \citet{ps93} fails for stellar 
masses above $\sim$10\,M$_\odot$  \citep{stahler00}.  These objects 
start core hydrogen burning before they finish accreting.  For such 
stars, the radiation pressure on the dust in the infalling material 
may halt accretion and thus limit the growth of the YSO through 
accretion (e.g. \citealt{yorke77, wolfire87, beech94}).  Consequently, 
there are two different schools of thought on the formation of massive 
stars.  Studies by \citet {bonnell98}, \citet{stahler00}, \citet{bally02}, 
etc. propose that the leading mechanism for the formation of massive
stars is the merger of low and intermediate mass objects.  An 
observational manifestation of this would be extremely poorly 
collimated outflows or no outflows.  The other scenario is  a direct 
extension of the mechanism for low-mass star formation, via accretion.  
This mechanism predicts that the outflows observed would be collimated, 
similar to those observed from low-mass YSOs, but at a larger scale.  
Studies by \citet{jijina96} and \citet{yorke02} conclude
that massive stars up to $\sim$40\,M${\odot}$ could form 
through infall and accretion irrespective of the large radiation 
pressure.  Numerical calculations of Krumholz, Klein \& McKee (2005) 
show that the cavities carved by outflows from massive stars will 
provide an optically thin region through which radiation can escape 
and, thereby, reduce the radiation pressure on the accreted matter 
and aid the growth of the star through accretion.  There is still 
difficulty in explaining the formation of extremely massive stars,
however.  \citet{bally05} gives a discussion on possible 
observational consequences of the different scenarios of 
massive star formation.

High mass star formation is usually clustered and is often
associated with other high, intermediate and low mass star 
formation.  Inadequate spatial resolution of the early studies,
mostly with single dish telescopes at radio and millimetre 
wavelengths, gave an impression that the outflows from massive 
young stars are less collimated when compared to those from 
their lower mass counterparts (e.g. \citealt{richer00, ridge01}). 

However, more and more results are being obtained from recent 
observational studies which suggest that massive stars also form 
through accretion.  \citet{zhang05} carried out a survey of 
several massive YSO outflows in CO (J=2-1).  Nearly half of 
the outflows detected by them exhibited spatially resolved 
bipolar features. The dynamical time-scales estimated by them 
for the outflows (without correcting for the angle of 
inclination) were $\sim$10$^{4}$--10$^{5}$\,years and the 
energy of the outflows was $\sim$10$^{46}$ ergs, similar to 
the turbulent energy in the core. Their large detection rate 
favoured an accretion process for the formation of HMYSOs.  
From the CO data, they estimated accretion rates of a few 
times 10$^{-4}$ M$_{\odot}$ year$^{-1}$ consistent with the 
value obtained by \citet{beuther02c}.  This is much higher 
than the average accretion rate derived from low-mass outflows 
(10$^{-7}$--10$^{-5}$ M$_{\odot}$ year$^{-1}$; 
\citealt{bontemps96}).  For HMYSOs, \citet{kim06} estimate an
accretion rate of 10$^{-4}$ M$_{\odot}$ year$^{-1}$, which 
is large enough to overcome the radiation pressure from the 
central stars with L$_{bol}$ $\leq$ 2$\times$10$^{4}$ L$_{\odot}$.  
They conclude that the outflows from the HMYSOs are well 
collimated, similar to the low-mass outflows.  Their comparison 
of the properties of the outflows from massive YSOs with those 
from low-mass YSOs shows that the collimation factors are 
similar (2.2$\pm$1.2) in the CO line, consistent with
a similar formative mechanism for the low- and high-mass YSOs. 
From CO maps with better spatial resolution, \citet{beuther02c} 
derived a collimation factor of 2.1 for HMYSO outflows.  From 
a large sample, \citet{wu04} derived average collimation 
factors of 2.8$\pm$2.2 and 2.1$\pm$1.0 for low- and high-mass 
outflows respectively.  These observations show that the 
outflows from low- and high-mass YSOs have similar 
collimation within error limits.

There are several massive YSOs for which collimated bipolar outflows 
have been observed in CO and in the near-IR (e.g. \citealt{cesaroni97,
beuther02a, beuther02c}; Beuther, Schilke \& Stanke 2003; 
\citealt{davis98, davis04, kumar02, shepherd00, todd06, zhang05}).  
Studying the luminous YSO IRAS~19410+2336, \citet{beuther03} found 
that the field hosts multiple bipolar outflows, which are as
collimated as the outflows from low-mass YSOs.  They also derived 
an accretion rate of $\sim$10$^{-4}$\,M$_{\odot}$year$^{-1}$, 
which is necessary to overcome the radiation pressure from the 
massive star in a disc-dominated accretion \citep{stahler00}.  
\citet{beuther02a} resolved the outflow from IRAS~05358+3543 
into three, with one of collimation factor $\sim$10, the highest 
observed among HMYSO outflows, similar to the largest collimation 
factor observed among outflows from low-mass YSOs.  From a study 
of CO outflows of 26 massive YSOs, \citet{beuther02c} concludes 
that massive star formation is a scaled up version of 
low-mass star formation.  From near-IR observations,  
\citet{davis04} and \citet{caratti08} also conclude that the 
outflows from the luminous YSOs IRAS~18151-1208 and IRAS~20126+4104, 
both with L$\sim$2$\times$10$^{4}$\,L$_{\odot}$, are like scaled-up 
versions of outflows from low-mass YSOs.

From our understanding of the low-mass star formation nearby and 
through theoretical studies \citep{pudritz83, shu94},  we expect 
that outflows require compact accretion discs.  Such discs have 
been proposed before \citep{cesaroni97, cesaroni99a, shepherd99}.
Evidence for rotating structures of radii of several 1000s of AU
have been presented by several investigators (\citealt{cesaroni97, 
cesaroni05}; Zhang, Hunter \& Sridharan 1998; \citealt{beltran05},
\citealt{zapata09}, etc).
Direct imaging of the disc from HMYSOs is challenging because of 
the very high angular resolution required; recently \citet{patel05} 
reported the detection of a dusty accretion disc around a 
$\sim$15\,M$_{\odot}$ protostar, using sub-millimetre interferometry.

Even with interferometric techniques, most of the millimetre and radio 
observations, which usually form the basis for the identification of HMYSOs, 
lack the spatial resolution required to probe these objects and their 
environments in sufficient detail.  Due to their location in dense molecular 
clouds confined mostly to the galactic plane and the resulting high extinction 
towards shorter wavelengths, most of these objects are not observable at 
optical wavelengths.  However, the near-IR regime offers a balance between the 
high extinction in the optical and the poor spatial resolution at millimetre 
and radio wavelengths.  The availability of highly sensitive infrared arrays 
on large telescopes,  which offer sub-arcsec resolution at IR wavelengths, and 
the presence of very good outflow tracers like the lines of molecular hydrogen, 
make the near-IR a very useful regime for studying embedded YSOs and their 
outflows.  The purpose of our study presented in this paper is to understand 
the formation of massive stars through near-IR imaging. We have obtained 
near-IR images of 50 high- and intermediate-mass YSO candidates with good 
spatial resolution as outlined in Table \ref{obslog}.  $K$-band and 
continuum-subtracted H$_2$ images are presented for all of the objects 
observed.  We also present the images obtained using the Br$\gamma$ filter 
for those objects which exhibited extended emission in Br$\gamma$.  

\subsection{Target Selection}

Emission lines of NH$_{3}$ and CO  
are good indicators of sites of high-mass star 
formation and the associated outflow activity.  The (1,1) and (2,2) 
lines of NH$_3$ are very good tracers of dense gas \citep{molinari96}.  
CO lines have been traditionally used to detect emission from cold 
gas in entrained molecular outflows from YSOs 
\citep{cabrit86, wb89, shepherd96a}.  

Maser emission from H$_{2}$O \& CH$_{3}$OH have also been considered 
as sign posts of massive star formation.  Some of the observations 
conclude that H$_2$O masers are formed in circumstellar discs 
\citep{torrelles96, torrelles98, goddi04}.  However, many studies 
(e.g. Felli, Palagi \& Tofani 1992; \citealt{torrelles98}) and recent
multi-epoch very high angular resolution VLBI observations 
(\citealt{patel00}; Moscadelli, Cesaroni \& Rioja 2000, 2005; 
\citealt{goddi05}) of the locations and proper motions of H$_2$O 
maser spots confirm that the H$_2$O masers associated with massive YSOs 
originate in the shocked regions of jets.

Observations of CH$_{3}$OH masers provide valuable information
about the formation of massive stars. Class I methanol masers are 
observed in high-mass star forming regions and are believed to be 
produced in the regions of interaction between the bipolar outflows 
from these YSOs and the dense ambient material or in cloud-cloud 
collision regions \citep{menten96}.  Observation of the
44\,GHz methanol maser emission by  Kurtz, Hofner \& \'{A}lvarez (2004)
support the idea that Class I methanol masers are mainly produced by
molecular outflows.  However, they posit that accretion shocks very 
near the central objects might also produce conditions conducive to 
the formation of Class I methanol masers.  Class II methanol masers 
are found closer to the YSOs than Class I masers and are seen to be 
radiatively excited in the dense warm molecular material surrounding 
the compact H{\sc{ii}} regions excited by YSOs \citep{menten91,
menten96} or in circumstellar discs (e.g. \citealt{nielbock07}), 
although there are examples where they appear to trace the directions 
of jets and outflows as well (\citealt{debuizer03,debuizer09}).

We have selected targets which exhibit both H$_{2}$O maser and NH$_{3}$ line 
emission from \citet{molinari96},  objects from \citet{sridharan02} which 
exhibit both H$_{2}$O \& CH$_{3}$OH emission, and HMYSOs with molecular 
outflows evident in high-velocity CO line wing emission from the list of 
Shepherd \& Churchwell (1996b) and Churchwell (1999).  Fourteen of the 
sources that we have selected from \citet{sridharan02} are also shown to 
have molecular outflows, mapped in CO by \citet{beuther02c}.

\subsection{Organization of the paper}

The details of observations and data reduction are presented in 
section 2.  Table \ref{obslog} summarizes the observations carried 
out.  A compilation of the previous observations and our own 
detection of outflows in H$_2$ is presented in Table \ref{sourceprop}.
The properties of the outflows deduced from our observations and the 
ZAMS spectral types corresponding to the luminosities of the sources 
are presented in Table \ref{outflowprop}.  The results and discussion 
on the 50 fields studied in this paper are  given in sub-sections 
A1--A50 of Appendix A.  The figure numbers of the sources agree with 
the corresponding sub-section numbers.  Each sub-section is  
self-contained so that those who are interested in any particular 
source can read that sub-section independently.   The outflows 
detected in this work are assigned identification numbers in 
the Catalogue of Molecular Hydrogen Emission-Line Objects (MHOs)
in Outflows from Young 
Stars{\footnote{http://www.jach.hawaii.edu/UKIRT/MHCat/}}
\citep{davis09b}, hosted by the Joint Astronomy Centre,
Hilo, Hawaii. Table \ref{MHO} of Appendix \ref{MHO-sect} 
gives a compilation of the `MHO' numbers.  Since the 
long-wavelength observations that are used along with our 
observations to draw conclusions on the 50 sources were obtained 
by different people using different telescopes,  we have given a 
compilation of the sources from which these are taken, the 
frequencies of observations,  telescopes used, the pointing 
accuracies and the spatial resolutions in Table \ref{resolutions} 
of Appendix \ref{resolutions-sect}.  Section 3 gives a discussion 
of the paper.  The near-IR colours of the YSO candidates 
identified in the 50 fields are discussed in sub-section 
3.1; their MSX and IRAS colours are discussed in 
sub-section 3.2.  Our inferences from the H$_2$ detections 
and near-IR colours are discussed in sub-sections 3.3 and 3.4.  
The summary and conclusions are given in section 4.  

\section{Observations}

Observations were carried out with the 3.8-m United Kingdom Infrared 
Telescope (UKIRT),  Mauna Kea, Hawaii, using the facility near-IR 
camera UFTI \citep{roche03}. UFTI is a 1--2.5$\mu$m camera with a 
1024$\times$1024 HgCdTe array giving a plate scale of 
0.091\,arcsec~pixel$^{-1}$ and a field of view of 
1.5$\times$1.5\,arcmin$^{2}$ at the f/36 Cassegrain focal plane
of the UKIRT.

Fifty sources were observed using narrow band filters centred 
at the wavelengths of H$_{2}$ ${\it{v}}$=1--0 S(1) (2.1218\,$\mu$m)
and Br$\gamma$ (2.166\,$\mu$m) lines and a broad-band $K$ 
(central wavelength=2.2\,$\mu$m; FWHM = 0.34\,$\mu$m) Mauna Kea 
Observatory (MKO) consortium filter.  Two different filters were used 
for the H$_{2}$ observations -  all observations before 2003 July were 
performed using a narrow-band filter named ``2.122'' with 
FWHM=0.02\,$\mu$m.  After this date, this filter was replaced with a 
filter ``2.122MK'' with an FWHM of 0.031\,$\mu$m micron, designed for 
the MKO consortium.  The central wavelength of both filters is 
2.119\,$\mu$m.  All Br$\gamma$ observations were done using an MKO 
consortium filter with a  central wavelength of 2.166\,$\mu$m and 
FWHM=0.022\,$\mu$m.  H$_{2}$ S(1) and Br$\gamma$ were selected to 
yield complementary information on the warm, entrained, molecular 
outflow and the collimated, high-velocity, high-excitation jet
respectively. The $K$-band images were used to continuum-subtract the 
line data, and to identify the likely outflow sources.

The observations were obtained by offsetting the telescope to nine 
positions on the sky around the object, separated by 20\,arcsec from 
the base position resulting in a field of view of 
2.2$\times$2.2\,arcmin$^{2}$ for the mosaics. Dark frames were 
obtained before each set of observations.  The data were flat fielded
using sky flats generated by median combining the observed object 
frames.  The dithered frames were combined, matching the centroids of 
stars after applying the dark and the flat field corrections.  For 
very crowded fields or regions with extended nebulosity, flat fielding 
was accomplished using the flat frames generated by observing less 
crowded regions nearby or using the flats from the object observation 
prior to or after the observation of the specific object of concern.
Typically, a total integration time of 180\,sec was used in $K$ and
594\,sec was used in H$_2$ and Br$\gamma$ filters.  At these 
integrations, we reached an average sensitivity of 5$\sigma$ at 19\,mag 
on point sources in $K$ and a per pixel sensitivity of 5$\sigma$ at 
1.3$\times$10$^{-18}$\,W~m$^{-2}$\,arcsec$^{2}$ in the narrow-band
filters.

H$_{2}\,$S(1) and Br$\gamma$ images, having their central wavelengths 
within the $K$ band, were continuum-subtracted using scaled $K$-band 
images observed before or after the narrow-band images, employing 
the methods adopted by Varricatt, Davis \& Adamson (2005).  For 
continuum subtraction, in each field, the integrated, sky-subtracted 
counts on many isolated point sources (without any IR excess) were 
measured in the narrow-band filters and in the $K$ filter using a 
circular aperture, typically $\sim$4 times the FWHM.  The ratios of 
the counts $K$/H$_2$ and $K$/Br${\gamma}$ were evaluated for these 
stars and the average values were taken.  The constant sky background 
was subtracted from the $K$-band images. This image was then divided 
by the above ratios and and the resulting images were subtracted 
from the sky subtracted narrow-band filter images.  The diffuse 
continuum was generally subtracted out well.  However, stars often 
showed residuals due to changes in PSF between the broad- and the 
narrow-band observations. Sometimes, continuum-subtracted images
exhibit large negative residuals on some objects because of 
reddening, caused by large extinction or IR excess or both.
Table {\ref{obslog}} gives a log of the 
observations and the average FWHM measured on point sources
from the $K$-band mosaics.  Figs. A1--A50 show the observed 
images
{\footnote{\bf The reduced images are available
for download (as fits files) at http://cdsarc.u-strasbg.fr/ftp/cats/J/MNRAS/vol/page
}}.

\begin{table*}
\begin{centering}
\caption{Log of observations}
\label{obslog}
\begin{tabular}{@{}llllllll}
\hline
No. &\multicolumn{2}{c}{Object Name}&UTDate&Filters               	&Sky Condition$^{2}$ &FWHM($K$)$^{3}$\\
    &(IRAS)	  &Mol$^{1}$&(yyyymmdd)	  &used		 	 	&	        &(arcsec)	\\
\hline
1   &00420+5530   &3      &20030915	  &$K$, 2.122MK, Br$\gamma$	&Thin cirrus	&0.56	\\
2   &04579+4703   &7      &20030417       &$K$, 2.122, Br$\gamma$	&Clear		&0.96 	\\
3   &05137+3919   &8      &20031012       &$K$, 2.122MK, Br$\gamma$	&Thin Cirrus	&0.42	\\
4   &05168+3634   &9      &20031012       &$K$, 2.122MK, Br$\gamma$	&Thin Cirrus  	&0.48	\\
5   &05274+3345   &10     &20031012 	  &$K$, 2.122MK, Br$\gamma$	&Thin Cirrus	&0.61	\\
6   &05345+3157   &11     &20030317       &$K$, 2.122, Br$\gamma$	&Clear 		&0.67	\\
7   &05358+3543   &       &20030317       &$K$, 2.122, Br$\gamma$  	&Clear		&0.62	\\
8   &05373+2349   &12     &20030317       &$K$, 2.122			&Clear		&0.63	\\
9   &05490+2658   &       &20030315       &$K$, 2.122, Br$\gamma$	&Clear		&0.59	\\
10  &05553+1631   &14     &20030315       &$K$, 2.122, Br$\gamma$	&Clear		&0.65	\\
11  &06061+2151   &16     &20030315       &$K$, 2.122, Br$\gamma$  	&Clear		&0.47	\\
12  &06584-0852   &28     &20030315       &$K$, 2.122, Br$\gamma$	&Clear		&0.59	\\
13  &18144-1723   &45     &20030528       &$K$, 2.122 			&Clear		&0.55	\\
14  &18151-1208   &46     &20030528       &$K$, 2.122, Br$\gamma$  	&Clear		&0.57	\\
15  &18159-1648   &49     &20031012       &$K$, 2.122MK, Br$\gamma$  	&Thin Cirrus	&0.66	\\
16  &18174-1612   &       &20030616       &$K$, 2.122, Br$\gamma$	&Clear		&0.52	\\
17  &18182-1433   &       &20031012       &$K$, 2.122MK, Br$\gamma$	&Thin Cirrus	&0.84	\\
18  &18264-1152   &       &20031023       &$K$, 2.122MK, Br$\gamma$     &Clear		&0.49	\\
19  &18316-0602   &62     &20030616       &$K$, 2.122, Br$\gamma$  	&Clear		&0.52	\\
20  &18345-0641   &       &20031009       &$K$, 2.122MK, Br$\gamma$     &Thin cirrus  	&0.87	\\
21  &18360-0537   &65     &20031009       &$K$, 2.122MK               	&Thin cirrus  	&0.90	\\
22  &18385-0512   &       &20030823       &$K$, 2.122MK           	&Clear		&0.7	\\
23  &18507+0121   &74     &20030527       &$K$, 2.122, Br$\gamma$       &Clear          &0.5    \\
24  &18517+0437   &76     &20030823       &$K$, 2.122MK, Br$\gamma$ 	&Clear	 	&0.63	\\
25  &19088+0982   &97     &20031006  	  &$K$, 2.122MK, Br$\gamma$    	&Clear		&0.75	\\
26  &19092+0841   &98     &20031006       &$K$, 2.122MK             	&Clear		&0.68	\\
27  &19110+1045   &       &20030926       &$K$, 2.122MK, Br$\gamma$	&Clear 		&1.12	\\
28  &19213+1723   &103    &20030926       &$K$, 2.122MK, Br$\gamma$  	&Clear 		&0.64	\\
29  &19217+1651   &       &20030512       &$K$, 2.122			&Thin cirrus?	&0.43 	\\
30  &19374+2352   &109    &20030512       &$K$, 2.122, Br$\gamma$ 	&Thin cirrus?	&0.61 	\\
31  &19388+2357	  &110    &20030926       &$K$, 2.122MK, Br$\gamma$     &Clear  	&0.72	\\
32  &19410+2336   &       &20030512       &$K$, 2.122, Br$\gamma$  	&Thin cirrus?	&0.57	\\
33  &20050+2720   &114    &20030518       &$K$, 2.122, Br$\gamma$	&Thin cirrus 	&0.50	\\
34  &20056+3350   &115    &20030518       &$K$, 2.122, Br$\gamma$  	&Thin cirrus 	&0.46	\\
35  &20062+3550   &116    &20030518       &$K$, 2.122, Br$\gamma$	&Thin cirrus 	&0.42	\\
36  &20126+4104   &119    &20030330       &$K$, 2.122, Br$\gamma$	&Thin cirrus	&0.83	\\
37  &20188+3928   &121    &20030330       &$K$, 2.122, Br$\gamma$  	&Thin cirrus	&0.74	\\
38  &20198+3716   &       &20030518       &$K$, 2.122, Br$\gamma$ 	&Thin cirrus 	&0.39	\\
39  &20227+4154   &124    &20030517       &$K$, 2.122, Br$\gamma$  	&Thin cirrus?	&0.62	\\
40  &20286+4105   &126    &20030517       &$K$, 2.122, Br$\gamma$ 	&Thin cirrus? 	&0.50	\\
41  &20293+3952   &       &20030516       &$K$, 2.122, Br$\gamma$	&Thin cirrus?	&0.49	\\
42  &20444+4629	  &131    &20031215	  &$K$, 2.122MK, Br$\gamma$	&Clear		&1.16	\\
43  &21078+5211   &133    &20030516       &$K$, 2.122			&Thin cirrus? 	&0.46	\\
44  &21307+5049   &136    &20030516       &$K$, 2.122, Br$\gamma$ 	&Thin cirrus?	&0.73	\\
45  &21391+5802   &138    &20030808       &$K$, 2.122MK, Br$\gamma$	&Clear		&0.62	\\
46  &21519+5613	  &139    &20031215	  &$K$, 2.122MK, Br$\gamma$  	&Clear		&0.99	\\
47  &22172+5549   &143    &20031215	  &$K$, 2.122MK, Br$\gamma$	&Clear		&0.99 	\\
48  &22305+5803	  &148    &20030808       &$K$, 2.122MK, Br$\gamma$    	&Clear 		&0.64 	\\
49  &22570+5912   &153    &20030915       &$K$, 2.122MK, Br$\gamma$  	&Thin cirrus	&0.62	\\
50  &23139+5939	  &       &20030915       &$K$, 2.122MK, Br$\gamma$  	&Thin cirrus  	&0.68	\\
\hline
\multicolumn{7}{l}{$^{1}$From \citet{molinari96}, $^{2}$A question mark is given when the presence of thin cirrus}\\
\multicolumn{7}{l}{during the observations is suspected,  $^{3}$FWHM of the average PSF measured from each reduced}\\
\multicolumn{7}{l}{$K$-band mosaic (the H$_2$, Br$\gamma$ and $K$-band data were obtained consecutively).}\\
\end{tabular}
\end{centering}
\end{table*}

\begin{table*}
\begin{centering}
\caption{Detections at different wavelengths.  Column1 gives the reference numbers (same as in Table \ref{obslog}, the subsections in 
Appendix A and in the accompanying figures), columns 2--6 list the detections in ammonia, water maser, methanol maser, CO and radio 
with references, and columns 7--8 give the kinematic distances and  the estimated luminosities with references.
``---'' implies that no published observation is available.  ``+'' implies that not all references are listed in the table.
Column 9 lists our detection in H$_2$.
``Y'' implies confirmed detection; ``N'' implies null detection; a ``?'' against ``Y'' shows that the detection is only tentative.}
\label{sourceprop}
{\small
\begin{tabular}{@{}llllllllll}
\hline
No  &NH$_3$ &H$_2$O      &CH$_3$OH    &CO line	 &radio (cm)	&d(kpc)   	&Luminosity	&H$_2$$^{2}$ \\[-0.6mm]
    &  	     &		 &	      &wings/outflow	&emission$^{1}$ &	&(10$^3$L$_\odot$)& \\[-0.6mm]
\hline
1   &Y(m8)  &Y(bb,m6,p1) &N(sh)	      &Y(bc,j1,w9)&Y(3.6-mb) 	&5.0(mb);7.72(m8)&12.4(mb);51.5(m8)	&{\bf Y(c)}	\\[-0.5mm]
2   &Y(m8)  &Y(bb,m6,w7,w8)&Y(w8)     &Y(w8);N(z3)&N(6-m9)	&2.47(m8)     	&3.91(m8)		&{\bf Y(c)}	\\[-0.5mm]
3   &Y(m8)  &Y(p1,m6)    &N(sh)       &Y(z3)   	&Y(3.6-mb)	&11.5(mc)   	&225(mc) 		&{\bf Y(c)}	\\[-0.5mm]
4   &Y(m8)  &Y(p1,m6)    &N(sa,sh)    &Y?(z3)$^{3}$ &N(6-m9)	&6.08(m8)       &24.0(m8)		&N         	\\[-0.5mm]
5   &Y(m8,z2)&Y(p1,g2,t1) &Y(sa,sh)   &Y(h5)    &Y(3.6-t1)	&1.55(m8)	&4.35(m8)  		&Y(c)	\\[-0.5mm]
6   &Y(m8,v2)&Y(p1,t2,v2,w8)&--- &Y(sc,w7,r1,z3)&Y(3.6-mb,t2)	&1.8(m8)	&1.38(m8)		&Y(c)+(f)\\[-0.5mm]
7   &Y(se)  &Y(se,b6)	 &Y(sh,se,g1,m5)&Y(b4) 	&N(3.6-se)	&1.8(se)	&3.8(se)       		&Y(c)+(f)\\[-0.5mm]
8   &Y(m8)  &Y(p1)       &N(sh)       &Y(z3) 	&Y(3.6-mb)	&1.17(m9)       &0.47(ma)  		&{\bf Y(c?)}\\[-0.5mm]
9   &---    &N(se)       &N(se)       &Y(sd) 	&N(3.6-se)	&2.1(sd)	&3.16(se);4.2(sd)	&Y	\\[-0.5mm]
10  &Y(m8)  &Y(bb,m1,p1) &N(se,m1)    &Y(w7,sd, &Y(1.3-s8;3.6-se,s6;&2.5(sd);	&6.5(sd);6.31(se);		&Y(c) 	\\[-0.5mm]
10  &---    &Y(s6,w8+)   &	      &se,s5) 	&6-h3) 		&3.04(m8)	&11.7(m8)		& 	\\[-0.5mm]
11  &Y(m8)  &Y(bb,p1)    &Y(g1,sh)    &Y(w7,s2)	&Y(3.6-k2)	&0.1(m8);2.0(c1)&0.0278(m8);4.0(c1)	&Y(c)	\\[-0.5mm]
12  &Y(m8)  &Y(bb,p1)    &---         &Y(w7,z3)	&N(2,6-m9)	&4.48(m8)	&5.67(ma)  		&N 	\\[-0.5mm]
13  &Y(m8)  &Y(p1)       &Y(sh,k3)    &N(z3)   	&N(6-m9)	&4.33(ma)       &21.2(ma)  		&Y(c?) 	\\[-0.5mm]
14  &Y(m8)  &N(b6,p1)  	 &Y(b6)       &Y(b5,se)	&N(3.6-se)	&3.0(se)        &19.95(se)		&Y(c) 	\\[-0.5mm]
15  &Y(m8)  &Y(p1)       &Y(w1)       &---     	&--- 		&2.5(m8)        &29.5(m8)  		&Y(c?)  \\[-0.5mm]
16  &Y(m3)  &N(m3,j2)    &Y(m5,ba,w2) &N(s3)  	&Y(2,6,21-f1)	&2.1(w6)    	&433(w6) 		&Y(f)	\\[-0.5mm]
17  &Y(se)  &Y(se,b6)    &Y(se,w2)    &Y(se,b5,b9)&Y(3.6,1.3-z1)&4.5,11.8(se) 	&20,125.9(se);50.1(w5)	&N      \\[-0.5mm]
18  &Y(se)  &Y(b6,se)    &Y(b6,se)    &Y(b5,se)	&Y(3.6,1.3-z1)	&3.5,12.5(se)	&10,125.9(se)		&Y(c)	\\[-0.5mm]
19  &Y(m8)  &Y(bb,k4)    &Y(w1,sa,sh) &Y(s2)   	&Y(3.6-k2,w2)	&3.17(m8)       &44.1(m8)  		&Y(c) 	\\[-0.5mm]
20  &Y(se)  &Y(b6,se,v1,sh)&Y(b6,sh,v1)&Y(b5,se)&Y(3.6-se)	&9.5(se)	&39.8(se)		&Y	\\[-0.5mm]
21  &Y(m8)  &Y(bb,p1)    &N(v1)	      &---  	&--- 		&6.28(m8)       &116(m8)        	&N      \\[-0.5mm]
22  &---    &Y(bb,b6,se) &N(b6,se,w2) &Y(se)	&Y(3.6-se)	&2,l3.1(se)	&5,199.5(se);4.0(w5)	&N	\\[-0.5mm]
23  &Y(m8)  &Y(bb,m7)    &Y(sh,s1)    &---     	&Y(2,6-m7;6-m9,s9)&3.87(m8)     &48.4(m8)		&N	\\[-0.5mm]
24  &Y(m8)  &Y(bb,c5,se) &Y(sh)       &Y(se)	&N(3.6-se)      &2.9(se)	&12.6(se)		&Y(f)	\\[-0.5mm]
25  &Y(m8)  &Y(bb,p1,m1) &N(m1,sh,v1)&Y(o1)   	&Y(6-m9)	&4.71(m8)	&29.9(m8) 	     	&N      \\[-0.5mm]
26  &Y(m8)  &Y(bb,p1)    &Y(m2,sh)    &N(z3)	&N(6-m9)	&4.48(m8,ma)    &9.2(ma)  		&N      \\[-0.5mm]
27  &---    &Y(h1)    	 &Y(c2,m5,sh) &Y(h6)    &Y(2,6-w6;23.4-v3)&6(v3);8.3(h6);9.7(w6)&330(v3);588.8(h6)&N      \\[-0.5mm]
28  &Y(m8)  &Y(bb,p1)    &N(sh,sb,v1) &Y(z3)	&Y(6-m9)$^{4}$	&4.12(m8)  	&28.2(m8)       	&Y(f)	\\[-0.5mm]
29  &Y(se)  &Y(se,b6)    &Y(se,b6)    &Y(se,b6,b8)&Y(3.6-se)&10.5(se)		&79.4(se)      		&N	\\[-0.5mm]
30  &Y(m8)  &Y(bb,p1)    &N(s1)       &Y(z3)	&Y(6-m9)&4.3(m8)		&26.7(m8)      		&{\bf Y(c?)}\\[-0.5mm]
31  &Y(m8)  &Y(bb,p1)    &Y(s1,sa)    &Y(z3)	&Y(6-m9,h4)&4.27(m8)		&14.8(m8) 		&Y(c)   \\[-0.5mm]
32  &Y(se)  &Y(b6,se)    &Y(b6,se,sh) &Y(se,b7)	&Y(3.6-se)	&2.1,6.4(se)    &10,100(se)		&Y(c)	\\[-0.5mm]
33  &Y(m8)  &Y(p1)       &---         &Y(b1,z3) &Y(6-w3)$^4$	&0.73(m8)  	&0.388(m8)  		&{\bf Y(c)}\\[-0.5mm]
34  &Y(m8)  &Y(bb,j1,p1) &---         &Y(z3)   	&Y(3.6-j1)	&1.67(m8)      	&4.0 (m8)  		&N	\\[-0.5mm]
35  &Y(m8)  &Y(bb,p1)    &Y(sb,sh,g1) &Y(z3)	&N(6-m9)	&4.9(mb)	&3.2(mb) 		&{\bf Y(c)}\\[-0.5mm]
36  &Y(m8+) &Y(e1,md,p1,t1+)&Y(e1,g1,k3,+) &Y(l1,s7+) &Y(3.6-h2)&1.7(c4)   	&10(c4)        		&Y(c)   \\[-0.5mm]
37  &Y(m8,a1,+)&Y(a1,bb,j1,p1)&N(s1)  &Y(l2,z3)&Y(3.6,1.95-j1;6-m9)&0.31(m8);3.91(p1)&0.343(m8);52.8(p1)&{\bf Y(c)} \\[-0.5mm]
38  &Y(d1)  &Y(f4,h1)    &Y(sh)       &Y(m4,s4)	&Y(6-w6) 	&0.9,5.5(d1)    &20,605(d1)		&Y(c)	\\[-0.5mm]
39  &Y(m8)  &Y(p1,bb)    &N(s1)       &Y(k1)	&N(2,6-w3)	&0.1(m8);3.39(p1)&0.00914(m8);9.59(p1)	&Y(c)	\\[-0.6mm]
40  &Y(m8)  &Y(p1,bb)    &---         &Y(z3)   	&N(6-m9)	&3.72(m8)       &39(m8)  		&{\bf Y(c)+(f)}\\[-0.5mm]
41  &Y(se)  &Y(se)       &N(se)	      &Y(b5,b8)	&Y(3.6-se)	&1.3,2(se)	&2.5,6.3(se)		&Y(c)+(f)\\[-0.5mm]
42  &Y(m8)  &N(p1,w8)    &N(m2)       &Y(d2,w7)	&Y(6-m9)	&2.42(m8)	&3.34(m8)  		&Y(f)	\\[-0.5mm]
43  &Y(m7,m8)&Y(p1,bb,c5,j1,+)&N(sh,sa)&Y(w7,b3)&Y(2,6-m7;6-m9)	&1.49(m8)	&13.4(m8)  		&N	\\[-0.5mm]
44  &Y(m8)  &Y(p1)       &Y(k3)       &Y(f3,z3) &N(3.6-mb)$^{4,5}$&3.6(mb)	&4(mb)			&{\bf Y(c?)}\\[-0.5mm]
45  &Y(m8)  &Y(f2,t1,p2) &Y(k3)       &Y(sg,w4,c6,b2)	&Y(3.6-b2)&0.75(m8)	&0.0939(m8);0.15(b2)  	&Y(c)	\\[-0.5mm]
46  &Y(m8)  &Y(c3,w8)    &N(m2)	      &Y(sf,w7,w8,z3)&N(6-m9)	&7.3(m8)	&19.1(m8)		&{\bf Y(c)}\\[-0.5mm]
47  &Y(m8)  &Y(c3,w8)    &---  	      &Y(w7,z3) &N(2,6-m9,6-mb)	&2.4(mb)	&1.8(mb)		&Y(c)	\\[-0.5mm]
48  &Y(m8)  &Y(p1)       &---         &Y(w7,z3) &Y(3.6-mb)	&5.4(m8)        &14.1(m8)  		&Y(c?)	\\[-0.5mm]
49  &N(m8)  &N(p1,se)    &N(sa,se,m2) &Y(se,b5) &Y(3.6-se)&2.92(m8);5.1(se)	&20.1(m8);50.1(se)	&{\bf Y(c)+(f)}\\[-0.5mm]
50  &---    &Y(g3,t1,b6) &Y(sh)       &Y(b5)	&Y(3.6-se,t1)&4.8(se)		&25(se)     		&Y?	\\[-0.5mm]
\hline
\multicolumn {9}{l}{$^{1}$column is formatted as Y/N(cm-reference); $^{2}$results from our H$_{2}$ line observations. ``(c)'' is given if the emission is collimated or is seen as a set of aligned knots;}\\[-0.5mm]
\multicolumn {9}{l}{a question mark against ``c'' implies that collimated emission in H$_2$ is suspected and ``f'' is given when a fluorescent component to the H2 emission is suspected.}\\[-0.5mm]
\multicolumn {9}{l}{Bold faces show new H$_2$ detections in this study;  $^{3}$CO outflow detected is away from the IRAS position; $^{4}$extended; $^{5}$emission is from a Supernova remnant.}\\[-0.5mm]
\multicolumn {9}{l}{References: a1-Anglada et al. (1997); b1-Bachiller et al. (1995); b2-Beltran et al. (2002); b3-Bernard et al. (1999); b4-Beuther et al. (2002a);}\\[-0.5mm]
\multicolumn {9}{l}{b5-Beuther et al. (2002c); b6-Beuther et al. (2002d); b7-Beuther et al. (2003); b8-Beuther et al. (2004a); b9-Beuther et al. (2006); ba-\citet{blaszkiewicz04};}\\[-0.5mm]
\multicolumn {9}{l}{bb-Brand et al. (1994); bc-Brand et al. (2001); c1-Carpenter et al. (1995); c2-Caswell et al. (1995); c3-Cesaroni et al. (1988); c4-\citet{cesaroni97};}\\[-0.5mm]
\multicolumn {9}{l}{c5-Codella et al. (1996); c6-Codella et al. (2001); d1-Dent et al. (1988); d2-Dobashi et al. (1995); e1-\citet{edris05}; f1-Felli et al. (1984); f2-Felli et al. (1992);}\\[-0.5mm]
\multicolumn {9}{l}{f3-Fontani et al. (2004); f4-Forster et al. (1978); g1-Galt (2004); g2-Goddi et al. (2004); g3-Goddi et al. (2005); h1-Hofner \& Churchwell (1996);}\\[-0.5mm]
\multicolumn {9}{l}{h2-Hofner et al. (1999); h3-Hughes \& MacLeod (1993); h4-Hughes \& MacLeod (1994); h5-Hunter et al. (1995); h6-Hunter et al. (1997); j1-\citet{jenness95};}\\[-0.5mm]
\multicolumn {9}{l}{j2-Johnson et al. (1998); k1-Kim \& Kurtz (2006); k2-Kurtz (1994); k3-Kurtz et al. (2004); k4-Kurtz \& Hoffner (2005); l1-\citet{lebron06}; l2-Little et al. (1988);}\\[-0.5mm]
\multicolumn {9}{l}{m1-\citet{macleod98a}; m2-MacLeod et al. (1998b); m3-Massi et al. (1988); m4-Matthews et al. (1986); m5-Menten et al. (1991); m6-Migenes et al. (1999);}\\[-0.5mm]
\multicolumn {9}{l}{m7-Miralles et al. (1994); m8-Molinari et al. (1996); m9-\citet{molinari98}; ma-\citet{molinari00}; mb-\citet{molinari02}; mc-\citet{molinari08};}\\[-0.5mm]
\multicolumn {9}{l}{md-\citet{moscadelli00}; o1-\citet{osterloh97}; p1-\citet{palla91}; p2-Patel et al. (2000); r1-Ridge \& Moore (2001); s1-Schutte et al. (1993);}\\[-0.5mm]
\multicolumn {9}{l}{s2-Shepherd \& Churchwell (1996a); s3-\citet{shepherd96b}; s4-Shepherd et al. (1997); s5-Shepherd et al. (1998); s6-Shepherd \& Kurtz (1999);}\\[-0.5mm]
\multicolumn {9}{l}{s7-\citet{shepherd00}; s8-Shepherd et al. (2004a); s9-Shepherd et al. (2004b); sa-Slysh et al (1990); sb-Slysh et al. (1999); sc-Snell et al. (1988);}\\[-0.5mm]
\multicolumn {9}{l}{sd-Snell et al. (1990); se-Sridharan et al. (2002); sf-Su et al. (2004); sg-\citet{sugitani89}; sh-Szymczak et al. (2000); t1-Tofani et al. (1995);}\\[-0.5mm]
\multicolumn {9}{l}{t2-Torrelles et al. (1992b);  v1-van derWalt et al. (1995); v2-Verdes-Montenegro et al. (1989); v3-Vig et al. (2006); w1-Walsh et al. (1997); w2-Walsh et al. (1998);}\\[-0.5mm]
\multicolumn {9}{l}{w3-\citet{wilking89}; w4-\citet{wilking90}; w5-Williams et al. (2005); w6-\citet{wc89b}; w7-Wouterloot \& Brand (1989);}\\[-0.5mm]
\multicolumn {9}{l}{w8-\citet{wouterloot93}; w9-\citet{wu01}; z1- \citet{zapata06}; z2-\citet{zhang02}; z3-\citet{zhang05}}\\[-0.5mm]
\end{tabular}
}
\end{centering}
\end{table*}

Astrometric corrections were applied to the coordinates of our images 
employing the starlink package GAIA, adopting the 2MASS positions 
as reference.  Coordinates of typically six isolated unresolved bright point 
sources spread throughout the 2.2$\times$2.2\,arcmin$^{2}$ field of each 
image were compared with those of 2MASS, and the required coordinate corrections 
were applied to our images to match the positions with those of 2MASS.  
After applying the corrections, the coordinates of the unresolved objects in 
our images agreed well with those of 2MASS.  For objects which 
were not resolved by 2MASS, but were resolved by us due to the higher 
spatial resolution of our data, the 2MASS positions would only be close to 
the centroids of the resolved components from our images.  Positional 
accuracy of our data after the astrometric corrections is $\sim$0.5\,arcsec, 
which is the positional accuracy of 2MASS.

The spatial resolution of our images are limited by the seeing.
The average FWHM measured in the $K$-band images of the fields observed 
are listed in Table \ref{obslog}.  These are much better than the spatial 
resolution of IRAS ($\sim$1-2\,arcmin), MSX ($\sim$18\,arcsec) and 
2MASS ($\sim$4\,arcsec).

\section{Discussion}

\subsection{The near-IR colours}

\begin{figure*}
\centering
\includegraphics[width=17.7cm,clip]{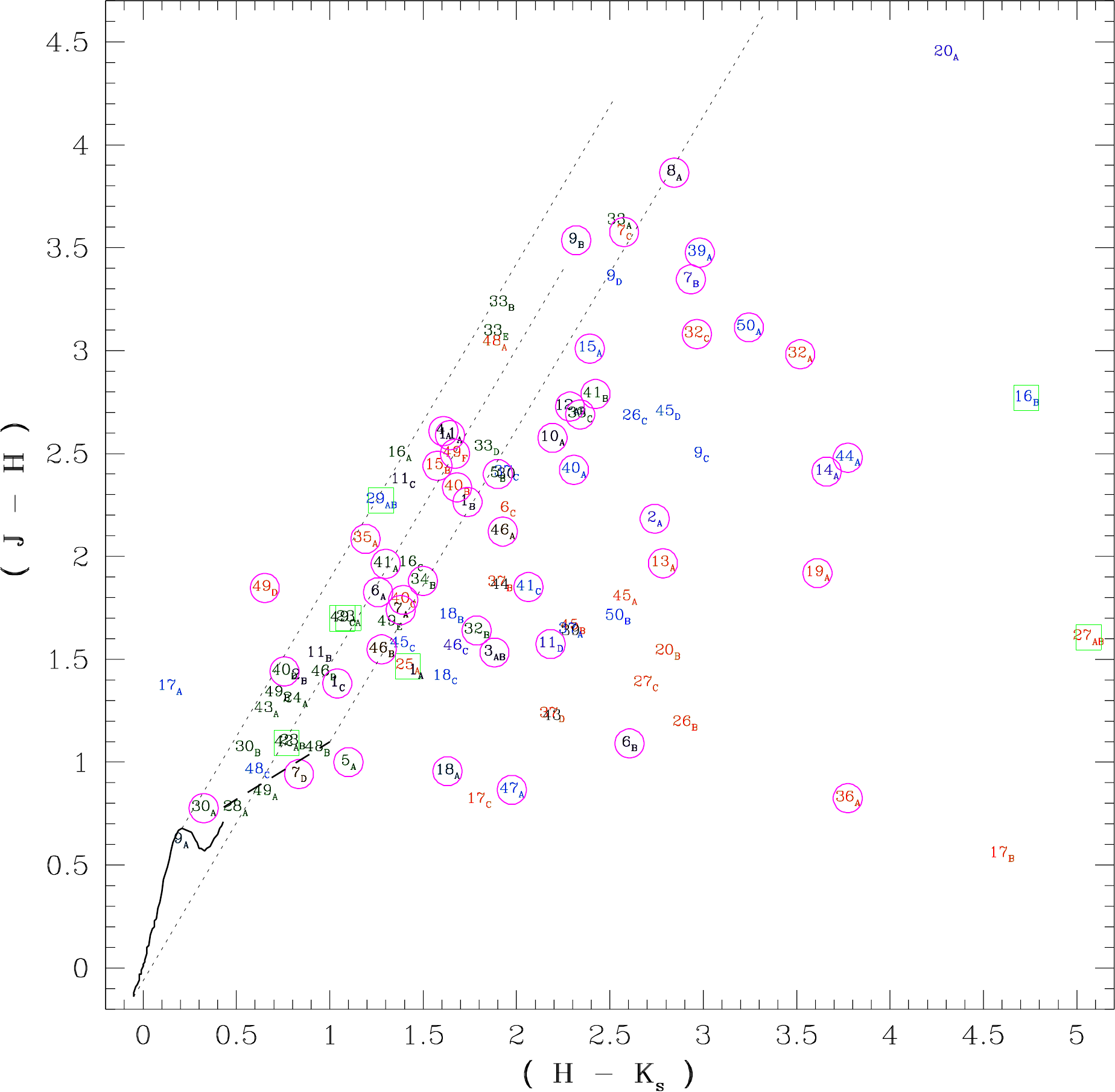}
\caption{Near-IR colour-colour diagram produced using the 2MASS 
$JHK_s$ magnitudes.  The continuous line denotes the locii of the 
main sequence stars, the dashed line shows the locii of CTTS 
and the three dotted lines show the reddening vectors up to A$_V$=30.   
The sources which we think are the IR counterparts of the YSOs 
driving the outflows detected in this work are circled.  Sources 
which are detected by 2MASS in all three bands are shown black in 
colour.  Limiting colours of those which are detected in only two 
bands are shown in blue; limiting colours of those detected in only 
one band are shown in red.   Objects which are not detected in $J$ 
due to extinction may be located further up.  Those which are not 
detected in $J$ and $H$ due to extinction may be located further 
up and right. Squares mark the near-IR counterparts of confirmed 
UCH{\sc{ii}} regions.}
\label{JHKcol}
\end{figure*}

Unless otherwise stated, 
2MASS $JHK_s$ magnitudes are used to derive the colours of the
YSO candidates indentified by us in the 50 fields studied here.
Fig. \ref{JHKcol} shows the ($J-H$) - ($H-K_s$) colour-colour diagram.
The continuous line shows the locii of the intrinsic colours of main 
sequence stars from \citet{koornneef83}.  The dashed line shows the
location of Classical T Tauri stars (CTTS) from 
Meyer, Calvet \& Hillenbrand (1997).  The dotted lines show the 
reddening vectors up to A$_V$=30 for main sequence stars and CTTS.  
We adopted a reddening law with R=A$_V$/(E(B-V))=5, which is 
typical of dense clouds \citep{cardelli89}.  The objects in between 
the reddening vectors for the main sequence stars are expected to 
be reddened main sequence stars.  The region below the lower 
reddening line for the main sequence stars is occupied by YSOs where 
CTTS, Herbig Ae/Be (HAeBe) stars and Luminous YSOs occupy different 
regimes \citep{lada92}.
 
The 2MASS colours of the near-IR sources of interest described 
throughout the text are plotted in Fig. \ref{JHKcol}.  Plot symbol 
adopted is the source number ``1--50'' (as in column 1 of 
Tables \ref{obslog} and \ref{sourceprop}, and in the subsections 
A1--A50 for the 50 fields in Appendix A) subscripted by ``A'', 
``B'', etc. for the near-IR objects identified in each field.  
The sources which we consider as the IR counterparts of the YSOs 
driving the outflows detected in H$_2$ are circled. 

Most of the objects in the diagram are located in the region of 
the near-IR colour-colour diagram occupied by YSOs.  The association 
of these objects with the MSX/IRAS, millimetre and radio continuum
emissions and their location central to the detected outflows suggest
that we are detecting the driving sources of the outflows in the near-IR.  
Even for several objects which are convincingly in the active outflow 
phase, we detect near-IR counterparts (e.g. IRAS~19410, 18151, 18316 
[henceforth, IRAS names of our sources are often truncated to the 
first five digits]).  The near-IR spectra available on some 
of these objects (e.g. IRAS~18151 - \citealt{davis04};  IRAS~06061, 
19110 - \citealt{hanson02}; IRAS~04579, 05137 - \citealt{ishii01}) 
exhibit steeply rising SEDs on which emission lines of molecular 
hydrogen, and often of CO and Br$\gamma$, are superposed.  This 
implies that in the near-IR we are mostly witnessing re-processed 
thermal emission from circumstellar gas+dust shells or spatially 
unresolved discs and therefore, not directly detecting the 
stellar photospheres.  

Among UCH{\sc{ii}}s and pre-UCH{\sc{ii}}s, we do not find a large
difference between their appearance in the near-IR,  although we 
would expect more of the UCH{\sc{ii}}s to be visible at $JHK$ 
wavelengths.  Even though many of the pre-UCH{\sc{ii}}s are detected 
in $JHK$, several UCH{\sc{ii}}s in our sample are not detected at 
NIR wavelengths (e.g. IRAS~18385, 19374, 21078); see also 
IRAS~18449-0115 of \citet{bik04}.  The detection in the near-IR 
of both these classes is highly influenced by the foreground 
extinction.  The near-IR counterparts identified for sources that 
are confirmed to be UCH{\sc{ii}}s (based on their radio fluxes) 
are enclosed in green squares in Fig. \ref{JHKcol}.  Three of the 
seven objects, IRAS~18174 (16$_B$), 19088 (25$_A$) and 19110 
(27$_{AB}$), show large excess.  However, 25$_A$ and 27$_{AB}$ 
are detected only in $K_s$ by 2MASS, so their colours are
highly uncertain.   For 27$_{AB}$,  two sources contribute to the 
2MASS $K_s$ magnitude,  one of which is likely to be in a 
pre-UCH{\sc{ii}} stage.  16$_B$ was detected only in $H$ and $K_s$ 
and has a large foreground extinction of A$_V$=40 \citep{nielbock07}.
The large extinction may imply a $J-H$ colour which is redder than 
the limiting colour; hence,  16$_B$ may be actually located further 
up in Fig. \ref{JHKcol},  closer to the reddening band.  The four 
remaining UCH{\sc{ii}}s are located either in the reddening band 
or very close to it.  This is indicative of an evolutionary trend 
which suggests that when the object moves to a UCH{\sc{ii}} stage,
more of the circumstellar matter which causes the excess and 
reddening is cleared and that the accretion and outflow will be 
at a lower rate when compared to pre-UCH{\sc{ii}} objects.  

\subsection{The MSX and IRAS colours} 

\begin{figure*}
\centering
\includegraphics[width=17cm,clip]{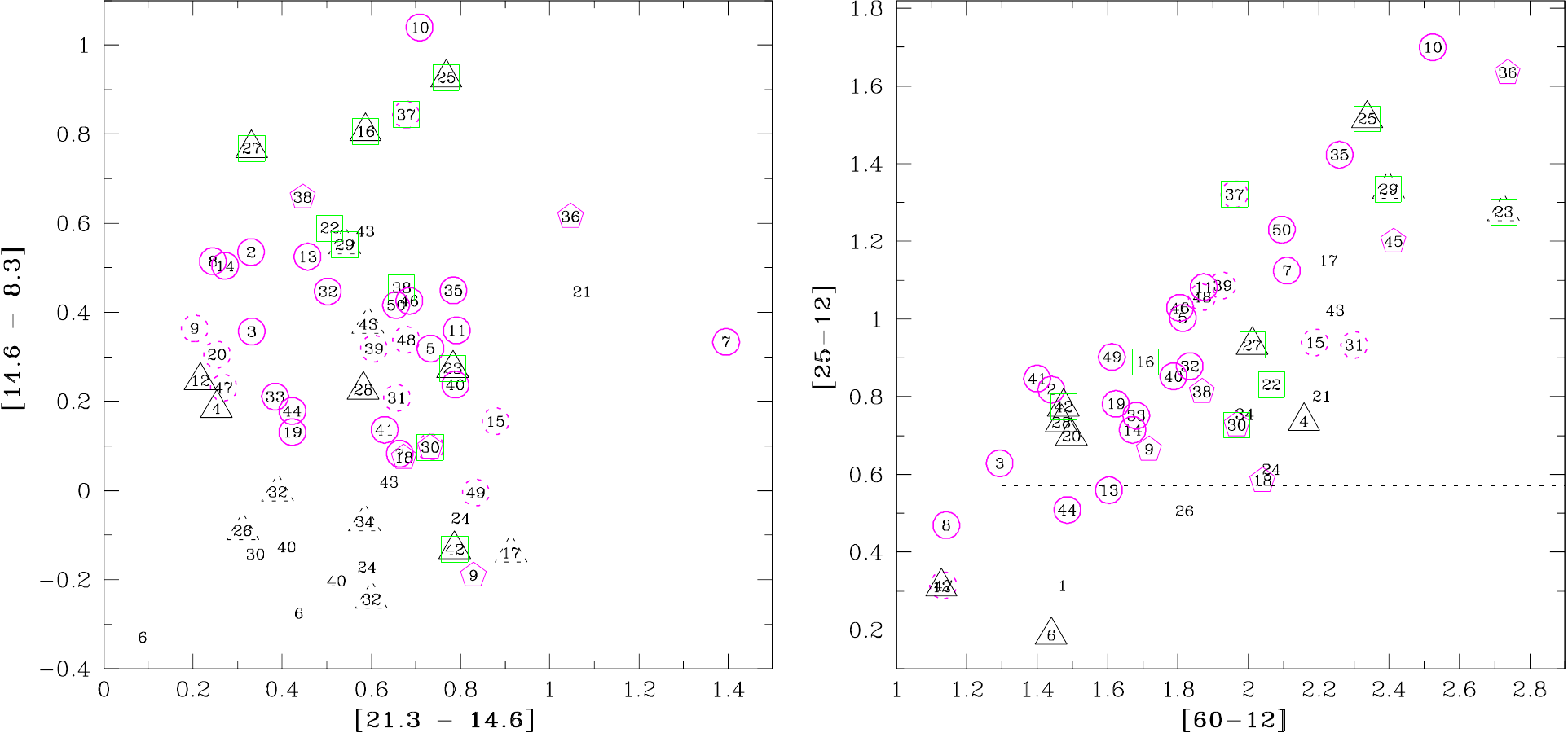}
\caption{The colour-colour diagrams produced from the MSX fluxes (left)
and the IRAS fluxes (right).  The sources are labelled according to their 
numbers 1--50 given in Table. \ref{obslog}.  Objects detected in $K$ 
and appearing to be associated with H$_2$ outflows are circled;  dotted 
circles show tentative identifications of the $K$-band conterparts.  When 
an H$_2$ outflow is detected, but its driving source is not visible in $K$, 
its MSX or IRAS counterpart is enclosed in a pentagon.  If no outflows are 
detected in $H_{2}$, but a YSO is identified in $K$ to be associated with 
the MSX or IRAS source, a solid triangle is shown; a dotted triangle is 
given when the $K$-band detection is only tentative. Infrared counterparts 
of confirmed UCH{\sc ii}s are enclosed in squares.   The dotted lines on 
the IRAS plot show the colour criteria of \citet{wc89a} for identifying 
UCH{\sc ii}s.}
\label{msxirascol}
\end{figure*}

The MSX and IRAS colours are derived by taking the logarithms 
of the ratios of fluxes at different wavelengths.  The MSX 
([21.3 - 14.6]) vs. ([14.6 - 8.3]) colour-colour diagram is 
shown in the left panel of Fig. \ref{msxirascol}.  The plot 
symbols adopted for the fifty sources are again ``1--50'', 
consistent with their labels in the tables, subsections A1--A50 
of Appendix A where they are discussed,  and in the 2MASS 
colour-colour diagram (Fig \ref{JHKcol}).  The objects 
reliably detected in $K$, from which we detected outflows at
2.122\,$\mu$m, are circled (when we have doubts about the 
identification of an outflow driving source in $K$, a dotted 
circle is used).  These objects have colours implying 
rising SEDs.  The driving sources of the outflows have a 
rough tendency to occupy intermediate [14.6-8.3] colours 
throughout the [21.3 - 14.6] colour regime.  MSX counterparts 
of H$_2$ outflow driving sources not detected in $K$
are enclosed in pentagons.  When a YSO candidate is 
detected in the near-IR at the expected position, with an 
MSX counterpart identified, but for which no outflow was 
seen in H$_2$, it is enclosed in a triangle.  When the 
near-IR detection was only suggestive of a YSO association,  
we have used a dotted triangle.  Some of these objects 
with negative colours are likely to be evolved objects.   
MSX counterparts of the positively identified UCH{\sc{ii}}s 
are enclosed in squares.  Most of the UCH{\sc{ii}}s,
especially the luminous ones, occupy the upper region 
of the [14.6 - 8.3] colour in Fig. \ref{msxirascol}, showing 
a steeper SED in this regime when compared to an average 
outflow driving source, agreeing with their location close 
to the reddening band in Fig. \ref{JHKcol}.

The right panel of Fig. \ref{msxirascol} shows the objects in a 
colour-colour diagram derived from the IRAS fluxes at  12, 25 and 60-$\mu$m.
The plot symbols used are the same as in Fig. \ref{msxirascol} - left.  
The IRAS colours of the H$_2$ outflow driving sources detected in $K$
are enclosed in solid circles and those which have only tentative
detection in $K$ are shown with dotted circles.  The IRAS colours of 
the outflow sources which do not have a $K$-band detection are 
enclosed in pentagons.  The objects for which YSO counterparts were 
detected close to the IRAS position in $K$, but for which no outflows 
are detected in H$_2$ are enclosed in triangles.  Confirmed 
UCH{\sc{ii}}s are shown enclosed in squares.

The dotted lines in Fig. \ref{msxirascol} (right) mark the IRAS 
colour criteria defined by \citet{wc89a} 
(log$_{10}$(F60\,$\mu$m/F12\,$\mu$m)$\geq$1.3 
and Log$_{10}$(F25\,$\mu$m/F12\,$\mu$m)$\geq$0.57),
where  UCH{\sc ii}s are expected to be located above and towards the 
right, respectively, of the two dotted lines. A major fraction of our
objects are located in this region.  Our results and the available
observations on these objects show that our sample is a mixture of
UCH{\sc ii} and pre-UCH{\sc ii} objects.  The radio detections, most 
of which are from VLA observations with good angular resolution and
positional accuracy, are compared with the locations of the MSX sources
and the IR counterparts of the YSOs identified in this work (based on 
their near-IR colours and location at the centroids of outflows). 
The MSX flux ratios in Fig. \ref{msxirascol} show that the outflow 
sources have rising SEDs in the MSX bands and the MSX detections 
are the counterparts of the IRAS sources in a major fraction of 
these regions.  Many of these are multiple sources embedded in 
clusters.  As described above, the radio detections are not
always from the YSOs driving the outflows; instead they are often 
from more evolved YSOs in the neighbourhoods of the driving sources
of the outflows.  The confirmed UCH{\sc{ii}}s are all located in the 
region defined by the Wood \& Churchwell colour criteria for 
UCH{\sc{ii}}s, but they are mixed with pre-UCH{\sc{ii}}s.  The colour 
criteria of \citet{wc89a} are, thus, more representative of intermediate 
and luminous YSOs (UCH{\sc{ii}}s and  pre-UCH{\sc ii}s) and the embedded 
clusters that contain these rather than just UCH{\sc ii}s.  Some of 
the previous studies have also noticed this ({e.g. \citealt{bik04}).

\subsection{Detection of outflows}

76\,\% (38 out of 50) of our objects display H$_2$ emission.
50\,\% (25 out of 50) of the objects exhibit collimated outflows
as inferred from the aligned H$_2$ emission knots;  including
the objects with suspected collimated emission (objects with
``?'' in the last column of  Table. \ref{sourceprop}),  this
fraction would be 62\% (31 out of 50).  Many of these are new
detections.  These numbers  should be treated as lower limits
only.  Even at 2.122\,$\mu$m,  the extinction will hamper the
the detection of several outflows.  Fig. \ref{gal_pos} shows
the distribution of the survey objects in Galactic coordinates,
where the objects are labelled by their distances from the Sun.
The 12 objects from which outflows are not detected in  H$_2$
(shown in red and enclosed in hexagons) are  located symmetrically 
above and below the galactic plane.  Most of these are located 
close to the galactic plane and are at large distances, 
indicating that the non-detections of outflows in many cases 
may be mainly due to extinction.

\begin{figure*}
\centering
\includegraphics[width=15cm,clip]{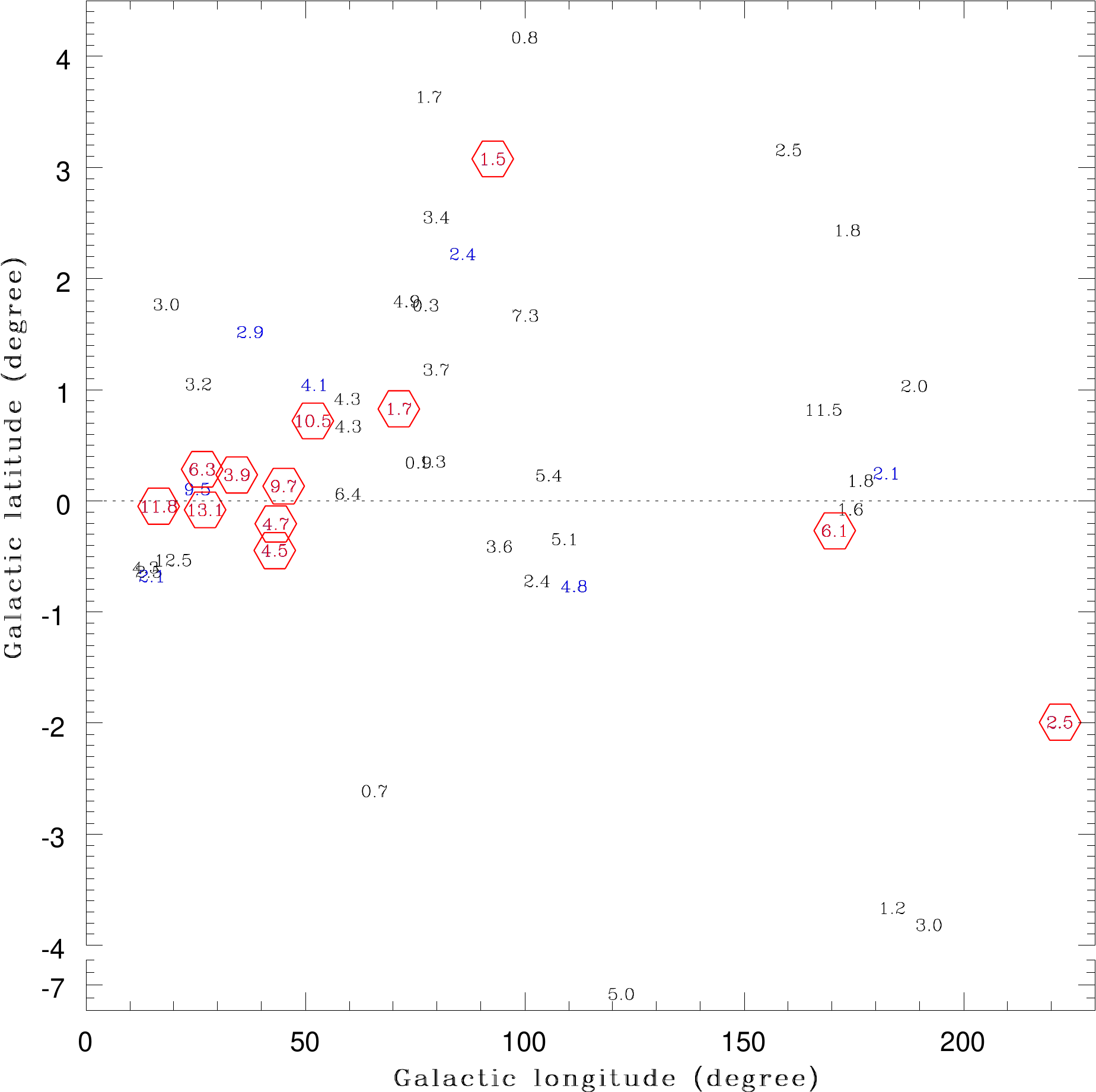}
\caption{The location of the survey objects in the galactic 
coordinates.  The fifty objects are labelled with their distances
from the sun.  Those from which outflows are positively detected
in H$_2$ are shown black in colour and those with only
a tentative detection of outflows in H$_2$ are shown in blue.
The objects from which no  H$_2$ outflows are seen in our images
are shown in red and are enclosed in hexagons.}
\label{gal_pos}
\end{figure*}

In most of the objects with outflows, we find good agreement between 
the centroids and directions of the outflows obtained from the 
aligned H$_2$ emission knots in our images and those deduced from 
published CO maps.  The aligned knots that we  see in H$_2$ 
emission are most certainly due to shocked emission from jets.
These observations show   that the outflows are mainly jet-driven 
as in low-mass stars and not wind-driven.  (Some examples of 
spectroscopic verifications of shocked emission from HMYSO jets 
are: IRAS~20126 - \citealt{caratti08}, IRAS~18151 - \citealt{davis04}, 
IRAS~11101-5928 - \citealt{gredel06}, IRAS~18316 - \citealt{todd06}).

\subsubsection {Do massive stars form through disk accretion?}

Collimation factors derived from our H$_2$ line observations
are listed in Table \ref{outflowprop}.  (Collimation factors are
derived by dividing the observed maximum lengths of the outflows from 
the driving sources by the widths,  both measured from our H$_2$ images.
We caution that collimation factors derived from H$_2$ will be affected 
by the large extinction at 2.122\,$\mu$m and should not be compared 
directly with those derived from CO maps).  Within the range of 
luminosities covered in our sample, the outflows are seen to be 
well collimated irrespective of the luminosities of the driving sources. 
The highest collimation ratio observed is $\sim$19 (for IRAS~05358+3543). 
The sample of H$_2$ jets presented in this paper seem to be as 
collimated as those from low-mass YSOs (a compilation of the
lengths and opening angles of outflows in Orion, derived from 
H$_2$ images, can be seen in \citealt{davis09a}).  The main sequence 
spectral types of single stars corresponding to the luminosities 
of the YSOs driving the outflows \citep{vacca96, morton68} are also 
listed in  Table \ref{outflowprop}.  From these observations, we 
infer that objects up to early B and late O-types produce collimated 
outflows, implying that disk accretion forms the main mechanism leading 
to their formation.   However, accretion rates may be highly 
variable, as parcels of matter (like clumps of gas and dust) 
fall from the disc on to the central source (e.g: source \#90 
in IRAS~05361+3539, from photometric variability of the central 
source and the twisted outflow observed in H$_2$ - \citealt{wpv05};  
IRAS~20126+4104, from the apparently large precession angle - 
\citealt{shepherd00}).   These large clumps may be capable of 
self-shielding against  the large radiation pressure to a 
significant level, thereby aiding accretion.  This also points 
to the possibility of highly clumpy nature of the accretion discs.

It is noteworthy that with the present calculations
(e.g. \citealt{yorke02}) the scenario of radiation pressure
halting the formation of massive stars by accretion is a problem
only for very massive stars. The outflows from HMYSOs will also
contribute to the reduction of radiation pressure on the accreted
matter and thereby aid accretion \citep{krumholz05}.  Our sample
does not include any extremely massive object.  The most luminous
objects in our sample with outflows observed in H$_2$ are
IRAS~18264 and 19410, which have L$\sim$10$^{5}$\,L$_{\odot}$.
They may have single star ZAMS masses of $\sim$38\,M$_{\odot}$
(M$_{evol}$ from \citealt{vacca96}).   However, IRAS~19410 is
confirmed to host multiple YSOs and more than one are suspected
to be present in IRAS~18264, which will  further reduce their masses. 
Even within a dynamical time of $\sim$10$^5$ years, with an accretion
rate of  $\sim$10$^{-4}$ M$_{\odot}$year$^{-1}$ \citep{zhang05},
these objects will accumulate only a maximum of a few 10's of 
\,M$_{\odot}$ more of matter.  This will not take any of the 50 
objects surveyed to the most massive limits, unless the accretion 
rate is significantly higher,  frequent captures of dense 
clumps of gas and dust occur which will increase the effective 
accretion rate on the long run, or we are underestimating the present
luminosity and mass.  Indeed, some of the recent studies
estimate accretion rates of $\sim$10$^{-3}$ M$_{\odot}$year$^{-1}$
(eg: \citealt{zapata08} in W51-North; \citealt{sandell05} in
NGC7538-IRS9).  

\subsubsection {Objects that exhibit H$_2$ line emission} 

Thirty eight objects in our sample exhibit H$_2$ line emission 
(Table \ref{sourceprop}). Twenty five of these (31 if we include the 
fields in which the detection of the  jet in H$_2$ is only tentative) 
exhibit collimated emission. (For two objects, IRAS~18517 and 20444, 
the H$_2$ emission is very faint and doesn't appear to be from outflows).  
The HMYSOs are subject to very high interstellar  extinction. If we 
integrate deeper, it is possible that we may detect H$_2$ emission from 
more. Except for a few cases,  near-IR counterparts were detected in our 
images for the YSOs driving the outflows.  Their association with 
the H$_2$ line emission features, association with other tracers of 
star formation like millimetre or radio emission and masers, large 
reddening and excess and the location in the region of the
near-IR colour-colour diagram (Fig. \ref{JHKcol}) occupied by YSOs 
suggest that we are detecting the driving sources of the outflows.  
Fig. \ref{msxirascol} shows that these objects occupy specific 
regions in the MSX and IRAS colour-colour diagrams,  which is 
indicative of the temperatures involved.  Co-ordinates of YSOs 
identified and suspected YSOs are given throughout the text to 
enable future spectroscopic and high angular resolution 
investigations of these objects and their close environments.

\subsubsection {Objects that do not exhibit any H$_2$ line emission} 

Twelve objects do not exhibit any H$_2$ emission at our level of 
sensitivity. These are: IRAS~05168, 06584, 18182, 18360, 18385, 18507, 
19088, 19092, 19110, 19217, 20056 and 21078.  For two other objects,
IRAS~18517 and 20444, the H$_2$ emission is very faint and appears 
to be due to fluorescence.   These 14 objects could be broadly classified 
into two different populations - the young and the old - based on whether 
they exhibit significant radio continuum emission or not.  IRAS~05168, 
06584, 18182, 18517, 19092, 20444 and 20056 did not exhibit any appreciable 
emission at radio frequencies, or the radio emission, when detected, was 
very weak. Among these, except for IRAS~05168, 06584 and 20444,  near-IR
counterparts of the YSOs are not positively identified in our images.  These 
could represent very young objects.  For IRAS~18517, it is possible that
the YSO responsible for the CO outflow is not detected in our $K$-band 
image since it is highly embedded;  for IRAS~18360, no radio observations
have been reported; faint radio emission was detected from IRAS~20444 and 
it is probably a YSO of  mid-B spectral type, in or approaching UCH{\sc ii} 
stage. The rest of the objects - IRAS~18385, 18507, 19088, 19110, 19217 
and 21078 - do exhibit radio emission.  Among these six objects,  only 
two (IRAS~18385 and 21078) do not have convincing IR counterparts.  For 
IRAS~19217, it is not clear if any of the NIR sources labelled on our 
image is the IR counterpart.  Hence, the objects which exhibit radio 
emission and have NIR counterparts may be more evolved YSOs that are in 
the UCH{\sc ii} stage.  The objects without any detected outflow in 
H$_2$ and without any significant radio emission are likely to be 
pre-UCH{\sc ii}s that are in the early stages of their formation.

\subsection {Do YSOs in the UCH{\sc ii} phase drive outflows?}

Thirteen objects are seen with strong radio emission (more than a few mJy).
Six of the strongest radio sources do not have any detection of jets in H$_2$
(IRAS~18385, 18507, 19088, 19110, 19217 and 21078).  IRAS~18345 probably
hosts an H$_2$ jet. The location of the radio source in this field is
not available.  However, our $K$-band image shows that there are possibly
two YSOs in this field.  For IRAS~19213 and 19374 the location of the radio 
sources do not agree with the possible locations of the YSOs identified 
from the IR colours.  The remaining four objects (IRAS~20188, 20198, 20293
and 22570) have jets detected in H$_2$.  However, the locations of the
radio sources are offset from the driving sources of jets identified
here or from the expected positions of the driving sources (the locations 
of the radio sources and the YSOs identified are labelled on the 
Figs. A1--A50).  Thus, none of the strong radio emitters are convincingly 
associated with H$_2$ jets.

\begin{table*}
\begin{centering}
\caption{Outflow properties from our H$_2$ observations}
\label{outflowprop}
\begin{tabular}{@{}llllllll}
\hline
No.  &Object Name$^1$   &Outflow direction$^2$&\multicolumn{2}{c}{Outflow length$^{3}$}&Collimation  &Log(L/L$_{\odot}$)$^4$ &ZAMS\\
     &(IRAS)            &(E of N)       &(arcsec)       &(parsec)$^{4}$ &factor &        &sp. type$^{4,5}$ \\
     \hline
2    &04579+4703        &125.5          &$>$29.5        &$>$0.35        &7.3    &3.59    &B1.5  \\
3    &05137+3919(1)     &19.1, 187      &38             &2.12           &8.5    &5.41    &O8    \\[-0.8mm]
     &05137+3919(2)     &166.5          &14.5           &0.81           &6.6    &        &      \\
5    &05274+3345(2)     &313            &               &               &??     &3.64    &B1.5  \\[-0.8mm]
     &05274+3345(1)     &49             &$\geq$33       &$\geq$0.25     &       &        &      \\
6    &05345+3157(I)     &132            &35.5           &0.31           &4.3    &3.14    &B3    \\[-0.8mm]
     &05345+3157(II)    &36, 229        &73             &0.64           &       &        &      \\
7    &05358+3543(I)     &130.4          &43.5           &0.38           &7.2    &3.58    &B1.5  \\[-0.8mm]
     &05358+3543(II)    &170            &83             &0.72           &18.8   &        &      \\[-0.8mm]
     &05358+3543(III)   &332.5          &$\geq$48       &$\geq$0.42     &6.8    &        &      \\
8    &05373+2349        &45.5           &20.5           &0.12           &2.55   &2.67    &B5    \\
10   &05553+1631        &84, 287        &66             &0.97           &1.9    &4.07    &B0.5  \\
11   &06061+2151        &128            &34             &0.33           &4.3    &3.60    &B1.5  \\
13   &18144-1723        &274            &18             &0.38           &10     &4.33    &B0.3  \\
14   &18151-1208(1)     &131            &$\geq$29       &$\geq$0.42     &$\geq$6.3	 &4.30 &B0.3  \\[-0.8mm]
     &18151-1208(2)     &35.5           &10             &0.15           &5.7    &        &      \\
18   &18264-1152        &69.2, 286      &$\geq$43.5     &$\geq$0.74, $\geq$2.64 &7  &4, 5.1 &B0.5, O7 \\
19   &18316-0602        &125, 315	&50             &0.77           &4.5    &4.64    &B0.5  \\
31   &19388+2357        &               &               &               &       &4.17    &B0.5  \\
32   &19410+2336(II)    &65             &18             &0.18, 0.56	&4.1    &4, 5     &B0.5, O7 \\[-0.8mm]
     &19410+2336(I)     &99.5           &31.5           &0.32, 0.98 	&3.8    &        &      \\
33   &20050+2720        &101            &$\geq$28.3     &$\geq$0.1      &4      &2.59    &B5    \\
35   &20062+3550        &43, 212 	&16             &0.38           &5.4    &3.51    &B2    \\
37   &20188+3928        &0 and/or 50    &               &               &       &2.54    &B5    \\
36   &20126+4104(I)     &122            &$\geq$11.5     &$\geq$0.09     &5      &4.00    &B0.5  \\
38   &20198+3716        &76.5 (and 61)  &               &               &??     &3.74    &B1    \\
39   &20227+4154(II)    &128            &$\geq$20       &$\geq$0.33     &8.3    &3.98    &B0.7  \\[-0.8mm]
     &20227+4154(I)     &239.5          &24             &$\leq$0.39     &4.5    &        &      \\
40   &20286+4105(1)     &274            &               &               &??     &4.59    &O9.5  \\[-0.8mm]
     &20286+4105(2)     &329            &7.5            &0.14           &??     &        &      \\
41   &20293+3952        &57             &               &               &       &3.8, 3.4 &B1, B2 \\
44   &21307+5049        &141            &               &               &       &3.6     &B1.5  \\
45   &21391+5802        &50             &38             &0.14           &6.5    &2.17    &B7    \\[-0.8mm]
     &21391+5802        &65             &$>$27		&$>$0.1		&$>$9	&        &      \\[-0.8mm]
     &21391+5802        &128            &9.9            &0.04           &4.6    &        &      \\
46   &21519+5613(1)     &48             &8.1            &0.29           &3.3    &4.28    &B0.3  \\[-0.8mm]
     &21519+5913(2)     &126.5          &7.8            &0.28           &5.5    &        &      \\
47   &22172+5549        &189            &               &               &       &3.26    &B2.7  \\
49   &22570+5912(1)     &149            &15             &0.37           &11.5   &4.7 	 &O9	\\[-0.8mm]
     &22570+5912(2)     &50.7           &7              &0.17           &4.5    &        &      \\
\hline
\multicolumn{8}{l}{$^1$When multiple outflows are detected, outflow numbers (labelled on Figs. A1--A50) are given in brackets;  $^2$two values}\\
\multicolumn{8}{l}{are given when the two lobes of the bipolar outflow deviate from a straight line containing the central source  identified;}\\
\multicolumn{8}{l}{$^3$lower limits only since these are not corrected for the (unknown) angle of inclination (this will also affect the}\\
\multicolumn{8}{l}{collimation factors derived); a ``$\geq$'' is shown against some sources since additional $H_2$ emission features could be}\\
\multicolumn{8}{l}{a part of the outflows which will inrease the outflow length.  These will in turn increase the collimation factors too;}\\
\multicolumn{8}{l}{$^4$two values separated by a comma are given when there is distance ambiguity; $^5$ZAMS spectral types corresponding to}\\
\multicolumn{8}{l}{the luminosites from \citet{panagia73} (for B3 and early spectral types) and from \citet{morton68} (for spectral}\\
\multicolumn{8}{l}{types later than B3)}\\
\hline
\end{tabular}
\end{centering}
\end{table*}

Among the 25 objects that are confirmed to be associated with H$_2$ jets,
nine objects did not exhibit radio continuum emission.  Out of the 
remaining 16, four objects have bright radio emission.  For three 
of these bright radio emitters, IRAS~20198, 20293 and  22570, the radio 
sources are convincingly not the driving sources of the H$_2$ outflows 
(Fig. \ref{20198_KH2}, \ref{20293_KH2}, \ref{22570_KH2BrG}) and for the 
fourth one  (IRAS~20188, Fig. \ref{20188_KH2}), the radio source appears 
to be  different from the driving source of the H$_2$ jet, though it remains
to be established.  The remaining 12 objects are faint in radio.  For some 
of these faint radio emitters, the radio sources detected are centred on 
the driving sources of the H$_2$ jets (eg. IRAS~05137, 05373, 05553, 
18316, 20126); some of these are interpreted as emission from ionized 
jets (eg.  IRAS~20126 - \citealt{hofner99}; IRAS~05553 - \citealt{shepherd99}).
Assuming that the strong radio sources are UCH{\sc{ii}}s, we conclude that
by the time the massive YSOs reach their UCH{\sc{ii}} phase, they
would have already accumulated a major fraction of their main sequence
mass, and thus, their accretion and outflow rates will be much lower
when compared to those in the pre-UCH{\sc{ii}} stage.  The location 
of the near-IR counterparts of UCH{\sc{ii}}s close to the reddening 
band in the colour-colour diagram (Fig. \ref{JHKcol}) also suggests 
the same.

A massive YSO may still be accumulating mass after it has 
developed an H{\sc{ii}} region.  Bulk rotation of molecular gas 
surrounding five massive UCH{\sc{ii}} objects have been detected 
recently by \citet{klaassen09}.  Ionized accretion have been proposed
as a significant contributor to the growth of massive YSOs which have 
formed H{\sc{ii}} regions around them (eg. \citealt{keto02, keto03, 
keto06});  accretion and outflows are detected in some HCH{\sc{ii}} 
objects (eg. \citealt{keto08}).  However, it remains to be seen if 
this can add a large fraction of the existing mass after the star 
has appeared as a UCH{\sc{ii}} object.

\section{Summary and Conclusions}

\begin{enumerate}

\item 76\,\% of our objects display H$_2$ emission.  
50\,\% exhibit collimated outflows; including the objects with 
suspected collimated emission,  this fraction would be 62\%.  
These numbers should be considered as lower limits only.  Even 
at 2.122\,$\mu$m,  the extinction will hamper the the detection 
of several outflows. 

\item From the good agreement that we find between the
centroids and directions of the outflows obtained from the
aligned H$_2$ emission knots in our images and those deduced from
the published CO maps, and from the available spectroscopic
results suggesting shock excitation of the H$_2$ emission
in jets, we conclude that the outflows are mainly jet-driven
as in low-mass stars and not wind-driven.

\item
Within the range of luminosities covered in our sample, the 
outflows are seen to be well collimated irrespective of the 
luminosities of their driving sources and are nearly as 
collimated as those from low-mass stars.  Considering the 
main sequence spectral types of single stars corresponding 
to the luminosities of the outflow driving sources 
(Table \ref{outflowprop}), we infer that objects up to 
early B and late O-types produce collimated outflows, 
implying that disk accretion forms the main mechanism 
leading to their formation.   However, accretion rates may be 
highly variable and non-uniform with the discs highly clumpy 
in nature and the individual clumps in the discs capable of
self-shielding against  the large radiation pressure to a 
significant level, thereby aiding accretion.
Accretion rates larger than 10$^{-4}$\,M$_{\odot}$year$^{-1}$
can produce even earlier spectral types.  Accurate estimates 
of accretion rates are requiered to understand the formation 
of massive stars at the very high end of the ZAMS mass limit.

\item 
For the objects that exhibited H$_2$ line emission in our survey,
except for a few cases,  near-IR counterparts were detected in our 
images for the YSOs driving the outflows. These objects occupy 
specific regions in the MSX colour-colour diagram,  which is 
indicative of the temperatures involved.  Co-ordinates of YSO
candidates identified are given throughout the text to 
enable future spectroscopic and high angular resolution 
investigations of these objects and their close environments.

\item 
Fourteen objects from which H$_2$ emission was either not detected
or, when detected, was suspected to be due to fluorescence
could be broadly classified into two different populations 
- the young and the old - based on whether they exhibit significant 
radio continuum emission or not.  The objects which exhibit radio 
emission and have NIR counterparts may be more evolved YSOs that 
are in the UCH{\sc ii} stage.  The objects without any detected 
outflow in H$_2$ and without any significant radio emission are 
likely to be pre-UCH{\sc ii}s that are in the early stages of their 
formation.  

\item  From the non-detection of strong H$_2$ jets from confirmed 
UCH{\sc{ii}}s,  we conclude that by the time the massive YSOs reach
their UCH{\sc{ii}} phase, they would have already accumulated a major 
fraction of their main sequence mass, and thus, their accretion and 
outflow rates  will be much lower when compared to pre-UCH{\sc{ii}}s.
The lack of high infrared excess implied by the location of the 
near-IR counterparts of UCH{\sc{ii}}s close to the reddening band in
the colour-colour diagram (Fig. \ref{JHKcol}) also suggests that they
have less circumstellar material than for pre-UCH{\sc{ii}}s objects.
Recent studies show that ionized accretion may significantly 
increase the mass of YSOs, which have already developed H{\sc{ii}} 
regions; it is to be understood how efficient it will be to add a
significant amout of mass after they appear as UCH{\sc{ii}}s.

\item As can be seen in Fig. \ref{msxirascol} (right panel), most of 
our objects, which are a mixture of UCH{\sc{ii}}s and pre-UCH{\sc{ii}}s,
are located within the region for UCH{\sc{ii}}s defined by 
\citet{wc89a}.   This shows that the objects identified by the colour 
criteria of  \citet{wc89a} are not always UCH{\sc{ii}}s. Instead, they 
represent luminous YSOs which are UCH{\sc{ii}}s and pre-UCH{\sc{ii}}s 
located in clusters (see also \citealt{bik04}).  

\item The close spatial locations of the objects resolved in our 
$K$-band images is suggestive of a very large fraction of binarity or 
multiplicity (confirmation of which may need long term spectroscopic 
monitoring and multi-colour high angular resolution photometry).  
Several new clusters are revealed;  multiple outflows 
are also observed in many of these clusters.  Our study shows that 
poor collimation factors derived in many (if not all) of the outflow 
maps in CO are due to unresolved multiple outflows (e.g. IRAS~05137, 
05345, 05358, 18151, 18316, 19410, 20227, etc., where we resolve 
multiple jets in H$_2$).  Some of these have recently been resolved 
in CO itself with high angular resolution interferometric mapping 
(e.g. 05358 - \citealt{beuther02a}).

\item As previously observed by many other investigators, intermediate
and high-mass stars form mostly in clusters and are  associated with 
lower mass star formation.

\item There are several candidate clusters in our sample which display 
a ring, arc, or flattened morphology (e.g. IRAS~05274, 06061, 20050, 
20056, 20188, 21519, 22172, 22570).  Another example outside this work 
is the ring cluster imaged by Kumar, Ojha \& Davis (2003).  The 
possible reasons for their morphologies need to be investigated.

\item Br$\gamma$ emission is not detected in any of the outflows; as is
the case with jets from low-mass YSOs, it remains a poor tracer of outflows
from young stars.  It is instead a better tracer of accretion process and
has been used to measure accretion rates in both low and high-mass YSOs
(e.g. \citealt{blum04, folha01, muzerolle98}).  Br$\gamma$ emission is
detected from some the outflow driving sources identified by us 
(e.g. IRAS~04579 - \citealt{ishii01} and  our Fig. \ref{04579_KH2};
IRAS~05137 - \citealt{ishii01};  IRAS~18151 - \citealt{davis04}).

\end{enumerate}

\section*{Acknowledgments}

The United Kingdom Infrared Telescope is operated by the Joint Astronomy 
Centre on behalf of the Science and Technology Facilities Council (STFC) 
of the U.K.  We have made use of 2MASS data obtained as a part of 
Two Micron All Sky Survey, a joint project of University of Massachusetts 
and the Infrared Processing and Analysis Centre/California Institute 
of Technology.  This research has also made use of IRAS and MSX data 
products, and SIMBAD database operated by CDS, Strasbourg, France.  
We thank the referee Thomas Stanke for carefully going through the
text and giving valuable inputs which have improved the quality
of the paper.


\appendix

\section{Results and Discussion on individual sources}

All of the YSO fields observed in the survey are discussed individually
below in fifty subsections (A1--A50). Table \ref{sourceprop} summarizes
the details on each source, including previous observations and our H$_2$
detections.  The order of presentation of the sources in the discussion
below is the same as in Tables \ref{obslog} \& \ref{sourceprop};
the subsection numbers of the sources agree with their reference numbers
in column 1 of the tables.  The figures are also named accordingly.
Conclusions from our images are discussed in context with existing
observations.  The $K$-band images are presented in all cases;  H$_2$ and
Br$\gamma$ images are shown only when line emissions are detected at 
these wavelengths.  H$_2$ emission line features are marked with
circles, or dotted  or dashed arrows and are numbered.  We use letters
for labelling point sources.  The coordinates of several sources of
interest detected in our images are listed throughout the paper.
The coordinates given are in the J2000 epoch. The convention adopted
is - the source label (``A, B'' etc.; RA($\alpha$)=hours:minute:seconds,
Dec($\delta$)=deg:arcmin:arcsec). The IRAS positions are marked on each
figure using a ``$\triangle$'' sign.  The positions of the objects
detected by MSX (except for source 1, which does not have MSX
observations available)  are shown with a ``+'' sign.  These are not
specifically mentioned anywhere else in the text except for Figs. 1 and 2.
Millimetre or sub-mm continuum peaks are marked on some of the images
with a ``*'' sign;  radio continuum positions are marked with ``x''.
When a millimetre or radio position from an additional source of data
is plotted on a figure, we use ``\#''.  Since the wavelengths and the 
literature from which the mm and radio observations were taken are
different for each source, these are listed in the figure captions
and in the text.  A compilation of the spatial resolutions and pointing
accuracies of the previous observations of these sources
is given in Appendix B (Table \ref{resolutions}) to help the readers
understand the possible association of the YSO candidates identified
by us in the near-IR with the YSOs identified in the long wavelength
studies before.  Since
north-east, south-west, etc. are used frequently in the text, we have
abbreviated these with NE, SW, etc.

\subsection{IRAS~00420+5530 -- {\it Mol 3}\\ ({\small \it d = 5.0; 7.72\,kpc, L = (12.4; 51.8$)\,\times$10$^3$\,L$_{\odot}$})}

IRAS~00420+5530 was identified as a candidate high mass protostar
based on its IRAS colours \citep{palla91, molinari96}.  A series
of survey papers show that the source is associated with NH$_3$
emission \citep{molinari96} and with H$_2$O maser emission
\citep{migenes99, palla91, brand94}. No 6.7-GHz CH$_3$OH maser was
seen by Szymczak, Hrynek, Kus (2000).  The lack of detected radio
continuum \citep{molinari98} combined with the other observed
properties suggests that  this source is young.  Jenness, Scott \&
Padman (1995) detected an 850-$\mu$m source at a flux level of
5.7\,Jy with a positional offset of 4.5\,arcsec east and 23\,arcsec
south from the IRAS position.  They detected $^{12}$CO and C$^{18}$O
emission from this region at -52 and -51.8\,km\,s$^{-1}$ respectively and
weak H$_2$O maser emission at -45\,km\,s$^{-1}$ close to the 850-$\mu$m
peak.  They did not detect any radio continuum emission at 3.6\,cm
above a flux density of 0.3\,mJy.  Attribution of the line shapes
to one or other of the properties of protostars (outflow, infall,
different velocity components within the beam) was not possible
with the sub-mm data. $^{13}$CO\,(J=1-0) was detected by
Wu, Wu \& Wang (2001), who report no evidence of any velocity
structure in their 55-arcsec beam.  \citet{brand01} detected extended
clumps of emission from  $^{13}$CO (2-1), C$^{18}$O (2-1), CS(3-2),
C$^{34}$S(3-2) and HCO$^+$(1-0) peaking 24\,arcsec south of the IRAS
position.  With the exception of the CS emission, which is more of a
tracer of dense gas from the core, all lines show red and blue shifted
components.  \citet{brand01} estimated a lower limit for the total
mass of the molecular material (from the $^{13}$CO emission) to be
2.8$\times10^3$\,M$_{\odot}$.

\begin{figure*}
\centering
\includegraphics[width=16.5cm,clip]{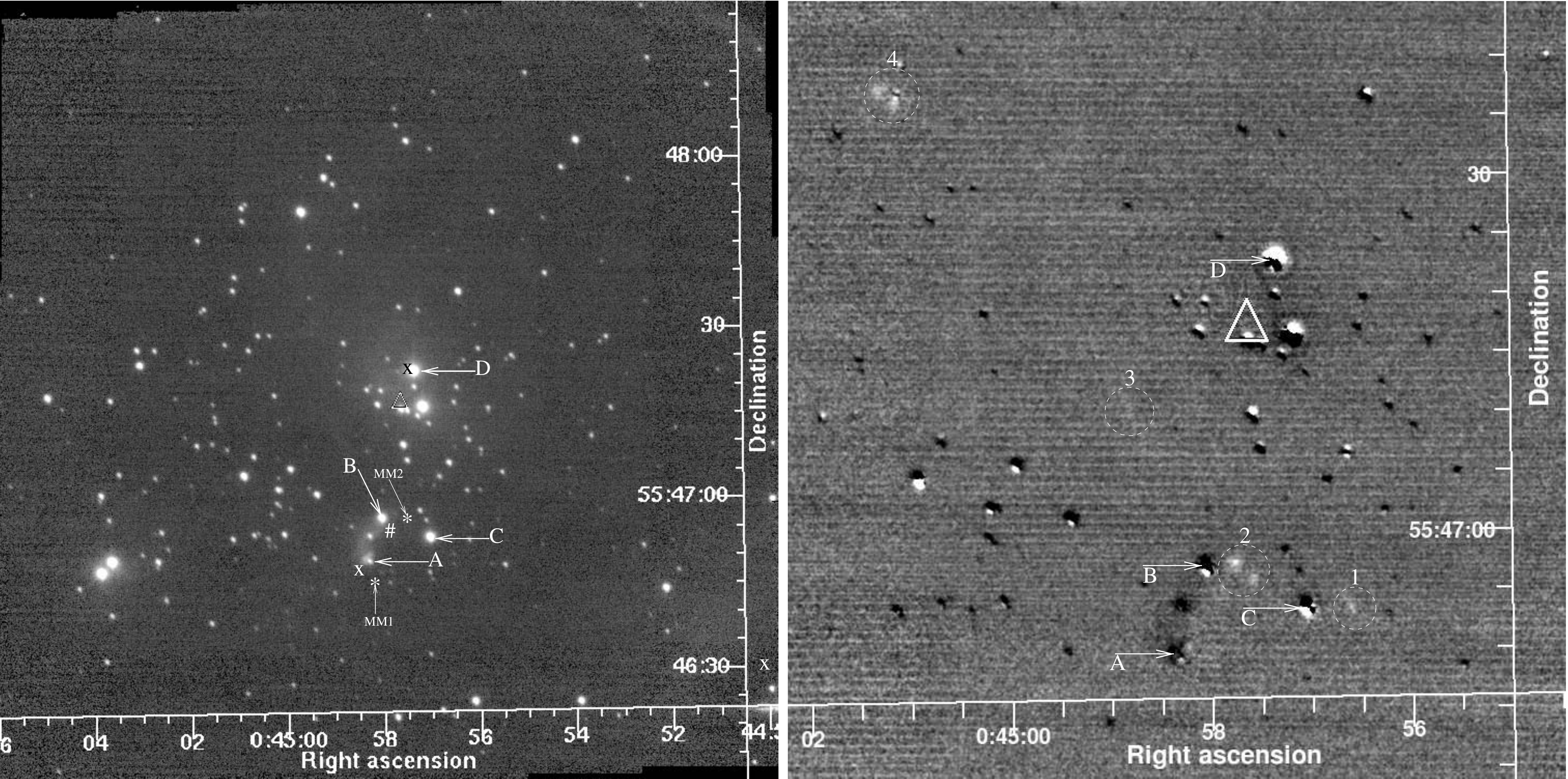}
\caption{The left panel shows the $K$-band image of IRAS~00420+5530. The
right panel shows the central part of the continuum-subtracted H$_2$ image.
``$\triangle$'' shows the location of the IRAS source.
The 850\,$\mu$m position of \citet{jenness95} is shown by ``\#''.
``*'' shows the two unresolved 3.4-mm continuum sources (MM1 and MM2)
and ``x'' shows the 3.6-cm emission peaks of \citet{molinari02}.}
\label{00420_KH2}
\end{figure*}

Our near-IR images (Fig. \ref{00420_KH2}) reveal a cluster of objects
within the 2.2$\times$2.2-arcmin$^2$ field.  The 2MASS colours of the
brightest of the near-IR sources located close to the IRAS position do
not exhibit any IR excess.  However, we see a set of IR-bright objects
located close to the peak of the sub-mm continuum and the line 
emissions.  Three objects are labelled here -
``A''($\alpha$=00:44:58.30, $\delta$=+55:46:49.7),
``B''($\alpha$=00:44:58.03, $\delta$=+55:46:57.2)
and ``C''($\alpha$=00:44:57.03, $\delta$=+55:46:53.8).
All three objects show IR excess in Fig. \ref{JHKcol},
with ``C'' having the lowest excess of the three.  ``A'' and ``B''
are apparently connected by nebulosity and there is a fainter
object embedded in it, which was not detected by 2MASS.
There is also a `cap-like' nebulosity seen in the east and north
directions  of ``A'' at a separation of $\sim$1 arcsec from the
star.  The nebulosity disappears in the continuum-subtracted H$_2$
image.  However, the H$_2$ image shows some faint line emission
features which are circled on Fig. \ref{00420_KH2} and labelled
``1--4''.  These are likely to be produced by an outflow
in the field.  At the present depth of integration, we cannot
say for sure if more than one outflow is involved in producing
the observed H$_2$ emission features.  The two unresolved
3.4-mm sources detected by \citet{molinari02},  the sub-mm source
of \citet{jenness95} and the water masers are all located in
the vicinity of the sources ``A--C''.  The molecular line
emissions mapped by \citet{brand01} are also peaked in this
region,  $\sim$2.1\,arcsec NE of ``A''.  This shows that this
is a cluster where there is on-going star formation.
\citet{molinari02} detected two faint point sources in this
region at  3.6\,cm (in addition to a brighter extended source
near the lower right region of our field); the brighter one
(0.32$\pm$0.03\,mJy) almost coincides with one of the bright
$K$-band sources near the IRAS position labelled
``D'' ($\alpha$=00:44:57.33, $\delta$=+55:47:23.1).  The 2MASS
colours of ``D'' do not exhibit any IR excess.  The fainter
radio source (0.17$\pm$0.03\,mJy) is located very close to ``A''.

The emission detected by IRAS could be from multiple sources.
The sub-mm, millimetre and CO emission could be from ``A'' or 
``B'' or some other embedded source in their vicinity.  It is 
not clear if any of these objects is the near-IR counterpart 
of the millimetre and sub-millimetre peaks and the driving 
source of a possible outflow implied by the H$_2$ emission 
features ``1--4''.  From its location, ``B'' is a strong 
candidate. Observations in CO rotational transitions and 
mid-IR continuum with better spatial resolution and deeper 
H$_2$ imaging will be useful.

\subsection{IRAS~04579+4703 -- {\it Mol 7}\\ ({\small \it d = 2.47\,kpc, L = 3.91$\times$10$^3$\,L$_{\odot}$})}

The $JHKLM$ photometric survey by Campbell, Persson \& Matthews (1989)
identified IRAS~04579+4703 as a possible luminous YSO. Multi-component
H$_2$O maser emission was detected from this source by several
investigators (\citealt{wb89}; Wouterloot, Brand \& Fiegle 1993;
\citealt{ brand94, migenes99}).  \citet{wouterloot93} also observed
12.2\,GHz methanol maser and CO emission.  However, the
CO observations of \citet{zhang05} did not reveal any molecular outflow
from IRAS~04579+4703.  A dense core was observed towards this
region in NH$_3$ emission by \citet{molinari96}.  The search by 
\citet{molinari98} did not detect any 6-cm radio emission from
this region. 

\begin{figure*}
\centering
\includegraphics[width=16.5cm,height=8.1cm,clip]{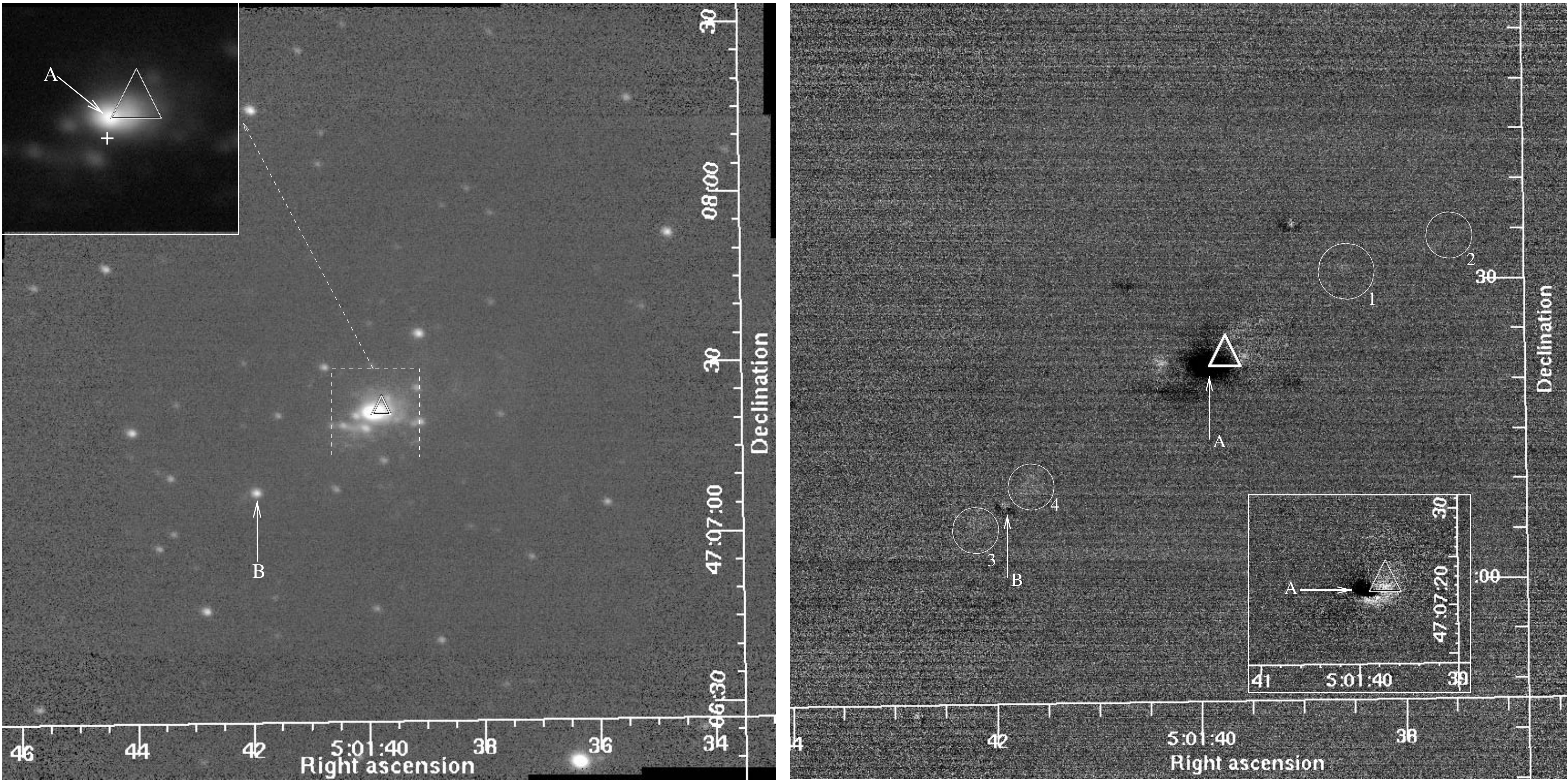}
\caption{Left: $K$-band image of IRAS~04579+4703.
An expanded view of the central source and its vicinity with better
contrast is shown in the inset.  Right: central region of the
continuum-subtracted H$_2$ image on which the continuum-subtracted
Br$\gamma$ image of the central source is shown in the inset.
``$\triangle$'' shows the location of the IRAS source and ``+''
shows the MSX source.}
\label{04579_KH2}
\end{figure*}

Our $K$, H$_2$ and Br$\gamma$ images (Fig. \ref{04579_KH2}) reveal a
compact cluster with a bright central source
(``A''; $\alpha$=05:01:39.91, $\delta$=47:07:21.6) located 2.35\,arcsec
SE of the IRAS position. ``A'' is associated with a cometary nebula seen
in $K$,  most of which disappears in the continuum-subtracted narrow-band
images.  Weak H$_2$ emission with a bipolar nature in the NW--east direction
is observed very close to ``A''.  The 2MASS position of ``A'' appears to be
centred off the source due to the presence of strong nebulosity in its
vicinity and its colours exhibit a large amount of reddening and excess
(Fig. \ref{JHKcol}).  Therefore, ``A'' is most probably the near-IR
counterpart of the IRAS source.  It should be noted that the MSX detection
is merely 1.6\,arcsec south of ``A''. There is a source 24.6\,arcsec to the
SW of ``A'', labelled ``B'' ($\alpha$=05:01:41.91, $\delta$=+47:07:7.7),
which displays a bipolar outflow-like H$_2$ emission features.  Both
lobe-like features are circled and labelled ``3'' and ``4'' respectively.
However, there are faint H$_2$ features in the diametrically opposite
direction  of ``3'' and ``4'' from ``A'', labelled ``1'' and ``2''.
Another very faint H$_2$ emission is detected  between ``4'' and ``A''.
It is also important to note that the 2MASS colours of ``B'' do not
exhibit any excess and place it well within the reddening band
(Fig. \ref{JHKcol}). Thus, even though morphologically ``3'' and ``4''
appear to be the two lobes of a bipolar outflow emanating from ``B'', it is
possible that these are parts of a major outflow from ``A'' in the NW-SE
direction at an angle of $\sim$125$^\circ$.5 in the sky plane.   The outflow
has a collimation factor of $\sim$7.3 if we include features ``3'' and ``4''.

As can be seen in the inset on the right half of Fig. \ref{04579_KH2},
there is Br$\gamma$ emission very close to ``A'', extending towards the
west.  Our detection of H$_2$ and Br$\gamma$ emission in the vicinity of
``A'' is consistent with the spectrum observed by \citet{ishii01}.
The H$_2$ emission seen in their slit must be coming from the bipolar
emission that we see close to the central source.  The Br$\gamma$ is
from a region closer to the source than the location of the H$_2$
knots.  Together, these observations show that this is a convincing
case of an intermediate-mass/high mass YSO driving a collimated outflow.

\subsection{IRAS~05137+3919 -- {\it Mol 8}\\ ({\small \it d = 11.5\,kpc, L = 225$\times$10$^3$\,L$_{\odot}$})}

IRAS~05137+3919 was first identified as luminous YSO based 
on its IRAS colours.  Modelling the SED, \citet{molinari08}
derived a luminosity of 2.55$\times$10$^5$L$_{\odot}$
at a distance of 11.5\,kpc.  The dense core was detected in 
ammonia emission by \citet{molinari96}.  Water maser emission 
has been observed towards this source \citep{palla91, migenes99},  
but no 6.7-GHz methanol maser emission was detected in the 
survey by \citet{szymczak00} and no SiO emission was detected 
by \citet{harju98}.  The VLA survey by \citet{molinari98} 
did not reveal any radio emission at 2\,cm and 6\,cm from 
IRAS~05137+3919.

\begin{figure*}
\centering
\includegraphics[width=16.5cm,clip]{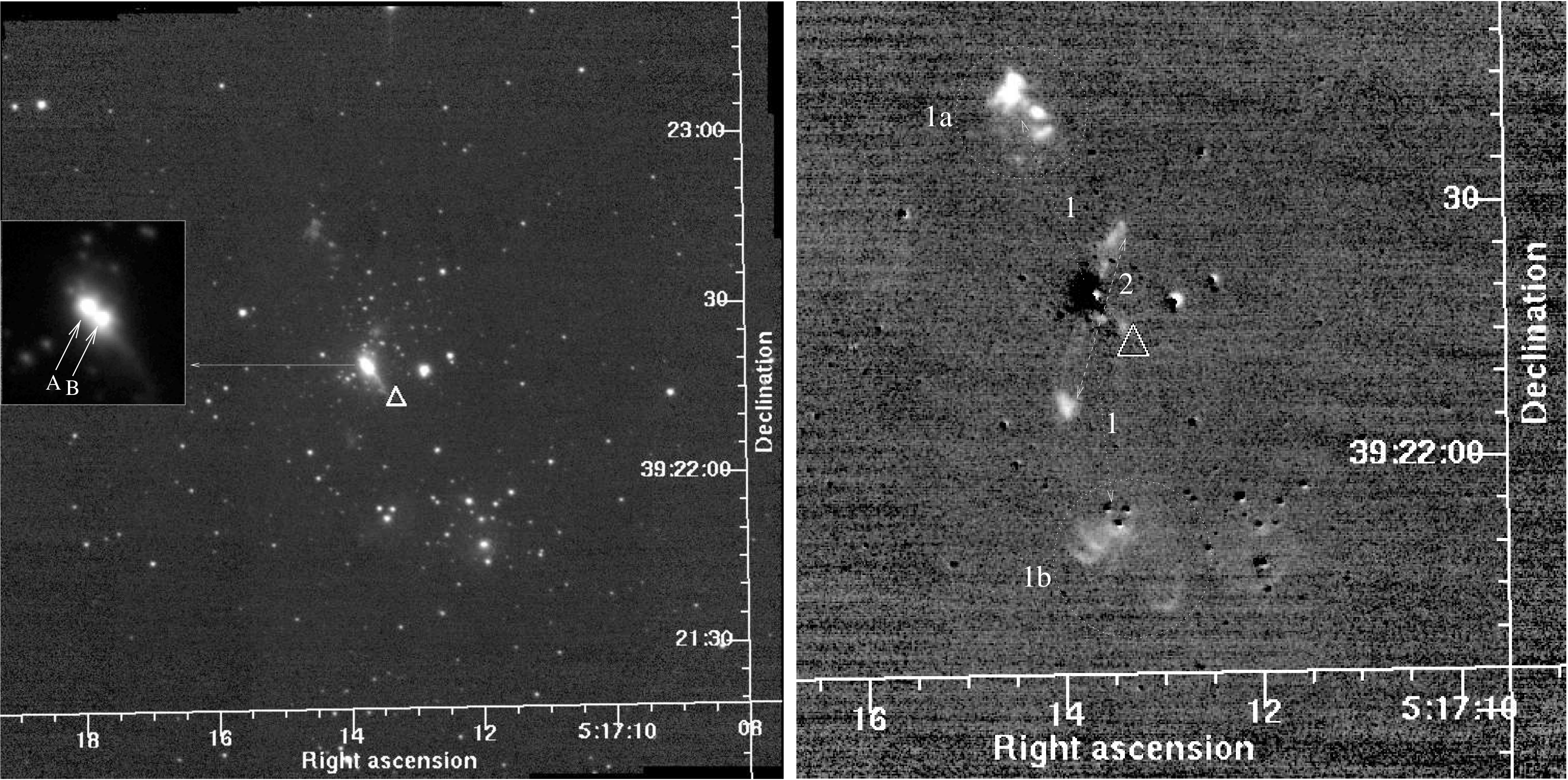}
\caption{Left: $K$-band image of IRAS~05137+3919.
The inset displays a magnified view of the central region 
to show the ``A,B'' pair resolved. ``*'' shows the 3.4-mm 
continuum peak (MM1) and ``X'' shows the 3.6-cm radio 
continuum peak of \citet{molinari02}.  Right: the central 
part of the continuum-subtracted H$_2$ image.  }
\label{05137_KH2}
\end{figure*}

Fig. \ref{05137_KH2} shows our $K$ and H$_2$ images. An expanded
view of the central region is shown in the inset on the $K$-band
image.  Through H$_2$ imaging, we discovered two spectacular
well-collimated outflows.  The brighter one is in the NE-SW direction
and the fainter one in the NW--SE direction. There are two
IR-bright sources embedded in nebulosity in the centre of
our images.  These sources are labelled
``A'' ($\alpha$=05:17:13.73, $\delta$=39:22:19.9) and
``B'' ($\alpha$=05:17:13.68, $\delta$=39:22:19.3).  They
are separated by 0.91\,arcsec at a position angle of 50$^\circ$.
The ``A,B'' pair is well detected by 2MASS, but not resolved.
The 2MASS near-IR colours of ``A'' and ``B'' combined place the
pair in the region of large excess emission, typical of luminous
YSOs, in the $JHK$ colour-colour diagram (Fig. \ref{JHKcol}).
The pair is located 7.25\,arcsec NE of the IRAS position.
MSX detected a source with a steep SED within one arcsecond
of ``A''.  We conclude that the two outflows discovered in our
H$_2$ images are produced by the sources ``A'' and ``B''.  On
the H$_2$ image, we have shown the directions of the two outflows
by the dotted and dashed arrows, labelled ``1'' and ``2'',
at angles 19.5$^{\circ}$ and 167.5$^{\circ}$ respectively.
The bow-shocks of outflow ``1'' are encircled and labelled ``1a''
and ``1b'' respectively.  Both ``1a'' and ``1b'' are composed
of multiple bow-shocks suggesting a possible precession of
the jet.  In H$_2$, the collimation factors could be $\sim$8.5
and $\sim$6.6 respectively for outflows ``1'' and ``2''.

\citet{molinari02} detected HCO$^+$ (1-0) emission from the central
source, and from locations close to the NE and SW lobes of the H$_2$
jet ``1'' described above, using the OVRO millimeter wave array.
They derived a position angle of $\sim$25$^{\circ}$ for the outflow,
which is comparable to a value of 19.5$^{\circ}$ we measure for jet
``1'' from the H$_2$ emission knots.  They detected 3.4-mm continuum
emission from the central object.  They also detected  faint radio
continuum emission from the central source at 3.6\,cm (at a flux
density of 0.33$\pm$0.03\,mJy) using the VLA.  From the  HCO$^+$(1-0)
spectrum of the central source, they suspect two distinct components
nearly along the line of sight which is supported by the binary stars
embedded in nebulosity as seen in our $K$-band image. \citet{wu01}
observed a dense core towards this region in $^{13}$CO (J=1-0).
\citet{brand01} detected two velocity components in their 
$^{13}$CO (J=1-0), HCO$^+$ and CS maps (at -25.5\,km\,s$^{-1}$ and 
-26.5\,km\,s$^{-1}$) with the redshifted component centred on the 
IRAS position and on the 3.4-mm and 3.6-cm continuum maps from 
\citet{molinari02}.

Previous infrared imaging by \citet{ishii02} did not resolve
``A'' and ``B''.  $K$-band spectroscopy by \citet{ishii01}
detected emission from H$_2$~${\it{v}}$=1-0~S(1), Br$\gamma$ and from
the ${\it{v}}$=2-0 overtone band of CO. The  H$_2$ emission  they
observed close to the central source must be due to the emission
from the H$_2$ jet in the NW-SE direction that we see in our image,
the base of which will be in their 2.4\,arcsec wide slit.
\citet{zhang05} also detected a very powerful outflow from this
source in CO, with the blue- and red-shifted lobes
roughly  in the direction of our flow ``1''.  These observations,
along with the detection of two jets in our H$_2$ image, confirm
that there are two outflows in this region and the jets that we see
in H$_2$  emission are powering the large scale outflows in
IRAS~05137+3919.

\subsection{IRAS~05168+3634 -- {\it {Mol 9}}\\ ({\small \it d = 6.08\,kpc, L = 24$\times$10$^3$\,L$_{\odot}$})}

IRAS~05168+3634 has water maser emission associated with it
\citep{palla91, migenes99}, but there was no detection of any
6.7-GHz methanol maser emission by either \citet{szymczak00} or
\citet{slysh99}. \citet{molinari98} detected 6-cm radio emission
from this region using the VLA.
However, the location of their radio source is 102\,arcsec away from
the IRAS source. Hence, the radio detection is not from the IRAS
source.  In addition, \citet{harju98} detected the SiO (J=2-1) line
at v$_{peak}$=-18.1\,km\,s$^{-1}$ and
Zinchenko, Henkel \& Mao (2000) observed C$^{18}$O  from this source.
\citet{zhang05} discovered a bipolar molecular outflow in CO.
The centroid of this outflow appears to be offset from the IRAS
position by a few arcminutes and therefore, may be outside our field
of view.  The dense core was mapped by Bronfman, Nyman \& May (1996)
in CS emission and by \citet{molinari96} in NH$_3$.

\begin{figure}
\centering
\includegraphics[width=8.10cm,clip]{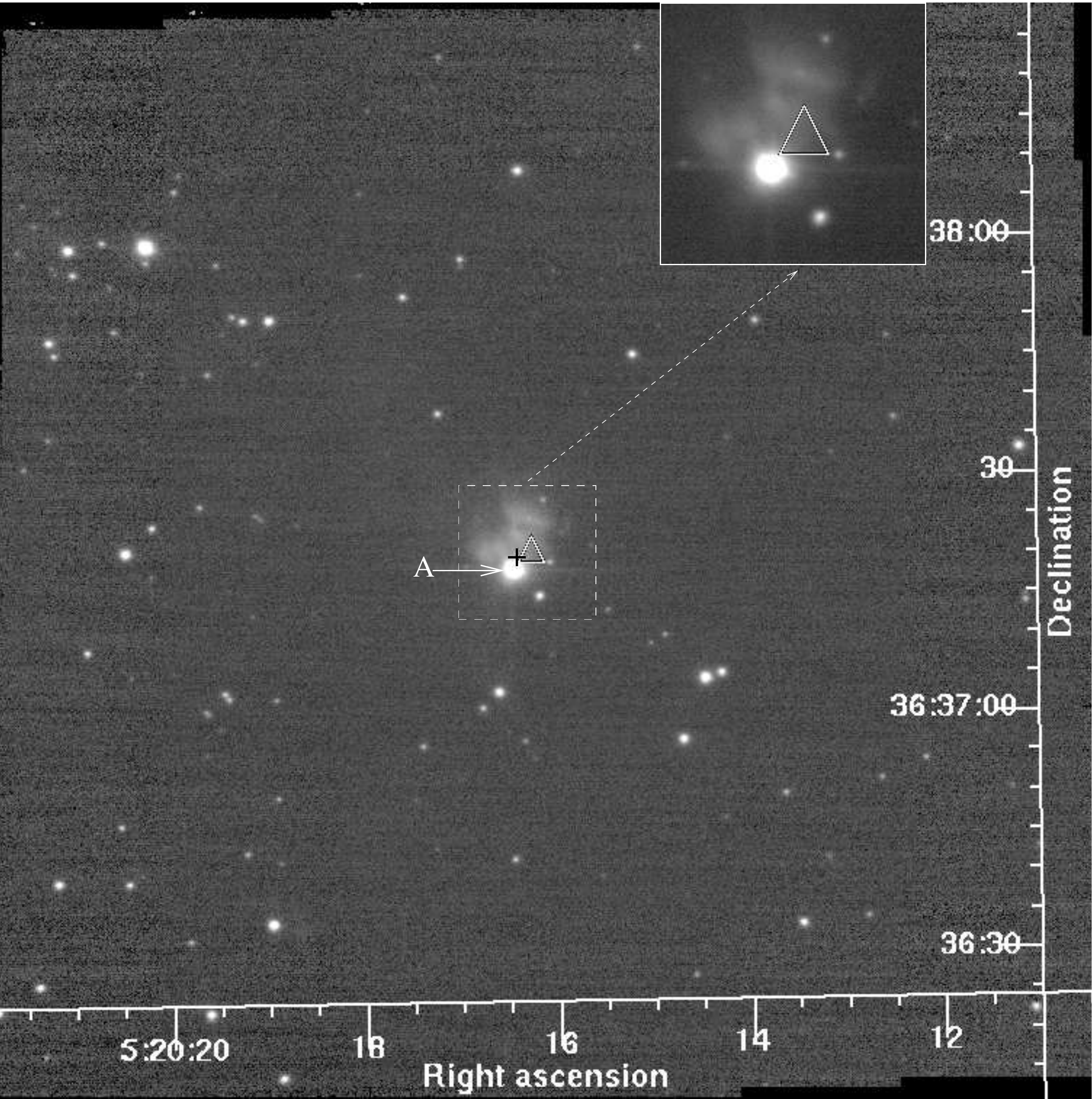}
\caption{$K$-band image of IRAS~05168+3634.  The inset shows
an expanded view of the central region.}
\label{05168_K}
\end{figure}

Fig. {\ref{05168_K}} shows our $K$-band image of IRAS~05168+3634.  The
$K$-band image exhibits nebulosity extending northward from a bright
infrared source labelled ``A'' ($\alpha$=05:20:16.44, $\delta$=+36:37:18.7),
located near the centre of the field.  This nebulosity disappears upon
continuum subtraction;  the subtracted H$_2$ and Br$\gamma$ images do
not reveal any line emission.   Hence these images are not presented
here.  The 2MASS colours of ``A'' do not exhibit any IR excess
(Fig. \ref{JHKcol}).  However, it should be noticed that ``A'' is only
poorly detected in $J$.  The IRAS source is 3.4\,arcsec NW of ``A'' 
and the MSX source is just 1.5\,arcsec NW of ``A'', which implies that 
``A'' is probably the luminous YSO.  The lack of any observed H$_2$ 
emission and the absence of any significant near-IR colour excess for 
``A'' imply that the source has probably passed the very early stages 
of its formation.  More accurate photometry is required to derive the
IR colours of ``A'' and its neighbours.

\subsection{IRAS~05274+3345 -- {\it Mol 10}\\ ({\small \it d = 1.55\,kpc, L = 4.35$\times$10$^3$\,L$_{\odot}$})}

\begin{figure*}
\centering
\includegraphics[width=16.5cm,clip]{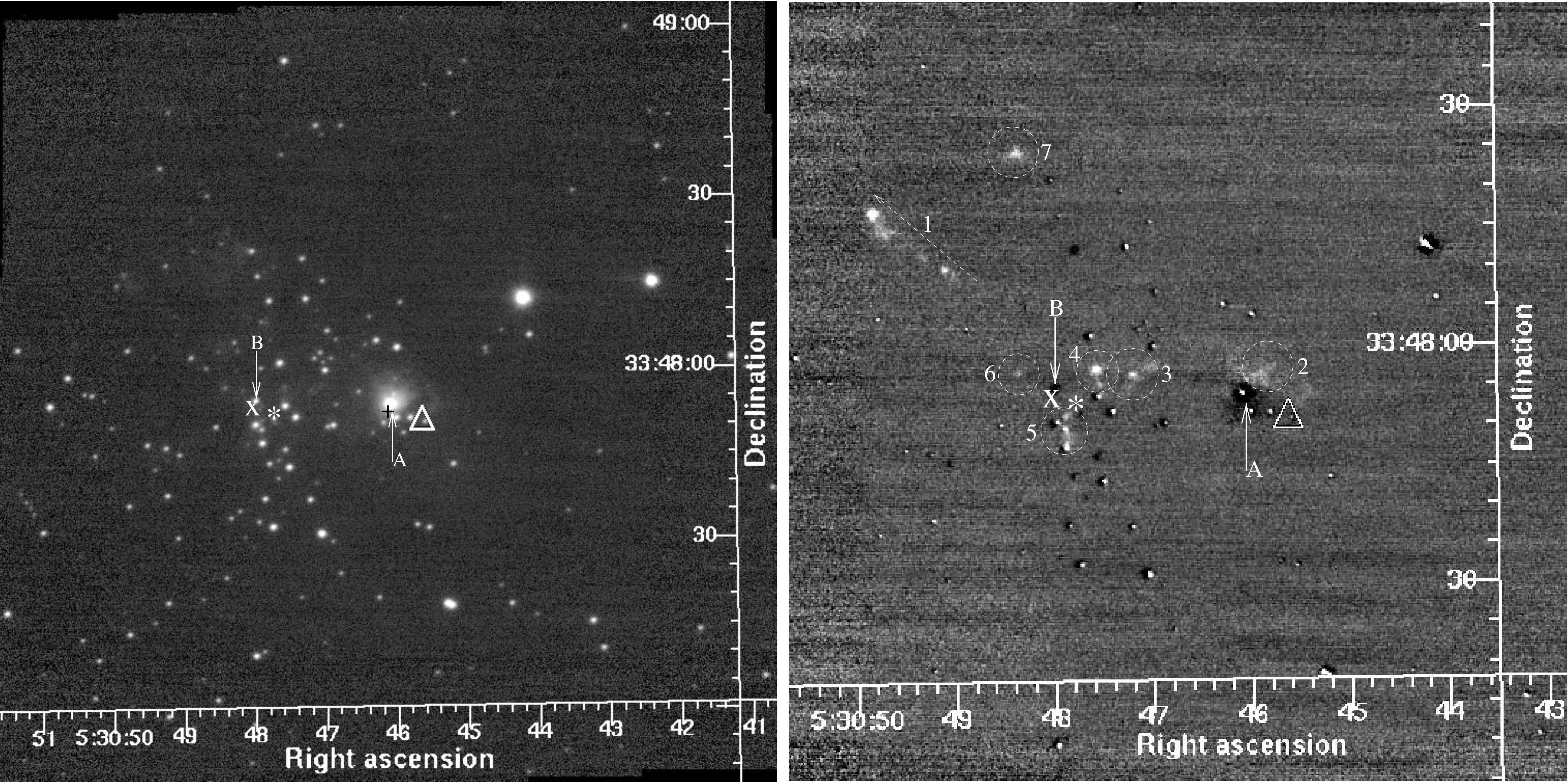}
\caption{Left: the $K$-band image of IRAS 05274+3345. Right:
the central part of the continuum-subtracted H$_2$ image.
The prominent H$_2$ emission features are labelled on the figure.
``*'' shows the 850-$\mu$m position of \citet{jenness95}.
``x'' shows the 3.6-cm position of \citet{tofani95}.}
\label{05274_KH2}
\end{figure*}

Towards IRAS~05274+3345 (also referred to as AFGL~5142), there
are two distinct sites of star forming activity.  The high density
core towards this source was detected in NH$_3$ emission by
\citet{molinari96}.  A radio continuum source detected in this
region (integrated flux of 0.83$\pm$0.12\,mJy at 3.6\,cm; 
\citealt{tofani95}) is offset $\sim$30\,arcsec east of the 
IRAS position, and is coincident with the peak millimetre 
emission from dust \citep{hunter99}, an ammonia core and a 
possible compact disc \citep{zhang02}.  The brightest 
infrared source in the previous $K$-band image from 
\citet{hunter95} is located near the centre of the error 
ellipse of the IRAS source position. However, the density 
distribution for the infrared sources in the field peaks 
close to the radio continuum peak. There is evidence for 
outflows centred on the radio position,  observed in $^{12}$CO 
and H$_2$ by \citet{hunter95} and in SiO and NH$_3$ by 
\citet{zhang02}, who deduce the presence of a disc associated 
with this source.  A sub-mm continuum peak was also detected 
by \citet{jenness95} at 850\,$\mu$m within the IR cluster 
towards the east of the IRAS source.  The H$_2$O maser 
emission from this source has been studied by many authors 
\citep{palla91, tofani95, goddi04}.  The masers are located 
in the close vicinity of the radio continuum source and on
either sides of it \citep{tofani95}.  Most recently, 
\citet{goddi04} determined the proper motions of 7 of the 12 
maser sources. The proper motions of a subset of these masers 
is consistent with Keplerian rotation in the disc posited by 
\citet{zhang02};  the remainder are associated with the outflow.
Detection of 6.7-GHz maser emission has been made by \citet{szymczak00}
and \citet{slysh99}.  An extensive survey for molecular line emission
associated with the H$_2$O maser activity in high mass YSOs by
Cesaroni, Felli \& Walmsley (1999b) detected $^{13}$CO~(2-1), 
HCO$^{+}$, HCN~(1-0), CH$_3$CN, C$^{34}$S and CS~(3-2) peaking 
at the position of the water masers, with a second, quiescent 
clump located 12\,arcsec south.  The southern clump is detected 
only in  C$^{34}$S and CS.  They propose that the northern 
clump associated with the water masers is at a later stage of 
star formation than the southern clump.

Fig. \ref{05274_KH2} shows the $K$ and H$_2$ images of IRAS~05274+3345.
The $K$-band image shows a cluster of IR sources with a bright object
embedded in nebulosity
(labelled ``A''; $\alpha$=05:30:46.07, $\delta$=+33:47:54.1) located
near the centre of the field. The nebulosity shows a mild cometary
nature extending NW.  There are many fainter objects close to ``A'',
the closest being 1.7\,arcsec NW of ``A'' and embedded in the
nebulosity.  Most of the other IR sources deteced in our image
are located towards the east of ``A''.  $JHK$ photometry of
\citet{hunter95} shows that ``A'' (their ``IRS2'') has reddening and
excess.  The 2MASS $K_s$ magnitude of ``A'' has large uncertainty.
It is possible that the 2MASS magnitudes of ``A'' are
contaminated by the nebulosity and the presence of nearby stars.
So we have used the $JHK$ photometry of \citet{hunter95} to plot it 
in the colour-colour diagram (Fig. \ref{JHKcol}).  ``A'' is located 
6.1\,arcsec NE of the IRAS position.  MSX detected a source with a 
steep SED within 1.1\,arcsec of ``A'', which suggests that ``A'' 
is probably a young source.

The continuum-subtracted H$_2$ image shows several emission features 
and generally agrees with the image presented by \citet{hunter95}. 
Prominent emission features are labelled in Fig. \ref{05274_KH2} (right panel).
This image is smoothed with a 2-pixel FWHM Gaussian to enhance the faint
emission. Features circled and labelled ``2--5'' are located in an
arc-like pattern extending from ``A''.  More conspicuous is the set of
aligned features in the direction of the arrow labelled ``1'' on
Fig. \ref{05274_KH2},  which appears to be a jet.  This jet is inclined
at an angle of 51$^{\circ}$.5 and appears to be originating from one of
the objects in the eastern cluster;  the mm, sub-mm and radio emissions
seen are from this cluster.  The 3.6-cm radio continuum source detected
by \citet{tofani95} is 24\,arcsec east of ``A''.  Many of the cluster
members show reddening and excess. Being located in a cluster with
several sources exhibiting reddening and excess, it is not sure if the
radio source, the mm source and the source driving jet ``1'' are all
the same.  It is noteworthy that the IR source labelled 
``B'' ($\alpha$=05:27:29.94, $\delta$=33:45:40.5;  ``IRS1'' of 
\citealt{hunter95}) shows large reddening and excess
(Fig. \ref{JHKcol}; colours are from \citealt{hunter95} since it is
detected only in K$_s$ by 2MASS) and is located very close to the 
radio source and the associated H$_2$O masers detected by \citet{tofani95}.
From the positional accuracies of the maser and radio sources
(Table \ref{resolutions}) and their locations, they appear to be 
associated with ``B''.  ``B'' is therefore a strong candidate for 
the driving source for an outflow in the direction of ``1''. 
``A'' (IRS2) is likely to be the driving source  of the large scale 
outflow mapped in CO(2-1) by \citet{hunter95}.  The location of the  
H$_2$ knot ``2'' is in the same direction from ``A'' as the 
blue shifted lobe of their CO outflow.  The compact outflow 
perpendicular to the large scale outflow,  mapped by Hunter et al., 
is roughly in the direction of ``1'' and is centred near ``B'', 
the radio source and the water masers.  They propose that the radio 
emission could be from an ionized wind.  There is an emission
feature labelled ``7'', which is probably part of another outflow.  
Overall, it looks like this region hosts more than one YSO,  with 
at least one well collimated outflow confirmed in H$_2$.  The 
multiple H$_2$ knots in different directions imply multiple outflows.

\subsection{IRAS~05345+3157 -- {\it Mol 11}\\ ({\small \it d = 1.8\,kpc, L = 1.38$\times$10$^3$\,L$_{\odot}$})}

Towards IRAS~05345+3157, better known as AFGL~5157,  \citet{snell88}
and \citet{ridge01} mapped an EW CO outflow at low spatial
resolution, centred on a dense molecular core extending roughly
in the NS direction that was mapped in NH$_3$ emission \citep{torrelles92a,
molinari96, verdes89}.  The CO outflow was also detected by
\citet{wb89}.  Multiple H$_2$O masers were detected from the region
close to the centre of the NH$_3$ core and CO outflow by several
investigators \citep{verdes89, palla91, wouterloot93, brand94,
torrelles92b}.  \citet{torrelles92b} and \citet{molinari02}
detected an H{\sc ii} region at 3.6\,cm.
The location of the radio emission is close to the centre of
the NH$_3$, CO and H$_2$O maser  emission.  \citet{chen99, chen03}
found an infrared cluster $\sim$1\,arcmin SW of the core - coincident
with the IRAS position - that is enveloped by diffuse, fluorescent
H$_2$ filaments (associated with a PDR).  They also observed a number
of compact H$_2$ knots,  implying the presence of multiple outflows
in the region.  \citet{chen03} show a schematic diagram summarising
the many observations of AFGL~5157.  The brightest 3.6-cm radio
continuum position of \citet{molinari02} agrees with the 6-cm
position given by \citet{molinari98}, which is towards the NW of
our images and is outside our field of view.  The second brightest
radio source of \citet{molinari02} (0.80$\pm$0.05\,mJy) is located
very close to an IR-bright source near the IRAS position and is
extended.  A fainter radio source (0.19$\pm$0.03\,mJy) was detected
by them, located close to the object labelled ``C'' in our image
(Fig. \ref{05345_KH2}).  This radio source appears to be the same as
the faint radio source detected by \citet{torrelles92b}
at 3.6\,cm and the locations of both agree within their
beam sizes (Table \ref{resolutions}).  
The positions of some of the
H$_2$O maser spots detected by \citet{torrelles92b} agree well with
that of ``C''. However, there are other H$_2$O maser spots observed
by \citet{torrelles92b} further SE of the ones near ``C'',  close to
the location of the H$_2$O maser observed by  \citet{verdes89}.  The
location of the NH$_3$ emission peak detected by \citet{verdes89}
is near ``C'', located between ``B'' and ``C'' and further north on
Fig. \ref{05345_KH2}.  However, both ``B'' and ``C'' fall within the
1.4\,arcmin beam of  \citet{verdes89}.

Fig. \ref{05345_KH2} displays our $K$-band and continuum-subtracted
H$_2$ images.  The $K$-band image reveals a cluster of IR-bright
sources towards the centre of the field.  The IRAS position is within
this cluster.  The image also shows emission in ridges around
this cluster and towards the NE. The H$_2$ image reveals a well
defined bipolar jet oriented SE-NW at an angle of 131$^{\circ}$.6, in
the direction of the dashed arrow labelled ``I''.  Emission features
circled and labelled ``1--4'' appear to be part of this jet.
Close inspection of the H$_2$ image shows that there is more than
one jet in this direction, or that the jet is precessing.  The
collimation factor of the jet would be $\sim$4.3 if it is a single jet.

Three objects are labelled on the figures -
``A'' ($\alpha$=05:37:49.83, $\delta$=+31:59:47.9),
``B'' ($\alpha$=05:37:50.07, $\delta$=+31:59:52.5) and
``C'' ($\alpha$=05:37:51.97, $\delta$=+32:00:04.2).
``A'' appears to be a point source close to the centroid of the
observed jet/s represented by ``I''.  It exhibits reddening and
excess.  ``B'' shows copious amount of H$_2$ emision. The central
region of ``B'' does not appear as a single point source in our
image and has multiple nebulous extensions.  The 2MASS $JHK_s$ colours
of ``B'' place it in the region occupied by YSOs in the colour-colour
diagram (Fig. \ref{JHKcol}).  The H$_2$ image shows several other
emission features, the prominent ones are circled on our H$_2$ image.
The featues ``5--9'' are located in a direction nearly
perpendicular to the direction of the jet ``I''.  These appear
to be part of another jet.  A dotted arrow is drawn in this
direction and labelled ``II''.   Two other knots (``10'' and ``11'')
are labelled on the H$_2$ image, with faint filementary features
extendening towards them from ``B'', the directions of which are
shown by dotted arrows labelled ``?''.  ``5--11'' could be
from multiple sources located in ``B'' or from a single wide angle
outflow from ``B''.

\begin{figure*}
\centering
\includegraphics[width=16.5cm,clip]{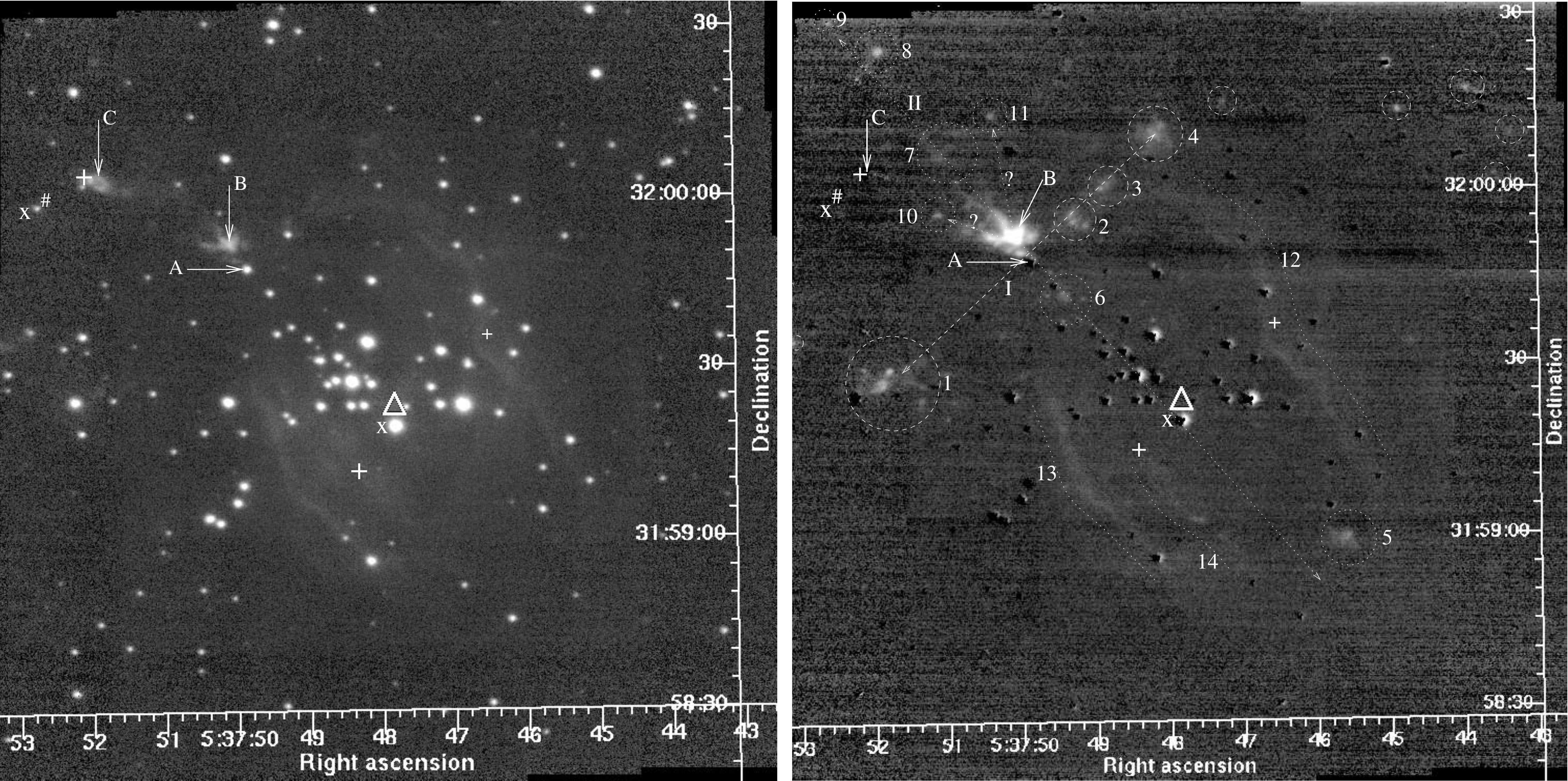}
\caption{Left: the $K$-band image of IRAS~05345+3157.  Right:
the continuum-subtracted H$_2$ image smoothed with a Gaussian of
FWHM=2 pixels.  ``x'' shows the locations of the faint radio
sources detected by \citet{molinari02}.  ``\#'' shows the
location of the 3.6-cm radio sources detected by
\citet{torrelles92b}.}
\label{05345_KH2}
\end{figure*}

Also prominent in the figure are filementary ridges of H$_2$ emission
around the IR cluster surrounding the IRAS position.  Dotted lines are
drawn on the figure close to these features to guide the eyes.
Labelled ``12--14'',  these are the fluorescent filaments of H$_2$
emission detected by \citet{chen03}. They are likely to be due to the
UV emission from the more evolved hot stars in the infrared cluster.

The MSX mission detected three sources in this field, the positions
of which are denoted by a ``+'' in our images.  The brightest one is
13.5\,arcsec SE of the IRAS position and is probably associated with
the IR cluster near the IRAS position.  The faintest of the three
is 20\,arcsec NW of the IRAS position. The third one is located
NE in our field and is very close to ``C''.  Source ``C'' has a
nebulous patch of emission which is seen in $K$, with faint point
sources in the vicinity; the nebula disappears upon continuum
subtraction.  It is detected only in $K_s$ by 2MASS and the magnitude
limits in the $J$ and $H$ bands imply strong excess.  The proximity
of the MSX source suggests that there is a deeply embedded YSO
located here.

Our H$_2$ image reveals bow shocks produced by outflows from the
deeply embedded core located NE of the IR-bright cluster.  The
locations of the radio continuum source, H$_2$O masers, NH$_3$
emission and the CO outflow all are near this core from which
the H$_2$ emission knots appear to be emanating.  The position of the
IRAS source is centred off this region and is towards the IR cluster.
However, an  MSX source is detected  near this region (only at
21\,$\mu$m), the location of which is close to the centre of the NH$_3$
core.  All these observations imply that there is multiple star
formation occuring within this embedded core and that the HH-type
features seen in the H$_2$ image are from a region which is much
younger than the IR cluster at the centre of our field.  With the
positional accuracy of the radio detection given by \citet{torrelles92b}
(Table \ref{resolutions}) and our near-IR images, it appears that 
the radio emission near ``C'' is caused by a source which is 
different from those driving the outflows in the direction of 
``I'' and ``II'' and is probably more evolved. Also, ``C'' does 
not appear to be the near-IR counterpart of the radio source.
From the positional uncertainty of the MSX, it is not clear if 
the MSX detection is from ``C'', ``B'',  the radio source,
or a combination of all.

Overall, our observations support the results
of \citet{chen03}.  It should be noted here that \citet{zhang05}
also mapped a bipolar outflow in this region,  the centroid of which
is offset from the IRAS position by 43\,arcsec each in RA and dec and
is close to the YSOs that we detect in the NE region of the field.

\subsection{IRAS~05358+3543\\ ({\small \it d = 1.8\,kpc, L = 3.8$\times$10$^3$\,L$_{\odot}$})}

IRAS~05358+3543 (S233IR) has been studied in the near-IR by a number
of groups (Porras, Cruz-Gonz\'alez \& Salas 2000; Kumar, Bachiller
\& Davis 2002; \citealt{khanzadyan04}). S233 is associated with two
embedded stellar clusters, the IRAS 12-$\mu$m position being coincident
with the central cluster (referred to as the south-western cluster by
Porras et al.).  Methanol maser emission has been detected by a
number of investigators \citep{menten91, szymczak00, sridharan02, galt04}.
Single dish observations of \citet{menten91} place the methanol maser
in the NE cluster.  Water maser emission observed in this region is from
the NE cluster \citep{sridharan02, beuther02d}.  In the CO maps of
\citet{beuther02a}, multiple molecular outflows have been resolved,
which are traced by the H$_2$ images shown by \citet{porras00},
\citet{kumar02} and \citet{beuther02a}.
The most prominent flow is oriented roughly NS and appears to be
associated with the NE cluster, the IRAS 100-$\mu$m position and a
density peak in the 1.2\,mm continuum maps \citep{beuther02a,beuther02b}.
The lobes of this outflow are capped by two prominent H$_2$ bow shocks,
N1 and N6 \citep{porras00, kumar02}.  The northern lobe is blue-shifted
and the southern red-shifted.  Additional outflows are traced in
both CO and H$_2$ \citep{beuther02a, khanzadyan04}.  Notably,
\citet{sridharan02} did not detect 3.6-cm emission, down to 1\,mJy,
from the NE cluster and the density peak, which suggests that the
driver of the north-south molecular flow \citep{beuther02a} is in a
pre-UCH{\sc{ii}} phase.

\begin{figure*}
\centering
\includegraphics[width=16.5cm,clip]{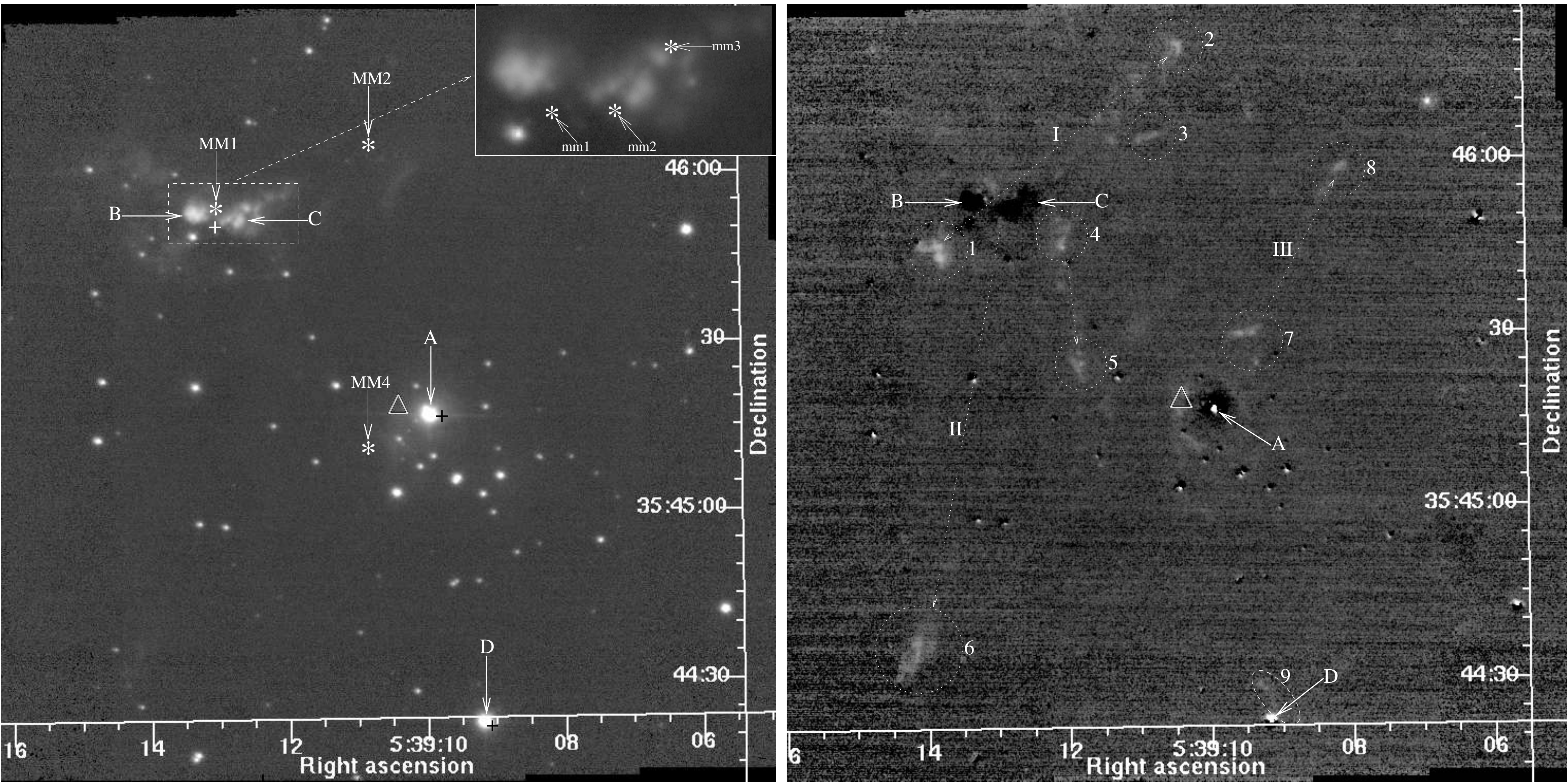}
\caption{The left panel shows our $K$-band image of IRAS~05358+3543.
``*'' on the main figure shows the location of the three 1.2-mm peaks
(``MM1, MM2, MM4'') within our field listed by \citet{minier05}. Labelled
by ``*'' on the inset and named ``mm1--mm3'' are the three massive
sub-cores resolved in this region by \citet{beuther02a} at 2.6\,mm.
The right panel shows the continuum-subtracted H$_2$ image smoothed with
a 2-pixel FWHM Gaussian.}
\label{05358_KH2}
\end{figure*}

Fig. \ref{05358_KH2} shows our $K$ and H$_2$ images.  Both the central
and the NE cluster are recorded in our images.  Prominent H$_2$ emission
features are circled in the diagram. Source
``A'' ($\alpha$=05:39:09.93, $\delta$=+35:45:17.2) is the IR-bright
object close to the IRAS 12-$\mu$m position and is most probably the
near-IR counterpart of the IRAS 12-$\mu$m source.  There is a ring-like
H$_2$ emission feature around it, which is probably due to fluorescence
in a shell \citep{porras00, kumar02} projected on to the sky plane.  ``A''
is the brightest member of the central cluster and it exhibits IR excess
in the colour-colour diagram (Fig. \ref{JHKcol}).  A jet-like feature
is emanating from this region;  the two knots labelled ``7'' and ``8''
are probably shock-excited H$_2$ emission along this jet in the direction
of the dotted arrow labelled ``III'' on Fig. \ref{05358_KH2}.  ``A''
appears to be the driving candidate for this jet. MSX detected two
bright sources in this region; the bluer of the two is
within 1.5\,arcsec of ``A''.

We label two objects in the NE cluster:
``B'' ($\alpha$=05:39:13.29, $\delta$=+35:45:53.5) and
``C'' ($\alpha$=05:39:12.66, $\delta$=+35:45:52.2). As can be seen in
the inset in Fig. \ref{05358_KH2}, these objects are not point
sources in our images and are resolved into multiple components.
Both these objects, which are at  similar reddening and are much more
reddened compared to ``A'', exhibit IR excess in our $JHK$ colour-colour
plot  (Fig. \ref{JHKcol}).  \citet{porras00} presented
$JHK$ colour-colour and colour-magnitude diagrams and concluded that the
NE cluster is much more reddened and is much younger than the central
cluster.  This is very much consistent with the difference in the
reddening that we see between ``A'', and ``B'' and ``C''.   The redder
of the two bright MSX sources is located between ``B'' and ``C''.

Several H$_2$ emission features are seen associated with the NE cluster
which provide evidence for multiple collimated outflows from this
region.  The feature labelled ``6'' is the southern bow shock of the
main outflow (in the direction of the dotted arrow labelled ``II'')
oriented NS in the figures of \citet{beuther02a} and \citet{kumar02}.
The northern bow shock is outside our field.  ``1'' and ``2'' are
probably the bow shocks from a bipolar outflow in the direction ``I''.  
``1'' appears double lobed and could be produced by more than one 
outflow, which has to be verified through velocity studies.  
``3'' also appears to be part of a different outflow.  There are 
two other features labelled ``4'' and ``5'',  which could be 
produced by a different outflow from an obscured source.

\citet{beuther02b} detected four 1.2\,mm sources in this region.  All
four are detected at 850\,$\mu$m and three of these are detected
at 450\,$\mu$m by  \citet{minier05}; they label the 1.2\,mm sources
as ``MM1--MM4''.   The location of the brightest mm peak,  ``MM1'', is
between our sources ``B'' and ``C''  and is probably the combined
emission from both of these.  (High-angular-resolution observations of
\citet{beuther02a} at 2.6\,mm in fact resolve ``MM1'' into three
components,  labelled ``mm1--mm3'' on Fig. \ref{05358_KH2}).  Modeling
by  \citet{minier05} shows that ``MM1'' is the most massive of the
four clumps detected here in the mm.  ``MM2'' is NW of ``B'' and ``C''
and is not detected in $K$.  ``MM3'' is outside our field.  ``MM4''
is close to ``A'', but is offset from it by 11.9\,arcsec towards
the SE.  This offset is comparable to a beamsize of 11\,arcsec
of the the 1.2\,mm observations of \citet{beuther02b}
(Table \ref{resolutions}).  It has to be noted that the 21-$\mu$m
MSX  emission plotted by \citet{minier05} appears to be peaking at
the locations of ``B'' and ``A''.  From our observations, it
appears that ``A'' drives an outflow in the direction of ``III'', 
which is detected in H$_2$.
It is possibly the near IR counterpart of the MSX and IRAS sources.
It needs to be verified through millimetre observations at high
angular resolution and positional accuracy if ``MM4'' is the mm
counter part of the detections here at lower wavelengths or is a
younger and more embedded source.  The near-IR source
``D'' ($\alpha$=05:39:09.19, $\delta$=+35:44:22.6) also appears
to be associated with an outflow.  An H$_2$ knot (``9'') is
detected close to it. ``D'' has a weak MSX detection.  It is well
detected by 2MASS and exhibits mild extinction and excess.

\vskip 10mm
\subsection{IRAS~05373+2349 - {\it Mol 12}\\ ({\small \it d = 1.17\,kpc, L = 0.47$\times$10$^3$\,L$_{\odot}$})}

IRAS~05373+2349 is one of the luminous YSOs selected by \citet{molinari96}
and has been extensively studied by that group.  This source is associated
with ammonia emission from gas with a kinetic temperature  of 21.2\,K
\citep{molinari96} and there is no detection of radio emission at 2\,cm or 6\,cm
\citep{molinari98}, from which it is concluded that no H{\sc ii} region has
developed and that the source is in the younger population of their sources
selected using IRAS colours. More recent, more sensitive, radio and
millimetre observations \citep{molinari02} detected IRAS~05373+2349 at
3.6\,cm and 3.4\,mm respectively.  The emission derives from the vicinity
of one of the two cores resolved in their HCO$^+$ observations. The radio
emission is faint (0.70$\pm$0.04\,mJy at 3.6\,cm) and is thought to arise
from an ionized wind. Searches for 6.7\,GHz methanol maser emission by
\citet{szymczak00} found none, even though water maser emission was
detected by \citet{palla91}. Sub-mm and mm observations at 0.35\,mm,
0.45\,mm, 0.8\,mm, 1.1\,mm, 1.3\,mm and 2.0\,mm were carried out by
\citet{molinari00}.  From a fit to the SED, a mass of 8.8\,M$_{\odot}$ is
derived for IRAS~05373+2349 and an H$_2$ column density of
5.9$\times 10^{22}{\rm cm}^{-2}$. This source has also been detected in
many molecular species: C$^{34}$S, HCN (1-0), $^{13}$CO~(2-1) and
CH$_3$CN~(8-7) \citep{cesaroni99b} and SiO \citep{harju98}.
\citet{zhang05} mapped a bipolar outflow from this object, oriented
roughly in the NE--SW direction.

\begin{figure}
\centering
\includegraphics[width=8.10cm,clip]{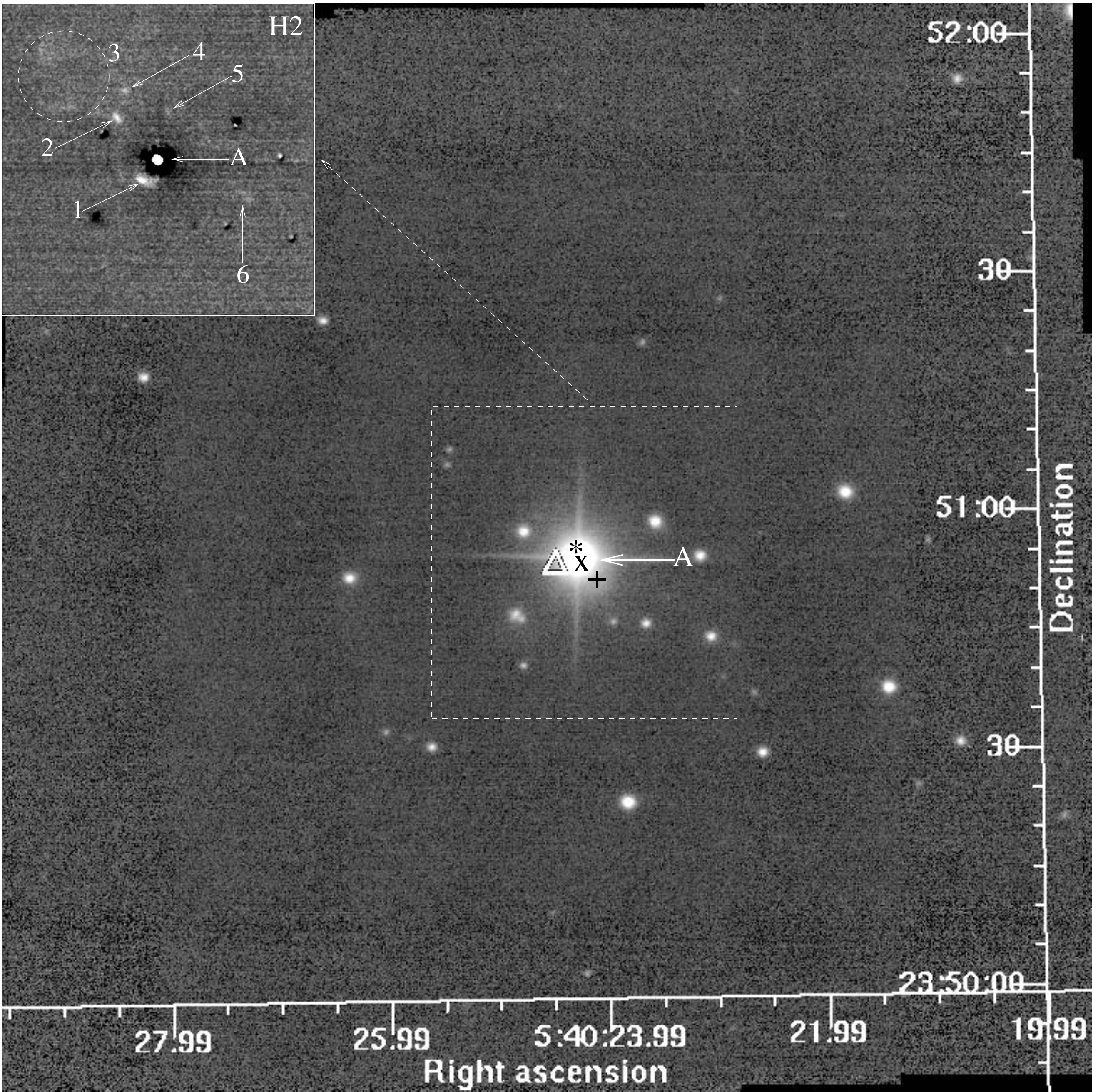}
\caption{$K$-band image of IRAS~05373+2349.  ``*'' shows  the location
of the 3.4-mm continuum peak and ``x'' shows the location of the 3.6-cm
continuum peak detected by \citet{molinari02}.  The inset shows the
central region of the continuum-subtracted H$_2$ image revealing the
faint line emission.}
\label{05373_KH2}
\end{figure}

Fig. \ref{05373_KH2} shows our $K$-band image of IRAS~05373+2349.
The central region of the continuum-subtracted H$_2$ image is
shown in the inset.  The field was not observed in the Br$\gamma$
filter.  A bright source, labelled ``A''
($\alpha$=5:40:24.33, $\delta$=23:50:54.6), is seen 2.8\,arcsec
NW of the IRAS position.  The MSX mission detected an object
2.4\,arcsec SW of ``A''.  The 2MASS colours of ``A'' exhibit IR
excess and very high reddening (Fig. \ref{JHKcol}), which is
typical of HMYSOs.  It is only weakly detected by 2MASS in $J$.
The continuum-subtracted H$_2$ image shows several faint emission
features around ``A'', the prominent ones of which are labelled
``1--6''. At the sensitivity of our imaging, these faint features
do not make a conclusive case for a well collimated bipolar 
outflow. However, the NE-SW alignment of the line emission 
features centred on ``A'' is suggestive of a bipolar outflow 
in that direction with a collimation factor of $\sim$2.55.  
Deeper integration is required for a better understanding of 
the H$_2$ emission in this region.  It is noteworthy that the 
H$_2$ knots are roughly aligned in the direction of the outflow 
mapped in CO by \citet{zhang05}.

The observations show that ``A'' is likely to be the luminous 
YSO in this region.  The 2.4\,mm peak of \citet{molinari02} 
is within 1.6\,arcsec of ``A''.  Their faint 3.6-cm radio 
continuum source nearly coincides with ``A''. The presence of 
the weak radio emission and the weakness of the outflow suggest 
that it has probably evolved out of the very early stages and 
is approaching the UCH{\sc{ii}} stage.

\subsection{IRAS~05490+2658\\ ({\small \it d = 2.1\,kpc, L = 3.16; 4.2$\times$10$^3$\,L$_{\odot}$})}

IRAS~05490+2658 is located $\sim$5\,arcmin east of the H{\sc ii}
region S242.  \citet{snell90} detected an outflow in CO. In their
low-resolution map, the blue- and red-shifted lobes of the outflow
were found pronounced towards the NW of the IRAS position with the peak
located $\sim$1 arcmin NW and nearly overlapping.  This high-velocity
emission seems to be coincident with an embedded cluster seen only at
near-IR wavelengths \citep{carpenter93}, which is separate from a
less-embedded cluster coincident with the IRAS position.  In the
1.2-mm maps of \citet{beuther02b}, knotty cloud cores are observed
towards these two clusters, which are embedded in a more diffuse
envelope that spreads over a 2$\times$3\,arcmin$^2$ field.
\citet{sridharan02} did not detect H$_2$O and CH$_3$OH masers
from this source.  From the IRAS HiRes, MSX and 1.2-mm photometry
they estimate a total luminosity for the region of
$\sim$10$^{3.5}$\,L${_\odot}$ at a distance of 2.1~kpc.

Fig. \ref{05490_KH2} shows our $K$ and H$_2$ images.  The 2MASS
colours of the bright star,
``A'' ($\alpha$=05:52:13.83, $\delta$=+26:59:40.8), place it
within the reddening band and it does not exhibit any IR  excess
(Fig. \ref{JHKcol}).  The IRAS position is $\sim$15 arcsec SW
of ``A''.  MSX detected an object within five arcsec of the IRAS
position and one of the two 1.2-mm continuum peaks observed by
\citet{beuther02b} is 7.8\,arcsec west and 4.4\,arcsec north of the
IRAS position. The positional offsets of the IRAS, MSX and mm
detections are in a similar direction from ``A'', confirming that
the YSO is a deeply embedded object and that ``A'' is just a
foreground star.  There are two faint IR sources close to the mm
position.  Accurate photometry of  the field is required to
investigate if any  of them could be the near-IR counterpart of
the YSO detected at longer wavelengths.

Most of the nebulosity in the field disappears upon continuum
subtraction.  The continuum-subtracted H$_2$ image shows some
emission features which are circled and labelled ``1'' and ``2''
in  Fig. \ref{05490_KH2}.  ``1'' could be produced by an outflow
from one or more sources near the centre of the image.  Our
Br$\gamma$ image does not show any significant emission from
the outflow.

\begin{figure*}
\centering
\includegraphics[width=16.5cm,clip]{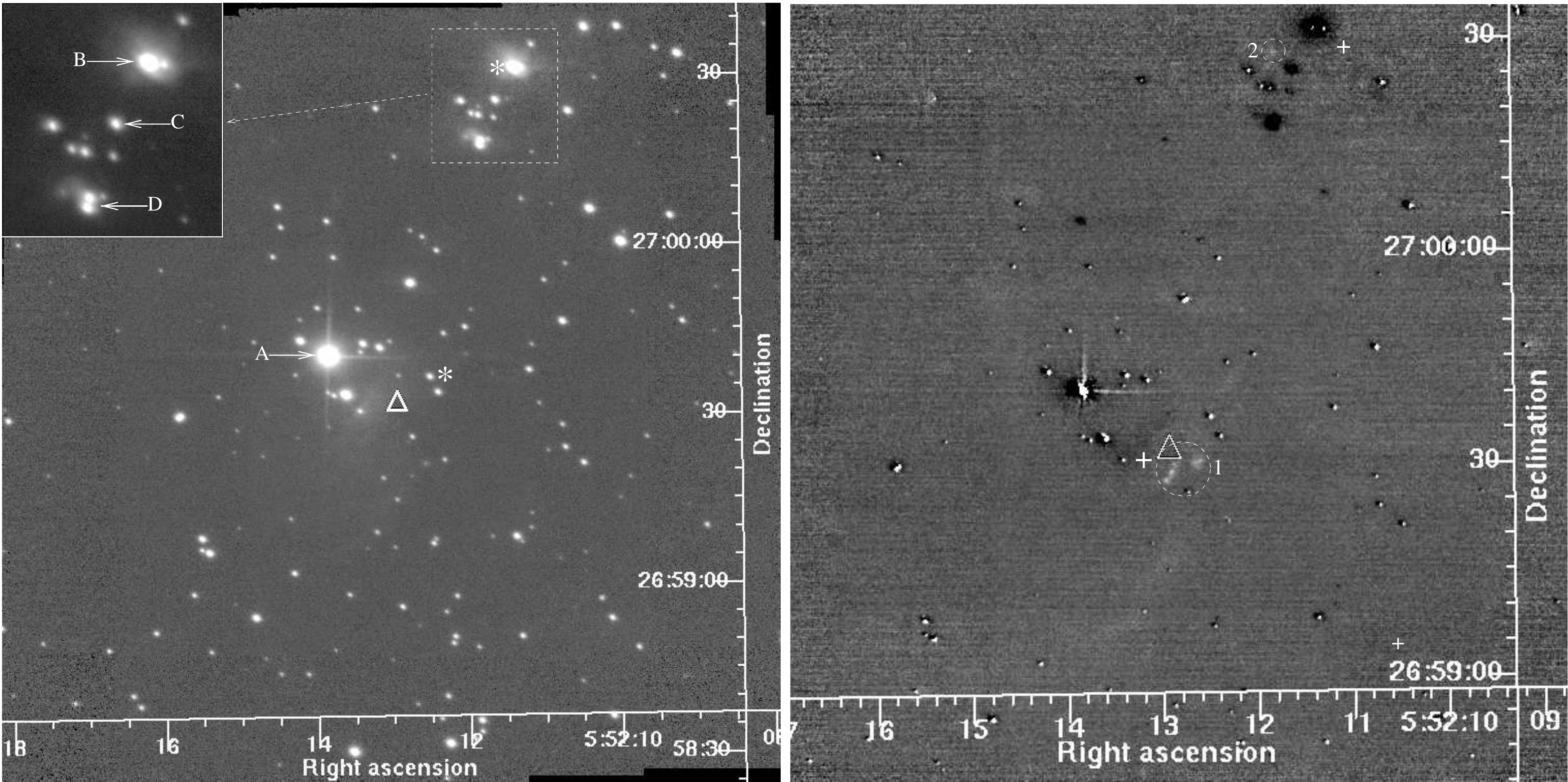}
\caption{The left panel shows the $K$-band image of IRAS~05490+2658.
The two 1.2-mm peaks detected by \citet{beuther02b} are shown by
``*''.  The right panel shows the
continuum-subtracted H$_2$ image.}
\label{05490_KH2}
\end{figure*}

There is a set of reddened sources (shown in the inset) in the region,
located nearly one arcmin NW of the IRAS position.  This is presumably
a second cluster, NW of the cluster associated with the IRAS position,
where the CO outflow discovered by \citet{snell90} is centred.
Three sources are labelled in this cluster; all are plotted in our
colour-colour diagram (Fig. \ref{JHKcol}).  The locations of ``A--D''
on the colour-colour diagram show that the cluster containing
``B--D'' is at a much larger extinction when compared to ``A''.
``B'' ($\alpha$=05:52:11.34, $\delta$=+27:00:31.5) is located at
the early-spectral-type limit of the reddening band.  Hence it shows
excess unless it is of very early spectral type.  It is surrounded
by nebulosity in our $K$-band image and is highly reddened. ``B''
has a fainter neighbour at a separation of $\sim$1.6\,arcsec.
``C'' ($\alpha$=05:52:11.59, $\delta$=+27:00:25.7) is also
reddened and shows IR excess.
``D'' ($\alpha$=05:52:11.79, $\delta$=+27:00:18.4) is embedded in
nebulosity in $K$.  It is resolved into a set of four objects with
two equally bright components at a separation of $\sim$1 arcsec
and two fainter components located on either sides of the pair.
The 2MASS colours of ``D'' (aggregate of the four objects) show IR
excess and its reddening is similar to that of ``B''.
The NW outflow seen in the CO maps of \citet{snell90}
is most likely to be produced by one or all of ``B, C, D''.
The brighter of the two 1.2-mm sources imaged by \citet{beuther02b}
is located 17.3\,arcsec west and 58.8\,arcsec north of the IRAS
position and is only $\sim$2.9\,arcsec east of  the brighter
component of ``B''. MSX observations reveal a second far-IR
source four arcseconds SW of ``B'' which appears only at longer
wavelengths.  The location of the MSX source agrees with that of
the mm source within the positional accuracy of MSX.
Hence, it is likely that ``B'' or the fainter star near it is the
YSO detected at longer wavelengths and is driving the CO outflow.
The lack of any 3.6-cm emission above the 1-mJy detection limit of
\citet{sridharan02} also suggests that the YSOs here are
very young.

\subsection{IRAS~05553+1631 - G192.16-3.82 -  {\it Mol 14}\\ ({\small \it d = 2.5; 3.04\,kpc, L = 6.31; 11.7$\times$10$^3$\,L$_{\odot}$})}

A massive core was mapped in this region in CS by
\citet{beuther02b}, \citet{bronfman96} and \citet{carpenter95}.
Carpenter et al. observed the region in the near-IR $JHK$ bands also.
The core was detected in NH$_3$ emission by \citet{molinari96}
and in 1.2-mm continuum emission by \citet{beuther02b}.  Observations
by  \citet{macleod98a} and \citet{sridharan02} did not reveal any
CH$_3$OH maser emission.  Multi-component H$_2$O maser emission was
detected at a range of velocities by several investigators
\citep{palla91,brand94, wouterloot93, macleod98a, shepherd99, shepherd04a}.

\begin{figure*}
\centering
\includegraphics[width=16.5cm,clip]{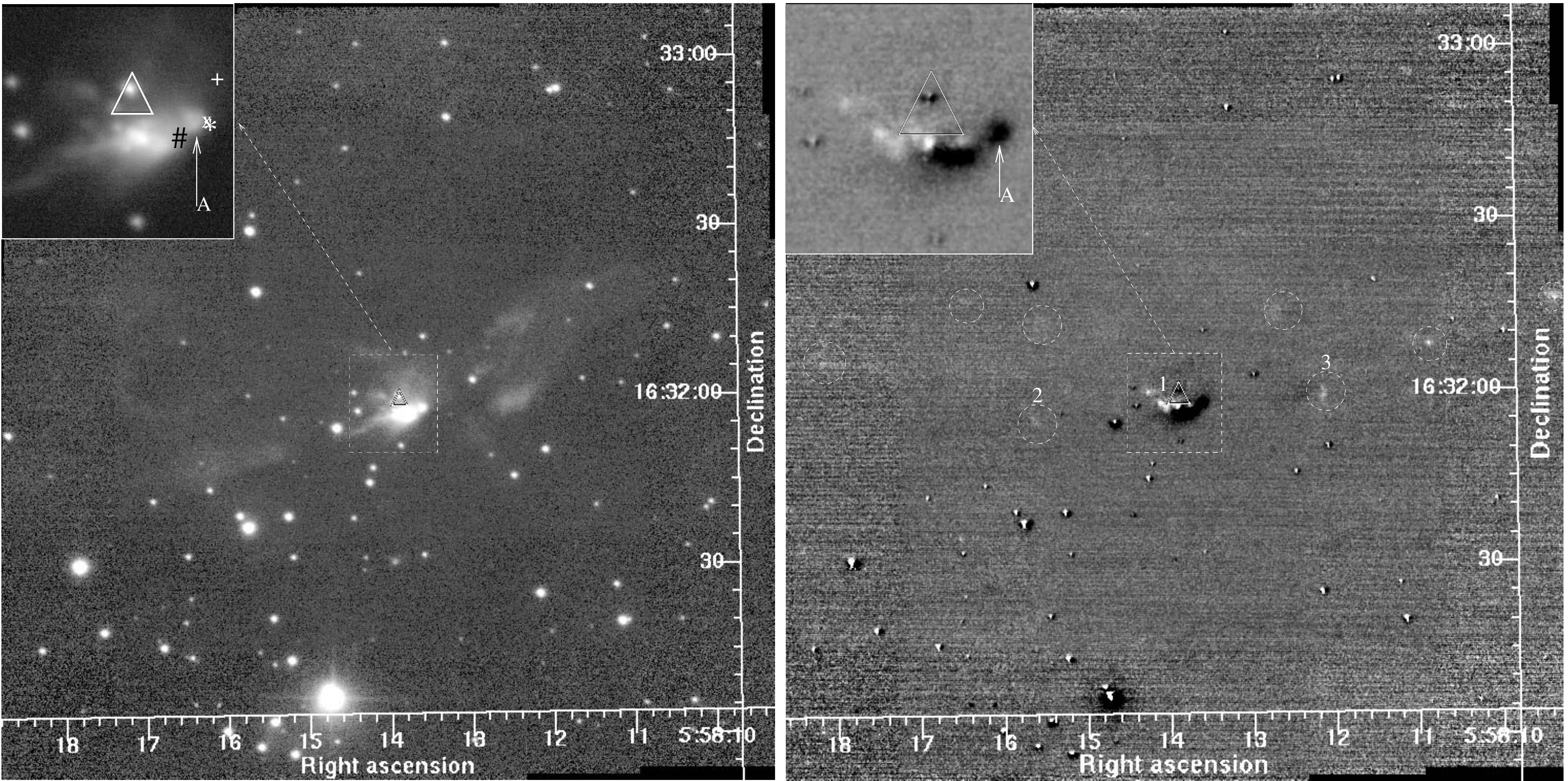}
\caption{The left panel shows our $K$-band image of IRAS~05553+1631.
The 2.7-mm continuum peak of \citet{shepherd98} is shown by ``*''
and the 1.2-mm continuum peak of \citet{beuther02b} is shown by ``\#''.
``x'' shows the 1.3-cm radio continuum position of \citet{shepherd04a}.
The right panel shows the continuum-subtracted H$_2$ image.  An
expanded view of the region enclosed in the box is shown in the inset.}
\label{05553_KH2}
\end{figure*}

CO line wing emission is detected towards this region by a
number of investigators \citep{wb89, sridharan02}.  The distance
estimated by \citet{wb89} (1.17\,kpc), and therefore the
luminosity, are somewhat lower than that derived by
\citet{molinari96} from the NH$_3$ observations.  \citet{snell90}
detected a bipolar outflow in CO with a spatial extent of
3\,pc (4\,arcmin).  OVRO observations by \citet{shepherd98}
at the wavelength of CO (J=1-0) line also revealed a well
defined bipolar outflow in this field, roughly in the EW
direction, originating from a B2-B3 star.  They also detected
shocked emission in the H$_2$ 2.122-$\mu$m line.

The position of the MSX, mm and cm sources all agree to
within 3.5\,arcsec of each other.  The 6-cm observations
of \citet{hughes93} showed emission with multiple structure
(with an integrated flux density of 1.55\,mJy), which was
interpreted by them as due to a number of B3 stars.
\citet{macleod98a} proposed a spectral type of B2.7 for the
exciting source. 3.6-cm observations by \citet{shepherd99}
revealed a total integrated flux density of 1.51\,mJy with
a central source of integrated flux density of 1.1\,mJy.
The source has also been detected at 3.6\,cm by
\citet {sridharan02} at 1.3\,mJy; they do not detect any H$_2$O
or CH$_3$OH maser from this region.  Sridharan et al. detected
emission in a number of molecular lines - SiO, CH$_3$OH and
CH$_3$CN - from this region.  From the spectral index derived
from their 3.6-cm radio emission along with the 6-cm emission
measured by \citet{hughes93}, \citet{shepherd99} proposed an
ionized jet. They suggested that the outflow from this source
is driven by the thermal jet and a wide angle wind, 
contributing $\sim$25\% and $\sim$75\% respectively to the
momentum.  From the observed radio fluxes, \citet{shepherd99}
and \citet{shepherd04a} derive a B2 spectral type for the
underlying protostar.  Williams, Fuller \& Sridharan (2005)
also estimated an early-B spectral type for the exciting source.

The locations and velocities of the 22-GHz H$_2$O masers
trace a 1000\,AU Keplerian disc detected at mm wavelengths
and an ionized jet detected at four wavelengths from 1.3 to 6\,cm
\citep{shepherd99, shepherd04a}.  Nevertheless, the high angular
resolution observations of \citet{shepherd04a} did not strongly
support or deny the disc origin of the masers.  The 7-mm
continuum observations of \citet{shepherd01}, at an angular
resolution of 40\,mas, detected an inner 130-AU accretion disc.
The orientation of the discs is roughly perpendicular
to the large scale molecular outflow mapped by \citet{shepherd98}.

Fig. \ref{05553_KH2} shows our $K$-band and continuum-subtracted
H$_2$ images.  The H$_2$ image has been smoothed with a 2-pixel 
FWHM Gaussian to enhance the appearance of the faint emission 
features.  The outflow source itself appears to be highly obscured.
In $K$ we see an extended nebulosity in the NW-SE direction, most
of which disappears in our continuum-subtracted H$_2$ image except for
a few emission features. The $K$-band image shows a faint point source
(``A'';  $\alpha$=5:38:13.59; $\delta$=+16:31:58.05), from which
a jet appears to emanate. The position of ``A'' agrees well with the
mm and the cm positions given by \citet{shepherd04a}. The 1.2-mm
continuum peak imaged by \citet{beuther02b} is 1.24\,arcsec SE from
``A'' and the 1.3-cm continuum source detected by \citet{shepherd04a}
is only 0.85\,arcsec from ``A''. The 2.7-mm continum peak of
\citet{shepherd98} is only 1\,arcsec away from ``A'' and it nearly
coincides with the 1.3-cm source. The IRAS position is $\sim$4.3\,arcsec
NE and the MSX position is $\sim$3.2\,arcsec NW of ``A''.  Probably
we are detecting the heated dust surrounding the YSO, rather than
the source itself in the near-IR.  The large negative residuals
on the continuum-subtracted H$_2$ and Br$\gamma$ images indicate 
the presence of a large amount of dust in the vicinity.  Our 
continuum-subtracted H$_2$ image shows a collimated set of knots 
(``1'') within 9.5\,arcsec, mostly located eastward, of the central 
source  ``A'' (Fig. \ref{05553_KH2}  right: inset).  In addition 
to these aligned emission knots close to the source, there are 
many other very faint emission features circled on the figure. 
These features are observed to be aligned roughly in the EW 
direction.  The H$_2$ knots (``1'') located close to ``A'',  and 
two of the other knots at $\sim$29.6\,arc SE (``2'') and 
21\,arcsec NW (``3'') from ``A'' were also detected by 
\citet{indebetouw03} (their Fig. 2).  Considering all the H$_2$
emission features  circled on Fig. \ref{05553_KH2}, the outflow has
a collimation factor of $\sim$1.9 and has a position angle of
$\sim$84$^{\circ}$ in the east and $\sim$287$^{\circ}$ in the west.

The 2MASS images do not appear to resolve the point source ``A'' well
from the nebulosity to the east. The 2MASS source is offset
2.7\,arcsec SE of ``A'',  is centred on the nebulosity, and shows
excess (Fig. \ref{JHKcol}).  Source ``A'' is well detected
in the $K$-band images of \citet{devine99} and \citet{indebetouw03}
and its coordinates derived by them agree well with ours.  As is
obvious from Fig. \ref{05553_KH2}, this position is very close to
the IRAS position and we identify ``A'' as the embedded YSO.

The direction of the outflow derived from CO maps \citep{shepherd98,
snell90} is consistent with the roughly EW alignment of the H$_2$ emission
features detected in our image (Fig. \ref{05553_KH2} - right panel).
The position angle of 84$^{\circ}$ for the outflow derived from our
H$_2$ observations is similar to 80$^{\circ}$ derived by \citet{shepherd98}
from their CO map.   The bright  H$_2$ line emission knots detected
by us close to ``A'' in the east are in the direction of the blushifted
lobe of the CO outflow.

Overall, IRAS~05553+1631 appears to be a luminous YSO driving outflow.
From the wide opening angle defined for the CO outflow, the H$_2$ line
emission knots and the ionized radio jet, \citet{shepherd98} and
\citet{shepherd99} proposed that the outflow is powered by a wide-angled
wind and a jet.  Eventhough the large scale distribution of the H$_2$
knots in Fig. \ref{05553_KH2} implies a wide angle, the arrangement of the
H$_2$ knots in the vicinity of ``A'' is suggestive of a well-collimated
jet.  From the current observations it is not clear if more than one YSO
is present.

\vskip 10mm
\subsection{IRAS~06061+2151 - {\it Mol 16}\\ ({\small \it d = 0.1; 2.0\,kpc, L = 0.0278; 4.0$\times$10$^3$\,L$_{\odot}$})}

\begin{figure*}
\centering
\includegraphics[width=16.5cm,clip]{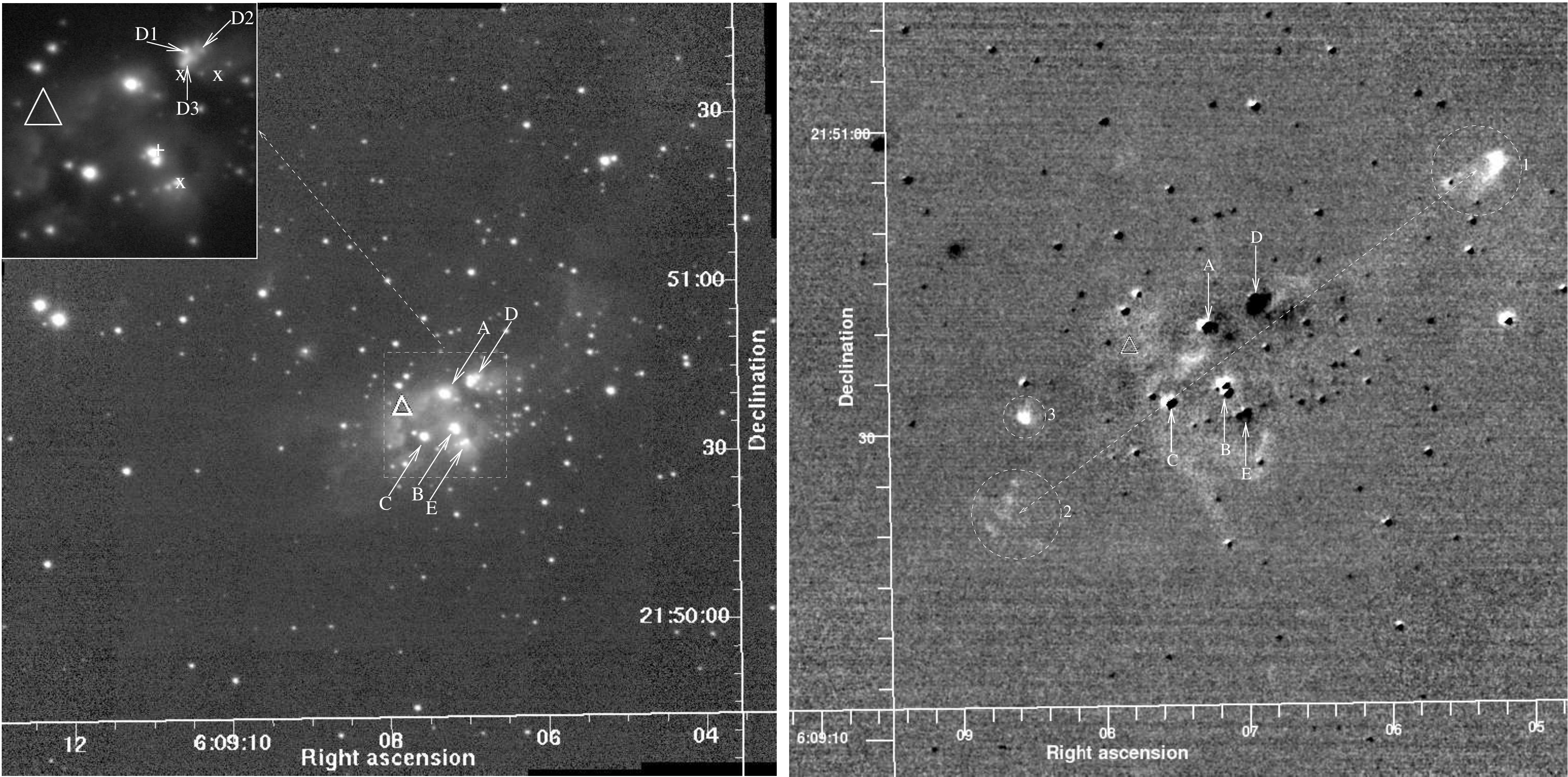}
\caption{Left: $K$-band image of IRAS~06061+2151.
The three radio positions listed in \citet{kurtz94} are shown by ``x''.
Right: central region of the continuum-subtracted H$_2$ image.}
\label{06061_KH2}
\end{figure*}

The dense core associated with IRAS~06061+2151 has been mapped in
NH$_3$ (\citealt{molinari96}; Jijina, Myers \& Adams 1999) and in
CS (J=2-1) (\citealt{bronfman96}; Carpenter, Snell \& Schloerb 1995).
The estimate of the distance towards this source (and thereby, the
luminosity) has considerable uncertainty.  \citet{molinari96} determined
a kinematic distance of 0.1\,kpc and derived a FIR luminosity of
27.8\,L$_{\odot}$, whereas \citet{carpenter95} assigned a membership
in the Gemini OB1 molecular cloud complex and derived a far-IR
luminosity of 4000\,L$_{\odot}$ at a distance of $\sim$2\,kpc.
Powerful H$_2$O maser emission has been detected from this region where
multiple (up to 6) maser spots were mapped at different radial velocities
\citep{brand94, palla91}. \citet{schutte93} did not detect any CH$_3$OH
maser,  whereas the detections by \citet{galt04} (observed in 1994) and
\citet{szymczak00} differ in intensity and radial velocity, showing the
variable nature of the CH$_3$OH maser emission from this region over a
period of a few years.  \citet{wb89} detected $^{12}$CO (J=1-0) with a
red wing and derived a kinematic distance less than 0.1\,kpc and a
FIR luminosity of 27 L$_{\odot}$.  \citet{shepherd96a} also detected
$^{12}$CO (J=1-0) emission.  Faint (0.6--3.4\,mJy) 3.6-cm continuum
emission was detected by Kurtz, Churchwell \& Wood (1994) from three
locations within 15\,arcsec of the IRAS position.  The source was also
detected at 1.3\,mm with a flux density of 4.2\,Jy by  \citet{chini86}.
Near-IR H$_2$ images of this object were recently published by
\citet{rao04}.  They proposed that the cluster hosts multiple YSOs,
the brightest being of early-B spectral type.  They also deteced two
knots of H$_2$ emission.

Fig. \ref{06061_KH2} shows our $K$-band and continuum-subtracted
H$_2$ images.  No extended Br$\gamma$ emission was detected,  so
this image is not shown.  Our $K$-band image unveils a rich cluster
embedded in nebulosity.  A major part of this nebulosity disappears
in the continuum-subtracted H$_2$ image.  Of the H$_2$ emission
features, the prominent ones are the NW and the SE knots detected
by \citet{rao04}, the locations of which are
``1'' ($\alpha$=6:9:5.22, $\delta$=21:50:6) and
``2'' ($\alpha$=6:9:8.69, $\delta$=21:50:22.8) respectively.
``2'' is resolved  into at least four components. There is another
feature labelled ``3'' ($\alpha$=6:9:8.55, $\delta$=21:50:31.7) 
located in the SE direction,  close to ``2''. The NW knot (``1'') 
is a bow shock.  The main jet,  composed of at least ``1'' and 
``2'', appears to be well collimated. These two features together 
give an outflow angle of 128$^{\circ}$ and a collimation factor 
of 4.3.  The collimation factor would be higher if the multiple
bow shocks seen in ``2'' are due to precession or multiple jets.  
The five prominent sources detected by \citet{rao04} are 
labelled on Fig. \ref{06061_KH2}, with  ``A--E'' representing 
the stars ``1--5'' discussed by them.  At our high spatial 
resolution, all five sources are resolved into multiple 
components as seen in the inset. 2MASS colours of
``A'' ($\alpha$=06:09:07.28, $\delta$=21:50:40.4) and
``B'' (centroid:  $\alpha$=06:09:07.14, $\delta$=21:50:34.2)
do not exhibit any excess. However, it should be noted that
``A'' has faint companions and ``B'' is composed of more than 
one object. Both ``A'' and ``B'' are poorly  detected by 2MASS 
in $K_s$. ``D'' (centroid: $\alpha$=6:9:6.91, $\delta$=21:50:42.8) 
appears to be the driving source of the main outflow (represented 
by ``1'' and ``2'') in this region.  It is resolved into multiple
components, three of which are labelled on the figure.  ``D3''
appears to be an extended feature directed away from ``D1/D2''
in the SE direction, but this feature gives strong negative
residuals upon continuum subtraction implying that it is probably
dominated by emission from dust.  The IRAS position is 13.5\,arcsec
SE of ``D'' and 10.2\,arcsec NE of ``B''.  The location of the
MSX source is  8\,arcsec SE of ``D'' and nearly coincides with
``B''.  It is possible that more than one source is contributing
to the IRAS and MSX fluxes.  ``D'' shows large excess in its
2MASS colours (not detected in $J$). Near-IR spectroscopy of
Hanson, Luhman \& Rieke (2002) detected 2.122-$\mu$m H$_2$
line emission from ``D'' or its vicinity.  2MASS detection near
``E'' only in $K$ and the coordinates centred off the
source that we label here.  So ``E'' is not shown in 
Fig. \ref{JHKcol}. Also obvious in Fig. \ref{06061_KH2} is
the line emission in the H$_2$ from within the cluster, many
of which appear filamentary and arc-like. Some of these could be
due to jets from other sources within the cluster or fluorescence
due to the emission from more evolved sources in the cluster.
Deeper imaging in H$_2$ is required to properly map these.
Overall, this cluster appears to host several YSOs.

\subsection{IRAS~06584-0852 {\it -- Mol 28}\\ ({\small \it d = 4.48\,kpc, L = 5.67$\times$10$^3$\,L$_{\odot}$})}

\begin{figure}
\centering
\includegraphics[width=8.10cm,clip]{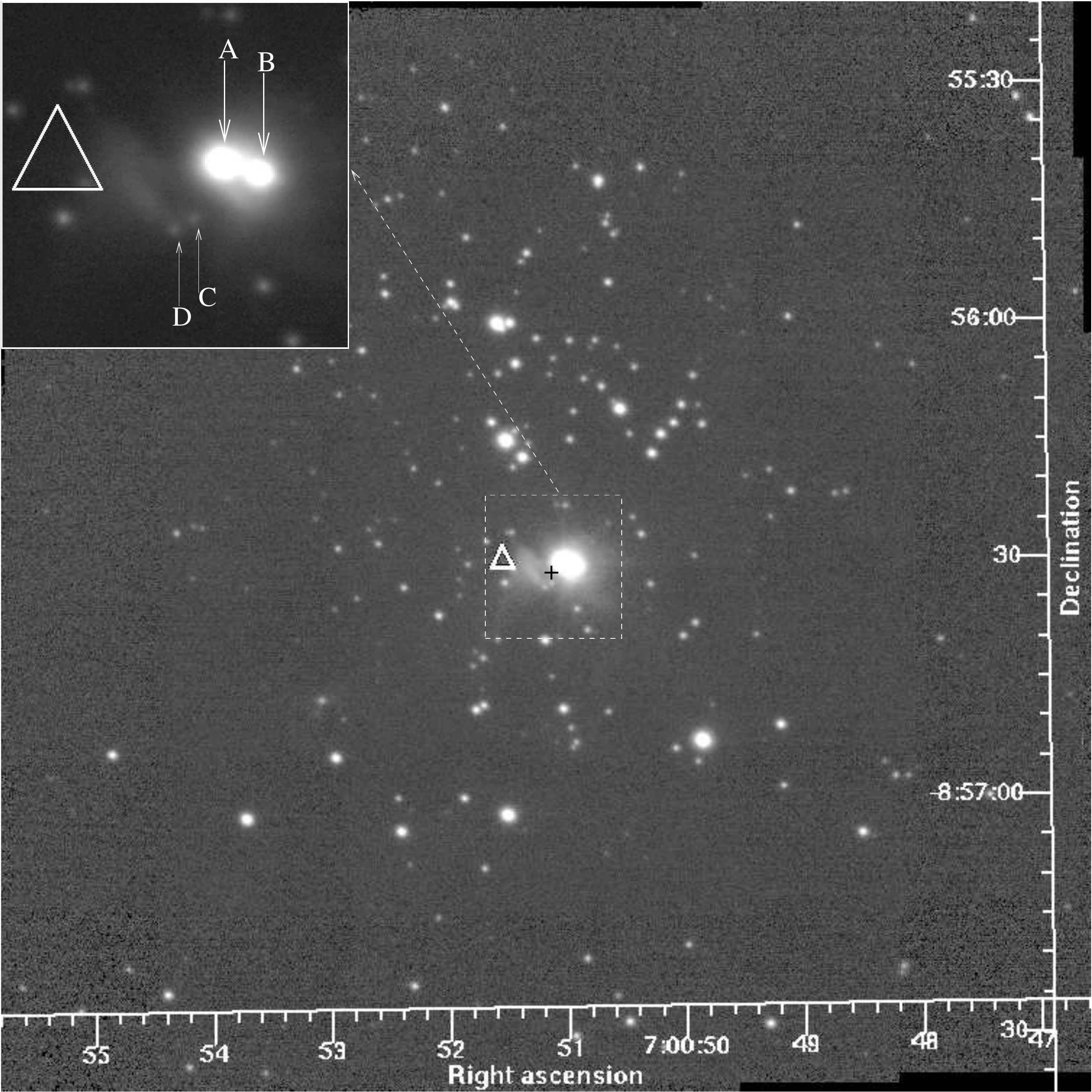}
\caption{$K$-band image of IRAS~06584-0852.  The inset shows
an expanded view of the central region showing the nebulosity
and the near-IR sources well resolved.}
\label{06584_K}
\end{figure}

\citet{molinari96} detected a  dense core at this location in
NH$_{3}$ emission.  They determined a kinematic distance of 4.48\,kpc
and derived a far-IR luminosity of 9.08 $\times 10^{3}$ L$_{\odot}$.
\citet{molinari00} obtained (sub)mm photometry and derived a bolometric
luminosity of 5.6 $\times 10^{3}$L$_\odot$ for the core.
\citet{ishii02} included IRAS~06584-0852 in their $JHK'$ survey of
clusters associated with luminous IRAS sources.  They detected a 
reflection nebula in this cluster.  The source was not detected by 
the VLA at 2 and 6\,cm \citep{molinari98}.  \citet{wb89}
detected red-shifted CO emission and \citet{zhang05} mapped a molecular
outflow in CO (J=1-0) line emission.    \citet{brand94} and
\citet{palla91} detected H$_2$O maser from this region.

Fig. {\ref{06584_K}} shows our $K$-band image of the region;  an
expanded view of the central region with higher contrast is shown in the
inset.  The near-IR image reveals a cluster of objects with two bright
stars (``A'' and ``B'') near the centre of the field separated by
$\sim$1.73\,\,arcsec, and located at
($\alpha$=07:00:51.01, $\delta$=-08:56:29.8)
and ($\alpha$=07:00:50.90, $\delta$=-08:56:30.3) respectively.
There are two stars located to the SE of this pair, that are much
fainter than ``A'' and ``B''.  These two objects located at
($\alpha$=7:00:51.10, $\delta$=-8:56:32.5 ) and
($\alpha$=7:00:51.16, $\delta$=-8:56:3) are labelled ``C'' and ``D''
respectively. ``D'' has a cometary nebula associated with it.
The continuum-subtracted H$_2$ and Br$\gamma$ images reveal no line
emission.  Hence these images are not shown here.  The bright central
pair is well detected by 2MASS, but the components are not resolved
in the 2MASS image. The colours of the two objects combined, as derived
from 2MASS photometry, exhibit large reddening and IR excess, placing
them in the region of reddened YSOs on the colour-colour diagram
(Fig. \ref{JHKcol}).  The IRAS source is located $\sim$8.5\,arcsec east
of the centroid of ``A'' and ``B'' and the MSX source is 2.3\,arcsec SE.
These observations show that ``A'' and ``B'' are probably the near-IR
counterparts of the YSO.  The source has probably past its very
early stage of formation.

\subsection{IRAS~18144-1723 -- \it{Mol 45}\\ ({\small \it d = 4.33\,kpc, L = 21.2$\times$10$^3$\,L$_{\odot}$})}

\begin{figure*}
\centering
\includegraphics[width=16.5cm,clip]{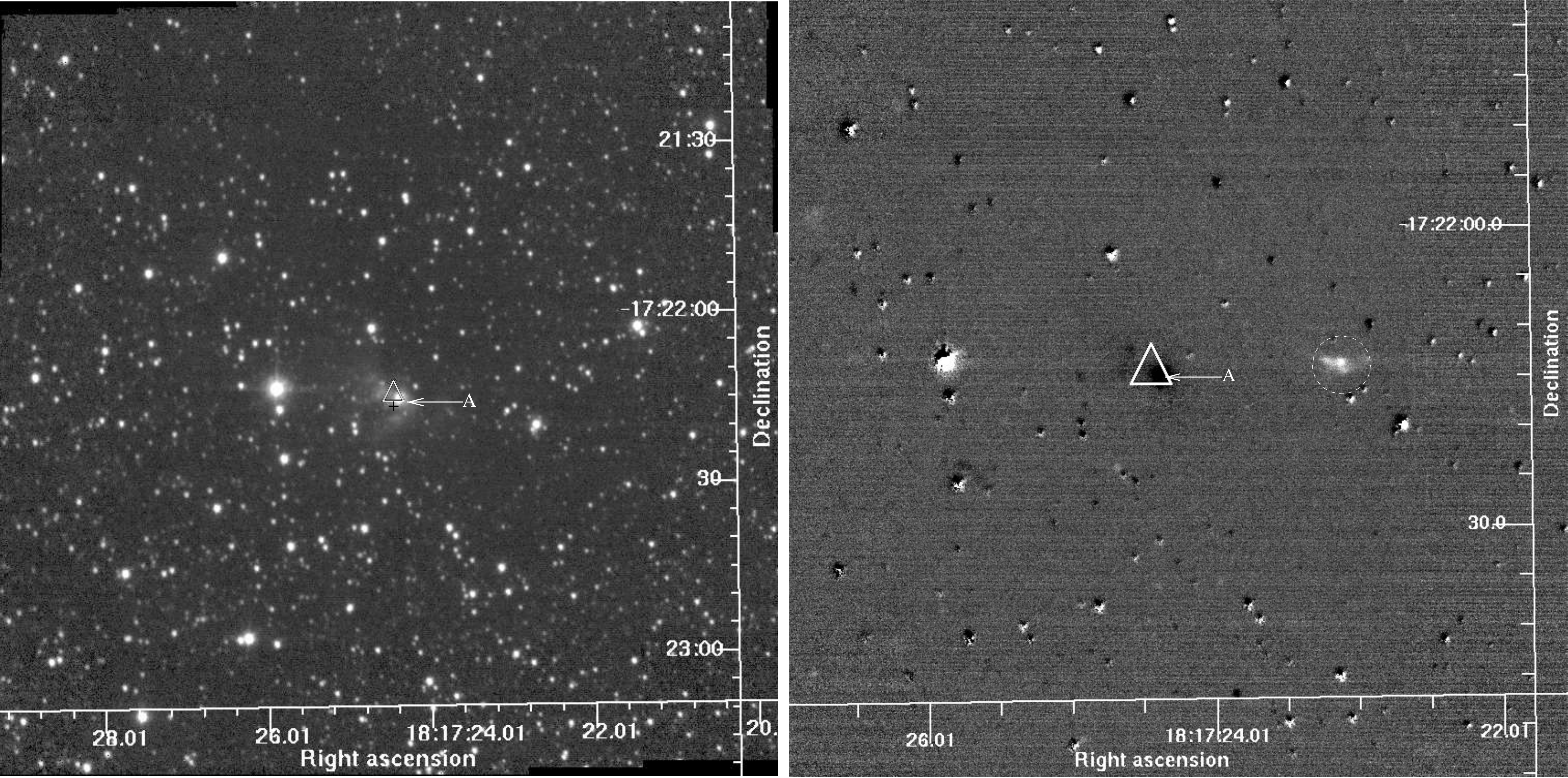}
\caption{Left: $K$-band image of IRAS~18144-1723.  Right:
The central portion of the continuum-subtracted H$_2$ image.}
\label{18144_KH2}
\end{figure*}

IRAS~18144-1723 is associated with an ammonia core \citep{molinari96} and
water and methanol masers \citep{palla91, szymczak00, kurtz04}.
\citet{molinari98} detected radio emission at 6\,cm from a region located
93\,arcsecs away from the location of the IRAS source and the masers.  Hence,
the radio emission does not appear to be associated with the IRAS source.
CO observations of this region by \citet{zhang05} did not reveal any outflow.

Fig. \ref{18144_KH2} shows our $K$-band and continuum-subtracted H$_2$ images
of IRAS~18144-1723.  There is a point source,
``A'' (at $\alpha$=18:17:24.38, $\delta$=-17:22:14.7), located within
1.3\,arcsec of the IRAS position.  This object was not detected in the 2MASS
$J$-band data, was only marginally detected in $H$, but is bright in $K_s$.
Our $K$-band image shows nebulosity around ``A'', most of which disappears
in the continuum-subtracted H$_2$ image.  The H$_2$ image reveals a
bow-shock like feature located at ($\alpha$=18:17:23.13, $\delta$=-17:22:13.5)
with an extent of over 2\,arcsec, mostly in the EW direction,  which is circled
in the figure.  If the  H$_2$ emission feature originates from ``A'',  the
outflow is at an angle of 274$^{\circ}$ with a collimation factor $\sim$10.
Notably, ``A'' is located within 0.4\,arcsec of the MSX position.  It exhibits
large infrared excess and is located in the region occupied by reddened YSOs
in the $JHK$ colour-colour diagram (Fig. \ref{JHKcol}).  ``A'' is the most
likely IR counterpart of the YSO producing the outflow in this region. The
region was not observed in Br$\gamma$.  The lack of strong radio emission
from ``A'' at 6\,cm \citep{molinari98} indicates that ``A''
is probably in a pre-UCH{\sc ii} stage.

\subsection{IRAS~18151-1208  -- {\it Mol 46}\\ ({\small \it d = 3.0\,kpc, L = 19.95$\times$10$^3$\,L$_{\odot}$})}

IRAS~18151-1208 is embedded in a high-density core detected in CS
and NH$_3$ emissions in the surveys carried out by \citet{bronfman96}
and \citet{molinari96} respectively. The IRAS source is bright 
and has FIR colours typical of UCH{\sc{ii}} regions
\citep{wc89a}.  However, the source was not detected at 3.6\,cm by
\citet{sridharan02}, which indicates the lack of a substantial H{\sc{ii}}
region.  Instead, the source may be in a pre-UCH{\sc{ii}} phase; its
double-peaked SED is typical of low-mass Class I YSOs, where the cold
dust component, peaking at $\sim$100\,$\mu$m, dominates over the warmer
dust component at $\sim$20\,$\mu$m. The bolometric luminosity of the
source, derived from IRAS HiRes photometry, is $\sim$20,000\,L$_\odot$
\citep{sridharan02}. The kinematic distance to
IRAS~18151-1208 is 3.0\,kpc \citep{brand93}.

A moderately collimated (collimation factor=2.1) molecular outflow 
was mapped in the $^{12}$CO (J=2-1) transition by \citet{beuther02c}.
\citet{beuther02d} observed methanol maser emission from this region,  
from a location very close to the luminous YSO that we identify
here.  They also detected faint H$_2$O maser emission.  However, the 
location of the water maser detected by them is far away from the 
YSO and is outside the field of view of our H$_2$ image.  Hence it is
not related to the YSO detected here.  The survey by \citet{palla91}
also failed to detect any H$_2$O maser emission from IRAS~18151-1208.

\begin{figure*}
\centering
\includegraphics[width=16.5cm,clip]{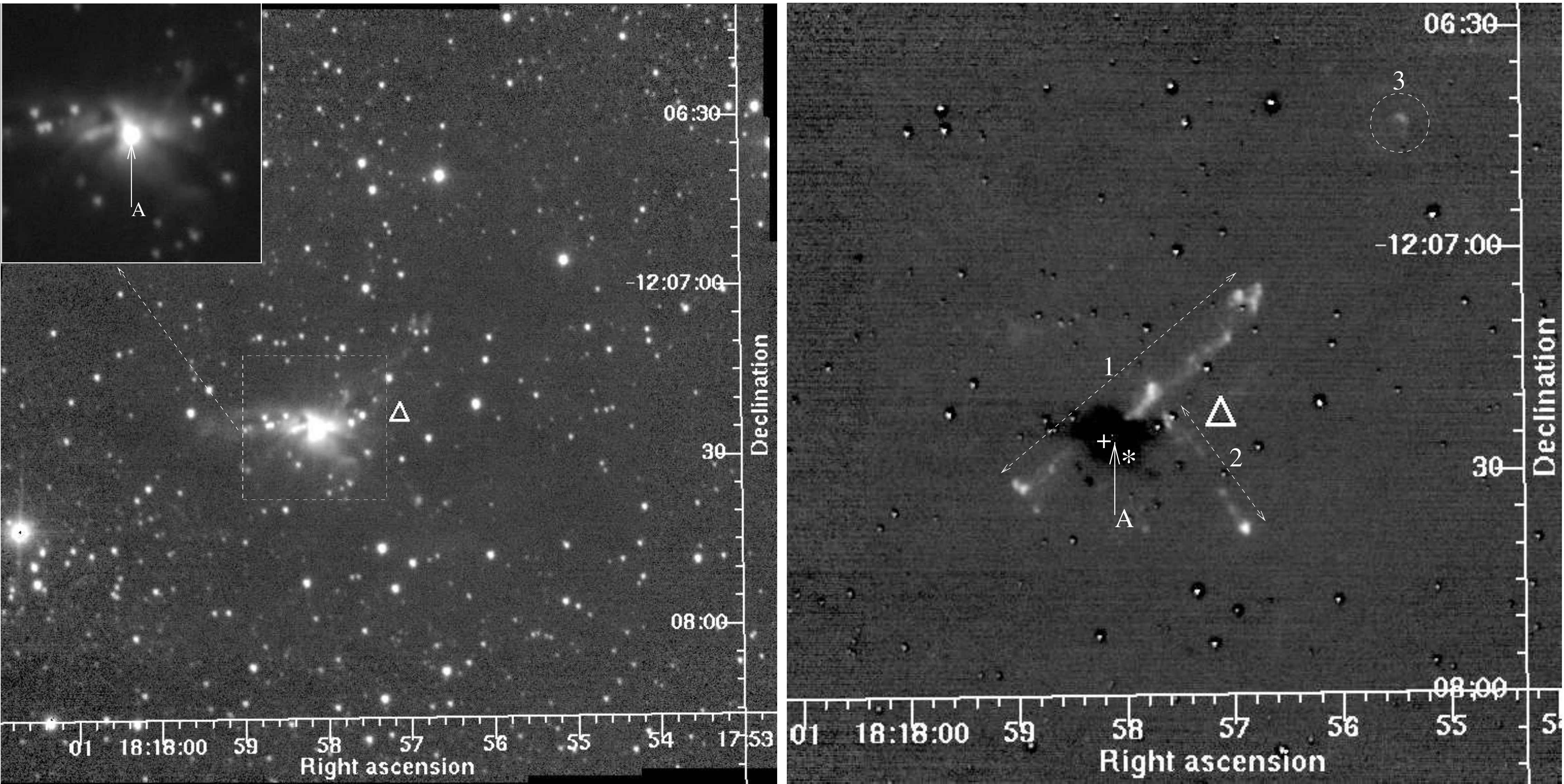}
\caption{The left panel shows the $K$-band image of IRAS~18151-1208
and the right panel shows the central portion of the
continuum-subtracted H$_2$ image.   The 1.2-mm continuum peak
observed by \citet{beuther02b} is shown by ``*''.  The H$_2$
image has been smoothed with a Gaussian of FWHM=2 pixels.}
\label{18151_KH2}
\end{figure*}

Our infrared images reveal a cluster of objects embedded in nebulosity
and two well-collimated outflows for the first time in this region.
Fig. \ref{18151_KH2} shows our $K$-band and continuum-subtracted
H$_2$ images. The central object labelled
``A'' ($\alpha$=18:17:58.12, $\delta$=-12:07:24.7) is deeply embedded.
2MASS gives only an upper limit in $J$; the $H-K$ colour shows
large reddening and excess, typical of a deeply embedded luminous YSO
(Fig. \ref{JHKcol}).  The H$_2$ image shows two jets, the directions
of which are shown by the two arrows on Fig. \ref{18151_KH2},
labelled ``1'' and ``2'' respectively. The brighter of the two
(``1'') is in the NW-SE direction,  at an angle 131$^\circ$ and has a
collimation factor of $\sim$6.3.  If we consider that the feature
circled and labelled ``3'' is  part of the same outflow, the
collimation factor could be as high as 12.8.  ``A'' is located at the
centroid of the NW-SE jet and is most probably the driving source
of the jet in the direction of ``1''.  The methanol maser detected by
\citet{beuther02d} appears to be associated with ``A''; it is located
only 0.75\,arcsec SW of ``A''.  This separation is within the 
positional accuracies quoted by Beuther et al. (Table \ref{resolutions}).
The 15.1 arcsec offset from the IRAS position could be due to the 
positional inaccuracy of the IRAS detection.  The MSX mission indeed 
detected a mid-IR source at
wavelengths above 8.28\,$\mu$m within 0.5\,arcsec of ``A'', and the
1.2-mm continuum peak observed by \citet{beuther02b} is only
2.21\,arcsec SW of ``A''.  The second outflow (``2'') appears to be
driven by another embedded source SW of ``A'' which is not detected
in the near IR.   This outflow is at an angle 35.5$^\circ$ and has
a collimation factor of $\sim$5.7.  A comparison of the directions
of the two jets with that of the CO outflow reveals that these two
well-collimated jets together drive the molecular outflow mapped
by \citet{beuther02c}.  If the driving source of the outflow ``2'' 
is deeply embedded and is located at the centroid  of ``2'', it is
likely to be at $\sim$15\,arcsec SW of ``A''.  This source
also may be contributing to the 1.2-mm continuum source detected
by \citet{beuther02b}.  Millimetre observations at high angular
resolution are therefore required to understand that.

Our discovery images prompted the detailed follow-up observations
of \citet{davis04}, who obtained near-IR echelle and Integral Field
spectra.  Their $HK$ spectroscopy revealed a steeply rising SED for
``A'' with Br$\gamma$ and CO overtone emission,  which are typical
of an accreting source with a disc.  High-resolution spectroscopy
at 2.122\,$\mu$m shows that the H$_2$ emission from the jet is
blue- and red-shifted on either sides of ``A''.  The IFU data also
reveal a collimated jet from ``A''.  \citet{davis04} found that
the general properties of the two outflows are similar to those of
low-mass outflows, and that these jets are very much like scaled
up versions of their low-mass counterparts.

\vskip 10mm
\subsection{IRAS~18159-1648 - {\it Mol 49}\\ ({\small \it d = 2.5\,kpc, L = 29.5$\times$10$^3$\,L$_{\odot}$})}

\begin{figure*}
\centering
\includegraphics[width=16.5cm,clip]{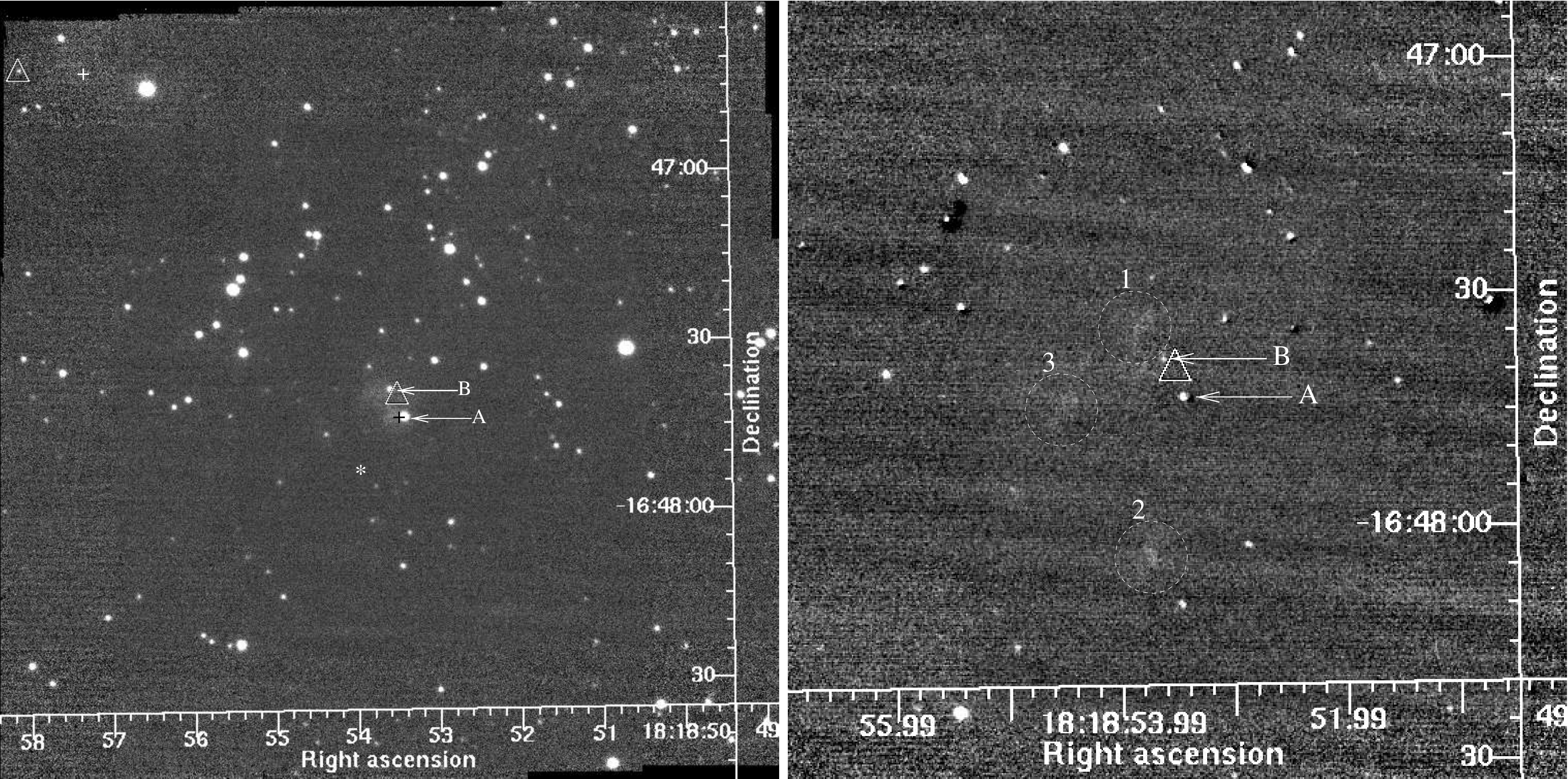}
\caption{Left: $K$-band image of IRAS~18159-1648. ``*'' shows the
1.2-mm peak position of \citet{faundez04}.
Right: Continuum-subtracted H$_2$ image of the central region.}
\label{18159_KH2}
\end{figure*}

IRAS~18159-1648 (GAL 014.33-00.64) was detected in CS(2-1) by
\citet{bronfman96}. \citet{molinari96} detected the dense core 
in NH$_3$ emission.  \citet{harju98} detected SiO (2-1) and 
(3-2), both at the same peak velocity. This is also consistent 
with the detection of two methanol maser spots at a velocity 
of 21\,kms$^{-1}$ by \citet{walsh97}, though this detection 
was not confirmed by \citet{szymczak00} or \citet{slysh99}, 
probably because their observations were less sensitive. 
Water maser emission was observed from this region by 
\citet{palla91}.

Fig. \ref{18159_KH2} shows our $K$ and H$_2$ images.
Two objects are labelled in the figure -
``A'' ($\alpha$=18:18:53.41, $\delta$=-16:47:43.1) and
``B'' ($\alpha$=18:18:53.58, $\delta$=-16:47:38.2) - both are 
embedded in faint nebulosity.  The 2MASS colours of ``A'' shows excess
and reddening, whereas ``B'' is located in the reddening band
(Fig. \ref{JHKcol}). However, it should be noted that ``B'' has only
upper limits in the 2MASS $J$ and $H$ and has poor S/N in $K_s$, whereas
``A'' has only upper limits in $J$ and has poor S/N in $H$.  The 1.2-mm
continuum source detected by \citet{faundez04} is only 12.25\,arcsec SE
of ``A''.  This separation is less than their beamsize of 24\,arcsec. 
The IRAS source is located 4.55\,arcsec NE of ``A'' and the MSX detection
is at 1.25\,arcsec east and 0.1\,arcsec north of ``A''.   From the
near-IR excess and reddening and the close proximity of the MSX, IRAS
and 1.2-mm sources,  it appears that ``A'' is the luminous YSO in
this field.  It remains to be investigated from better photometry if
``B'' also is a YSO.

There is a tentative detection of H$_2$ line emission close to the
centre of the field. The H$_2$ image (Fig. \ref{18159_KH2} - right panel) 
is smoothed with a 2-pixel FWHM Gaussian to enhance the very faint 
emission features.  Three features detected are circled and labelled 
``1--3'' on the H$_2$ image.  A deeper integration in H$_2$ is required
if we are to comment further on these possible outflow features.

\subsection{IRAS~18174-1612 - {\it G15.04-0.68, M17-UC1}\\ ({\small \it d = 2.1\,kpc, L = 433$\times$10$^3$\,L$_{\odot}$})}

\newcounter{subfigure}
\renewcommand{\thefigure}{A\arabic{figure}\alph{subfigure}}
\setcounter{subfigure}{1}
\begin{figure*}
\centering
\includegraphics[width=16.5cm,clip]{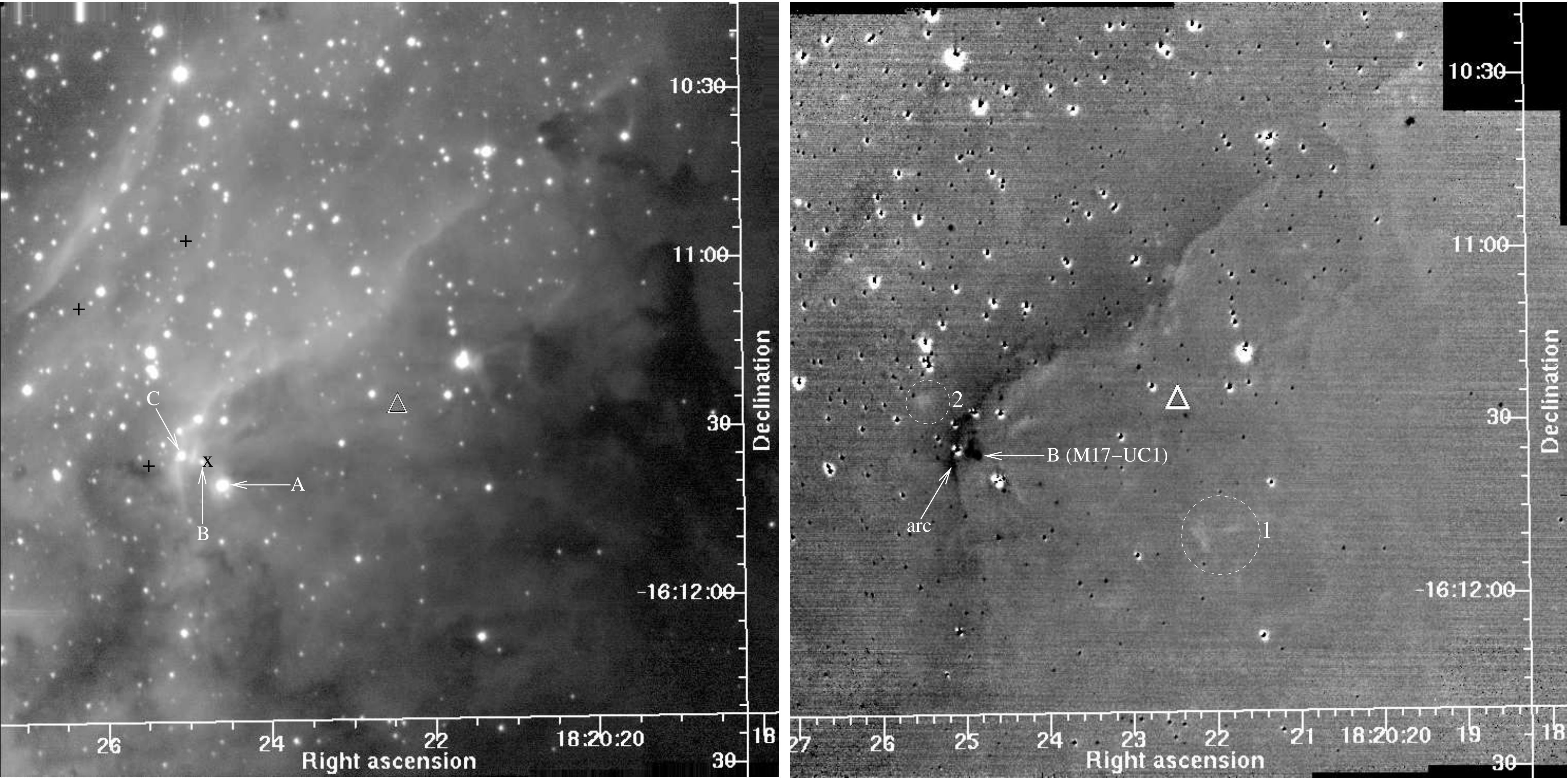}
\caption{The left panel shows the $K$-band image of IRAS~18174-1612
and the right panel shows the continuum-subtracted H$_2$ image
smoothed with a Gaussian of 2-pixel FWHM.  ``x'' shows the radio
position of the UCH{\sc{ii}} given by \citet{wc89b}.}
\label{18174_KH2}
\end{figure*}

Commonly known as M17-UC1, IRAS~18174-1612 is a Hyper Compact H{\sc ii}
(HCH{\sc ii}) region (Johnson, Depree \& Goss, 1998; \citealt{sewilo04})
located near the centre of the southern bar of the M17 (NGC~6618) molecular
cloud, identified by Felli, Johnston \& Churchwell (1980).  The object was
resolved at radio wavelengths by Felli, Churchwell \& Massi (1984) using the
VLA at high angular resolution, who showed that the M17-UC1 is a tiny shell of
gas (diameter $\sim$0.4\,arcsec or 1.4$\times$10$^{16}$\,cm) ionized from
within by a B0--0.5 YSO.  Their observations revealed M17-UC1 as a
powerful radio emitter with 147$\pm$7, 118$\pm$7, 28$\pm$3 and $<$\,7\,mJy at
1.2, 2, 6 and 21\,cm respectively. They interpret this object as the result of
shock-induced star formation in the H$^+$-H$_2$ interface.  An ``arc''-type
emission nebulosity was discovered $\sim$4\,arcsec east of the UCH{\sc{ii}}
at radio wavelengths.  OH masers \citep{caswell97, forster99} and
Class II methanol masers \citep{menten91, walsh98} have been detected within an
arcsec of M17-UC1.  Methanol masers have been detected by several other
investigators too. \citet{blaszkiewicz04} reported the variable
nature of the 6.7\,GHz methanol maser emission from this region.
However, H$_2$O maser detections are away from it by $\sim$30\,arcsec
or more; a group of them are located NW, $\sim$6\,arcsec north of the
IRAS position and another group $\sim$30\,arcsec SW of M17-UC1
(Massi, Felli \& Churchwell 1988, \citealt{forster99}).  H$_2$O maser
spots observed by \citet{johnson98} are also offset NW and SW of M17-UC1
(with some of them located closer than those detected by \citealt{massi88}).
NH$_3$ maps of this region show the clumpy nature of the molecular cloud,
with the NH$_3$ clumps, the HCH{\sc ii} M17-UC1 and the IR and radio
continuum emission of the ionization front all well separated.  They suggest
that the clumps detected in NH$_3$ may be producing low-mass stars.

Through near- and mid-IR imaging and spectroscopy, \citet{nielbock07}
proposed that M17-UC1 is surrounded by a disc of cool dust.  The
photometric distance towards this object (2.2 $\pm$ 0.2\,kpc)
derived by  Chini, Els\"aesser and Neckel (1980) is in agreement with
the kinematic distance.  A search by \citet{shepherd96b}  in CO found
no evidence of a bipolar molecular outflow from this region. 

\addtocounter{figure}{-1}
\setcounter{subfigure}{2}
\begin{figure}
\centering
\includegraphics[width=8.1cm,clip]{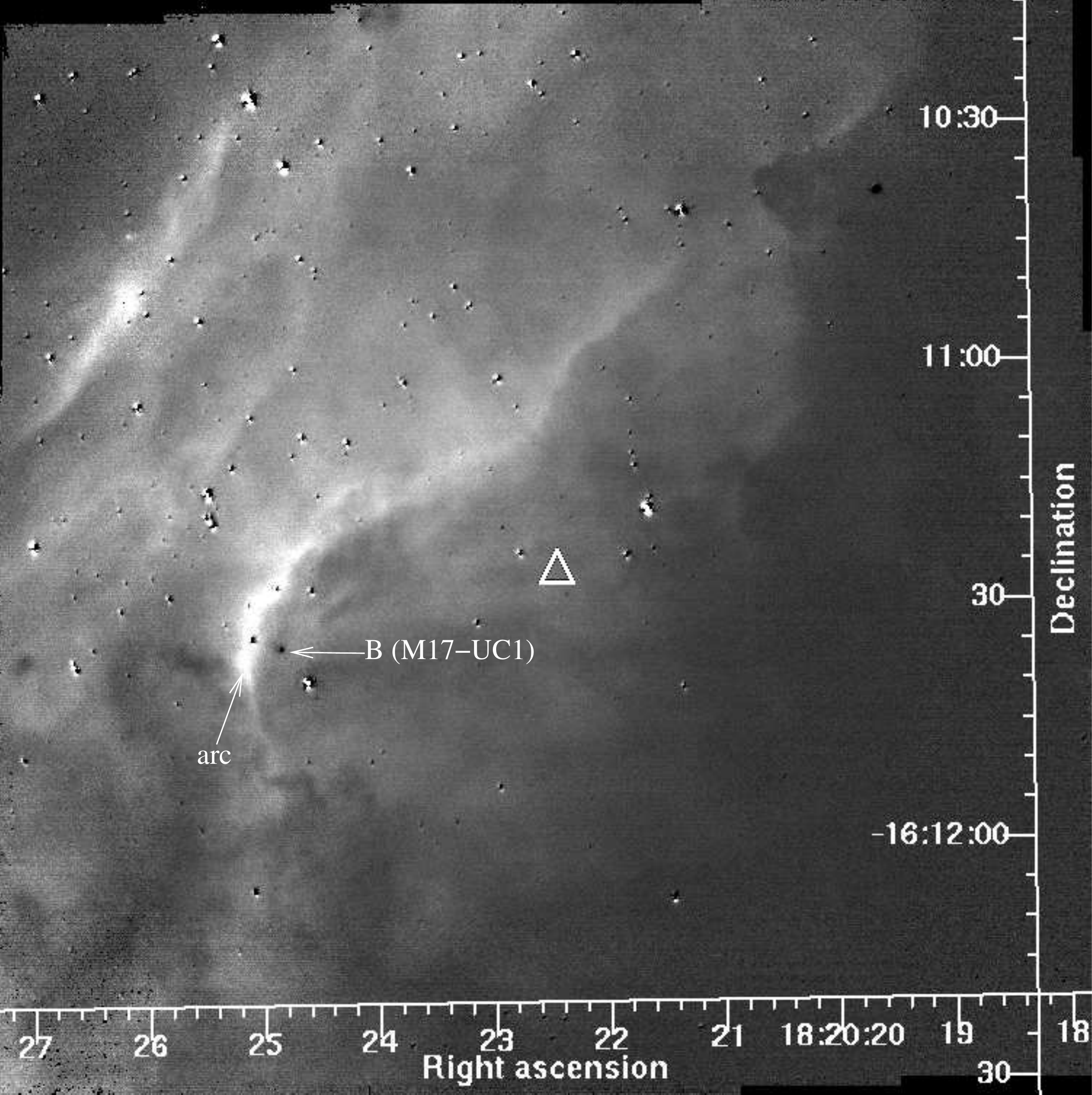}
\caption{Continuum-subtracted Br$\gamma$ image of IRAS~18174-1612.}
\label{18174_BrG}
\end{figure}
\renewcommand{\thefigure}{A\arabic{figure}}

The 2.3$\times$2.3\,arcmin$^{2}$ field observed by us is rather complex
and rich in nebulosity.  Fig. \ref{18174_KH2} shows the $K$-band and
continuum-subtracted H$_2$ images observed by us and Fig. \ref{18174_BrG}
shows our continuum-subtracted  Br$\gamma$ image. Three sources imaged
previously at high angular resolution in the near-IR  \citep{nielbock07}
are labelled in Fig. \ref{18174_KH2}:
``A'' ($\alpha$=18:20:24.60, $\delta$=-16:11:39.5),
``B'' ($\alpha$=18:20:24.83, $\delta$=-16:11:35.3 ) and
``C'' ($\alpha$=18:20:25.08, $\delta$=-16:11:34.1 ).  ``B'' is the
near-IR counterpart of M17-UC1.  It is not detected in the 2MASS images. 
Nielbock et al. detected it in $H$ and $K_s$ only.  The location of 
M17-UC1 in Fig. \ref{JHKcol} using the $H$ and $K_s$ magnitudes of 
Nielbock et al. and using their $J$-band detection limit shows that 
it has large reddening and IR excess.  However, for an A$_V$=40 
derived by Nielbock et al., the $J$ magnitude of ``B'' will be much 
fainter than the detection limit, which will move it further up
and place it in a region of larger reddening and lesser excess in 
Fig. \ref{JHKcol}.  ``A'' and ``C'' are identified with IRS5 and 
B273 respectively of \citet{nielbock07}.  ``A'' has a faint 
companion located $\sim$0.35\,arcsec north and $\sim$0.95\,arcsec 
west of the brighter one.   The physical association of these two 
stars is not certain at this stage.  The 2MASS colours of ``A''
are dominated by the brighter component and do not exhibit any 
IR excess.  ``C'' exhibits slight IR excess. Both ``A'' and ``C'' 
exhibit large reddening.  Part of the southern bar of M17 is seen 
in emission in Br$\gamma$, diagonally across the image.  
M17-UC1 (``B'') and the emission ``arc'' are labelled on the image.  
As can be expected for an ionization front, the ``arc'' is very 
bright in the continuum-subtracted Br$\gamma$ image.

Our H$_2$ image does not reveal any major collimated outflow in this
region.  Two faint bow-shock-like H$_2$ line emission features are
circled and are labelled ``1'' and ``2'' in Fig. \ref{18174_KH2} .  
There is also very faint H$_2$ emission seen from the inner edge of 
the Br$\gamma$ ``arc'' and at the boundary of the bright Br$\gamma$ 
emission ridge and the H$_2$ cloud, near the centre of the field.  
It is not clear if these are due to shocks or fluorescence.  The 
massive protostellar disc proposed by \citet{chini04} is further SE 
of ``B'' and is not related to any of the sources considered here.

The IRAS source is located $\sim$36\,arcsec NW of M17-UC1 and is 
probably not related to the radio source.  Three MSX sources are 
present in the field.  One of the sources is in the vicinity of 
M17-UC1.  There are no near-IR counterparts identified for the 
other two MSX sources.  None of the MSX sources is in the vicinity 
of the IRAS source,  which is probably very young and not very 
bright in the near- and mid-IR.  The lack of any convincing 
signatures of powerful outflow in H$_2$ and CO and the presence of 
strong radio emission show that M17-UC1 has probably
passed its main accretion and outflow stage and is actively 
moving towards a UCH{\sc ii} stage via its present HCH{\sc ii}
phase.

\subsection{IRAS~18182-1433\\ ({\small \it d = 4.5, 11.8\,kpc, L = (20, 125.9; 50.1)\,$\times$10$^3$\,L$_{\odot}$})}

This object was detected in CS emission by \citet{bronfman96} and
\citet{beuther02b}.  \citet{walsh98} detected CH$_3$OH maser emission
in this region, located NE of the IRAS position.  \citet{beuther02d}
resolved five H$_2$O maser spots, very close to the location of the
CH$_3$OH maser and the single 1.2-mm peak of \citet{beuther02b}.
Observations of \citet{sridharan02} revealed the presence of H$_2$O
and CH$_3$OH masers and emission by SiO, CH$_3$CN and NH$_3$.  They
also reported the detection of CO emission line wings implying
the presence of a molecular outflow.  Beuther et al. (2002c) mapped
the blue- and red-shifted lobes of the outflow and derived a
collimation factor of 1.7.  \citet{beuther06} performed
high-angular-resolution observations of IRAS~18182-1433 in 1.3-mm
continuum emission and in a number of molecular species including
$^{12}$CO\,(2-1).  In the CO data obtained using the SMA and IRAM,
they detect two outflows, inclined $\sim$90$^{\circ}$ with each
other.  Their SiO data (obtained using the VLA) is suggestive of
the possibility of an additional third outflow,  emanating towards
the north.  The southern lobe of this outflow is not detected in
their SiO data. Most of the molecular species peak $\sim$1-2\,arcsec
SE of the 1.2-mm peak and appear to coincide with one of the
nearby radio sources detected by \citet{zapata06}.
Williams, Fuller \& Sridharan (2004) imaged a core at
450 and 850\,$\mu$m.  By modelling the 850-$\mu$m data,
\citet{williams05} proposed a core with isothermal density profile
producing an O9 star.

The VLA observations of \citet{sridharan02} did not find any radio
emission at 3.6\,cm from this region, suggestive of a pre-UCH{\sc{ii}}
source.   Later observations of \citet{zapata06} revealed two faint
radio sources in this region at 1.3 and 3.6\,cm.  The first radio
source (``b''), which is interpreted as due to a thermal jet, is
located near, but offset by $\sim$1.9\,arcsec SE from, a 7-mm
continuum source (``a'') detected by them.  The 7-mm source is
also detected at 1.2\,mm by \citet{beuther02b} and later at
1.3\,mm at high angular resolution using the SMA by
\citet{beuther06}.  The cm emission is considered by
\citet{beuther06} as due to an unresolved companion to the mm
emitting source or due to a thermal jet.  A second radio source
detected by \citet{zapata06} (``c'') is 9.7\,arcsec SE of the first
and is considered to be due to strong shocks.  The labels
``a--c'' are given to be consistent with those of \citet{zapata06}.

Fig. \ref{18182_K} shows our $K$-band image of~IRAS 18182-1433. 
Since no extended emission was detected in H$_2$ or Br$\gamma$,
those images are not shown here.  Four objects are labelled on
the $K$-band image. The object labelled
``A'' ($\alpha$=18:21:9.10, $\delta$=-14:31:49.1) is located
within one arcsecond of the H$_2$O maser position of
\citet{beuther02d}.  Its 2MASS colours place it in the region of
evolved stars (Fig. \ref{JHKcol}).  However, the 2MASS $K_s$
magnitude of this object is only an upper limit.  Using a rough
estimate of the $K$-magnitude from our image (comparing with
brighter nearby objects), the location of the object comes closer
to the reddening band.  The object
``B'' ($\alpha$=18:21:9.70, $\delta$=-14:31:35.5) appears with a
large IR excess. Similarly, ``C'' ($\alpha$=18:21:9.48, $\delta$=-14:31:56.1)
also displays large excess.  However, ``B'' and ``C'' are detected
only in $K_s$ by 2MASS;  they are not detected in  $J$ and $H$
and have only upper limit magnitudes.  Deeper $JH$ imaging of the
field is therefore required.  ``C'' has a fainter companion
located 1.4\,arcsec NW and both are embedded in faint nebulosity
in our $K$-band image.  The coordinates given here are  those of
the brighter component.  This source also  exhibits IR excess. 
Finally, ``D'' ($\alpha$=18:21:8.83, $\delta$=-14:31:54.95) is a
very faint object that is not detected by 2MASS.  It is associated
with an arc-like feature in $K$, which is located 1.7\,arcsec NE
of a faint point source and has an extent of $\sim$2.5\,arcsec
in the NE direction.  There is no H$_2$ emission associated with this.

One of the two radio sources detected by \citet{zapata06} at 1.3 and 3.6\,cm
(their 18182-1433b), which is in the vicinity of the millimetre source,
is located very close to ``A''.   This source also is associated with the
CH$_3$OH maser detected by \citet{beuther02b} and is interpreted by
\citet{zapata06} as due to free-free emission from a thermal jet or a
stellar wind.  The brightest of their radio sources (18182-1433c), the
origin of which is suspected by them as due to strong shocks, is located
very close to the bright near-IR source in ``C''.  They did not detect
any 7-mm  emission associated with this.  The location of the MSX source
is between ``C'' and ``D'', closer to ``D'' (2.3\,arcsec SE of ``D'').  
The mid-IR observations of \citet{debuizer05} resolved two sources in
this region, the fainter one is located close to the location of the water
masers and a brighter, more extended one is located further SE.  The fainter
one also agrees with a detection in the near-IR by \citet{testi94}.  We do
not see any near-IR source at the location of either of these mid-IR sources.
However, applying a translation of $\sim$2.4\,arcsec towards the north
to the mid-IR positions of  \citet{debuizer05}, the fainter mid-IR source
agrees with ``A'' and the brighter and extended mid-IR source agrees with
the binary ``C'', with the direction of extension of similar to that of
the direction of orientation of the binary components.  This gives
justification in identifying ``A'' and ``C'' as two candidates
for the near-IR sources of the YSOs in this region.  The mutual
association of the sources detected at different wavelengths
in this region needs to be thoroughly investigated.

\begin{figure}
\centering
\includegraphics[width=8.10cm,clip]{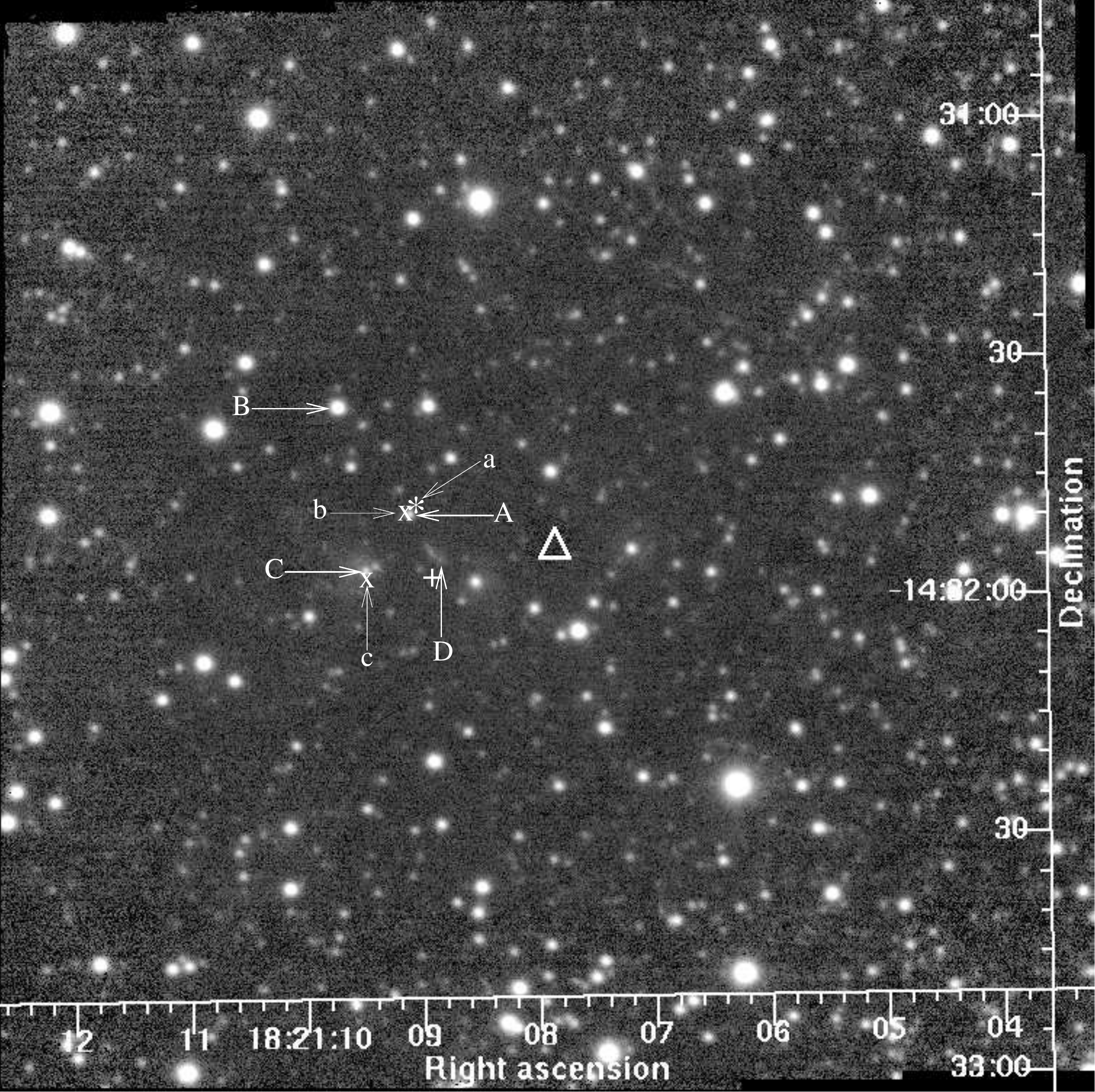}
\caption{$K$-band image of IRAS~18182-1433.  ``* (a)'' shows the location
of the 7-mm peak of \citet{zapata06} and the 1.3-mm peak of \citet{beuther06}. 
The 3.6-cm sources of \citet{zapata06} are shown by ``x (b and c)''.}
\label{18182_K}
\end{figure}

\subsection{IRAS~18264-1152\\ ({\small \it d = 3.5, 12.5\,kpc, L = (10, 125.9)\,$\times$10$^3$\,L$_{\odot}$})}

\citet{bronfman96} and \citet{beuther02b} detected CS emission
from this source.  \citet{sridharan02} observed CO emission line
wings implying the presence of bipolar outflows; observations by
\citet{beuther02c} indeed mapped the red- and blue- shifted lobes
of the CO outflow, which are located close to the 1.2-mm emission
peak of \citet{beuther02b}.  \citet{sridharan02} and
\citet{beuther02d} observed H$_2$O and CH$_3$OH maser emission
from this region,  from locations very close to the millimeter
peak.  The location of the 1.2-mm peak and the masers were
found to be north and west of the MSX position by $\sim$8.5
and 1\,arcsec respectively.  However, this offset is within the
spatial resolution of the MSX data.  \citet{sridharan02}
also discovered line emissions in NH$_3$, SiO, CH$_3$OH,
and CH$_3$CN from this region in addition to H$_2$O and
CH$_3$OH masers.  Their observation at 3.6\,cm using the
VLA resulted in a null detection ($<$ 1mJy).

\begin{figure*}
\centering
\includegraphics[width=16.5cm,clip]{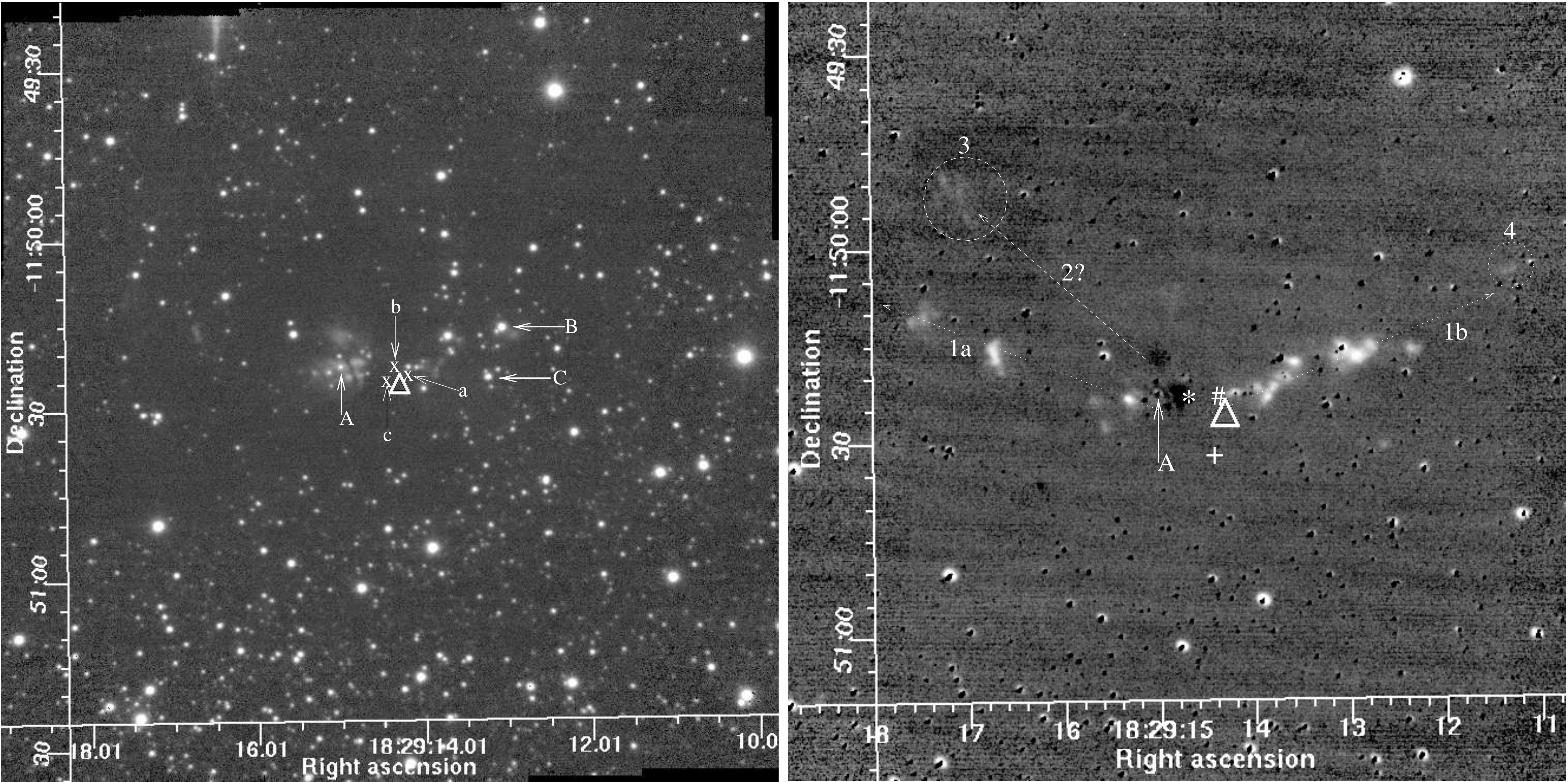}
\caption{The left panel shows the $K$-band image of IRAS~18264-1152 and
the right panel shows the continuum-subtracted H$_2$ image smoothed with
a 2-pixel FWHM Gaussian. The arrows show the directions of the outflows.  
The ``*'' sign shows the location of the 1.2-mm centroid of 
\citet{beuther02b}. The 1.3-cm positions of \citet{zapata06} are shown 
by ``x'' (labelled ``a--c'' to be consistent with their original naming);  
their 7-mm position is shown by ``\#''.}
\label{18264_KH2}
\end{figure*}

Our H$_2$ observations reveal a spectacular bipolar outflow from this
source (Fig. \ref{18264_KH2}).  The continuum-subtracted H$_2$
image shows an outflow roughly in the EW direction, originating from a
source somewhere near the millimetre position, the near-IR counterpart
of which is not detected here.  The direction of the outflow traced
by the  H$_2$ emission is in agreement with the direction inferred
from the CO map of \citet{beuther02c}.  The 1.2-mm continuum peak of
\citet{beuther02b} is located near the centre of the H$_2$ lobes.
One or both of the lobes of this bipolar outflow, the directions of
which are shown by the arrows labelled ``1a'' and ``1b'' at angles
63.5$^{\circ}$ and 291$^{\circ}$ respectively,  appear bent northward.
The outflow has a collimation factor of $\sim$7.  At this stage,  we 
also cannot rule out the possibility that ``1a'' and ``1b'' represent 
two different outflows rather than being a bent outflow as described 
above.  Detailed radial velocity studies are required to ascertain 
this.  There is another H$_2$ emission feature towards the west, 
circled and labelled ``4'',  which is possibly part of the outflow 
lobe in the direction of ``1b''.  There is also some faint H$_2$ 
emission feature shaped like a bow-shock, circled  and labelled ``3''.   
This is very faint and we are not sure at this stage if this feature 
reveals the bow-shock of a separate outflow or fluorescent emission.  
If it is due to an outflow, it is possibly in the direction of the 
arrow labelled ``2'' at an angle of $\sim$48$^{\circ}$.  Observations 
by \citet{zapata06} at 7\,mm and 1.3\,cm resolved the YSO as a possible 
triple system, with components separated by $\sim$2\,arcsec and the 
brightest peak at 7\,mm coinciding with the brightest at 1.3\,cm.  
Hence it is possible that ``3'' is the bow shock of another outflow.  
No emission was seen in our Br$\gamma$ image; hence it is not displayed 
here.

Three objects located close to the IRAS position are labelled on
the Fig. \ref{18264_KH2}.
``A'' ($\alpha$=18:29:15.01, $\delta$=-11:50:22.5) is 11\,arcsec NE
of the IRAS position and is near the geometric centre of the two lobes.
However, it is not clear if this is the driving source of the H$_2$
outflow.  The objects labelled ``B'' and ``C'' also show excess, but
they are far away from the apparent centre of the flow and are not
the driving sources of the outflow/s traced by ``1a''  and ``1b'' (they
are located along the western branch of the H$_2$ outflow).  All three
sources are shown in the 2MASS colour-colour diagram (Fig. \ref{JHKcol})
and do show reddening and excess.  Many of the fainter objects detected
in our $K$-band image are not detected by 2MASS.  Hence we do not have
any colour information on them.
Overall, IRAS~18264-1152  appears to be young, in
a phase of active accretion and outflow. Detailed near-IR studies
of this object and its multiplicity are underway.

\subsection{IRAS~18316-0602 -- {\it Mol 62}\\ ({\small \it d = 3.17\,kpc, L = 44.1$\times$10$^3$\,L$_{\odot}$})}

\begin{figure*}
\centering
\includegraphics[width=16.5cm,clip]{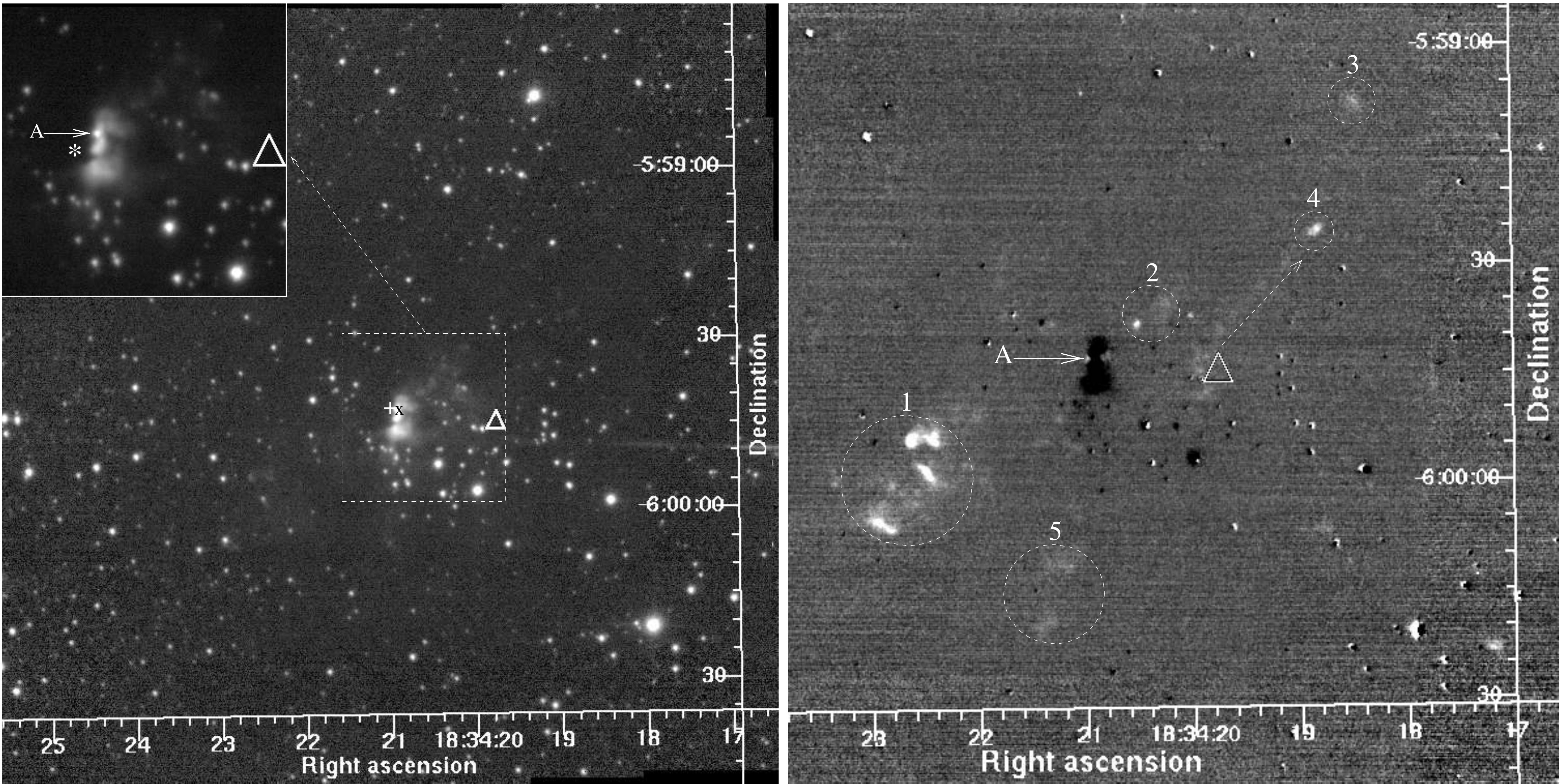}
\caption{The left panel shows the $K$-band image of IRAS~18316-0602.
The inset shows an expanded view of the central object.  ``x'' marks
the 3.6-cm radio continuum position of \citet{kurtz94}; ``*'' marks
the 450-$\mu$m position of \citet{walsh03}.  The right panel
shows the continuum-subtracted H$_2$ image smoothed with a
2-pixel FWHM Gaussian.}
\label{18316_KH2}
\end{figure*}

IRAS~18316-0602 (also known as G25.65$+$1.05 or RAFGL~7009S) is
associated with an irregular compact radio source,  first
identified at 3.6\,cm by \citet{kurtz94} with an integrated flux
density of 3.8\,mJy.  This is close in position to the radio source
detected at 3.6\,cm by \citet{walsh98} with a peak flux density
of 1\,mJy.  The radio peak of \citet{kurtz94} is coincident with an
unresolved infrared source, identified as a young B1V star with a
large $K$-band excess \citep{zavagno02}.  It is also closely
associated with NH$_3$ emission \citep{molinari96} and strong
CH$_3$OH \citep{walsh97,slysh99,szymczak00} and H$_2$O maser
\citep{brand94, kurtz05} emissions.  Sub-mm continuum observations
at 350\,$\mu$m by \citet{hunter00} and at 450\,$\mu$m and
850\,$\mu$m by \citet{walsh03} are all peaked at the position of the
radio continuum and maser sources. Observation of the CS (2-1)
line by \citet{bronfman96} shows good agreement with the observed radial
velocity from the masers (40.8-42.4\,km\,s$^{-1}$, \citealt{walsh03})
and the line emission (41.4\,km\,s$^{-1}$) indicating a strong link
between the dense gas, the maser sources and the massive star.
This also implies that the outflow is probably inclined close to
the sky plane.  \citet{shepherd96a} observed G26.65+1.05 in their
survey of high-velocity CO; they detected an energetic bipolar
outflow centred on the radio source.  \citet{todd06} have recently
presented high angular resolution images of this source in the
$K$ band and the {\it{v}}=1-0 S(1) line of H$_2$.  The source has an
extremely complex morphology and jet-like features abound. They 
conclude that these features arise from a single source centred
on the IRAS position.

Fig. \ref{18316_KH2} shows our $K$ and H$_2$ images.  The $K$-band
image reveals a cluster with a bright point source
(``A''; $\alpha$=18:34:20.92, $\delta$=-05:59:42.3) located towards
the centre of the field.  ``A'' is embedded in a compact nebula
extending roughly NS, most of which disappears in the
continuum-subtracted H$_2$ image and leaves a large negative residual
implying a cold environment, probably dominated by dust.  The
continuum-subtracted H$_2$ image reveals emission features NW and SE
of ``A'', which are encircled and labelled ``1--5'' on the figure. 
The IRAS source is located  17.6\,arcsec SW of ``A'', but the MSX
source location is very close to ``A'' (1.2\,arcsec east and
0.15\,arcsec north).  The 3.6-cm continuum source observed by
\citet{kurtz94} nearly coincides with ``A''.  This source was not
detected  by 2MASS in $J$ and $H$.  The colours derived from the upper
limits place it in the region of high reddening and excess in the
$JHK_s$ colour-colour diagram (Fig. \ref{JHKcol}), typical of YSOs.
The 2MASS colours would also be affected by the presence of nebulosity
close to ``A''.  We identify ``A'' as the HMYSO here, which drives the
outflow traced by the H$_2$ emission features and is the most likely
counterpart of the IRAS  and the MSX sources. \citet{zavagno02}
proposed the possibility of a disc associated with ``A''.  If the
features labelled ``1--3'' are produced by an outflow from ``A'',
we get a position angle of 130$^{\circ}$ (125$^{\circ}$ in the SE
and 315$^{\circ}$ in the NW) and a collimation factor of $\sim$4.5.
However, if the multiple bow shock like features in ``1'' are due to
precession of the jet, the collimation factor would be higher.

The locations of the H$_2$ emission features indicate the probable
presence of more than one outflow in this region.  Considering the
direction of ``4'' as seen by the line emission traced in the
direction of the dotted arrow shown on the H$_2$ image, one would get
an impression that ``4'' and ``5'' may well be due to an outflow from
a second  source in the vicinity of the IRAS position. This needs to
be examined with observations at longer wavelengths at high spatial
resolution.  Nevertheless, the 450-$\mu$m image obtained by
\citet{thompson06} (FWHM=8\,arcsec) shows only a single peak at ``A''.

The 3.6-cm radio continuum map of \citet{kurtz94} is suggestive of
a fainter radio companion at $\sim$2\,arcsec SW of the bright source.
Their map does not extend to the expected position of the driving 
source of the second jet seen in our H$_2$ image.
It remains to be investigated through multi-wavelength high angular
resolution studies if ``1--5''  originate from a single
(wide angle or precessing) outflow, or from two
different outflows which is  more likely.

\subsection{IRAS~18345-0641\\ ({\small \it d = 9.5\,kpc, L = 39.8$\times$10$^3$\,L$_{\odot}$})}

IRAS~18345-0641 is one of the high mass YSOs surveyed by
\citet{sridharan02} and \citet{beuther02b,beuther02c,beuther02d}.
Their $^{12}$CO (2-1) map shows a single, moderately collimated
(collimation factor$\sim$1.5) outflow, extending over 1.84\,pc and
centred close to the location of the peak mm-emission
\citep{beuther02c} which is also at the IRAS position. The estimated
mass of the outflow is 143\,M$_{\odot}$.  They \citep{sridharan02,
beuther02d} also detected water and methanol masers from this source. 
Comparison of the methanol maser observations of \citet{vanderwalt95}
with their own led \citet{szymczak00} to report that the maser emission
from this source is highly variable and was dominated by two main
components at the time of their observations. 

Fig. \ref{18345_KH2} shows our $K$-band image of IRAS~18345-0641 on
which the continuum-subtracted H$_2$ image of the central region is
shown in the inset.  Two sources are labelled ``A''
($\alpha$=18:37:7.00, $\delta$=-06:38:24.5) and ``B''
($\alpha$=18:37:6.91, $\delta$=-06:38:30.7) in the figure, with ``B''
6.5\,arcsec SW (nearly south) of ``A''.  This region does not exhibit
much H$_2$ emission except for a very faint feature circled and labelled
``1'' in the inset,  which is located at 
($\alpha$=18:37:17.41, $\delta$=-06:38:30.4).   The direction of the
blue-shifted lobe of the bipolar outflow mapped in  CO by \citet{beuther02c}
is roughly in the same direction as the H$_2$ emission feature ``1'' if
it is produced by an outflow from ``A''.  Deeper images in H$_2$ are
clearly warranted to ascertain any association of the H$_2$ emission
with the CO outflow.  Both ``A'' and ``B'' are detected by 2MASS. ``A''
has only an upper limit magnitude in $J$ and for ``B'',  2MASS gives only
upper limits in both $J$ and $H$.  Their 2MASS near-IR colours exhibit
large reddening and excess, with ``A'' much more reddened than
``B'' (Fig. \ref{JHKcol}).  The IRAS source is 8.5\,arc SW of ``A''
and 2.4\,arc SW of ``B''.  However, the location of the MSX source is
within 0.5\,arcsec of ``A''.   The peak of the 1.2-mm continuum imaged
by \citet{beuther02b} is only 1.65\,arcsec SE of ``B'' and the two
CH$_3$OH maser spots detected by \citet{beuther02d} are close to ``B''. 
However, both ``A'' and ``B'' are within the 11\,arcsec beam of
the 1.2\,mm observations of \citet{beuther02b}(Table \ref{resolutions}).
\citet{sridharan02} reported a 3.6-cm radio continuum source in this
region, with a flux density of 27\,mJy.  The position of the radio source
is not available to look for an association with any of the near-IR sources.
Possibly both ``A'' and ``B'' are YSOs with one of them in a more advanced
(UCH{\sc{ii}}) stage than the other.

\begin{figure}
\centering
\includegraphics[width=8.10cm,clip]{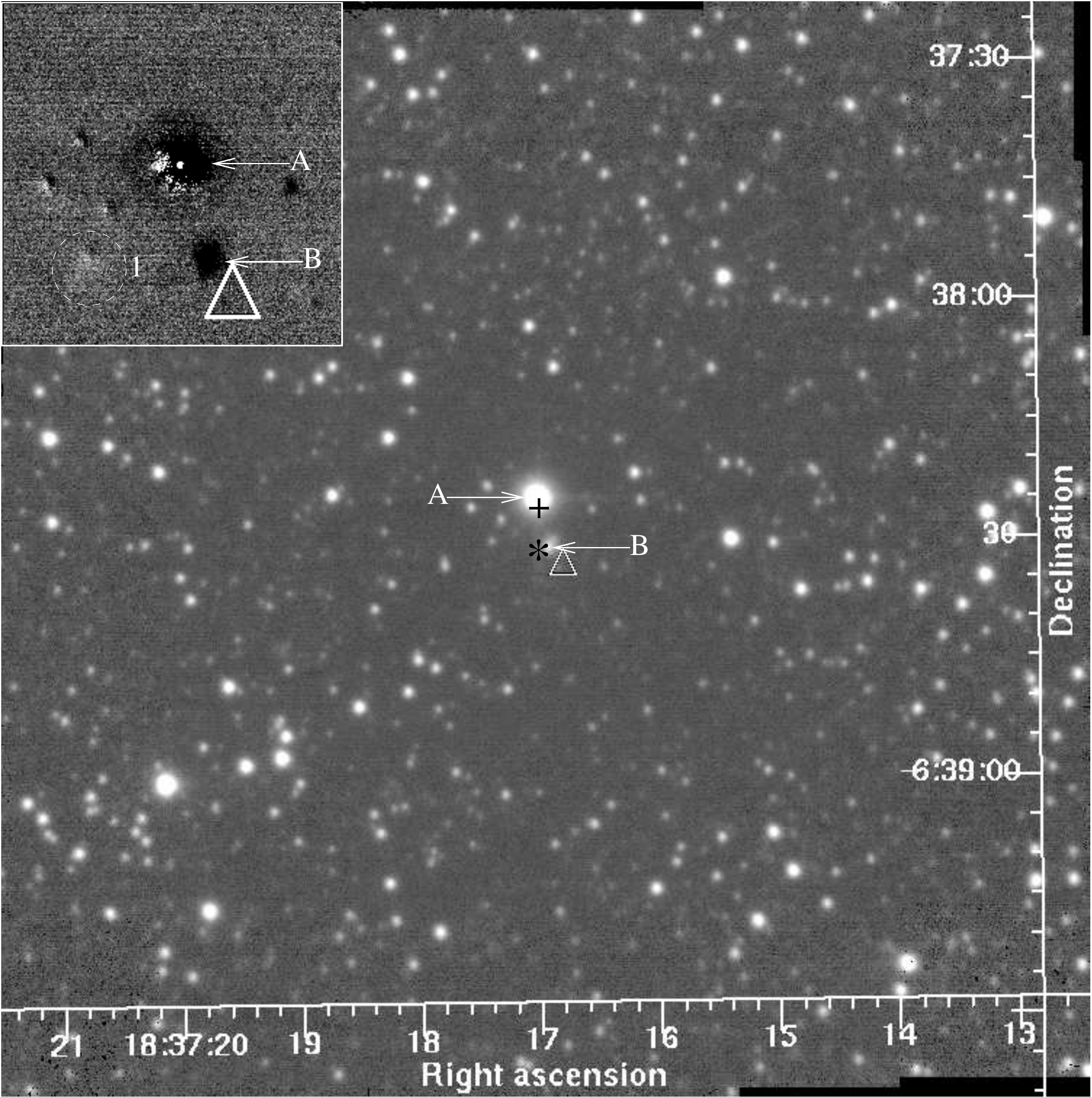}
\caption{$K$-band image of IRAS~18345-0641.  The continuum-subtracted
H$_2$ image of the central region is shown in the inset.  ``*'' shows the
location of the 1.2-mm continuum peak of \citet{beuther02b}.}
\label{18345_KH2}
\end{figure}

\subsection{IRAS~18360-0537 --{\it Mol 65}\\ ({\small \it d = 6.28\,kpc, L = 116$\times$10$^3$\,L$_{\odot}$})}

\citet{brand94} and \citet{palla91} detected water masers associated
with IRAS~18360-0537; Brand et al. found two spots with peak velocities
104.36\,kms$^{-1}$ and 105.01\,kms$^{-1}$ respectively.  Searches for
6.7-GHz methanol maser emission by \citet{vanderwalt95} found no emission.
A dense ammonia core associated with this source was reported by
\citet{molinari96} and CS (2-1) was detected by \citet{bronfman96}.
The peak velocities of the molecular species were 102.3\,km s$^{-1}$
and 101.6\,km s$^{-1}$ respectively, confirming the observed link between
emission from the dense molecular gas and masers.

\begin{figure}
\centering
\includegraphics[width=8.10cm,clip]{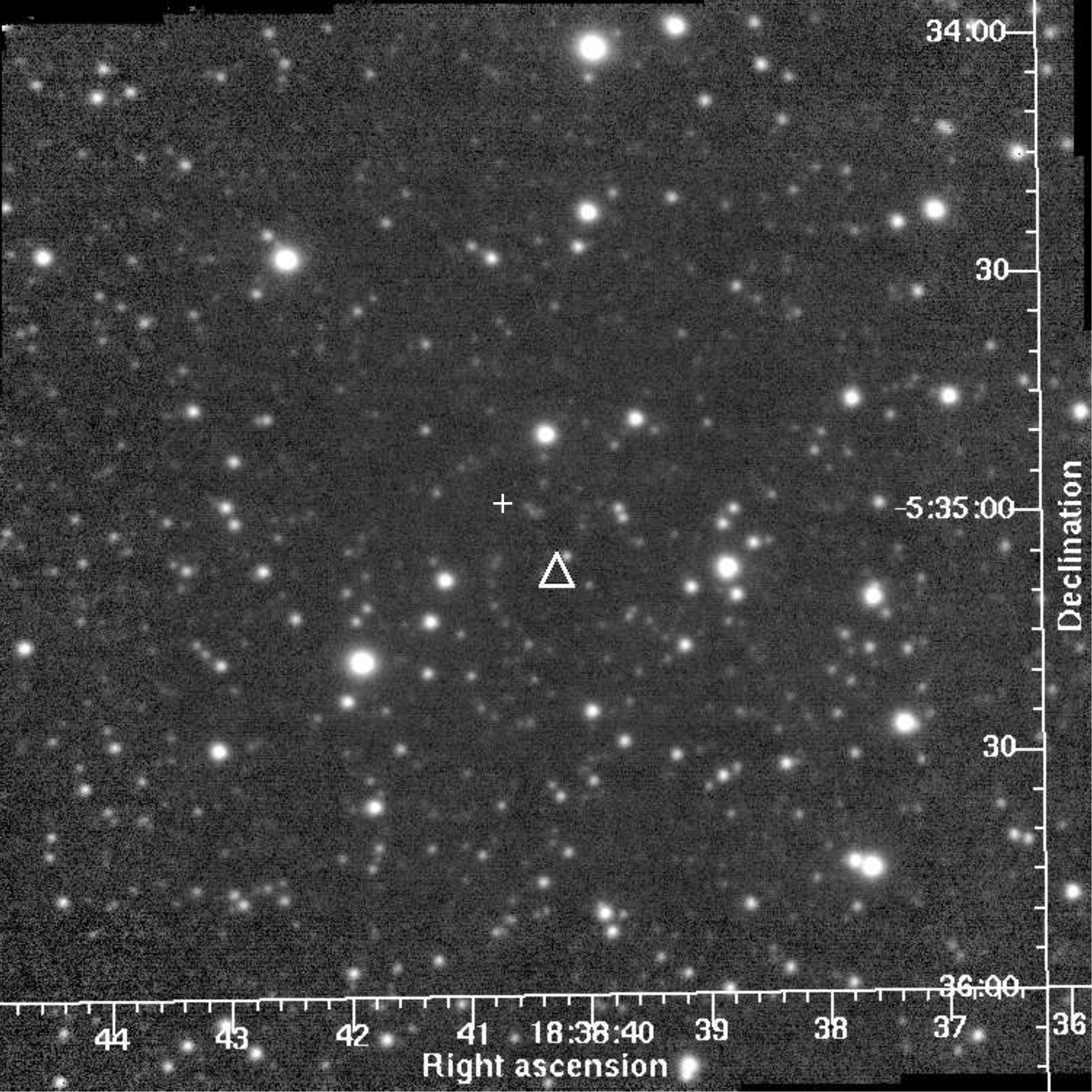}
\caption{$K$-band image of IRAS~18360-0537}
\label{18360_K}
\end{figure}

Fig. \ref{18360_K} shows our observed $K$-band image.  The
continuum-subtracted H$_2$ image did not reveal any line emission. 
Hence, it is not shown here. Since the region did not exhibit any
extended features, we did not observe it in Br$\gamma$.  The location
of the MSX object is $\sim$10.8\,arcsec NE of the IRAS position.
From Fig. \ref{18360_K}, we cannot identify any near-IR counterpart
of the HMYSO.  It is likely to be very young.

\subsection{IRAS~18385-0512\\ ({\small \it d = 2, 13.1\,kpc, L = (5, 199.5)\,$\times$10$^3$\,L$_{\odot}$})}

\citet{brand94}, \citet{sridharan02} and \citet{beuther02d} observed
H$_2$O maser emission from this region.  Searches for CH$_3$OH emission
by \citet{walsh98}, \citet{sridharan02} and \citet{beuther02d} did not
yield a detection.  Sridharan et al. detected CO emission
line wings  giving the indication of bipolar outflow.  They also
detected a 3.6-cm free-free continuum source, at a flux density of
29\,mJy, near the millimetre source and the masers. 

\begin{figure}
\centering
\includegraphics[width=8.10cm,clip]{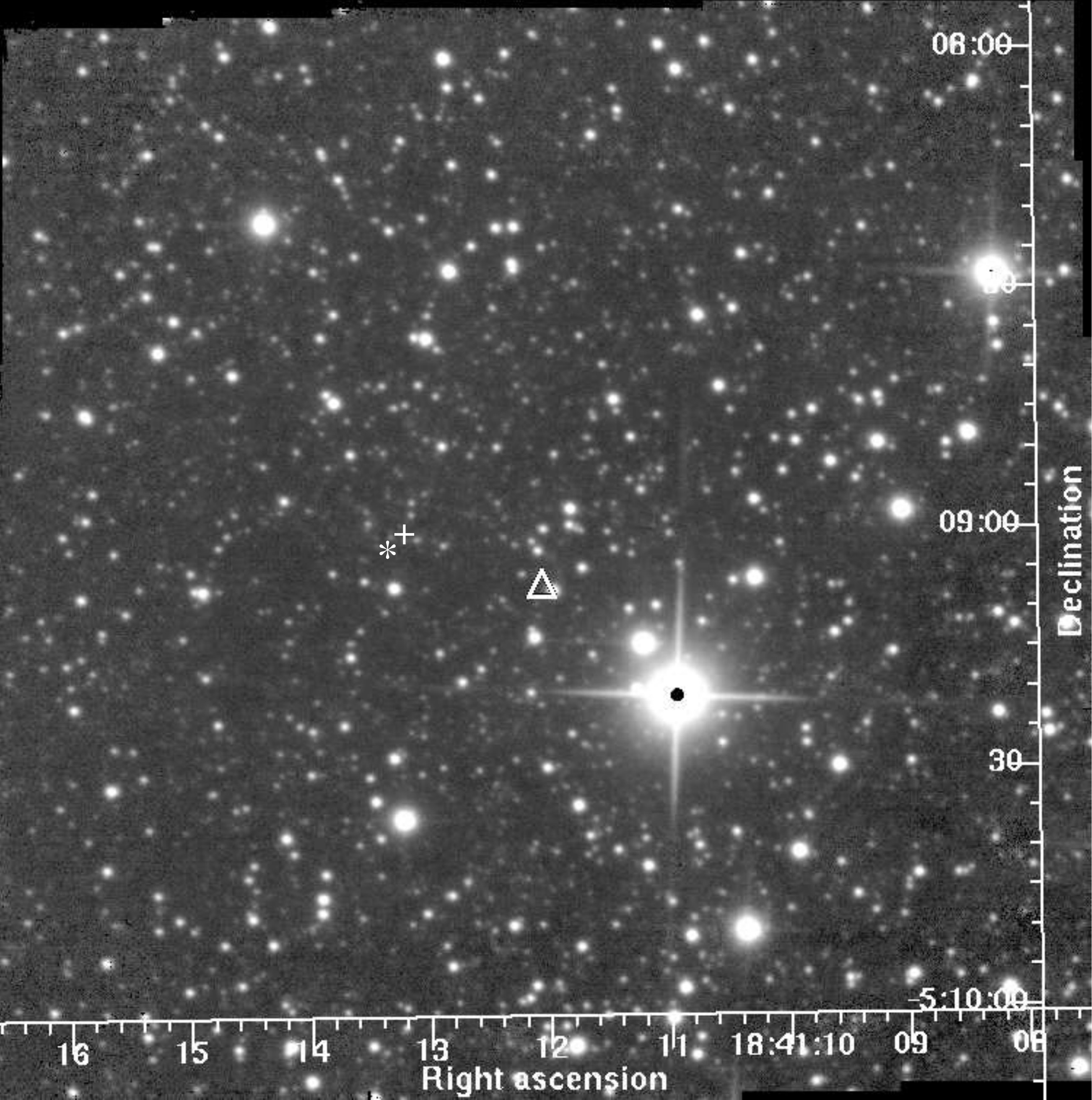}
\caption{$K$-band
image of IRAS~18385-0512.  The location of the brightest 1.2-mm peak of
\citet{beuther02b} is shown by ``*''.}
\label{18385_K}
\end{figure}

No near-IR observations of this object have been published.
Our $K$-band image shows a crowded field with a very bright source
(2MASS $K_s$=4.97 mag) located at ($\alpha$=18:41:10.93, $\delta$=-5:09:20.8),
$\sim$21.7\,arcsec SW of the IRAS position.  This object does not show
any significant IR excess. It is likely  to be a foreground object. 
Hence it is not included in Fig. \ref{JHKcol}.  No Br$\gamma$ observations
were acquired since we did not detect convincing signs of outflow in
the H$_2$ image.  Hence, only the $K$-band image is displayed
here (Fig. \ref{18385_K}).

\citet{williams04} detected a sub-mm core at 450\,$\mu$m and 850\,$\mu$m; 
modelling of the radial density profile by \citet{williams05} shows a
core with L$\sim$10$^4\,$L$_\odot$, which could be the pre-cursor of a 
B1 star. It should be noted that the 1.2-mm continuum source, the 
location of the H$_2$O masers and the 3.6-cm emission shown by 
\citet{beuther02d} all appear to be offset NE of the IRAS position, 
with the centroid of the centimetre emission offset from that of the 
millimetre emission by 3.4\,arcsec (\citealt{sridharan02}).  The 
H$_2$O masers are offset by only $\sim$1\,arcsec \citep{beuther02d} 
and the MSX position is only 2.8\,arcsec away \citep{sridharan02} 
from the 1.2-mm peak.  These offsets are within the positional 
uncertainty of the 1.2-mm detection.  The offsets of the IRAS and 
MSX positions from the mm and radio peaks are within their positional 
uncertainties. Both H$_2$O maser and 3.6-cm observations have good 
positional accuracy ($\sim$1\,arcsec; Table \ref{resolutions}).  
A  separation of $\sim$3.1\,arcsec between them and the close proximity 
of the H$_2$O masers to the 1.2-mm source \citep{beuther02d} 
suggests that there may be two YSOs here, one in a pre-UCH{\sc ii} 
stage detected at at 1.2\,mm and responsible for the H$_2$O maser
emission, and the other in a more evolved UCH{\sc ii} stage, 
detected in the radio.  The offsets of the IRAS and MSX positions 
from the mm and radio peaks are within their positional 
uncertainties.  2MASS does not detect most of the faint objects 
in the $K$ image, which makes identification of any near-IR 
counterpart difficult.  Deep mid-infrared photometry of this region 
is therefore required.

\subsection{IRAS~18507+0121 -{\it Mol 74}\\ ({\small \it d = 3.87\,kpc, L = 48.4$\times$10$^3$\,L$_{\odot}$})}

A detailed set of maps of this source has been made recently by
Shepherd, N\"{u}rnberger \& Bronfman (2004b) observing in the
radio (H$^{13}$CO$^{+}$ and SiO), millimetre (3-mm continuum)
and near-IR ($JHK'$).  These are the highest angular resolution
measurements, pre-dating those presented here. The H$^{13}$CO$^+$
maps show two distinct cores, separated by 40\,arcsec.  The more
southerly peak is 3\,arcsec NW of a UCH{\sc{ii}} region seen in
6-cm (12.24$\pm$0.1\,mJy) radio continuum emission 
\citep{molinari98}. The southern radio source was detected by
\citet{shepherd04b} at 6\,cm with an integrated flux density of
9\,mJy; a marginal detection (0.7\,mJy) was made by them at the
position of the northerly core.  Miralles, Rodriguez \& Scalise (1994)
also detected an unresolved radio source at 2 and 6\,cm
(integrated flux densities of 9.9$\pm$0.2 and 9.1$\pm$0.1\,mJy,
respectively) located very close to the 6-cm peak of
\citet{molinari98}.  The 3-mm continuum observations of
\citet{shepherd04b} show a single source at the position of
the northerly core seen in H$^{13}$CO$^+$. Previous studies in
molecular lines have been made by \citet{miralles94}, 
\citet{harju98}, \citet{bronfman96} and 
Ramesh, Bronfman \& Deguchi (1997). Although the ammonia maps of
Miralles et al. are consistent with that from H$^{13}$CO$^+$,
the SiO emission seen by Harju et al. was not detected by Shepherd
et al., probably due to the smaller beam size of their
interferometric observations. \citet{shepherd04b} detected 146 infrared
sources in their images, the brightest of which is in a young
cluster on the western edge of the southern HCO$^+$ peak. There is
evidence for circumstellar material around 50\% of the NIR
sources. There is strong and variable methanol emission from this
region, measured by \citet{szymczak00} and \citet{schutte93}
at velocities of 55--64\,km\,s$^{-1}$. Copious water maser
emission is also seen at these velocities \citep{brand94, miralles94}.
\citet{shepherd04b} derived spectral types of B2 and B0.5 for the
northern and the southern cores, respectively.

\begin{figure}
\centering
\includegraphics[width=8.10cm,clip]{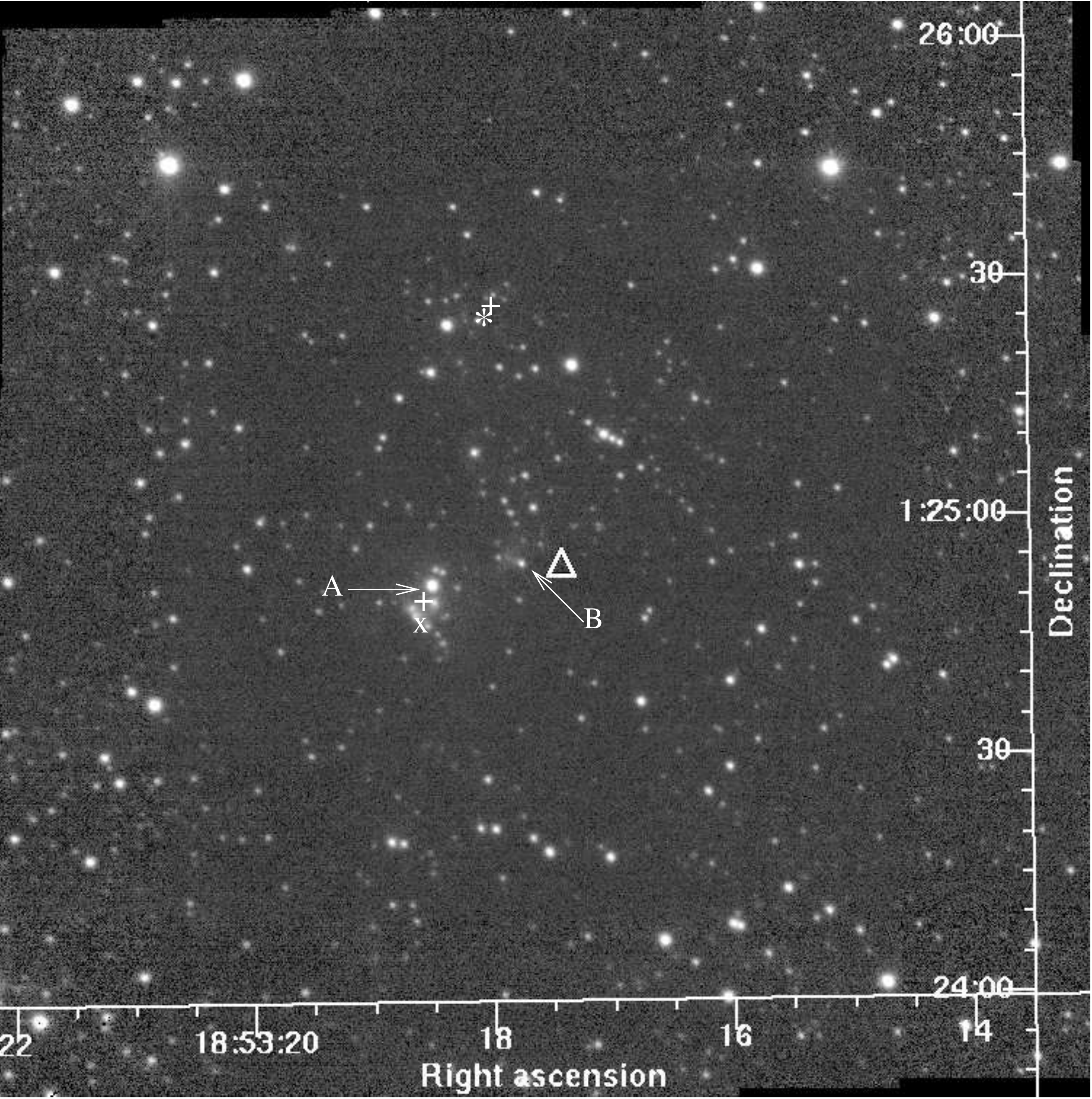}
\caption{$K$-band image of IRAS~18507+0121. The location of the
3-mm continuum peak imaged by \citet{shepherd04b} is shown by ``*''. 
The 2- and 6-cm radio continuum peak of \citet{miralles94} is shown by ``x''.
A second 6-cm source marginally detected by \citet{shepherd04b} coincides
with their 3-mm source in the north.}
\label{18507_K}
\end{figure}

Fig. \ref{18507_K} shows our $K$-band image.  H$_2$ and Br$\gamma$
images did not reveal any line emission, hence they are not displayed
here.  The source labelled ``A'' (``\#54'' of Shepherd et al. 2004b;
$\alpha$=18:53:18.49, $\delta$=+01:24:51.8) is located $\sim$16.4\,arcsec
SW of the IRAS position.  ``A'' is well detected by 2MASS in $K_s$ only.
The 2MASS magnitudes of this object are affected by the neighbouring
sources.  Hence, we used the magnitudes given by \citet{shepherd04b}
in Fig. \ref{JHKcol}.  The $JHK'$ colours exhibit reddening, but no
excess.  ``A'', and some fainter stars in its vicinity which are not
detected by 2MASS, appear to be embedded in a faint nebulosity.  The
fainter neighbours are detected by Shepherd et al. in $K'$ only and
not at shorter wavelengths.  There are two MSX sources detected in
this field, the brighter one is located within 2.5\,arcsec of ``A''. 
It is possible that one of the neighbours of ``A'' is the near-IR
counterpart of the YSO detected by MSX.  Another object ``B''
(``\#57'' of Shepherd et al. 2004b; $\alpha$=18:53:17.75, $\delta$=+01:24:54.8)
is located close to the IRAS position;  it is associated with a faint
nebulosity in $K$.  It was detected by 2MASS only in one band.  Hence
we have used its $JHK'$ colours measured by \citet{shepherd04b} to
plot it in  Fig. \ref{JHKcol}.  ``B'' exhibits reddening and mild
excess.  This could be another YSO in the field.  The second MSX source,
which appears to be much cooler than the first (it is detected only
at 21.34\,$\mu$m), is located $\sim$33\,arcsec NE of the IRAS  position.
The location of ``A'' and its sorrounding stars agree well with that
of the UCH{\sc{ii}} identified at 6\,cm \citep{miralles94, molinari98,
shepherd04b}, the southern H$^{13}$CO$^{+}$ core with no millimetre
continuum detected \citep{shepherd04b} and the brighter MSX source
in the field.  The second YSO located northward was detected in
mm continuum and H$^{13}$CO$^{+}$ emission \citep{shepherd04b},
while a fainter 6-cm emission peak located northward coincides well
with the second MSX source (which appears only at longer wavelength). 
There are no IR sources identified with this source and in general,
we agree with the conclusion of Shepherd et al. (2004b) that the
mm source coinciding with the cooler MSX source is younger than the
one associated with the UCH{\sc{ii}}; it is not certain from the current
study whether any of the IR sources identified near the location
of the UCH{\sc{ii}} is the IR counterpart of the massive YSO. More
observations are underway.

\subsection{IRAS~18517+0437 --  {\it Mol 76}\\ ({\small \it d = 2.9\,kpc, L = 12.6$\times$10$^3$\,L$_{\odot}$})}

\begin{figure}
\centering
\includegraphics[width=8.10cm,clip]{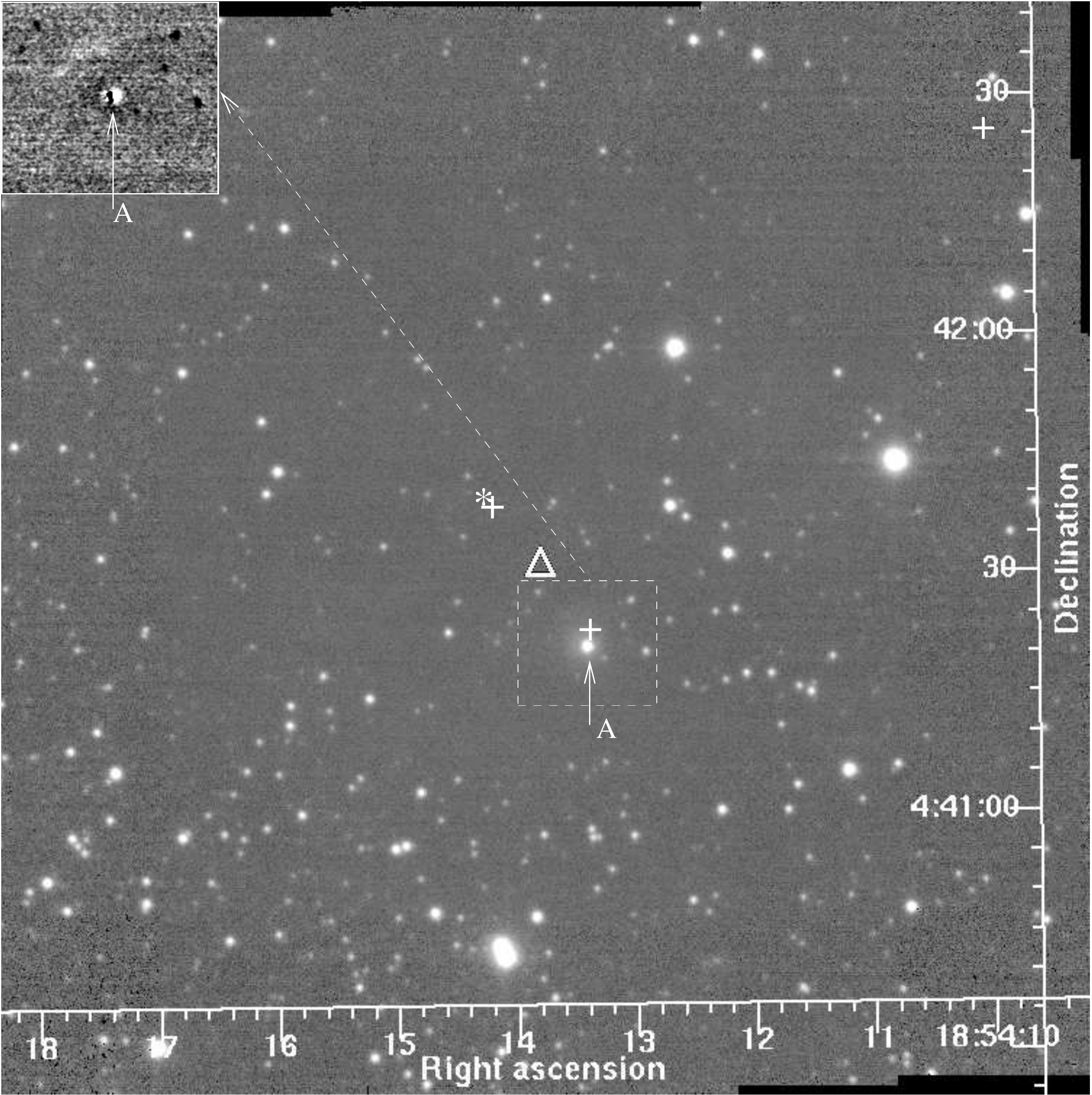}
\caption{$K$-band image of IRAS~18517+0437. The inset shows
the region of the continuum-subtracted H$_2$ image around source
``A'' (smoothed by a 2-pixel FWHM Gaussian) showing the faint
H$_2$ emission in the vicinity. The location of the bright
1.2-mm continuum peak mapped by \citet{beuther02b} is shown by ``*''.}
\label{18517_KH2}
\end{figure}

Towards this object, \citet{molinari96} discovered a dense core in
NH$_3$ emission.   \citet{szymczak00} detected strong CH$_3$OH maser
emission. \citet{brand94}, Codella, Felli \& Natalle (1996) and
\citet{sridharan02} observed H$_2$O maser emission.  A search for
3.6-cm continuum free-free emission by \citet{sridharan02} did not
yield any solid detection.  They discovered CO line wings indicating
the presence of bipolar outflows from this region.   However, CO
observations by \citet{zhang05} did not reveal an outflow.

No other near-IR observations of this region have been published.
Fig. \ref{18517_KH2} shows our $K$-band image and the central region of
the continuum-subtracted H$_2$ image. There is a source embedded in
faint nebulosity, located $\sim$11.5 arcsec SW of the IRAS position,
labelled ``A'' ($\alpha$=18:54:13.40, $\delta$=+4:41:21.1).  It
is well detected by 2MASS in all three bands and the $JHK$ colours
place it within the reddening band. Thus, there is no significant amount
of IR excess.   No collimated emission is detected in H$_2$; the only
H$_2$ line emission seen is a very faint arc-like feature
$\sim$5\,arcsec NE of ``A''.  

The MSX mission detected three objects in this field, with two of them
located towards the central region,  NE and SW of the IRAS position.
The IRAS source is likely to represent combined emission from these two.
The brighter of the three MSX sources is 5.4\,arcsec west and 9.1\,arcsec
south of the IRAS position and is only 1.98\,arcsec north of ``A''. 
The 1.2-mm continuum core mapped by \citet{beuther02b} is NE of the IRAS
position,  by $\sim$8\,arcsec in both RA and Dec.  The redder of the three
MSX sources is  8.8\,arcsec NE of the IRAS position.  It is indeed only
2.1\,arcsec SW of the 1.2-mm peak and is likely to be the MSX counterpart
of the millimetre source.  It is not detected in the near-IR. 
\citet{beuther02b} also discovered CS line emission from this region.

All these observations imply that the HMYSO is a very young
object, possibly in a pre-UCH{\sc{ii}} phase; the IR source
``A'' is unrelated and is likely to be a more evolved object. 

\subsection{IRAS~19088+0902 --  {\it Mol 97}\\ ({\small \it d = 4.71\,kpc, L = 29.9$\times$10$^3$\,L$_{\odot}$})}

IRAS~19088+0902 is one of the UCH{\sc ii} candidates from
\citet{palla91}, that is associated with H$_2$O maser emission (see
also \citealt{brand94, macleod98a} for water maser detection from
this source).  Searches for CH$_3$OH maser emission from this region
did not yield any detection \citep{macleod98a,szymczak00,vanderwalt95}.
The water maser emission peaks are associated with the CO emission
observed by Osterloh, Henning \& Launhardt (1997). An additional CO
peak has no corresponding maser emission, but broad line wings are
evident.  \citet{osterloh97} also present $K'$ images,  but those of
IRAS~19088+0902 are of low quality and do not reveal any extended,
diffuse emission.  The dense core was detected in ammonia emission
\citep{molinari96} and in CS (2-1) emission \citep{bronfman96}.

\begin{figure}
\centering
\includegraphics[width=8.10cm,clip]{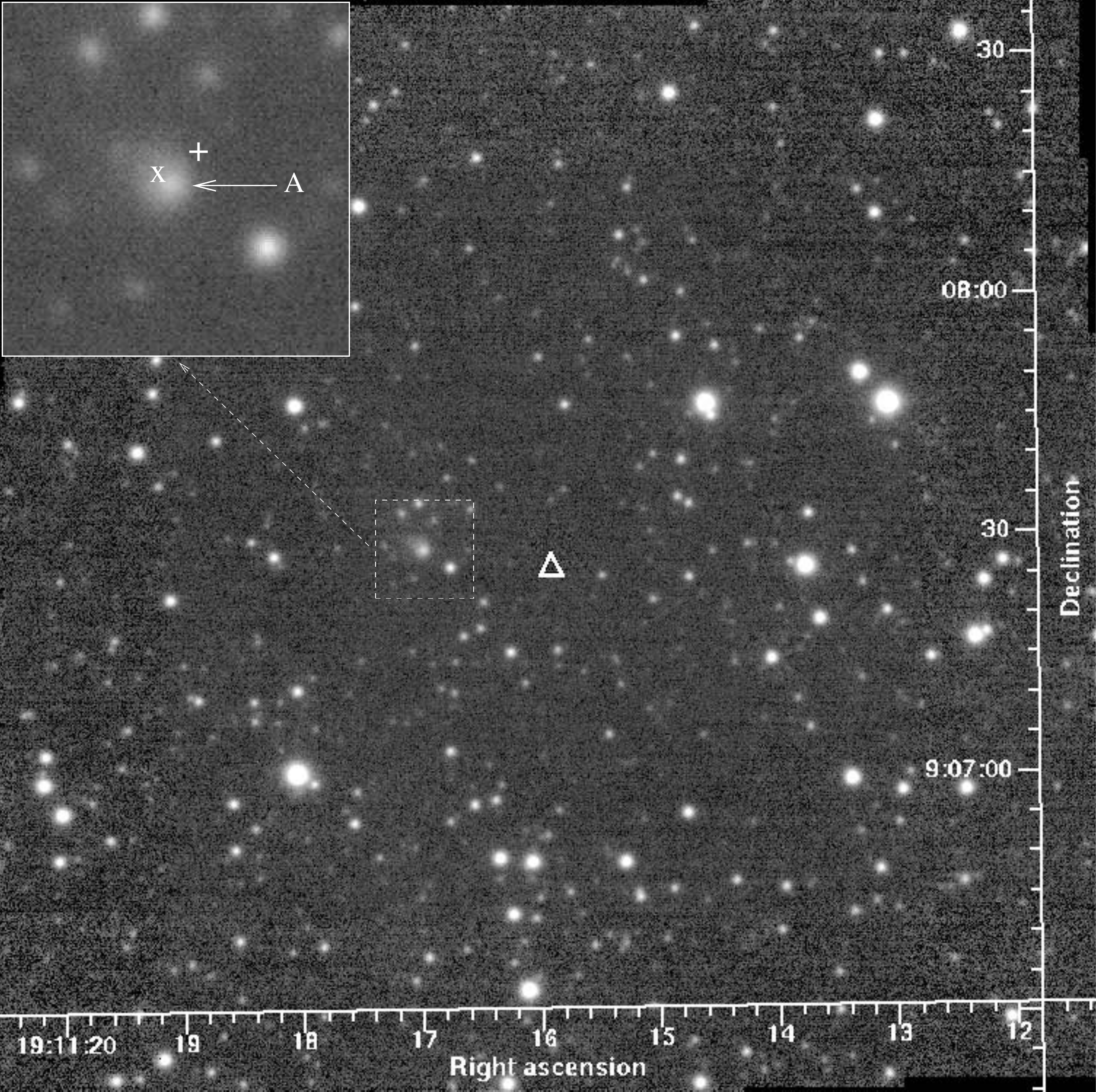}
\caption{$K$-band image of IRAS~19088+0902. An expanded view
of the vicinity of the proposed YSO is shown in the inset.
``x'' shows the location of the 6-cm radio source detected
by \citet{molinari98}.}
\label{19088_K}
\end{figure}

The VLA observations of \citet{molinari98} at 6\,cm reveal an
unresolved source with an integrated flux density of 10.78\,mJy,
located 17\,arcsec away from the IRAS position.  From the 20-cm
radio flux (observations by \citealt{zoonematkermani90}),
\citet{macleod98a} derived a spectral type earlier than B0.8.

Fig. \ref{19088_K} shows our $K$-band image. No line
emission was detected either in the H$_2$ or Br$\gamma$ image.
Hence these are not shown here.  A near-IR source embedded in faint
nebulosity in $K$ is labelled on the figure
(``A''; $\alpha$=19:11:16.97, $\delta$=09:07:28.6).  The IRAS source
is located 15.9\,arcsec SW of ``A''. However, the MSX source is just
1.7\,arcsec NW of ``A''. This object is well detected in $K_s$ in 2MASS,
but has only upper limits in $J$ and $H$.  The derived colours show
excess (Fig. \ref{JHKcol}).  The 6-cm radio source detected by
\citet{molinari98} is within 0.9\,arcsec of ``A''.  These observations
show that ``A'' is certainly a candidate for the HMYSO in this field.
It is probably in its UCH{\sc ii} phase.

\subsection{IRAS~19092+0841 -- {\it Mol 98}\\ ({\small \it d = 4.48\,kpc, L = 9.2$\times$10$^3$\,L$_{\odot}$})}

IRAS~19092+0841 is considered to be a good candidate
pre-UCH{\sc ii} object \citep{molinari96}.
Water maser emission was detected by \citet{brand94} and \citet{palla91}
and an ammonia core was found at the same velocity ($\sim$58 km s$^{-1}$;
\citealt{molinari96}).  Methanol and Hydroxyl masers were observed close
to this velocity by \citet{macleod98b}, with the methanol maser confirmed
by \citet{szymczak00}. The source strength ($\sim$10\,Jy) is consistent
between the two methanol maser observations and so this methanol maser
appears to be non-variable.  \citet{molinari98} detected a source at
6\,cm from this region, but offset from the IRAS position by
110\,arcsec in the NE direction.  Hence the radio detection by
\citet{molinari98} is not from the IRAS source.  A search by
\citet{zhang05} failed to detect any  CO outflow from this region.

Fig. \ref{19092_K} shows our $K$-band image.  No outflows are detected
in H$_2$ in the 2.25$\times$2.25\,arcmin$^{2}$ field.  There is a faint
object labelled ``A'' ($\alpha$=19:11:37.79, $\delta$=8:46:41.2) located
$\sim$12.3\,arcsec NE of the IRAS position,  which appears to be surrounded
by very faint nebulosity in $K$. It is located 4.25\,arcsec SW of the
brightest of the three MSX sources detected within this field.  (The other
two MSX sources are far from the centre and are not detected at longer
wavelengths).  This object is not detected by 2MASS and is deeply embeded.  
At this stage, it is not clear if ``A'' is the near-IR counterpart of the 
HMYSO.  The 1.1-mm peak reported by \citet{molinari00} is offset 10\,arcsec 
south and west of the IRAS position, ie., offset from the IRAS position 
in a direction opposite to that of ``A''.  Deep IR imaging would be 
required to obtain its colours.  Two other objects closest to the IRAS 
position, detected by 2MASS with an appreciable amount of IR 
excess (Fig. \ref{JHKcol}) are labelled 
``B'' ($\alpha$=19:11:38.61, $\delta$=+8:46:32.6) and
``C'' ($\alpha$=19:11:37.65, $\delta$=+8:45:45.5).  All three objects
are away from the mm position.  The YSO in this field may be in  a
pre--UCH{\sc ii} stage.  From the current observations, we cannot
identify any of the near-IR objects in the field as an IR counterpart.
Deeper IR imaging and longer wavelength observations with better
positional accuracy are required.

\begin{figure}
\centering
\includegraphics[width=8.10cm,clip]{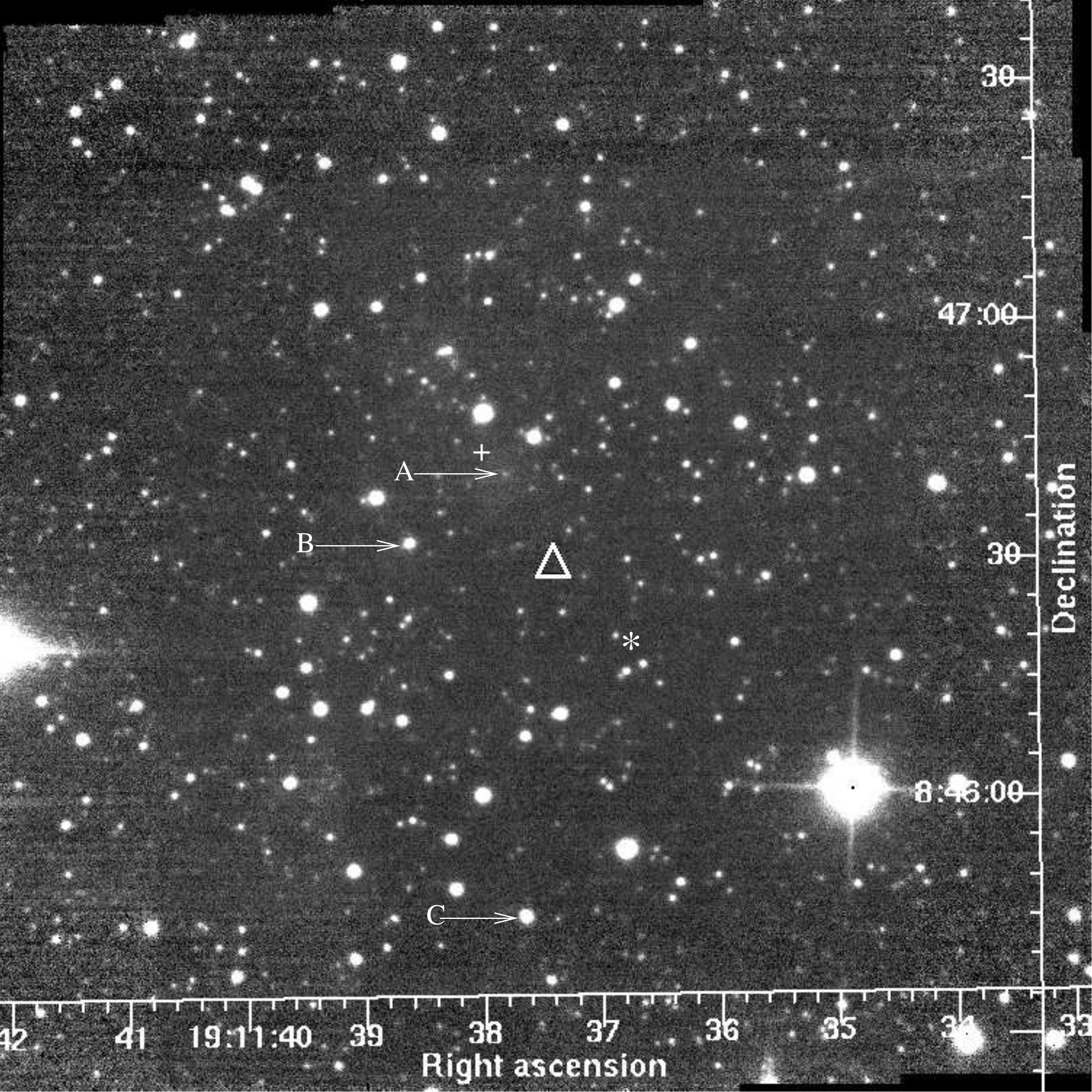}
\caption{$K$-band image of IRAS~19092+0841. ``*'' shows the 1.1-mm
position of \citet{molinari00}.}
\label{19092_K}
\end{figure}

\subsection{IRAS~19110+1045 -- (G45.07+0.13)\\ ({\small \it d =6; 8.3; 9.7\,kpc, L = (330; 588.8)\,$\times$10$^3$\,L$_{\odot}$})}
\begin{figure}
\centering
\includegraphics[width=8.10cm,clip]{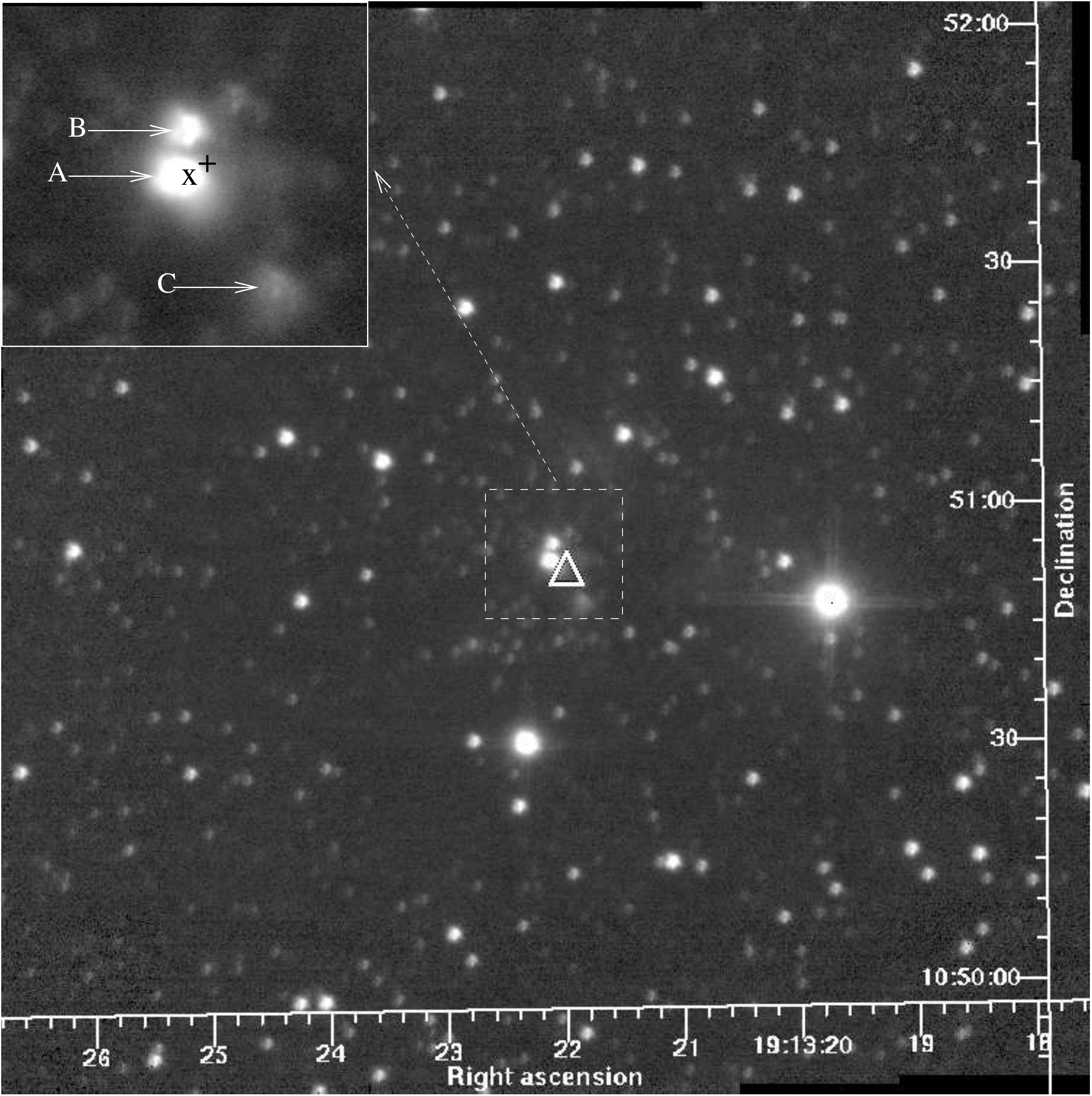}
\caption{$K$-band image of IRAS~19110+1045.  ``x'' shows the peak position
of the radio source detected by \citet{wc89b}.}
\label{G45p07_K}
\end{figure}

IRAS~19110+1045, also referred to as G45.07+0.13, is a known UCH{\sc{ii}}
region \citep{wc89b}.  Hunter, Phillips \& Menten (1997) were the first
to detect outflows from this source.  Their CO (J=6-5) map shows an
unresolved bipolar outflow (beam size 10\,arsec) whose origin is well
centred on the radio position of the UCH{\sc{ii}} region.  The estimated
mass of the outflow is 45\,M$_{\odot}$ and the length is 0.3\,pc.
Based on the relatively small extent of the outflow,  the existence of
H$_2$O maser emission from this source \citep{hofner96}, thought to
trace the earliest stages of star formation, and on the compactness of
the source (appeared to contain a single core),  \citet{hunter97} argue
that G45.07+0.13 is a younger source than the nearby G45.12+0.13.
The single core has been resolved into three sources in the
mid-infrared images of \citet{kraemer03} and previous near-infrared
imaging and spectroscopy towards this source by \citet{hanson02} 
showed it to be highly reddened (A$_K$=8).  They detected very faint
Br${\gamma}$ and H$_2$ 1-0 S(1) emission from this source.  6.7-GHz
methanol maser was detected by \citet{caswell95}, \citet{menten91} 
and \citet{szymczak00} with varying intensities suggesting a 
variable nature of the maser emission.

Fig. \ref{G45p07_K} shows our $K$-band image. Our data did not
reveal any significant emission in H$_2$ or Br$\gamma$, so
those images are not shown here.  The source labelled
``A'' ($\alpha$=19:13:22.09, $\delta$=10:50:53.4) and a
neighbouring source located $\sim$2.15\,arcsec NW
(``B'': $\alpha$=19:13:22.07, $\delta$=10:50:55.5) exhibit large
large negative residuals in the continuum-subtracted H$_2$ image.  
location of the IRAS source is 2.2\,arcsec SW of ``A'';  the the MSX
source is 1.4\,arcsec NW of ``A'' and 2.05\,arcsec SW of ``B''.  These
two sources are not resolved in the 2MASS images and the aggregate has
only upper limit magnitudes in $J$ and $H$.  Their combined colours
exhibit IR excess and place the pair in the region of reddened
YSOs (Fig. \ref{JHKcol}).  It appears that both ``A'' and ``B'' are YSOs.
These two objects are, in fact, the near-IR counterparts of the two
bright mid-IR sources ``2'' and ``3'' of \citet{debuizer05} which are
associated with the H$_2$O and OH masers.  The brighter component
``A'' (``2'' of De Buizer et al.) is associated with the unresolved
(at 2\,cm and 6\,cm) radio continuum source of \citet{wc89b} with
integrated flux densities of 594.2 and 141.9\,mJy respectively.
There is a fainter object in the vicinity, ``C'' ($\alpha$=19:13:21.83,
$\delta$=+10:50:48.2), which also exhibits negative residuals on
the continuum-subtracted images.  This object is reddened and is
detected by 2MASS only in $K_s$.  This corresponds to the mid-IR
object ``1'' of \citet{debuizer05}, which is not associated with
any maser emission.  The MIR sources were also detected at 3\,mm
by \citet{hunter97}.  \citet{vig06} detected low-frequency radio 
emission from the vicinity of these sources using the Giant 
Metrewave Radio Telescope (GMRT).  They estimate a total luminosity
of 3.3$\times$10$^{5}$ L$_{\odot}$ by integrating the observed SED
(adopting a distance of 6\,kpc from \citealt{araya02}).

With the brightness in the near-to-mid infrared and its association
with the compact radio source,  ``A'' appears to be the leading YSO
in this region.   It is probably in a UCH{\sc{ii}} stage.  ``B''
and ``C'' are also probably YSOs.  It remains to be investigated 
which one of these is the driving source of the CO outflow mapped
by \citet{hunter97}.

\subsection{IRAS~19213+1723 -- {\it Mol 103}\\ ({\small \it d = 4.12\,kpc, L = 28.2$\times$10$^3$\,L$_{\odot}$})}

IRAS~19213+1723 is one of the large sample of CO outflow sources
assembled by \citet{wu04}.  The source is reported by them as having
a bipolar outflow 1.1\,pc in length with a collimation factor 1.44.
From CO (2-1) observations, \citet{zhang01} derived the mass and
momentum of the outflow as 3.9\,M$_{\odot}$ and
24.8\,M$_{\odot}$~kms$^{-1}$ respectively.  The centre of the CO
emission peak, derived by \citet{zhang05}, is close to the IRAS
position.

IRAS~19213+1723 was included in the SiO survey of \citet{harju98},
but no detection was made.  \citet{bronfman96} do find CS (2-1)
associated with this source at a velocity of 42.1\,kms$^{-1}$,
close to the velocity at which the NH$_3$ emission was detected
by \citet{molinari96} (41.7\,kms$^{-1}$).  PAH emission from
IRAS~19213+1723 has been studied by \citet{zavagno92} in the
mid-infrared at 7.7\,$\mu$m and 11.3\,$\mu$m and by
Jourdain de Muizon, D'Hendecourt \& Geballe (1990) at 3.4\,$\mu$m.
Methanol maser emission was sought by \citet{szymczak00},
\citet{slysh99} and \citet{vanderwalt95}, though none was detected.
\citet{palla91} and \citet{brand94} detected strong H$_2$O maser
emission from this region at v$_{peak}$$\sim$-27 km s$^{-1}$,
blue-shifted considerably with respect to the core velocity
(v$_{LSR}$$\sim$+42 km s$^{-1}$).  The masers are probably
excited in a face-on flow.  The CO map of \citet{zhang05}
shows the red- and blue-shifted wings overlapping.

\begin{figure}
\centering
\includegraphics[width=8.10cm,clip]{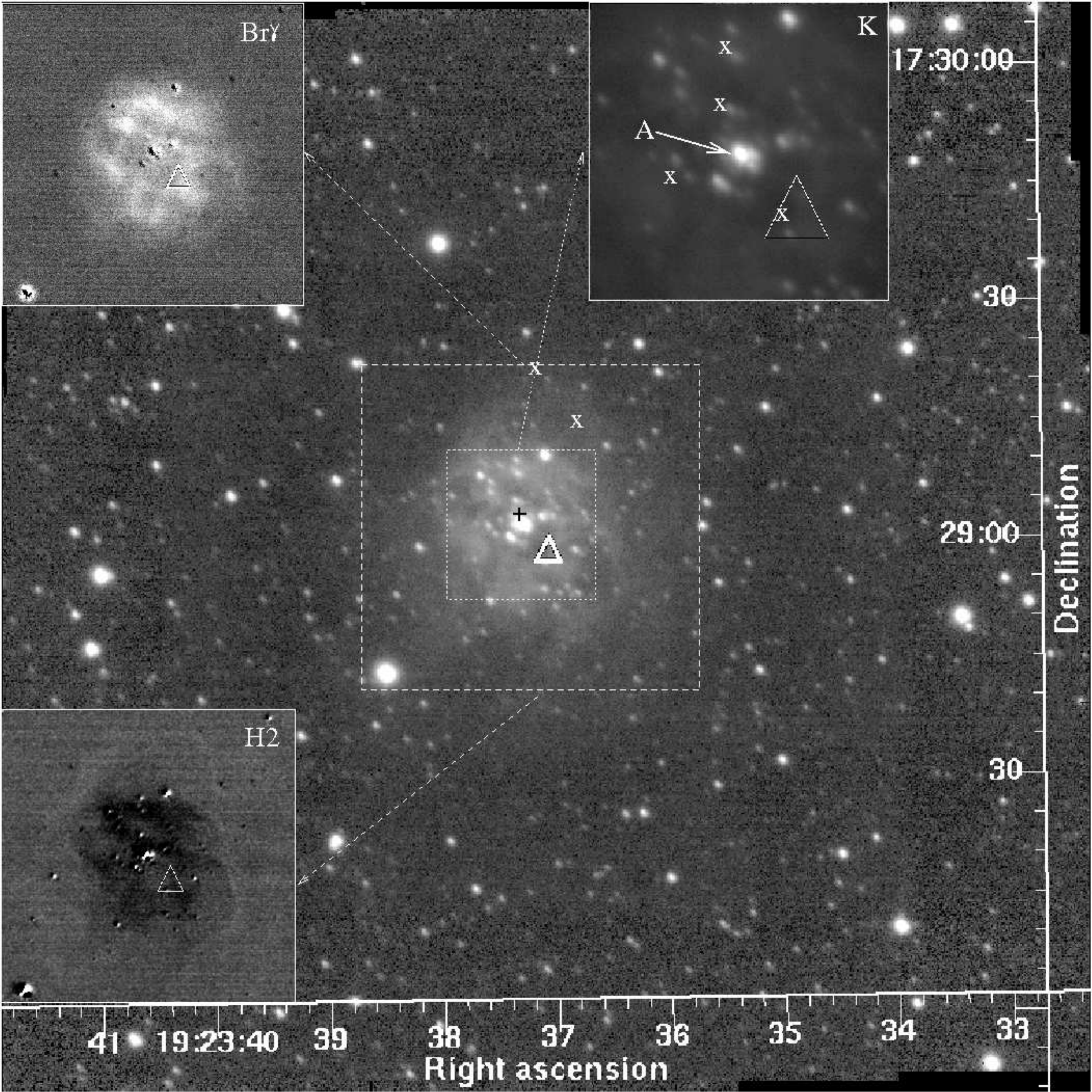}
\caption{$K$-band image of IRAS~19213+1723. The continuum-subtracted
Br$\gamma$ and H$_2$ images of the central region are shown inset
top-left and bottom-left respectively.  An expanded view of the
central region of the $K$-band image is shown inset top-right.
The six radio peaks detected at 6\,cm by \citet{molinari98} are
shown by ``x''.}
\label{19213_KBrG}
\end{figure}

Fig. \ref{19213_KBrG} shows our $K$-band image. The central
regions of the $K$ and the continuum-subtracted Br$\gamma$ and H$_2$
images are shown in the insets.  The bright source close to the IRAS
position is labelled ``A'' ($\alpha$=19:23:37.29, $\delta$=17:29:02.5).
The location of the IRAS source is $\sim$4.8\,arcsec SW of ``A'';
the MSX source is only 0.5\,arcsec NE of ``A''.   However, the 2MASS
colours of ``A'' do not exhibit any significant amount of IR excess.
It should be noticed that ``A'' has many fainter close companions
(Fig. \ref{19213_KBrG}), which would not have been resolved by 2MASS.
The $K$-band image shows copious amount of nebulosity around the
central region.  The Br$\gamma$ image shows line emission in the
nebulosity, distributed around the central source.  The H$_2$ image
does not reveal an outflow, but shows only faint filamentary emission
surrounding the region exhibiting the Br$\gamma$ emission.  This
emission is probably due to fluorescence; spectroscopy is required
to confirm whether the H$_2$ emission originates in a PDR or in a
wide angle wind.  Six spatially-resolved 6-cm radio continuum
emission cores within an extended halo of dimension
30$\times$30\,arcsec${^2}$ were detected by \citet{molinari98},
with an integrated flux density of 496.25\,mJy.  The dimensions
of the halo of radio emission roughly agrees with the diameter
of the extended emission in Br$\gamma$  ($\sim$26\,arcsec) observed
in our data. It is possible that this cluster hosts stars at different
stages of youth, with some in a UCH{\sc ii} stage.  From the current
images, we cannot identify the YSO responsible for the outflow
detected in CO.   More multi-wavelength observations are warranted.

\subsection{IRAS~19217+1651\\ ({\small \it d = 10.5\,kpc, L = 79.4$\times$10$^3$\,L$_{\odot}$})}

IRAS~19217+1651 has been studied in detail at radio and mm
wavelengths (\citealt{sridharan02, beuther02b, beuther02d}; Beuther,
Schilke \& Gueth 2004a).  \citet{sridharan02} \& \citet{beuther02d}
detected H$_2$O and CH$_3$OH masers, 3.6-cm radio continuum emission
and line emission from CO, SiO and CH$_3$CN. This source is associated
with both methanol and water masers.  The dense core mapped in CS
\citep{bronfman96, beuther02b} is also coincident with the mm peak.
High angular resolution CO (2-1) and SiO (2-1) maps of
\citet{beuther04a} show an energetic, moderately
collimated outflow (collimation factor = 3) which is interpreted as
a single flow, originating at the millimetre dust continuum peak.
At an angular resolution of 1.5\,arcsec, their 1.3-mm continuum
emission appears with a single peak $\sim$5\,arcsec north of the
IRAS position, almost  coincident with the 3-mm peak and the
32\,mJy 3.6-cm free-free continuum source.
The outflow mass in the red wing is
50.4 M$_\odot$ with 24.4 M$_\odot$ in the blue wing; the momentum is
2210 M$_\odot$ kms$^{-1}$ and the total extent, 1.6\,pc.
\citet{beuther04a} conclude that their observations allow the driving
mechanism to be the same as for low mass sources and that massive
stars may form via accretion.

Fig. \ref{19217_K} presents our $K$-band image of IRAS~19217+1651
on which an expanded view of the central region is shown in the inset.
The field does not exhibit any significant amount of emission in H$_2$
or Br$\gamma$, so these images are not shown here.  Four sources are
labelled ``A--D'' in the figure, which are located close to the
IRAS position.  MSX detected a single source 3.6\,arcsec NE of the
IRAS position (1.88\,arcsec SE of ``B'' and 3.7\,arcsec SE of ``A'').
In the field shown in the inset, the 2MASS detection is centred between
``A'' and ``B'' and thus represents the combined light from the two
objects;  it does not exhibit any IR excess (Fig. \ref{JHKcol}).
However, our continuum-subtracted H$_2$ image shows large negative
residuals for ``B'' ($\alpha$=19:23:58.80, $\delta$=16:57:40.9),
which suggests that this could be a source with a steeply rising SED
in the $K$ band.  ``C'' comprises two sources connected by an arc-like
nebulosity. ``D'' ($\alpha$=19:23:58.64, $\delta$=16:57:39.2)
is a faint nebulous patch.  ``C'' and ``D'' are not detected by 2MASS.
The centroids of the 1.3-mm, 3-mm and 3.6-cm emissions shown by
\citet{beuther04a} appear to coincide with the location of ``B''
to better than an arcsecond.  However, with the apparent offsets of
the centroids from each other, it is difficult to ascertain the
association of these with any of the near-IR sources discussed here.
It is not clear if the millimeter and centimetre emissions are from
the same YSO or represent objects at different ages.

\begin{figure}
\centering
\includegraphics[width=8.10cm,clip]{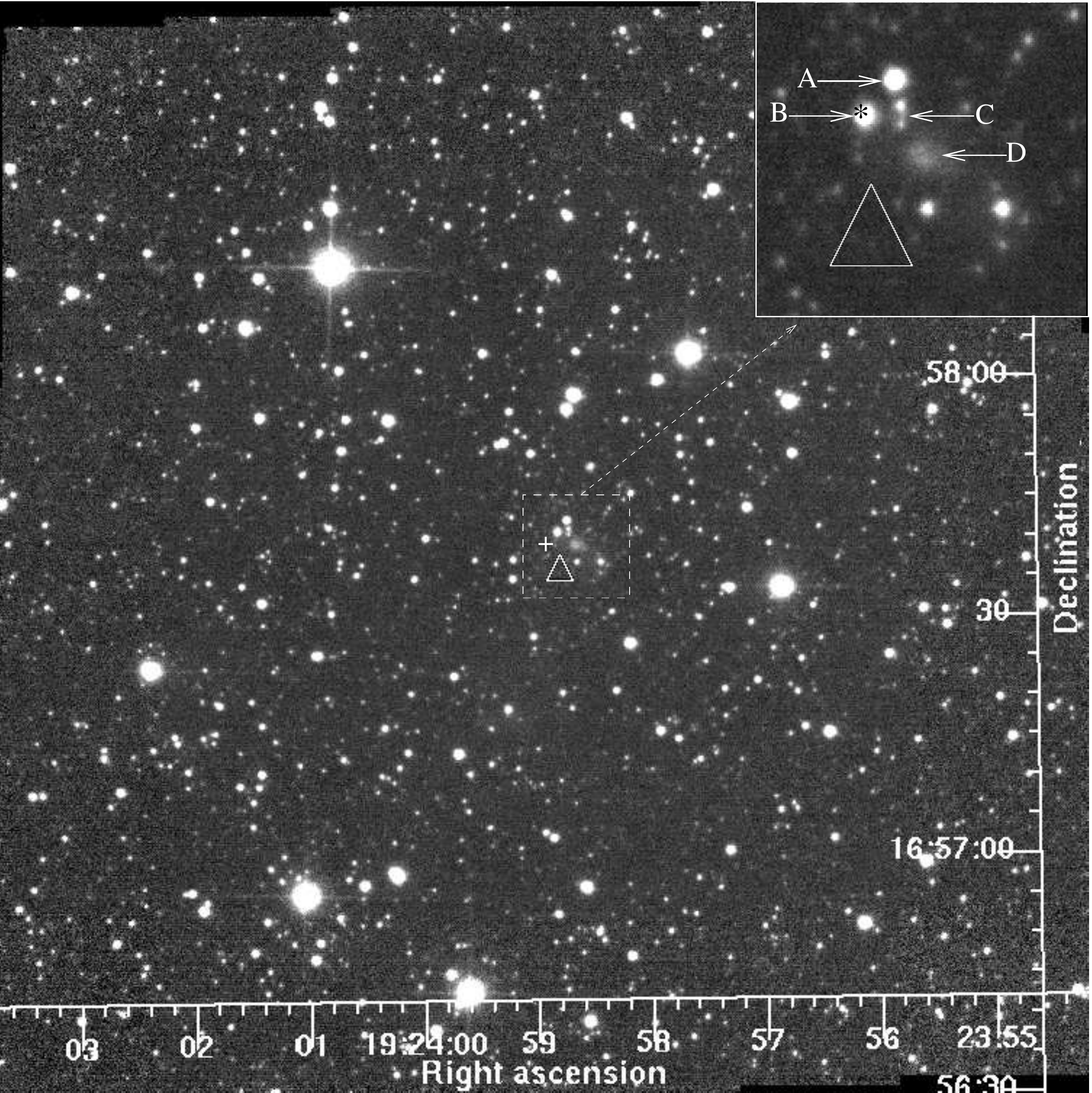}
\caption{$K$-band image of IRAS 19217+1651; an expanded view of the 
central region is shown in the inset. ``*'' shows the 3-mm peak 
position of \citet{beuther04a}; the locations of the 3.6-cm radio
continuum source,  the 22-GHz H$_2$O maser and the 6.7-GHz CH$_3$OH
maser agree with that of the mm peak.}
\label{19217_K}
\end{figure}

\subsection{IRAS~19374+2352 - {\it Mol 109}\\ ({\small \it d = 4.3\,kpc, L = 26.7$\times$10$^3$\,L$_{\odot}$})}

\begin{figure*}
\centering
\includegraphics[width=16.5cm,clip]{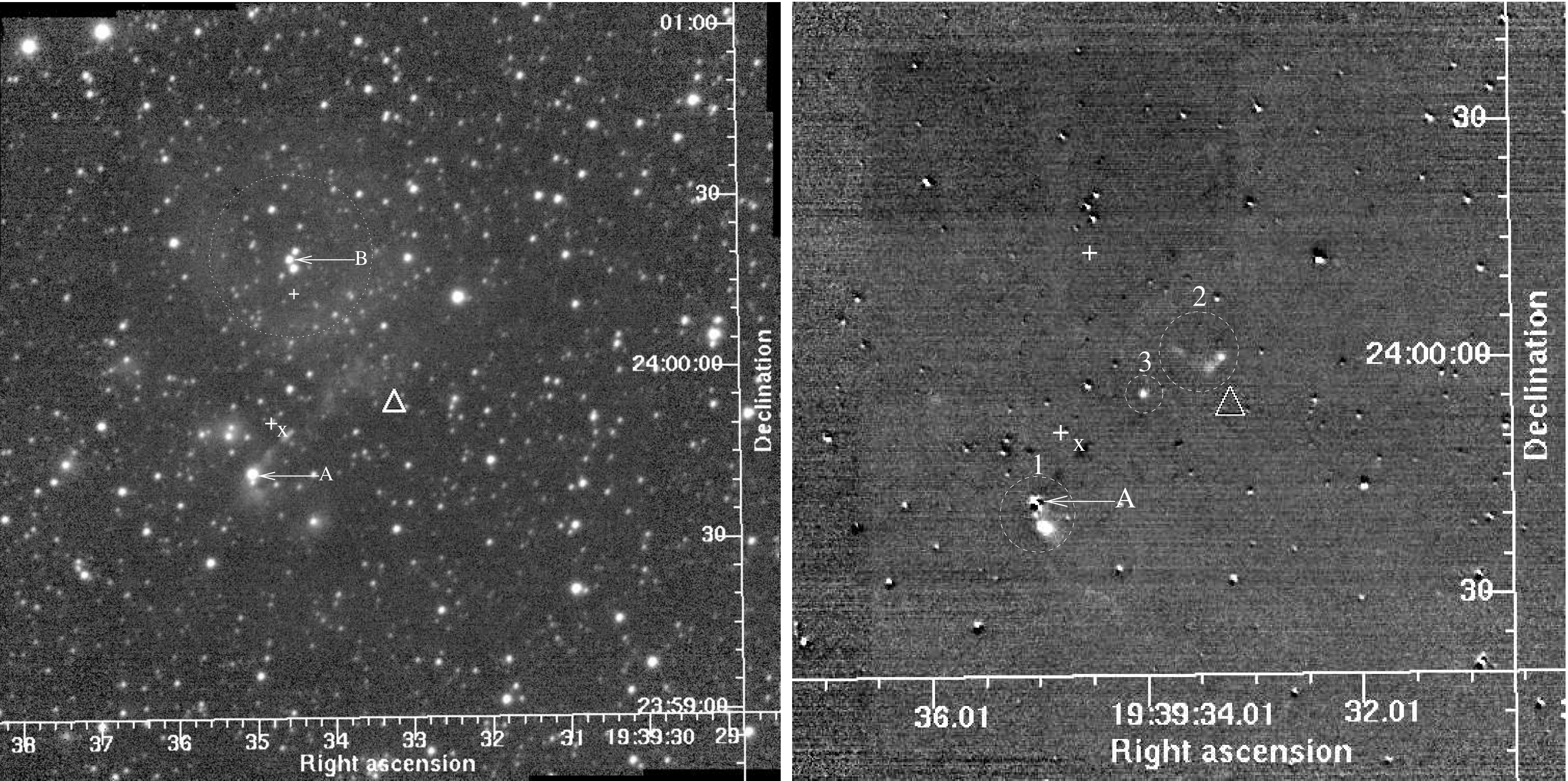}
\caption{Left: $K$-band image of IRAS~19374+2352.
Right: Continuum-subtracted H$_2$ image. The 6-cm position
of \citet{molinari98} is shown by ``x''.}
\label{19374_KH2}
\end{figure*}

IRAS~19374+2352 (G59.60+0.92) is associated with a dense core
detected in NH$_3$ emission by \citet{molinari96} and in CS by
\citet{bronfman96}.  An H{\sc ii} region was detected towards 
this source at 6\,cm by \citet{molinari98}.  H$_2$O maser 
emission was observed by \citet{palla91} and \citet{brand94}.  
\citet{schutte93} did not detect any 6.6-GHz CH$_3$OH maser
emission from this region at a 3$\sigma$ level of 3\,Jy.
\citet{watson03} estimated a kinematic distance of 4.3\,kpc
using H$_2$CO observations.  Through CO (J=2-1) observations,
\citet{zhang05} mapped a molecular outflow.

Fig. \ref{19374_KH2} shows our $K$-band image and the central
portion of our H$_2$ image. The $K$-band image shows a ring of very
faint nebulosity,  outlined by the dotted circle on the image.
The combined 2MASS colours of the three bright sources located
(the centroid is labelled ``B'') at the centre of the ring do not
exhibit any IR excess.  The fainter of the two MSX sources in this
field (26\,arcsec NE of the IRAS position) is located within this
ring.   This ring of nebulosity disappears in the
continuum-subtracted H$_2$ image and there is no line emission
detected within the ring.  However, we see some H$_2$
line emission features near the central region of the image,
which are circled on the H$_2$ image in Fig. \ref{19374_KH2}
and labelled ``1--3''.  ``1'' is observed to be closely
associated with a pair of stars at a separation of 1.5\,arcsec,
labelled ``A'' ($\alpha$=19:39:35.04, $\delta$=23:59:42.1) in
the figure.  The combined 2MASS colours of ``A'' do not exhibit
any IR excess (Fig. \ref{JHKcol}).  However, note that one of
the components of ``A'' is much fainter than the other and the
2MASS magnitudes will be dominated by the brighter component.
The brighter of the two MSX sources in the field, located
22\,arcsec SE of the IRAS position, is 9.5\,arcsec NE of ``A''.
This MSX detection is only 3.2\,arcsec NE of the 72.5-mJy 6-cm
radio continumm source of \citet{molinari98}.  Our $K$-band image
shows a faint patch of nebulosity very close to this radio source.
This nebulosity disappears in the continuum-subtracted images
and leaves a large negative residual, implying the possible
presence of dust.  ``2'' is split into two separate blobs of emission.

The H$_2$ features, along with the MSX sources, the radio source
and the CO map of \citet{zhang05}, suggest that there is more than
one outflow source in this region.  From the available photometry
we are not able to associate the outflow drivers with any
of the near-IR sources at this stage.  Even so, a collimated outflow
is inferred from our H$_2$ data.  Spatially resolved colours of ``A''
and its companion need to be obtained to establish if either is the
driving source of the outflow traced by the H$_2$ emission near ``A''.

\subsection{IRAS~19388+2357 -- {\it Mol 110}\\ ({\small \it d = 4.27\,kpc, L = 14.8$\times$10$^3$\,L$_{\odot}$})}

IRAS~19388+2357 is associated with H$_2$O maser emission 
\citep{palla91, brand94}, a radio source \citep{hughes94, 
molinari98} and dense molecular gas traced in NH$_3$ 
\citep{molinari96} and CS \citep{bronfman96}.  CH$_3$OH maser 
was also detected from this source \citep{schutte93,
slysh94}.  No near-IR  observations were found in the 
literature.  \citet{zhang05} mapped a CO outflow.  The 
centroid of their CO emission is 29\,arcsec south of the 
IRAS position.

\begin{figure}
\centering
\includegraphics[width=8.10cm,clip]{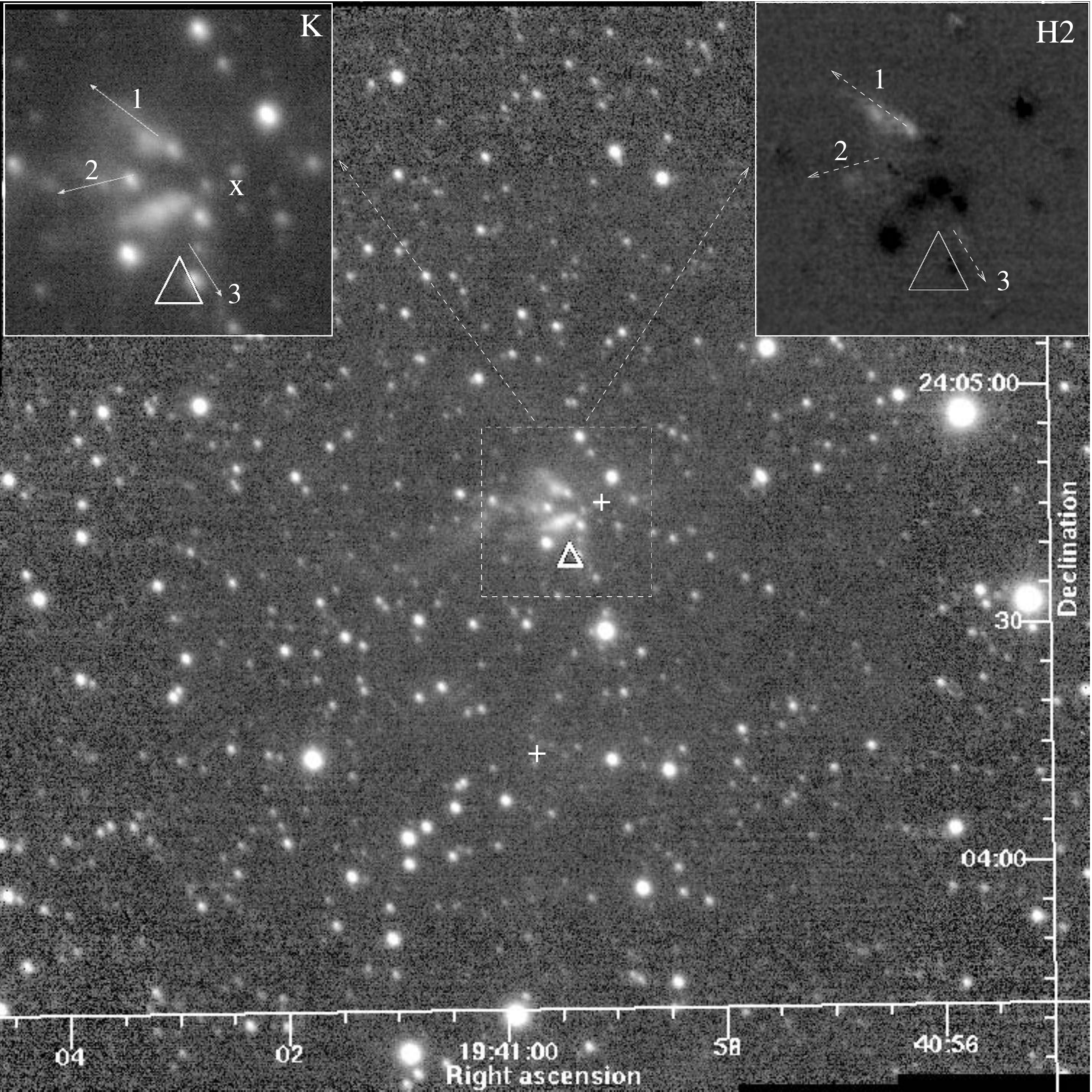}
\caption{$K$-band image of IRAS~19388+2357. The inset on the top-left
shows an expanded view of the central region of the $K$-band image, and
that on the top-right shows the central region of the continuum-subtracted
H$_2$ image.  ``x'' shows the 6-cm position of \citet{molinari98}.}
\label{19388_KH2}
\end{figure}

Fig. {\ref{19388_KH2}} shows our $K$-band and continuum-subtracted
H$_2$ images.  We see a set of aligned features embedded in nebulosity
in the $K$-band.  Due to the presence of the nebulosity, 2MASS does not
appear to detect the sources close to the centre; therefore the 2MASS
data are not included in Fig. \ref{JHKcol}.  Our H$_2$ image shows a 
set of three emission features;  three arrows are drawn in the 
directions of these features and are labelled ``1'', ``2'' and ``3''.  
As seen from the continuum-subtracted H$_2$ image, the sources
close to the centre appear very much reddened, since they show
large negative residuals upon continuum-subtraction.  It is possible
that some of these are the near-IR counterparts of the YSOs
responsible for the outflows from this region implied by the H$_2$
jets.  The MSX mission detected three objects in our two arcminute
field of view; the one closest to the IRAS position (7.85\,arcsec NW)
is the brightest and shows a rising SED,  typical of YSOs, and is
located well within the central cluster.  The second MSX source is
24.7\,arcsec SE of the IRAS position (24.3\,arcsec south, 3.8\,arcsec
east), close to the centre of the CO emission peak of \citet{zhang05}.
However, this MSX source is weak and is not detected in the 21.34-$\mu$m
band of MSX.  Hence it remains to be investigated if the CO outflow
is produced by a different source located south of the IRAS position,
as implied by the 29\,arcsec offsets of the centroid of the outflow
mapped by \citet{zhang05} (note that this offset is is only of the
order of their beamsize [Table \ref{resolutions}]) and the weak MSX
detection, or by one or more of the sources in the central cluster
itself as suggested by our H$_2$ image.

The 6-cm radio source reported by \citet{molinari98} is faint
(2.96$\pm$0.06\,mJy) and is separated by only 7\,arcsec from the
IRAS position.  This region appears to host more than one YSO, some
of which are in a pre-UCH{\sc ii} phase.  More observations,
especially in the mid-IR, are required to positively identify the 
sources driving the outflows.

\subsection{IRAS~19410+2336\\ ({\small \it d = 2.1, 6.4\,kpc, L = (10, 100)\,$\times$10$^3$\,L$_{\odot}$})}

\begin{figure*}
\centering
\includegraphics[width=16.5cm,clip]{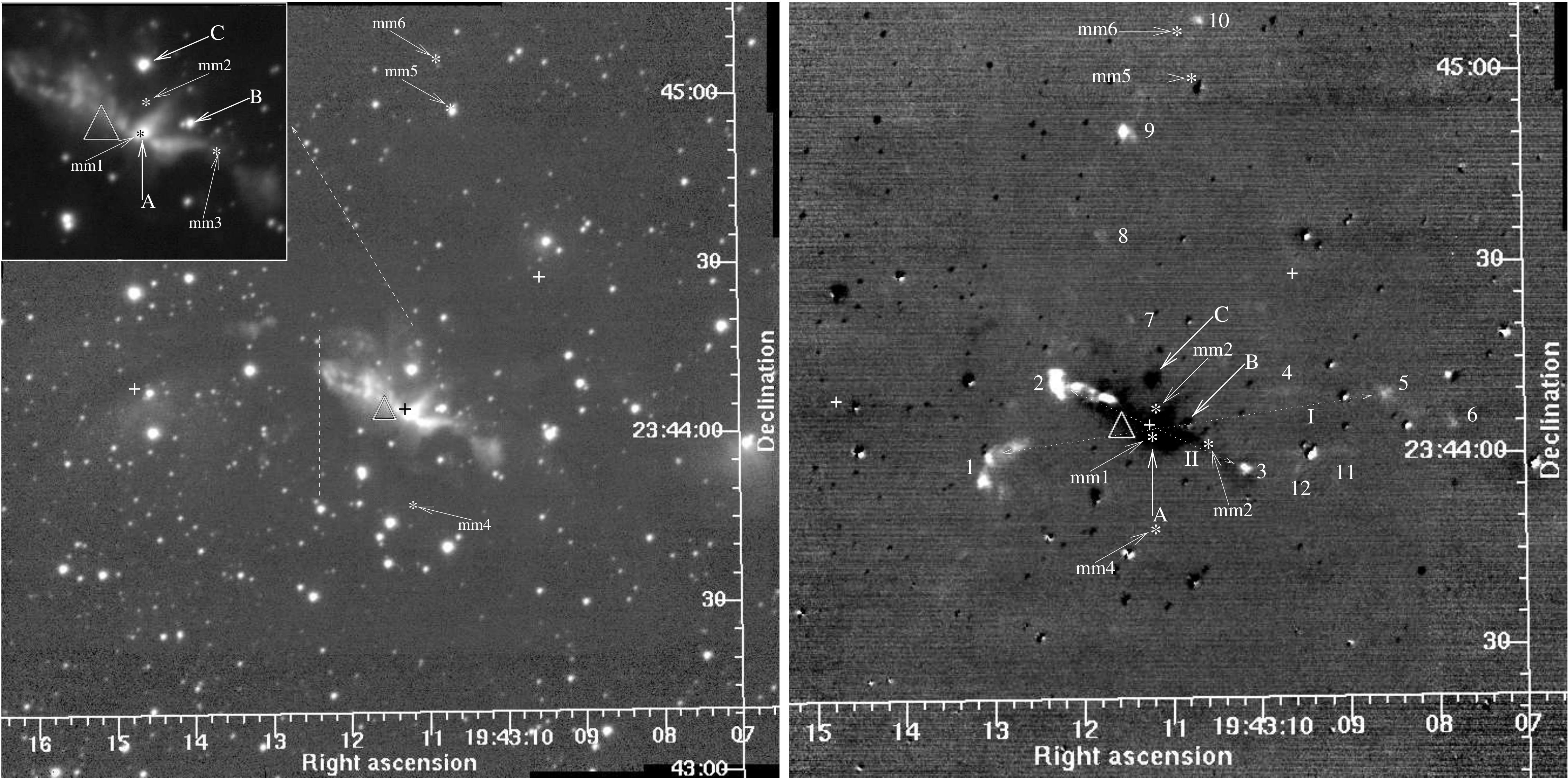}
\caption{The left panel shows the $K$-band image of IRAS~19410+2336 and the
right panel shows the continuum-subtracted H$_2$ image.  The H$_2$ image is
smoothed with a Gaussian of 2-pixel FWHM. The 2.6-mm positions (``mm1--mm6'')
of \citet{beuther03} are shown by ``*''.}
\label{19410_KH2}
\end{figure*}

IRAS~19410+2336 is a good example for HMYSOs driving
well-collimated outflows.  \citet{bronfman96} detected a dense core
towards IRAS~19410+2336 in the CS(2-1) transition and
\citet{sridharan02} detected NH$_3$ emission.  The observations of
Sridharan et al. and \citet{beuther02d} revealed water maser emission.
Strong 6.7-GHz methanol maser emission has been observed by many
\citep{szymczak00, sridharan02, beuther02d}.   Sridharan et al. also
detected CO emission line wings implying outflows;  the 3.6-cm radio
continuum emission detected by them was very faint (1\,mJy) implying
a possible pre-UCH{\sc{ii}} stage for this source.  Beuther, Schilke
\& Stanke (2003) observed this object in $^{12}$CO (J=1-0) using
Plateau de Bure Interferometer (PDBI), as well as in H$_2$ and $K'$.
Their observations resolved seven (or possibly nine) bipolar outflows.
They estimate an accretion rate of the order of
10$^{-4}$ M$_\odot$ year$^{-1}$ for these YSOs, which is more than
the accretion rates of typical low-mass stars.  CS observations at
high angular resolution using PDBI \citep{beuther04b} shows CS
tracing some of the outflows mapped in CO.

Fig. \ref{19410_KH2} shows our $K$ and H$_2$ images.
Three sources are labelled on the figure,
``A'' ($\alpha$=19:43:11.20, $\delta$=23:44:03.9),
``B'' ($\alpha$=19:43:10.81, $\delta$=23:44:04.9) and
``C'' ($\alpha$=19:43:11.19, $\delta$=23:44:11.8).
Both ``A'' and ``B'' are resolved and are found to have fainter companions
located $\sim$0.9\,arcsec SE of the brighter ones. The coordinates given
here are those of the brighter components.  ``A'' and ``B'' are separated
by only 5.5\,arcsec and would not be resolved by either IRAS or MSX.
``A'' is deeply embedded and is well detected by 2MASS, though only in $K_s$.
The derived colours using the magnitude limits show that it has a large
amount of reddening and IR excess (Fig. \ref{JHKcol}).  ``B'' is detected
by 2MASS in all three bands with large reddening and IR excess (although
less than those of ``A'').  The brightest MSX source is located 1.8\,arcsec
NE of ``A'';  the IRAS position is 5\,arcsec NE of ``A''.  Twelve H$_2$
emission features are labelled on the H$_2$ image.  The labels of the
H$_2$ emission features ``1--10'' are chosen to be consistent with those
used by \citet{beuther03} for ease of comparison.  The location of ``A''
close to the centroid of the observed H$_2$ emission features ``2'' and
``3'', its high reddening and excess and the close association with the
MSX source imply that this is the most likely candidate for the luminous
YSO driving the jet, traced by ``2'' and ``3''.  This jet, shown by the
dotted arrow and named ``II'' on Fig. {\ref{19410_KH2}}, is at an angle
of 65$^{\circ}$ with a collimation factor of $\sim$4.1.  Similarly, ``B''
and one of the components of ``A'' are other strong candidates for the
YSO driving the outflow revealed by the H$_2$ features ``1'',  ``4--6''
and ``11--12''.  The lateral spread in the locations of these features
shows that they represent more than one bipolar jet or a precessing jet.
If ``1'' and ``4--6'' are due to an outflow from ``B'' in the direction
of the dotted arrow labelled ``I'' on  Fig. {\ref{19410_KH2}}, the outflow
direction is $\sim$99.5$^{\circ}$ with a collimation factor of $\sim$3.8.
``C'' is also a deeply embedded object with large reddening and excess.
It is located 8.1\,arcsec north of ``A''. This object  is well detected
only in $K_s$ by 2MASS and exhibits deep reddening and a significant excess.

In addition to the line emission features mentioned above, there
are several other features. Some of the brighter ones are labelled
``7--10''.  These features could be driven by other YSOs in the
field,  like ``C''.  We have not attempted to explain all the observed
H$_2$ emission features in this complex region, which will instead be
addressed in a forthcoming paper.

Single dish 1.2-mm continuum observations of \citet{beuther02b}
revealed two massive cores, each associated with a bipolar molecular
outflow as revealed in their maps in the CO(2-1) and in the CO maps at
high angular resolution observed by \citet{beuther03}.  The blue- and
red-shifted lobes of the CO emission around the southern core roughly
trace the direction of the bright H$_2$ features ``1'' and  ``4--6''
showing that the outflow mapped in CO is driven by the jet detected
in H$_2$.  The southern core appears centred close to the IRAS position
and is resolved into four mm sources at 2.6\,mm by \citet{beuther03}.
We have plotted the positions of these four 2.6-mm peaks near the
sourthern core, ``mm1--mm4'', on Fig. \ref{19410_KH2}. The position
of ``mm1'' agrees well with that of ``A'', the object producing the 
outflow defined by the H$_2$ features ``2'' and ``3'' in the direction
``II''.  Except for one methanol maser spot
which is located 1.6\,arcsec SW of ``A'', the locations of the
remaining one methanol maser and the two water maser spots given by
\citet{beuther02d} are within an arcsecond of ``A'' (their 
positional offsets are less than 1\,arcsec; Table \ref{resolutions}) 
implying that these are produced by ``A'' or its companion.

\subsection{IRAS~20050+2720  -- {\it Mol 114}\\ ({\small \it d = 0.73\,kpc, L = 0.388\,$\times$10$^3$\,L$_{\odot}$})}

IRAS~20050+2720 is associated with an extremely high velocity
molecular jet, a dense core and a compact cluster. The outflow
in this region appears to be the superposition of more than one,
all of which radiate outward from the IRAS position (Bachiller,
Fuente, \& Tafalla 1995).  \citet{chen97} obtained $JHKLM$
photometry of the cluster, resolving a cluster population of
over a hundred sources.

The dense core has been mapped in various gas and dust tracers
(NH$_3$ - \citealt{molinari96}; CS - \citealt{bronfman96};
450, 850\,$\mu$m \& 1.3\,mm - \citealt{chini01}). Within a
5$\times$5\,arcmin$^2$ region, \citet{ridge03} resolved a
number of cores enveloped by a diffuse cloud that extends roughly
north-south.  The brightest CO emission peak mapped by them
coincides with the IRAS position. From the IRAS data, Ridge et al.
derived a FIR luminosity of $\sim$230\,L$_{\odot}$, adopting a
distance of 700\,pc.  IRAS~20050+2720 is also associated with
water maser emission \citep{palla91}.  No strong compact radio
emission is detected in this region.  The radio observations
at 6\,cm by \citet{molinari98} reveal no sources within
120\,arcsec of the IRAS position.  \citet{wilking89} detected a
faint 1.4\,mJy extended source at 6\,cm.  The outflow from this
region was mapped in CO by \citet{zhang05}.  It appears to be
oriented roughly EW, although, from the morphology, more than
one bipolar outflow appears to be contributing to their CO maps.
The centroid of their outflow is offset 29\,arcsec west and
north from the IRAS position.  Similarly, the
high-velocity component of the outflow mapped in CO by
\citet{bachiller95} is oriented roughly EW.  Their map also
is suggestive of the presence of multiple outflows in this region.

\begin{figure*}
\centering
\includegraphics[width=16.5cm,clip]{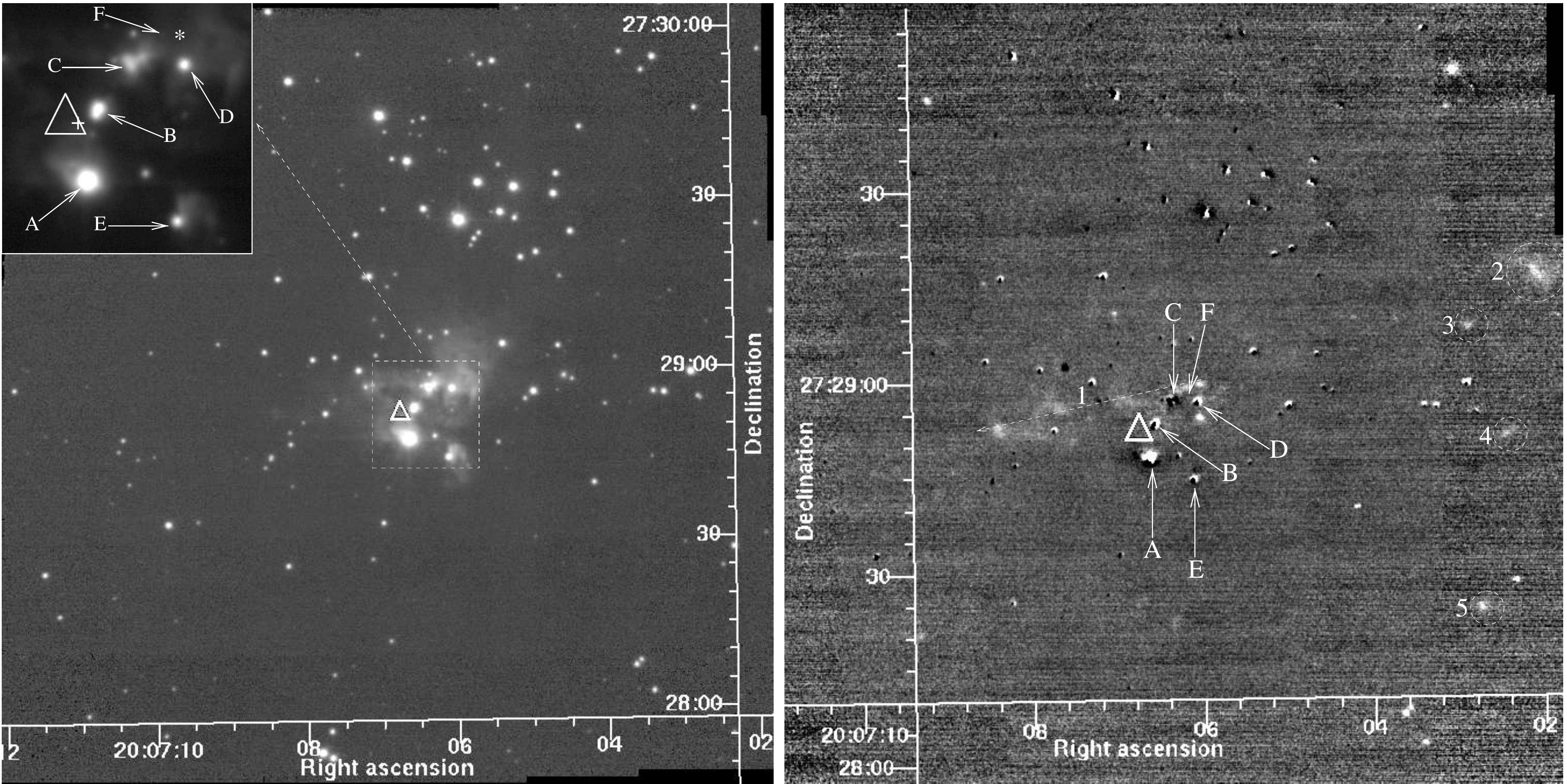}
\caption{Left: $K$-band image of IRAS~20050+2720.  ``*'' shows
the mm and sub-mm continuum peak ``MMS1'' of \citet{chini01}.
Right: Continuum-subtracted H$_2$ image, smoothed with a 2-pixel
FWHM Gaussian.}
\label{20050_KH2}
\end{figure*}

Fig. \ref{20050_KH2} displays our $K$-band image and the central region
of our H$_2$ image.  The continuum-subtracted H$_2$ image reveals at least
one well collimated outflow, the direction of which is shown by the arrow
on the image (labelled ``1'').  This outflow is roughly oriented EW
with a position angle of 101$^{\circ}$; it has a collimation factor of
$\sim$4.  The direction of the H$_2$ outflow is consistent
with the flow direction inferred from the CO maps.  Additionally,
there are some faint features detected in the H$_2$ image, circled and
labelled ``2--5'' in Fig. \ref{20050_KH2}. ``2'' and ``3'' are possibly
part of the same outflow detected here, but we can't say for sure
from the H$_2$ images alone. If these two features are part of the
same outflow described above, the collimation factor would be much higher.
The presence of the other features support the presence of
multiple outflows in this field.

Six reddened sources, located near the centre of the field and embedded
in nebulosity, are labelled ``A--F'' on Fig. \ref{20050_KH2}. All these
sources are not detected well in all three bands of 2MASS.  So, we have
used the $JHK$ magnitudes of \citet{chen97} to plot them in the
colour-colour diagram (Fig. \ref{JHKcol}).  Source ``F'' is detected by
Chen et al. only in $L$ and $M$.  Hence, that object is not shown in
Fig. \ref{JHKcol}.  Sources ``A--E'' show large and comparable reddening.
The source ``A'' (``2'' of Chen et al.; $\alpha$=20:07:06.63,
$\delta$=27:28:47.7) exhibits mild IR excess.  Both ``B'' and ``E''
(``5'' and ``1'', respectively, of Chen et al.) do not exhibit IR
excess.  ``B'' is resolved into two components  in our image.
``D'' (``7'' of Chen et al.; $\alpha$=20:07:06.05,  $\delta$=27:28:56.6)
exhibits mild excess.  
``C'' (``6'' of Chen et al.; $\alpha$=20:07:06.35, $\delta$=27:28:56.9)
has the highest IR excess among ``A--E''.  ``C'' is resolved into at least
four components in our image.  Chen et al. detected an extremely red
source (their ``8'') which was seen only in their $L$ and $M$. We have
a weak detection at its location;
``F'' ($\alpha$=20:07:06.18, $\delta$=27:28:59.1) is probably the $K$
counterpart of ``8''.  The IRAS position is 5.3\,arcsec NE of ``A'' and 
the MSX source is only 1.1\,arcsec away from the IRAS position.
From the location of the outflow detected in H$_2$ and the locations 
of the IRAS, MSX and $K$-band sources, ``C'', ``D'' or ``F'' could 
be the driving source of the outflow. The continuum source mapped by
\citet{chini01} at sub-mm and mm wavelengths is only 4.15\,arcsec NW 
of ``C'',  2.3\,arcsec NE of ``D'' and nearly coincides with the position 
of ``F''.  All three IR sources fall within their beam-size of 
8.3--10.7\,arcsec at these wavelengths.  From our H$_2$ image, 
``F'' and ``C'' are the most probable candidates for the driving
source of the jet in the direction of ``1''.

\subsection{IRAS~20056+3350 - {\it Mol 115}\\ ({\small \it d = 1.67\,kpc, L = 4.0\,$\times$10$^3$\,L$_{\odot}$})}

A number of groups detected H$_2$O maser emission from this source
\citep{palla91, brand94, jenness95}. No CH$_3$OH maser emission
has been reported. The dense core was detected in NH$_3$ emission
\citep{molinari96} and in CS(2-1) emission \citep{bronfman96}.
VLA radio continuum observation by \citet{molinari98} did not
detect any emission from this source, although \citet{jenness95}
report very faint emission at 3.6\,cm (from two locations separated
by $\sim$1.35\,arcsec with flux densities 0.93\,mJy and 0.6\,mJy,
respectively).  \citet{jenness95} detected the source at 450\,$\mu$m
and 800\,$\mu$m as a single core with a faint extension in the SW
direction, the peak position matching reasonably well with the
location of the H$_2$O maser and their radio sources and located
close to the IRAS position.  They also detected the source in
C$^{18}$O (J=2-1) at the same v$_{LSR}$ as the H$_2$O maser emission.
IRAS~20056+3350 was also detected in $^{13}$CO by \citet{casoli86} 
at a similar radial velocity.  A bipolar outflow centred on the 
IRAS source has been mapped in this region by \citet{zhang05}.

\begin{figure}
\centering
\includegraphics[width=8.10cm,clip]{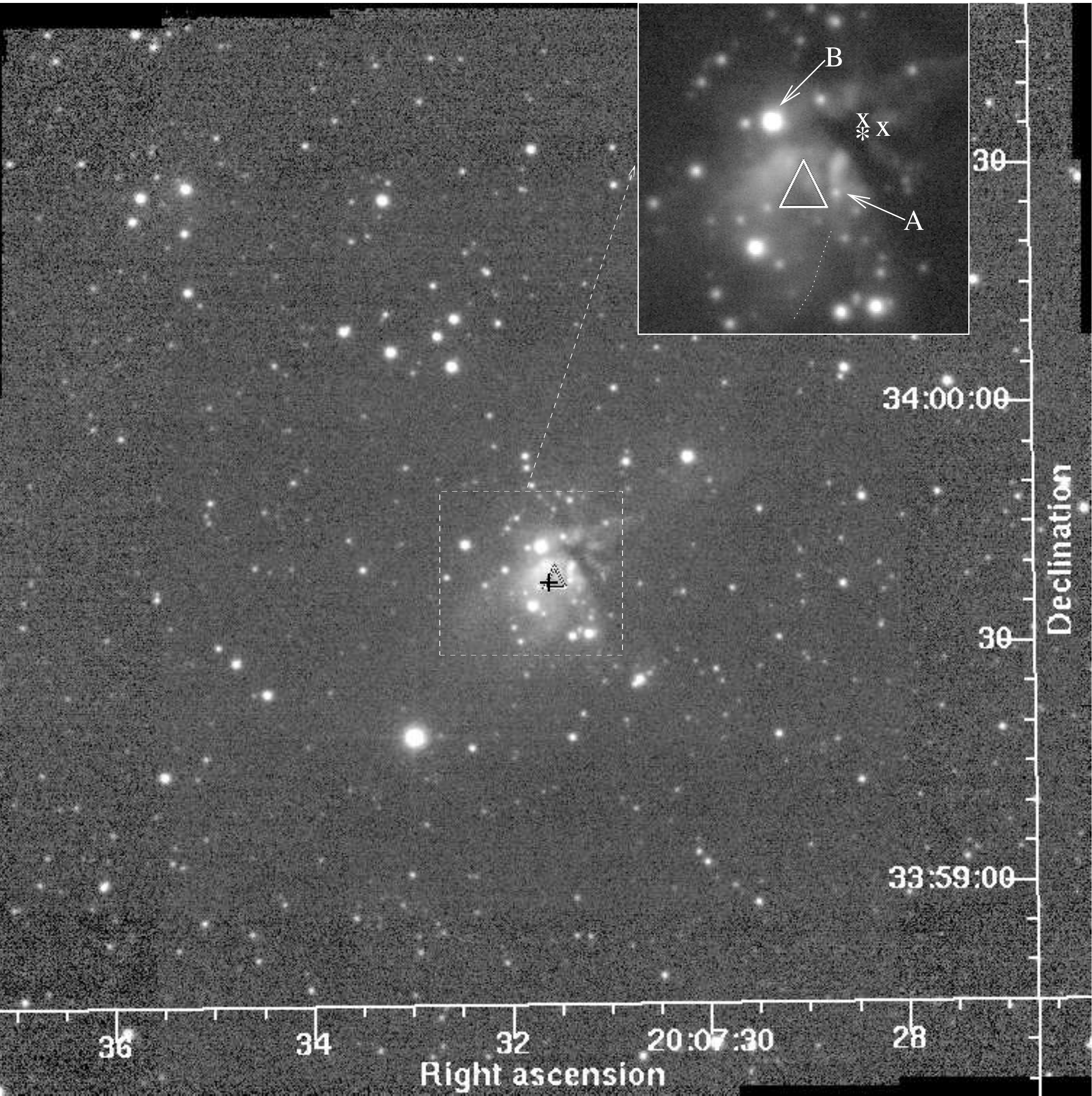}
\caption{$K$-band image of IRAS~20056+3350.  An expanded view of the
central region of the $K$-band image is shown in the inset.  3.6-cm
sources of \citet{jenness95} are shown by ``x''. The peak position of
the 800-$\mu$m source detected by them is show by ``*''.
}
\label{20056_K}
\end{figure}

Fig. \ref{20056_K} shows our $K$-band image of the region.  The image
reveals an IR cluster towards the centre of the field and a significant
amount of nebulosity. Most of the nebulosity disappears in the
continuum-subtracted narrow-band images.  The inset in Fig. \ref{20056_K}
shows an expanded view of the central region with better contrast.
The H$_2$ and Br$\gamma$ images did not show any significant amount of
line emission. Hence, these images are not shows here.  There is a
tentative detection of very faint H$_2$ emission along the dotted line
in the inset of Fig. \ref{20056_K}. This needs to be verified through
deeper imaging in H$_2$.  The object labelled
``B'' ($\alpha$=20:07:31.68, $\delta$=35:59:42.4) is well detected by
2MASS. It shows a small level of IR excess in our colour-colour diagram
(Fig. \ref{JHKcol}).  Most of the other bright stars embedded in the
nebulosity detected by 2MASS have only upper limits of magnitudes listed
in the catalogue.  A major fraction of the fainter objects were not
detected in the 2MASS survey.  Especially noteworthy among those objects
is the one labelled ``A''($\alpha$=20:07:31.38, $\delta$=35:59:38.2).
This object is located $\sim$2\,arcsec SW of the IRAS position and has
a ``coma-shaped'' nebulosity associated with it.  The centroid of this
structure is $\sim$1.3\,arcsec north
of ``A'' and is $\sim$2 arcsec in extent. It appears that none of the
bright objects are related to the IRAS source.  The near IR counterpart
of the IRAS source could be one of the fainter objects not
detected by 2MASS, or an object which is deeply embedded and is not
detected here.  Note that the 3.6-cm sources detected using
the VLA \citep{jenness95} are located close to the dark lane in
Fig. \ref{20056_K}.
This object appears to be an intermediate mass YSO.
We do not detect any signs of collimated outflows in our H$_2$ image.

\subsection{IRAS~20062+3550 -- {\it Mol 116}\\ ({\small \it d = 4.9\,kpc, L = 3.2\,$\times$10$^3$\,L$_{\odot}$})}

A dense core was mapped in this region using the emission lines of
NH$_3$ \citep{molinari96} and CS \citep{bronfman96}.  \citet{palla91}
and \citet{brand94} detected H$_2$O maser emission.  The source shows
variable CH$_3$OH maser emission \citep{slysh99, szymczak00, galt04}.
Search for 6-cm radio emission  by \citet{molinari98} did not yield
any detection.  The region hosts a well defined bipolar outflow,
which was detected in CO by  \citet{zhang05}.

\begin{figure*}
\centering
\includegraphics[width=16.5cm,clip]{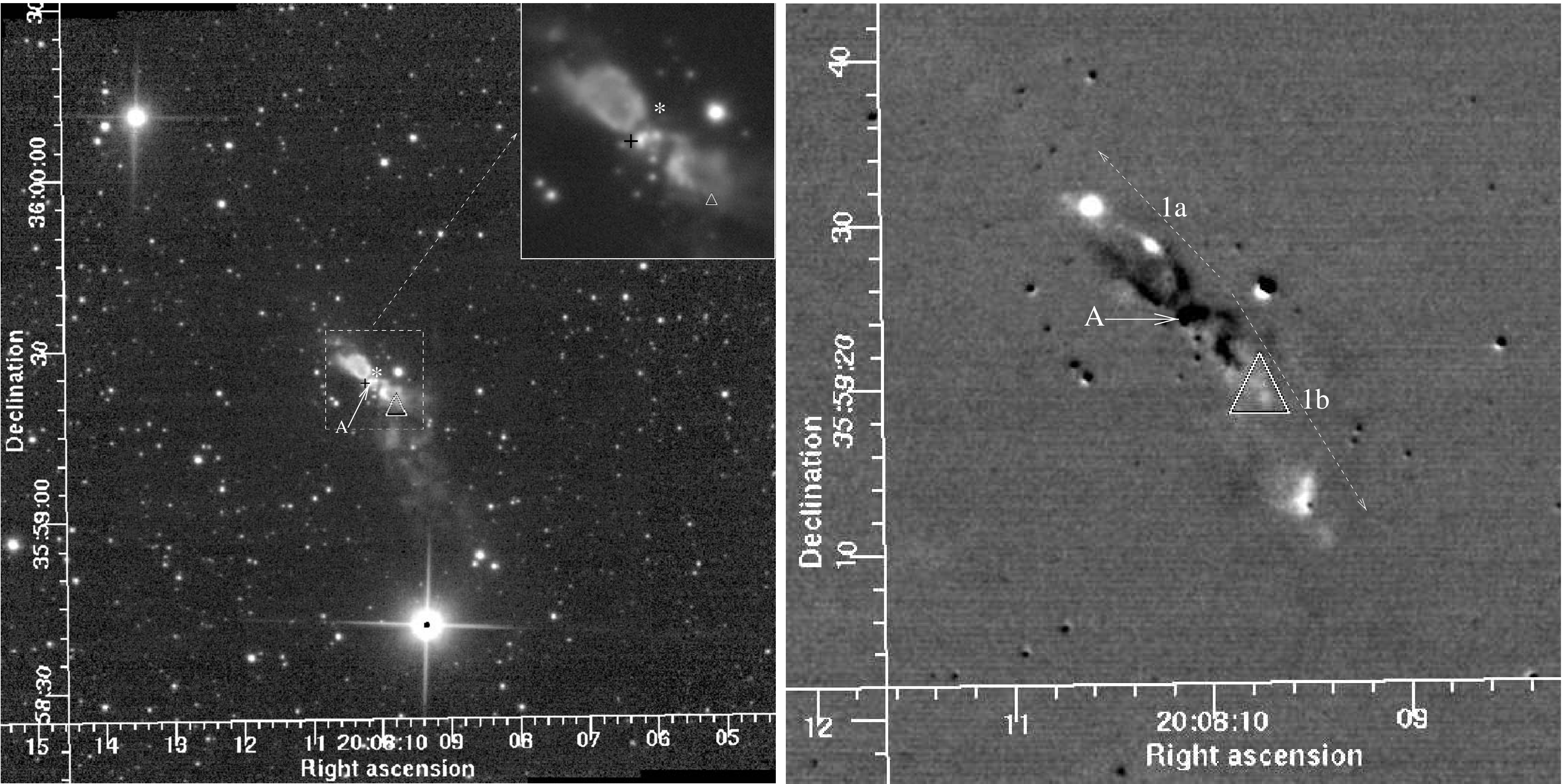}
\caption{The left panel shows the $K$-band image of IRAS~20062+3550.
``*'' shows the 3.4-mm continuum position of \citet{molinari02}.
The right panel shows the continuum-subtracted H$_2$ image.}
\label{20062_KH2}
\end{figure*}

Fig. \ref{20062_KH2} shows our $K$ and H$_2$ images.  Our near-IR
images unveil a spectacular outflow in the NE-SW direction from
this source.  The outflow appears to emanate from an infrared source
``A'' ($\alpha$=20:08:10.13, $\delta$=35:59:24.2), which is located
$\sim$6\,arcsec NE of the IRAS position.  This offset is consistent
with the NE offset of the centroid of the outflow given by
\citet{zhang05}. ``A'' is embedded in nebulosity and consists of two
components at a separation of 1.1\,arcsec and an angle of
$\sim$106$^{\circ}$ from the fainter to the brighter component;  
the coordinates given here are of the brighter of the two.
We also see very faint objects in the vicinity of the pair.
Multiwavelength imaging with higher spatial resolution is
required to understand if more than two components are present.
``A'' is detected only in $K_s$ in the 2MASS observations.  
In our $K$-band image,  the outflow
is roughly hourglass-shaped.  The NE lobe of the  outflow, in the
direction of the dashed arrow labelled ``1a'' in the 
continuum-subtracted H$_2$ image, is at an angle of 43$^\circ$
and extends up to 10\,arcsec from ``A''.  The southwest
lobe of the outflow (in the direction of ``1b'')
is inclined at 212$^\circ$ east of north, extends
up to 16\,arcsec in the H$_2$ image and shows a
bow shock.  The NE lobe also shows a bright blob in H$_2$ at its NE
extreme, which could be the emission from the apex of the bow shock.
The direction of the jet revealed here in H$_2$ agrees with the
direction of the CO outflow mapped by \citet{zhang05}.

\citet{molinari02} observed this object in HCO$^+$(1-0)
and resolved four distinct cores with a complex velocity structure.
They also detected two of these cores in H$^{13}$CO$^+$ and one in
continuum emission at 3.4\,mm.  Upon continuum-subtraction, our
H$_2$ image exhibits large negative residuals on ``A'',  along the
edges of the NE lobe of the hourglass, and in some regions of the
SW lobe.  This is likely due to emission from dust dominating in
these regions, which results in a steeply rising SED within the
pass band of the $K$ filter.

No emission was detected from the outflow or from the central
core in the Br$\gamma$ image.  Most of the nebular emission in
$K$ towards the South of the bipolar emission is subtracted out
upon continuum subtraction.  The MSX position is 0.82\,arcsec
SE of ``A''.  These observations suggest that ``A'' is the
near-IR counterpart of the YSO.

\subsection{IRAS~20126+4104 -- {\it Mol 119}\\ ({\small \it d = 1.7\,kpc, L = 10.0\,$\times$10$^3$\,L$_{\odot}$})}

IRAS~20126+4104 is a well studied luminous YSO.  Still, the
intricacies of this field are far from being unveiled.  It is 
embedded in a massive core of 230\,M$_{\odot}$, which was 
detected in NH$_3$ by \citet{estalella93} with a diameter 
of $\sim$0.4\,pc (see also \citealt{molinari96}).  The core 
was also detected in CS \citep{bronfman96}.

A NW-SE bipolar jet has been detected in H$_2$ $v$=1-0 S(1)
\citep{ayala98, cesaroni97, caratti08} and SiO (2-1) \citep{cesaroni99a}
lines.  A compact molecular outflow, the direction of which agrees very
well with that of the jet traced in H$_2$ and SiO ($\sim$120$^{\circ}$),
has been mapped in HCO$^{+}$  \citep{cesaroni97}. \citet{zhang99} discovered
shock-excited NH$_3$ emission tracing the jet revealed in SiO and H$_2$.
However, the large scale outflow mapped in CO is oriented roughly
north-south ($\sim$171$^{\circ}$;  \citealt{shepherd00}).  Recently, high
velocity bipolar outflows have been mapped in CO by \citet{lebron06}.
The direction of the CO outflow  is roughly traced by the large
scale arrangement of all the H$_2$ knots detected throughout the
region covered by the CO maps, with the innermost H$_2$ knots
tracing the SiO jet \citep{shepherd00}.  Detailed near-IR spectroscopic
study of the jet by \citet{caratti08} showed that the outflow
is like a scaled-up version of the outflows from low-mass YSOs.

The continuum emission from the sources on which the outflow is
centred has been detected from infrared to millimetre wavelengths
(e.g. \citealt{sridharan05, debuizer07, cesaroni99a, shepherd00}).
A Keplerian circumstellar disc was detected in  NH$_3$, CS and
CH$_3$CN \citep{zhang98, cesaroni97, cesaroni99a, cesaroni05}
with the disc plane oriented roughly perpendicular to the direction
of the jet/outflow detected in SiO, H$_2$, and HCO$^{+}$ by
\citet{cesaroni99a} and \citet{cesaroni97}.

The source exhibits methanol maser (e.g. \citealt{slysh99,szymczak00},
Minier, Conway \& Booth 2001; \citealt{galt04,kurtz04})
and water maser emission (e.g. \citealt{comoretto90, palla91,
tofani95, moscadelli00, moscadelli05}).  The location of the water
maser spots mapped by \citet{tofani95} have good positional agreement
with the millimetre continuum source of \citet{cesaroni97},
\citet{cesaroni99a} and \citet{shepherd00};  they are located
central to the jet and the millimetre source and are aligned
in the direction of the jet of \citet{cesaroni97, cesaroni99a}.
High-angular-resolution VLBA observations \citep{moscadelli00,
moscadelli05} show that the water maser spots move along the
surface of a conical jet.  \citet{edris05} also detected OH,
H$_2$O and 6.7-GHz methanol masers from this source with the
OH and methanol masers tracing a Keplerian disc and the water
maser the outflow.  However, the 44-GHz methanol maser spots
imaged by \citet{kurtz04} using the VLA trace the shocked H$_2$
and SiO emission from the jet \citep{cesaroni05}.

\begin{figure*}
\centering
\includegraphics[width=16.5cm,clip]{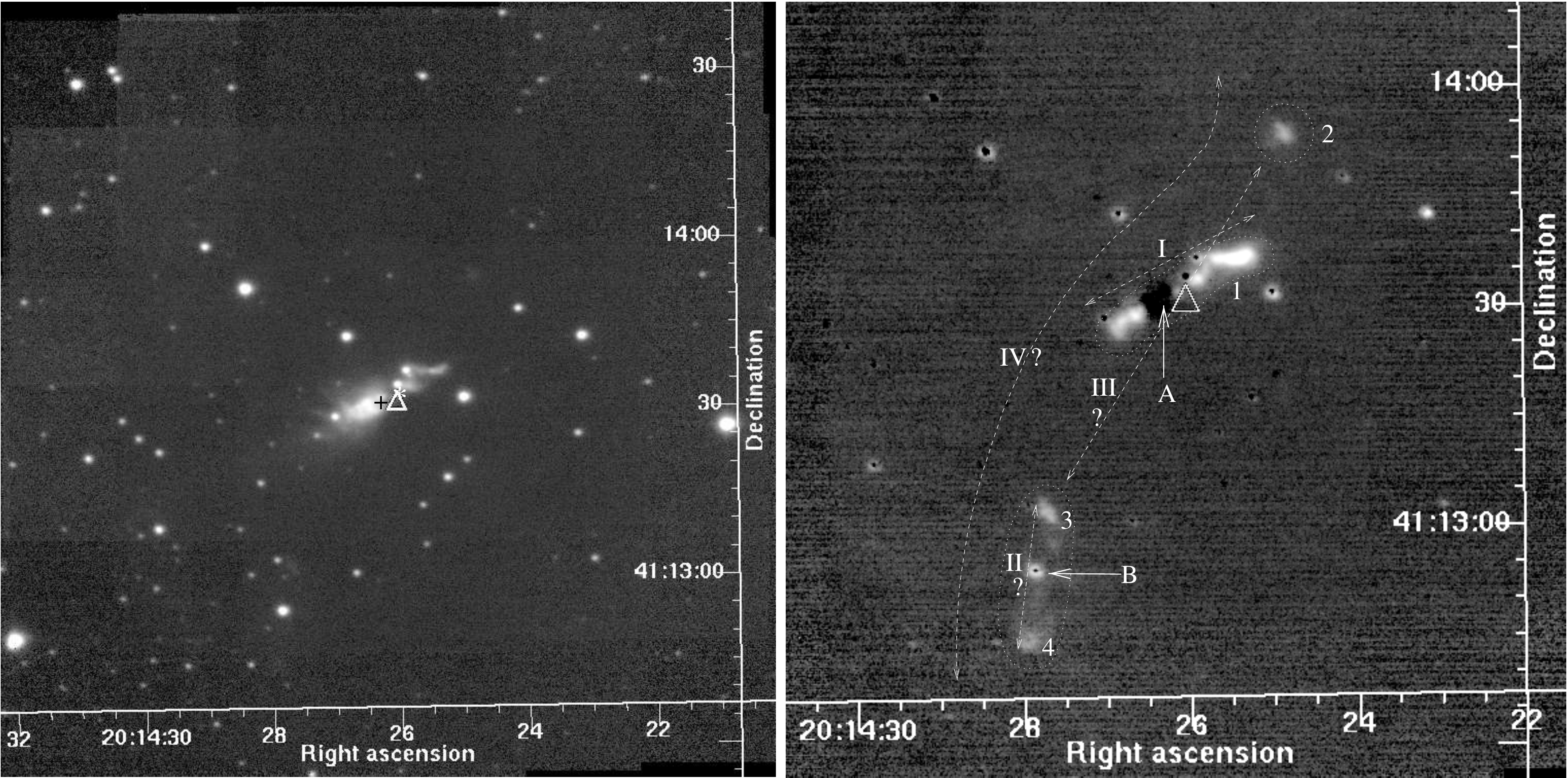}
\caption{The left panel shows the $K$-band image of IRAS~20126+4104.
The 7-mm continuum peak of \citet{hofner99} is shown by ``*''.
The right panel shows the continuum-subtracted H$_2$ image of the
central region smoothed with a 2-pixel FWHM Gaussian.}
\label{20126_KH2}
\end{figure*}

The outflow axis, as derived from the SiO jet, is inclined only
slightly with respect to the sky plane ($i \sim$10$^{\circ}$;
\citealt{cesaroni99a}).  However, the large scale CO outflow is found
to have $i \sim$45$^{\circ}$ \citep{shepherd00}.  The differences
in the angle of inclination with respect to the sky plane and in
the directions of the H$_2$/SiO jet and the CO outflow are
interpreted as due to a precession of the jet by $\sim$45$^{\circ}$
\citep{shepherd00, cesaroni05}.  The precession could be due to
the presence of a binary in a non-coplanar orbit or to
anisotropic accretion events \citep{shepherd00}.

A binary or a multiple system for the YSO has been proposed by
several investigators.  \citet{hofner99} carried out high angular
resolution observations of IRAS~20126+4104 at 7\,mm and 3.6\,cm
using the VLA.  They detected a single unresolved peak at 7\,mm,
coincident with the 3-mm peak of \citet{cesaroni97}.  Their
3.6-cm continuum emission consisted of two distinct elongated
sources  (separated by nearly an arcsec in declination and with flux
densities of $\sim$0.2 and 0.1\,mJy for the northern and southern
components respectively), with the major axes appearing to be
aligned and oriented at an angle $\sim$117$^{\circ}$,  which is
in the same direction as the moleculear
outflow imaged in HCO$^{+}$(1-0) by \citet{cesaroni97} and the jet
traced in SiO and H$_2$.  The H$_2$O maser sources resolved by
\citet{tofani95} are located along the northern 3.6-cm
continuum source and are also aligned in the same direction as
the northern jet. \citet{hofner99} interpreted the two 3.6-cm
continuum sources as due to two ionized jets from two YSOs, both
contributing to the outflow observed in this field.  The 1-mm
and 3-mm peaks of \citet{shepherd00}
are coincident with the northern ionized jet of  \citet{hofner99}.
The southern ionized jet was not detected at 1\,mm by \citet{shepherd00}.
However, their 3-mm source exhibits an extension to the south, in agreement
with the presence of the southern cm source.
\citet{cesaroni97} proposed a pre-UCH{\sc{ii}} stage for
IRAS~20126+4104.

Through imaging in the $K$, $L'$ and $M'$ bands,
\citet{sridharan05} proposed that the central object is a
binary/disc system, with the components of the binary system
separated by $\sim$0.5\,arcsec and located nearly in the disc 
plane. They estimated the thickness of the disc to be 
$\sim$850 AU for radii $\lesssim$1000 AU.  However, mid-IR 
observations by \citet{debuizer07} suggest that the near- and 
mid-IR emission could be produced by either multiple YSOs or 
by scattered and direct emission from dust in the cavity walls 
of an outflow from a luminous YSO.

Fig. \ref{20126_KH2} shows our $K$-band and H$_2$ images of IRAS~20126+4104.
The main outflow traced by the H$_2$ emission feature ``1'' (shown in the
direction of the arrow ``I'') is oriented  in the NW--SE direction at an
angle of $\sim$122$^{\circ}$ and has a collimation factor of $\sim$5. The source
driving this outflow is not resolved in $K$ and appears to be obscured by
the disc.  The point sources detected by IRAS and MSX are within 3 arcsec
of each other and are located close to the centroid of the H$_2$ jet feature.
The twisted nature of the  H$_2$ emission in ``1'' gives strong indication
of a precessing jet (see also \citealt{caratti08}).  This H$_2$ emission
is in very good agreement with the SiO jet discovered by \citet{cesaroni99a}
and traces the direction of the HCO$^+$ outflow imaged by \citet{cesaroni97}
as was noted by previous investigators.

There are additional features unveiled in H$_2$, which were also observed
by previous investigators.  The prominent ones are labelled  ``2--4''.
Four different scenarios for the outflow are indicated on the figure, the
directions of which are shown by lines labelled ``I--IV''.  Of these,
only ``I'' could be treated as conclusive, the rest require confirmation
with additional observations.  ``2'' and ``3'' in the direction of ``III''
are nearly equidistant from the centre of the flow ``I'' (at ``A'') and
therefore could be interpreted as arising from a second object located
near the driving source of ``I''.  ``3'' and ``4'' are located on either 
sides and are nearly equidistant from a point source
``B'' ($\alpha$=20:14:27.87, $\delta$=+41:12:54.4).  This source is
well detected by 2MASS.  As pointed out by \citet{shepherd00} (from the
photometry given by \citealt{cesaroni97}) and confirmed by the 2MASS
photometry, ``B'' does not exhibit any excess (Fig. \ref{JHKcol}) and 
it appears to be a foreground object.  However,  ``3'' and ``4'' do
resemble bow shocks and are directed as if produced by an outflow
(represented by ``II'') emanating from ``B'' or an embedded source
located close to ``B''.  There is also faint emission in H$_2$ between
``B'' and ``3'' and ``4''.  It has been suggested that the features
``1--4'' are caused by a single precessing jet \citep{shepherd00,
cesaroni05}, the direction of which is shown by ``IV''.  However,
their observations could not rule out multiple outflows produced by
more than one YSOs.  It remains to be investigated through further
observations which of these scenarios is a true representation of
this interesting object.

\vskip 15mm
\subsection{IRAS~20188+3928  --{ \it Mol 121}\\ ({\small \it d = 0.31; 3.91\,kpc, L = (0.343; 52.8)\,$\times$10$^3$\,L$_{\odot}$})}

\begin{figure*}
\centering
\includegraphics[width=16.5cm,clip]{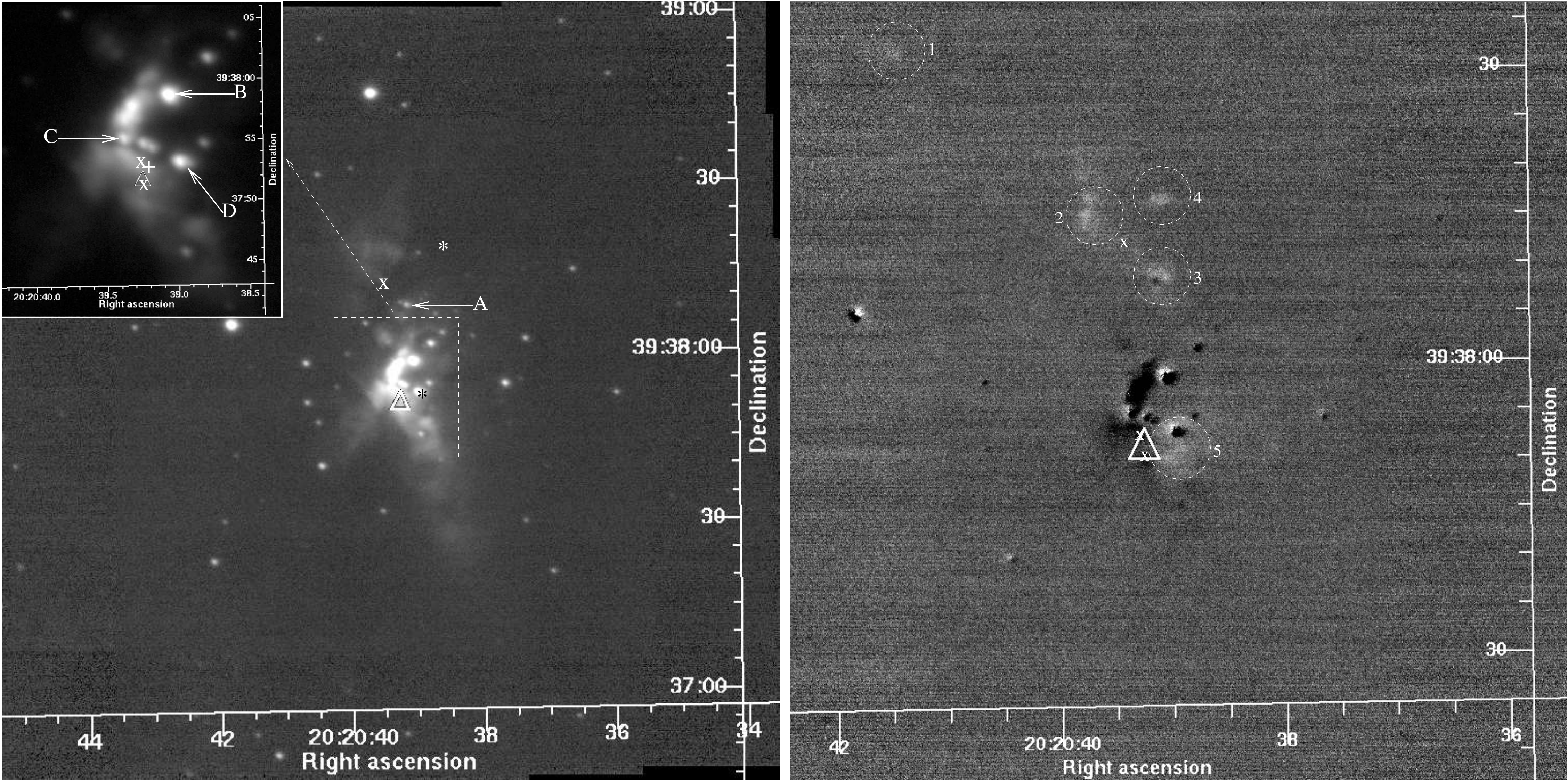}
\caption{Left: our $K$ band image of IRAS~20188+3928.
The inset shows an expanded view of the central region with higher
contrast. Right: the continuum-subtracted H$_2$
image of the central region. ``*'' shows the  850-$\mu$m positions
and ``x'' shows the 3.6-cm peaks of \citet{jenness95}.}
\label{20188_KH2}
\end{figure*}

\citet{little88} mapped IRAS~20188+3928 in HCO$^+$ and CO emission
lines and discovered a dense bipolar molecular outflow associated
with this source in the N-SW direction.  \citet{bronfman96} detected
the source in CS(2-1) line emission.  Several observers detected
NH$_3$  emission  (\citealt{molinari96}; Jijina, Myers \& Adams 1999;
Anglada, Sepulveda \& Gomez 1997) implying that the source is deeply
embedded. NH$_3$ observations by \citet{anglada97} showed a velocity
gradient indicating entrainment by high velocity gas.  They also
detected a variable H$_2$O maser $\sim$1 arcmin NW of the IRAS position.
H$_2$O maser emission was detected by several other investigators
\citep{palla91, brand94, jenness95}, some from more than one location
in the field.  \citet{mccutcheon95} detected continuum emission at
450, 800 and 1100\,$\mu$m. \citet{jenness95} imaged two sources at
450 and 800\,$\mu$m with the brighter and the fainter ones located NW
of the IRAS position by 4.2\,arcsec and 28.4\,arcsec respectively.
They detected three-component emission at 3.6\,cm, with the brightest
one (24.1\,mJy) and the faintest (2.2\,mJy) closer to the IRAS position
than another faint source (3.4\,mJy) 20.7\,arcsec north and 1.95\,arcsec
east of the IRAS position.  The brightest 3.6-cm source was also detected
by them at 1.95\,cm.  C$^{18}$O emission was also observed to be
peaking in its vicinity.  \citet{molinari98} detected two-component 6-cm
emission using the VLA, with the stronger component (29.95\,mJy) within
one arcsec of the IRAS position and agreeing with the position of the
brightest radio source of \citet{jenness95} and the weaker one (3.82\,mJy)
located $\sim$21\,arcsec NE, and agreeing in position with the northern
radio source of Jenness et al.  The distance estimated to this source
varies from 0.31 to 4 kpc (\citealt{molinari96, palla91, little88}; etc).
Accordingly the luminosity also is highly uncertain.

Our near IR images (Fig. \ref{20188_KH2}) show a cluster of deeply
embedded objects towards the centre of the field, very close to the IRAS
position.  The MSX mission detected mid-IR emission from this region at
8.28\,$\mu$m and above, within 1.4\,arcsec of the IRAS position.  The
good agreement of these positions indicates that the outflow source is
located within this cluster.  Most of the nebulosity disappears in the
continuum-subtracted H$_2$ image, leaving faint line emission features
that are labelled ``1--5'' in Fig. \ref{20188_KH2}. Features
``1--3'' appear aligned, as do ``3--5'', though in a
different direction.  The faint radio peak $\sim$20\,arcsec north of
the IRAS position detected by \citet{jenness95} and \citet{molinari98}
is located between ``2'' and ``3''.  This radio emission is likely to
be from a highly embedded source, although it is not clear if it is
the source which drives an outflow seen in H$_2$ emission.  There is a
pair of faint stars (``A'') located close to the centre of the feature
labelled ``3''.  With the radio positions of \citet{jenness95} and
\citet{molinari98} agreeing very well and separated from ``A'',  it
is unlikely that any of these two stars is the IR counterpart of the
YSO seen in radio emission. The 2MASS measurement of this source is
contaminated by the surrounding nebulosity and therefore we can't
say if the source shows any excess.  Likewise, the IR magnitudes of
the stars ``B'', ``C'', ``D'' etc. near the IRAS position are also
contaminated by the presence of nebulosity.  However, the presence
of the radio sources and the outflows implies that there are
multiple YSOs in the region, which are likely to be at different
stages of evolution.

\citet{zhang05} observed a bipolar outflow in CO centred on the IRAS
source and roughly in the NS direction.    This direction is consistent
with the H$_2$ emission features ``3--5''.  They note that the outflow
in CO is better centred on the illuminating source $\sim$6\,arcsec north
of the IRAS position, in agreement with the location of the deeply
embedded source northward of the IRAS position proposed by \citet{yao00}
from near-IR polarization studies. However, the presence of the H$_2$ 
emission features ``1--2'' is suggestive of an additional outflow in 
the field.  High angular resolution CO observations of the field is 
required to understand that.

\subsection{IRAS~20198+3716\\ ({\small \it d = 0.9, 5.5\,kpc, L = (20, 605)\,$\times$10$^3$\,L$_{\odot}$})}

\renewcommand{\thefigure}{A\arabic{figure}\alph{subfigure}}
\setcounter{subfigure}{1}
\begin{figure*}
\centering
\includegraphics[width=16.5cm,clip]{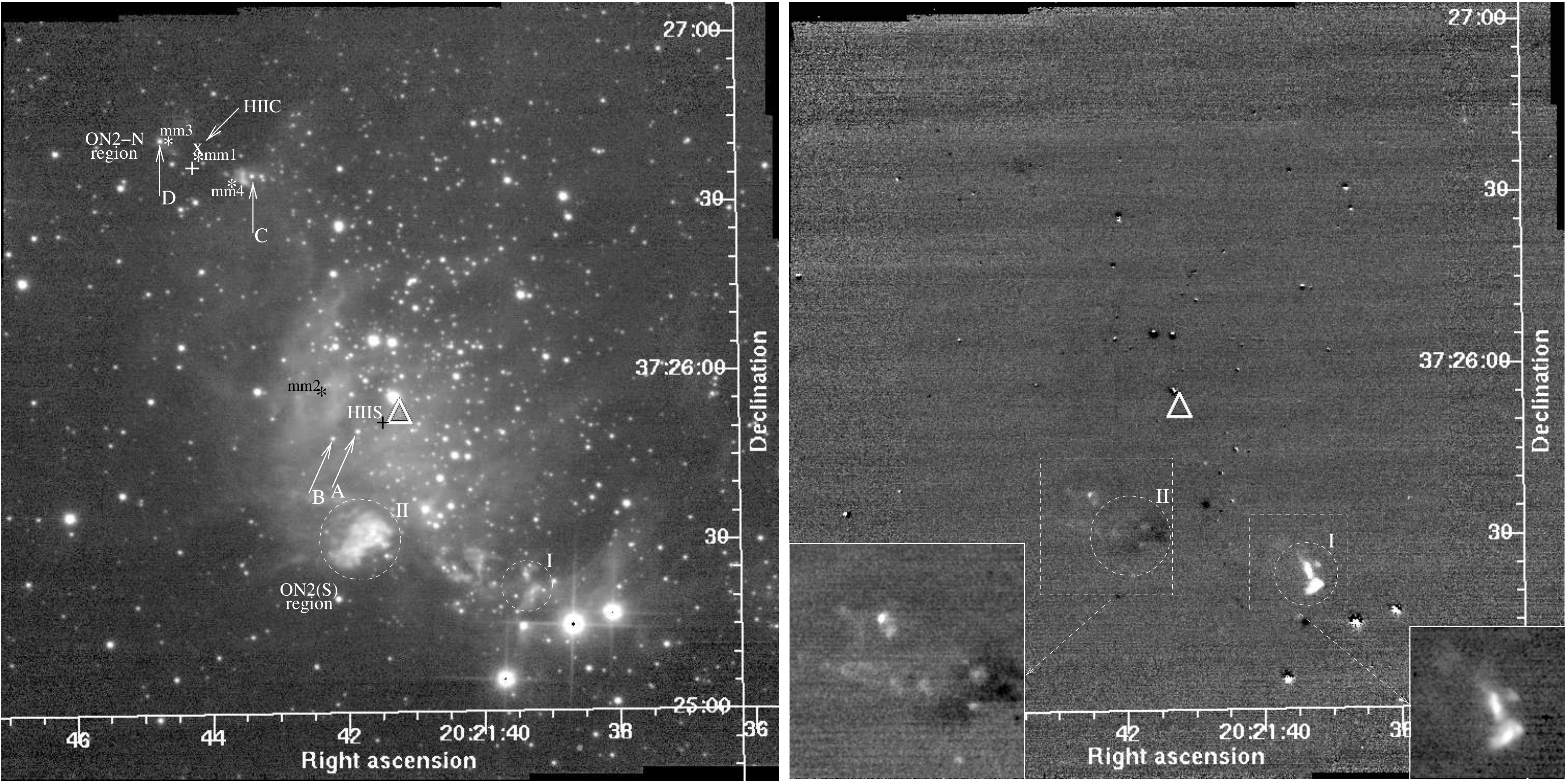}
\caption{The left panel shows the $K$ band image of IRAS~20198+3716.
The location of the millimeter continuum peaks of \citet{shepherd97}
are shown by ``*'' and labelled ``mm1--mm4''.  ``x'' shows the
location of the 6-cm peak emission of \citet{wc89b}.  The right
panel shows the continuum-subtracted H$_2$ image on which expanded
views of the regions shown in the boxes are overlaid as inset.
The regions in the insets are Gaussian smoothed with an FWHM=2 \
pixels to improve the contrast against the background noise.}
\label{20198_KH2}
\end{figure*}

IRAS~20198+3716 (ECX6-7b) coincides with the extensive ON2 star
forming region, which extends over roughly 15~pc at a distance of
5.5~kpc.  ON2-N (the northern region in our image, referred to as
ON2-N by Shepherd, Churchwell \& Wilner 1997 and as ON2(C) by
Matthews, Anderson \& MacDonald 1986) contains the  UCH{\sc{ii}}
region G75.78+0.34 while the southern cloud contains the
H{\sc{ii}} region G75.77+0.34.  The location of the latter is
close to the IRAS position.  The location of the ON2(S) given by
\citet{matthews86} is further SE of the IRAS position by
$\sim$30\,arcsec, which is close to their spatial resolution.

ON2 has been studied in considerable detail by \citet{shepherd97}
in 3-mm continuum emission, CO, SiO, H$^{13}$CO$^+$ and SO$_2$ at
high spatial resolution. They find evidence for at least four
outflows.  \citet{matthews86} present CO observations of the region
and Dent, MacDonald \& Anderson (1988) present NH$_3$ and HCO$^{+}$
observations.  Water and Methanol masers are observed at various
locations throughout the region (Hofner \& Churchwell 1996;
Szymczak et al. 2000).  \citet{forster78} found water maser emission
closely associated with the northern source.  \citet{dent88} derive
a late O spectral type for the southern H{\sc{ii}} region H{\sc{ii}}S
(G75.77+0.34) (ON2(S) of Matthews et al.) and an early B spectral 
type for H{\sc{ii}}C (G75.78+0.34; ON2(C) of Matthews et al.; ON2-N
of \citealt{shepherd97}).

\citet{comeron01} present low-resolution $JHK$ images of the
region, together with colour-colour diagrams. They point out that
because the near-IR cluster is offset SW of the high-density
regions mapped by \citet{shepherd97}, the cluster may be a blister
H{\sc ii} region where newly formed stars are disrupting the dense
molecular cloud that still composes the core of ON2.

Fig. \ref{20198_KH2} and \ref{20198_BrG} show our $K$, H$_2$ and
Br$\gamma$ images.  Our observations centred on the IRAS position
reveal a rich cluster of IR sources embedded in nebulosity.  2MASS
detects only a few of the bright objects in the field and many of the
detections have poor quality flags, probably due to the clustering and
the nebulosity.  Hence, we cannot draw solid conclusions about the
colours of many of the candidate YSOs.  A few objects and regions of
interest are labelled on the figure. The bright source near the IRAS
position does not exhibit any excess in the 2MASS colours.  However,
it is not well detected in their $J$ band.  The MSX mission detected
two objects in this field.  The brighter of the two is $\sim$4\,arcsec
SE of the IRAS position, located in between the IRAS position and
source ``A''.  Source ``B'' shows IR excess (Fig. \ref{JHKcol}).
There is a fainter MSX source, $\sim$56\,arcsec NE of the IRAS
position and located between the two sources in the NE
detected by \citet{comeron01}.  We label these sources
``C'' ($\alpha$=20:21:43.34, $\delta$=37:26:35.5) and
``D'' ($\alpha$=20:21:44.68, $\delta$=37:26:41.9).

\citet{shepherd97} resolved four millimeter continuum sources in this 
region.  The positions of these are indicated by ``*'' on 
Fig. \ref{20198_KH2} and labelled ``mm1--mm4'', consistent with ``1--4'' 
of Shepherd et al.  The brightest of these (``mm2'') is located within a 
few acrseconds of ``A'' and ``B'', but it does not coincide with either
of these or with the MSX source near ``A''.  ``mm1'', ``mm3'' and 
``mm4'' are located between ``C'' and ``D'',  with ``C'' and ``D'' 
located close to ``mm4'' and ``mm3'',  respectively.   The second 
brightest mm source (``mm1'') is within 2\,arcsec of the MSX source 
(towards the NW corner in Fig. \ref{20198_KH2}) and could be the 
mm counterpart of the MSX source.  However, with an $\sim$18\,arcsec 
spatial resolution of the MSX, more than one mm source could be 
contributing to the MSX detection.  H$_2$O \citep{forster78, hofner96} 
and OH \citep{hardebeck71} masers detected in this region are 
located close to ``mm1''.  

\citet{wc89b} detected a cometary UCH{\sc{ii}} region located between 
``C'' and ``D'' and very close to ``mm1'', at an integrated flux 
density of 40.4\,mJy at 6\,cm. 7-mm continuum observations at high 
angular resolution \citep{carral97} resolve ``mm1'' into two sources; 
one source coincides with the UCH{\sc{ii}} region and the second one 
is located $\sim$2\,arcsec SW, agreeing well with the location 
of the H$_2$O masers.  \citet{kurtz05a} discusses the possibility of the
7-mm source coinciding with the H$_2$O masers being in an
HCH{\sc{ii}} phase.

The ``Central'' and   ``Western'' outflows mapped by Shepherd et al.  
appear to be centred near ``mm3'' and ``mm1'' respectively;  
observations of outflows at higher spatial resolution are required 
to confirm any association.  They have not mapped any CO emission 
around the IRAS source and ``mm2''.  Both ``C'' and ``D'' have 
nebulosity around them in the $K$ band and there is a clumpy 
emission feature located $\sim$1.6\,arcsec west of ``C'', which 
disappears in the continuum-subtracted narrow-band images.  Our 
H$_2$ imaging did not detect any line emission from the northern 
cluster close to ``C''and ``D'' where they detected the outflow.  
However, from the southern region, we detect some well-collimated
outflows.

Two regions of interest in the southern region are circled in
Fig. \ref{20198_KH2}. Region ``II'' shows faint IR sources hidden
in strong nebulosity;  most of the nebulosity disappear in the
continuum-subtracted images.  The H$_2$ image shows several
bow shocks, which appear to be emanating from region ``II'', implying
the presence of one or more YSOs within ``II''.  The H$_2$ features
from region ``II'' reveal at least two outflows.  Region 
``I'' shows some very bright H$_2$ emission features which appear 
to be a bow-shock of the jet from a YSO in the IR cluster or 
could be the bow-shock/s of counter jet/s from the sources
producing the outflows seen in region ``II''.  Both H$_2$ and 
Br$\gamma$ images show negative residuals in this region upon 
continuumm subtraction, which implies that there could
be thermal emission from dust here, making the SED steep.  As is seen
from Fig. \ref{20198_BrG},  there is no Br$\gamma$ emission emanating
from the regions ``I'' and ``II''.  However, the Br$\gamma$ image
(Fig. \ref{20198_BrG}) shows a copious amount of line emission from
the nebulosity in which the IR cluster is embedded.   The location
of the H{\sc{ii}} region G75.77+0.35 is within this cluster.  With
the positional uncertainties of the IRAS and the MSX detections,
it is difficult to say if region ``II'' hosts the IRAS source and 
the MSX object near it,  or if it is separate.  Note that the IRAS 
and MSX positions are closer to ``mm2''  than to ``II''.

Together, these observations imply that this is a region very active in
star formation.  The YSOs present here drive the outflow that we observe
in H$_2$, while more evolved, luminous objects ionize the circumstellar
medium.  Since it is not clear which are the driving sources of the
outflows traced by these H$_2$ emission features, we do not list the
outflow angles and the collimation factors.  It should be noted that
the ON2(S) outflow observed by \citet{matthews86} is centred close
to ``II''.  It is not clear if ``I'' is produced by counter jet/s
originating from the sources producing the outflows in ``II'', or
if it is from a source responsible for the millimeter continuum source
``mm2'' and the MSX and IRAS sources.  What is important is that
\citet{dent88} derived a late O-type for the driving source of the
outflow in the ON2(S), from which, presumably, we see the strongly 
emitting bow-shock in the H$_2$ image.

\addtocounter{figure}{-1}
\setcounter{subfigure}{2}
\begin{figure}
\centering
\includegraphics[width=8.10cm,clip]{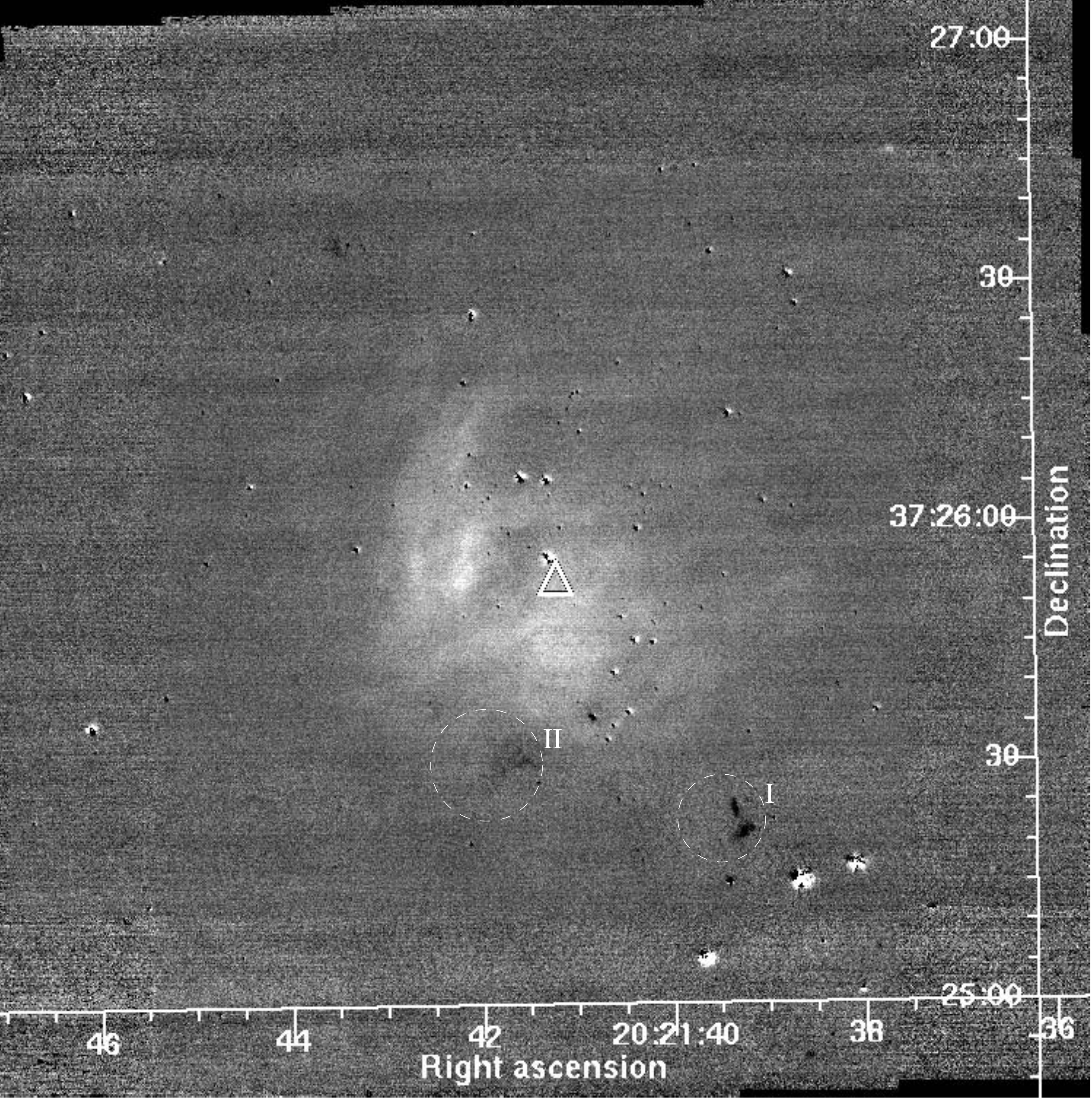}
\caption{The continuum-subtracted Br$\gamma$ image of IRAS~20198+3716.}
\label{20198_BrG}
\end{figure}
\renewcommand{\thefigure}{A\arabic{figure}}

\subsection{IRAS~20227+4154 -- {\it Mol 124}\\ ({\small \it d = 0.1; 3.39\,kpc, L = (0.00914; 9.59)\,$\times$10$^3$\,L$_{\odot}$})}

\begin{figure*}
\centering
\includegraphics[width=16.5cm,clip]{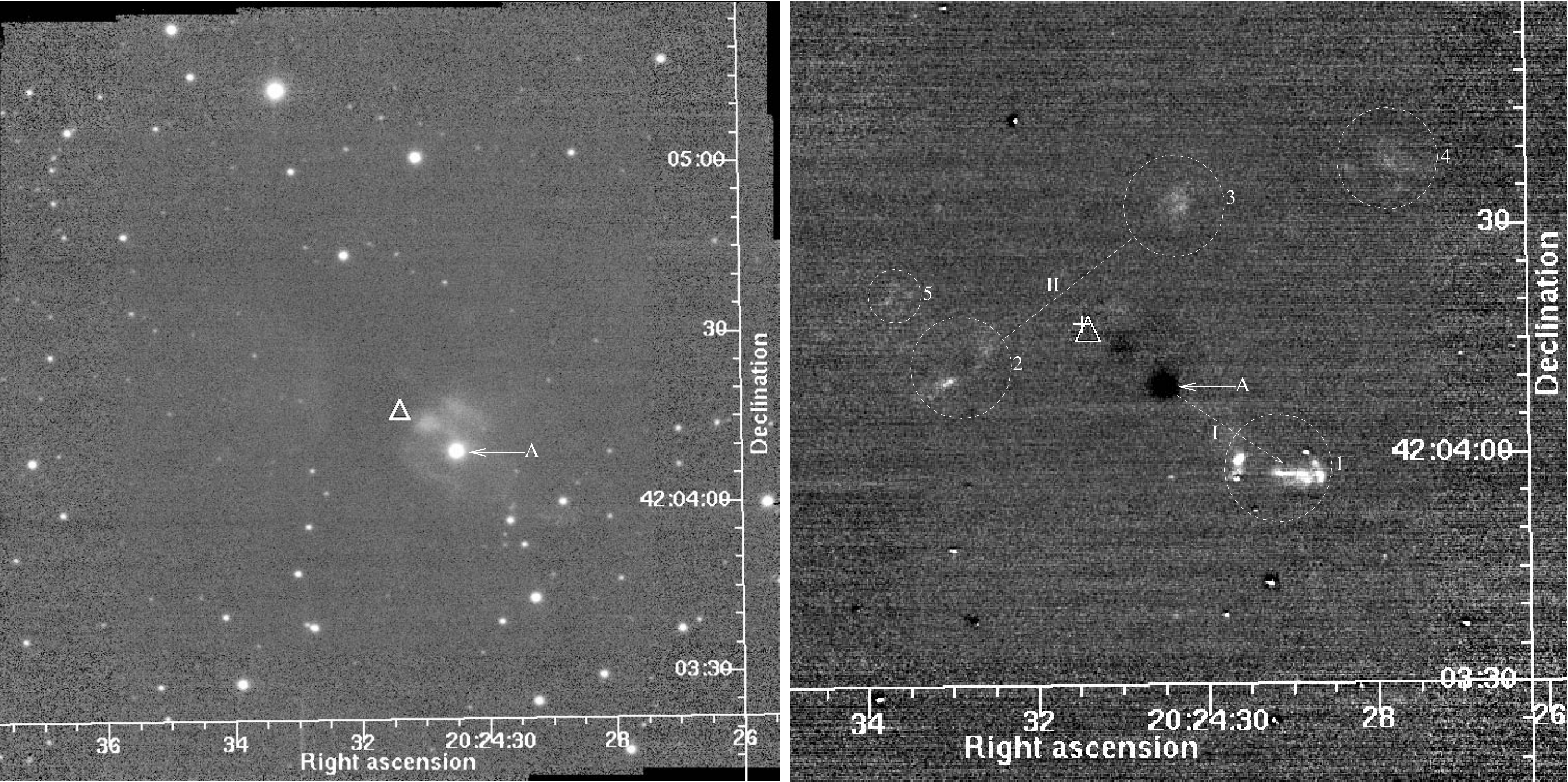}
\caption{Left: $K$-band image of IRAS~20227+4154.
Right: continuum-subtracted H$_2$ image of the
central region.}
\label{20227_KH2}
\end{figure*}

There is large uncertainty in the distance and, thereby, in the
luminosity of this object. \citet{palla91}, using their estimate
of 3.39\,kpc for the kinematic distance, calculated a FIR luminosity
of 9.59$\times$10$^3$L$_\odot$,  whereas \citet{molinari96}, through
observations of NH$_3$ emission, estimated a kinematic distance of
0.1\,kpc and a luminosity of 9.14\,L$_\odot$.  The v$_{LSR}$ of the H$_2$O
maser emission detected by \citet{palla91} and \citet{brand94} are
comparable (v$_{peak}$ at -2.06 and -1.99\,km~s$^{-1}$ respectively),
whereas these values differ significantly from the v$_{LSR}$ of the
NH$_3$ emission lines (+5.8/9\,km~s$^{-1}$; \citealt{molinari96}),
CS emission (+5.8\,km~s$^{-1}$; \citealt{bronfman96}) and 1665 and
1667\,MHz OH absorption line (+5.6\,km~s$^{-1}$; \citealt{slysh94}).
At 1667\,MHz, Slysh et al. also detected an OH emission component at
-10.4\,km~s$^{-1}$.  A search for 6.7-GHz methanol maser emission by
\citet{schutte93} did not yield any detection at a 3$\sigma$ level
of 3\,Jy.  \citet{wilking89} detected $^{12}$CO and $^{13}$CO line
emission from this region at v$_{LSR}$=+5.5 kms$^{-1}$.  Their
observations set upper limits of  48\,mJy, 0.8\,mJy and 0.6\,mJy
for flux densities at 2.7\,mm,  2\,cm and 6\,cm, respectively.

Fig. {\ref{20227_KH2}} shows our $K$-band and H$_2$ images.  The
near-IR point source close to the IRAS position is labelled
``A'' ($\alpha$=20:24:30.49, $\delta$=42:04:9.3) in the figure.
This is an extremely red object with IR excess (Fig. \ref{JHKcol}).
It has a $K_s$ magnitude of 10.39 and is not detected in the $J$-band
by 2MASS.  The H$_2$ image shows at least one well defined bow shock,
which is circled and labelled ``1'' on the figure.  It appears to be
directed away from ``A''.  This bow shock appears bright and we do not
see any emission from the counter jet.  Hence, this outflow could be
highly inclined with respect to the line of sight.  The bow-shock is
likely to be from the blue lobe of the jet.  From the very red IR colours,
``A'' appears to be a YSO.  However, ``A'' is located $\sim$12.3\,arcsec
SW of the IRAS position.  The MSX detected a source, separated by only
1.9\,arcsec from the IRAS position, at wavelengths 8.28 $\mu$m and above.
With the MSX and IRAS positions agreeing reasonably well and located
away from our near-IR source,  ``A'' is unlikely to be the near-IR
counterpart of the object detected by IRAS and MSX.  The mid-IR
source could be a more deeply embedded one.  Our $K$-band image does
not reveal any near-IR source close to the IRAS position.

The H$_2$ image reveals two emission features circled and labelled
``2'' and ``3'' in Fig. {\ref{20227_KH2}}.  The centroid of these 
two features falls close to the IRAS position.  These two could be 
produced by shocked emission from a bipolar outflow from the 
far-IR source.  There are two other emission features, which are 
labelled ``4'' and ``5'' on the diagram.  Together, these observations 
reveal at least two different outflows in this region.  The recent 
CO observations by \citet{kim06} confirm this result - the two outflows
detected by them in CO are roughly in the directions of the two
outflows that we label ``I'' and ``II'' on the H$_2$ image. From the
two different ranges of distance estimates discussed above,  it is
possible that the two different YSOs in the field are at different
distances from us, with IRAS and MSX detecting a more luminous
object farther from us, which is not detected in the near-IR.  However,
it should be noted that we have to be careful with ascribing the
differences in the v$_{LSR}$ of different species to different distances -
CS and NH$_{3}$ emission probe the core \citep{bronfman96, molinari96}
whereas H$_2$O maser emission traces the jet \citep{moscadelli05}.
More observations are required to understand if the two outflows are
from objects at different distances. Assuming that the H$_2$ emission
feature ``1'' is produced by ``A'' and ``2'' and ``3'' are produced
by the IRAS/MSX source not detected in the near IR, we measure a
collimation factor and outflow angle 4.5 and 239.3$^{\circ}$,
respectively, for outflow ``I'', and 8.3 and 128$^{\circ}$,
respectively, for outflow ``II''.  A poor collimation is inferred
for outflow ``II'' if ``4'' and ``5'' are part of it.  If the outflow 
``I'' also originate from a source near the IRAS position instead 
of from ``A'', its collimation factor would be as high as 6.9.

\subsection{IRAS~20286+4105 -- {\it Mol 126}\\ ({\small \it d = 3.72\,kpc, L = 39\,$\times$10$^3$\,L$_{\odot}$})}

IRAS~20286+4105 is associated with a dense core detected in NH$_3$
\citep{molinari96} and in CS \citep{bronfman96}.  \citet{palla91}
and  \citet{brand94} detected  H$_2$O maser from this region.  The
IRAS source has the colours of a UCH{\sc{ii}} region and, in optical
images, lies at the centre of a ring-shaped nebula \citep{odenwald89}.
However, \citet {molinari98} did not detect any radio emission at 6\,cm.
Comer\'on \& Torra (2001) present low-resolution $JHK$ images of this
source, together with near-IR colour-colour diagrams. They also
estimate a distance of 1~kpc to this star forming region, which differs
considerably from the kinematic distance of 3.72\,kpc estimated by
\citet{molinari96} from NH$_3$ observations.  \citet{molinari96}
determined a luminosity of 3.9$\times$10$^4$\,L$_\odot$.  
\citet{zhang05} detected a CO outflow in the CO (J=2-1) line.

\begin{figure*}
\centering
\includegraphics[width=16.5cm,clip]{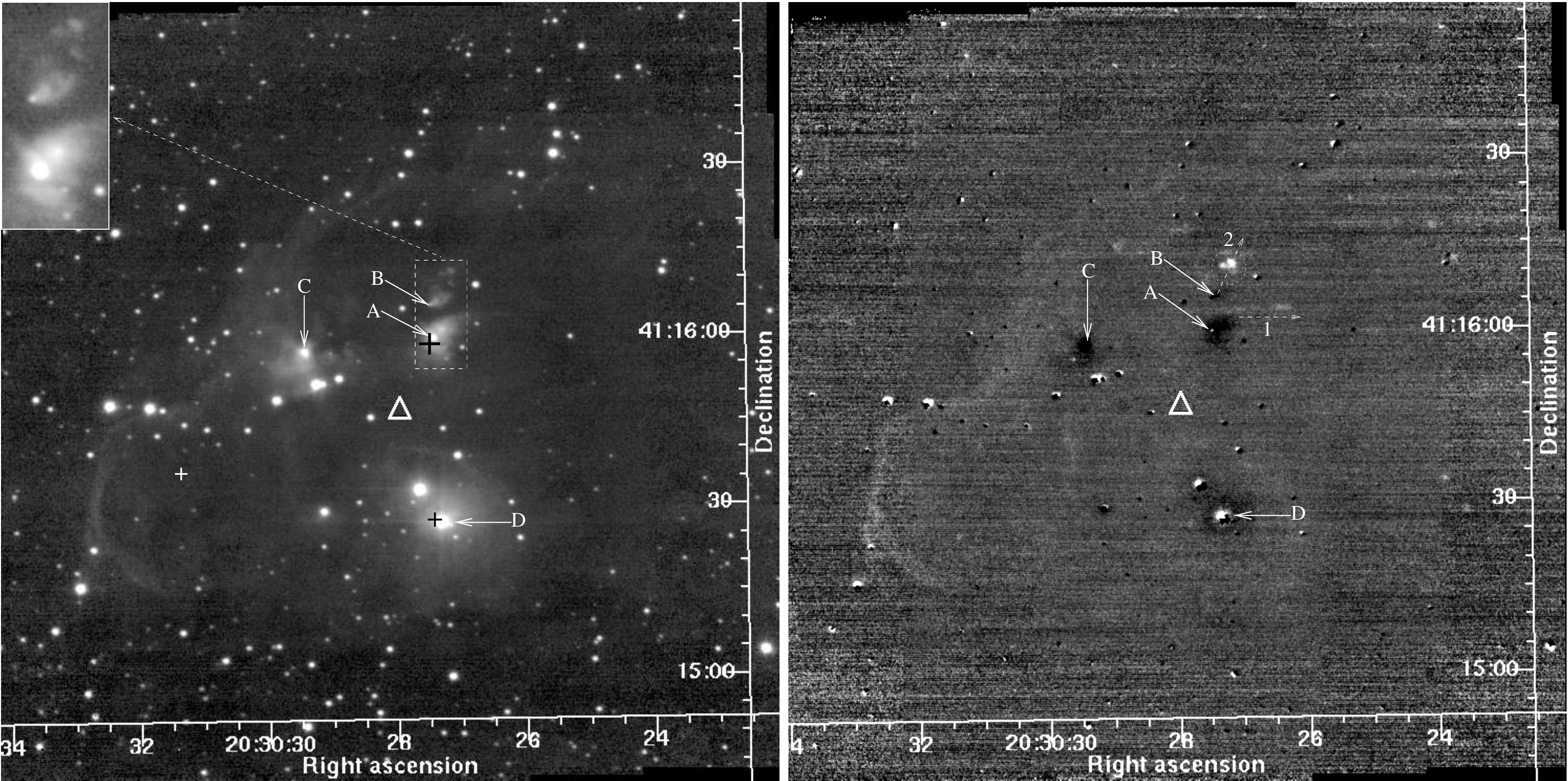}
\caption{Left: $K$-band image of IRAS~20286+4105.  An expanded view of
the region around ``A'' and ``B'' is shown in the inset.
Right: Continuum-subtracted H$_2$ image.}
\label{20286+4105_KH2}
\end{figure*}

Four objects surrounded by nebulosity are labelled in the figures -
``A'' ($\alpha$=20:30:27.41, $\delta$=41:15:59.7),
``B'' ($\alpha$=20:30:27.45, $\delta$=41:16:5.6),
``C'' ($\alpha$=20:30:29.41, $\delta$=41:15:57.5), and
``D'' ($\alpha$=20:30:27.32, $\delta$=41:15:27.3).
``A'' is associated with an extended cometary nebulosity in $K$ and
is detected by the 2MASS only in $H$ and $K_s$.  There appears to be
a faint source deeply embedded in this nebulosity and located
$\sim$2.2\,arcsec NW of ``A''.  Another faint source located
2.7\,arcsec SW of ``A'' is also associated with a cometary nebula.
The continuum-subtracted H$_2$ image shows aligned emission feature,
the direction of which is denoted by the dashed arrow (labelled ``1'')
on the figure.  From the direction of ``1'',  it appears to be
produced by the outflow from a deeply embedded source located
towards the north of ``A'', which is not detected in our image.
There is a  brighter H$_2$ emission feature, which is circled
and labelled ``2'' in Fig. \ref{JHKcol}.  It is located at the
apex of the cometary nebulosity seen to be associated with ``B''
in $K$ and is probably the bow-shock of an outflow from ``B'' in the
direction of the dashed arrow shown from ``B''.
``C'' also is embedded in strong nebulosity.  ``B'' and ``C'' are
detected by 2MASS only in $K_s$.  The 2MASS colours of all three
objects exhibit reddening and excess (Fig. \ref{JHKcol}), though
their colours could have large uncertainty. All three objects appear
to be YSOs.

MSX detected three objects in this field,  the positions of which
are denoted by ``+'' on the $K$-band image.  The brightest MSX
object is within an arcsec of ``A'';  the second brightest is
within an arcsec of ``D''.  The faintest object towards the east is
poorly detected by MSX and does not show any obvious near-IR
counterpart.  The 2MASS colours of ``D'' exhibit reddening, but do not
exhibit any IR excess.  The continuum-subtracted H$_2$  image
exhibits a very faint shell of H$_2$ emission around ``D'',  which
is marginally detected here.  ``D'' is probably a more evolved
object.  However, ``D'' has very faint neighbours which are not
detected by 2MASS.

The direction of the outflow as implied from the direction of
``1'' is roughly in the EW direction.  It is also noteworthy that
the location of the centre of the outflow imaged in CO by
\citet{zhang05} is offset northward from the IRAS position, which
is consistent with the location of the H$_2$ features and the
reddened objects with excess in our image northward of the IRAS
position.  The offsets of the CO and H$_2$ features from the
IRAS position are different by $\sim$10\,arcsec,  but agree within
the positional accuracy of \citealt{zhang05}.  Overall, the region
appears to be active in star formation, with more than one YSO
present.  We positively identify ``B'' as the YSO responsible for
the outflow in the direction of ``2''.  A deeply embedded YSO
located north of ``A'' is likely to be responsible for the
outflow in the direction of ``1'';  high angular resolution
imaging at longer wavelengths is required to understand that.

There are also other fainter H$_2$ emission features detected, the
prominent ones being the extended filementary structures seen towards
the east and NE in the H$_2$ image.  This is probably fluorescent
emission either from the periphery of materials pushed away by
more than one bright stars near the centre of the field or from
externally illuminated cloud boundaries.

\subsection{IRAS~20293+3952\\ ({\small \it d = 1.3, 2\,kpc, L = (2.5, 6.3)\,$\times$10$^3$\,L$_{\odot}$})}

\citet{beuther02b} detected a dense core in CS and 1.2\,mm dust
continuum emission $\sim$23\,arcsec to the east and 3.4\,arcsec
north of the IRAS position. They also detected a much fainter peak
located NW of the IRAS position.  The bright mm continuum peak was
resolved into three at 1.3\,mm and 3\,mm using high angular resolution
interferometric observations using PDBI \citep{beuther04a,beuther04b}.
High resolution CO and SiO maps of \citet{beuther04a} resolve four
outflows in this region.  A water maser is detected towards the dust
core and outflow source, while a resolved 3.6-cm radio continuum
source with an integrated flux density of 7.6\,mJy is detected
toward the IRAS position suggesting the presence of an H{\sc ii}
region \citep{sridharan02, beuther02d, beuther04a}. Sridharan et al.
also detected the hot-core tracers CH$_3$OH and CH$_3$CN towards the
IRAS source.  \citet{sridharan02} did not find any evidence for the
presence of 6.7-GHz methanol maser.  They detected strong H$_2$O
maser, located close to the dust continuum peak.

\citet{kumar02} present near-IR images of
IRAS~20293+3952; they detect H$_2$ emission coincident with the
blue-shifted CO lobe \citep{beuther02c}, as well as compact
features to the NE and SE.  They also observed a ring of H$_2$
emission surrounding one of the two bright near-IR sources in
the cluster near the IRAS position.

\begin{figure*}
\centering
\includegraphics[width=16.5cm,clip]{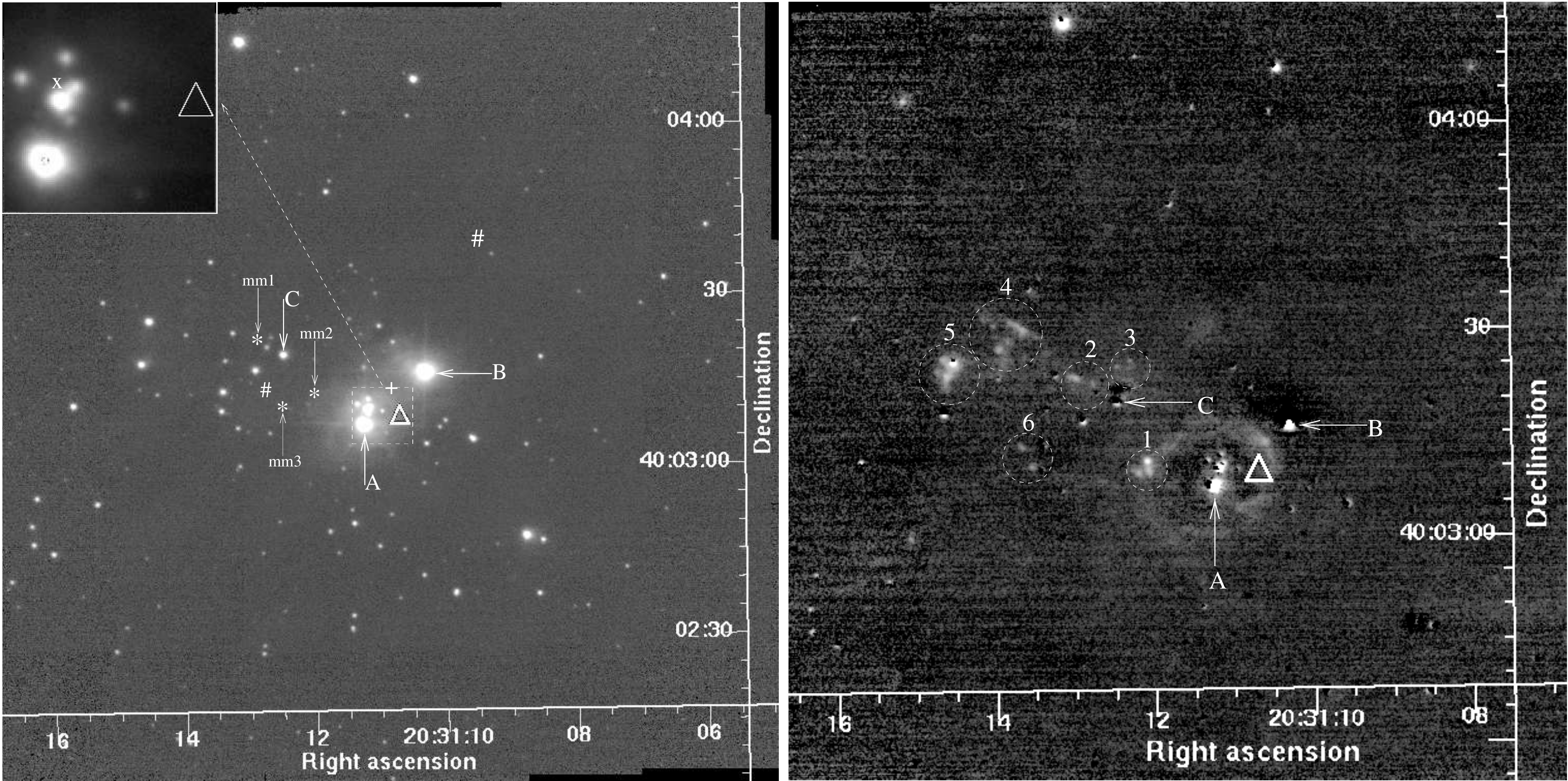}
\caption{The left panel shows the $K$-band image of IRAS~20293+3952.
The 1.2-mm continuum peaks (from IRAM 30-m telescope) of \citet{beuther02b}
are shown by ``\#''.  ``*'' represents the peaks (``mm1--mm3'') of
the higher resolution 1.3-mm maps shown by \citet{beuther04a}
observed using PDBI. ``x'' shows the location  of the radio source
\citep{sridharan02, beuther04a}. The right panel shows the
continuum-subtracted H$_2$ image of the central region.}
\label{20293_KH2}
\end{figure*}

Three sources are labelled on our IR images (Fig. \ref{20293_KH2}) -
``A''($\alpha$=20:31:11.26, $\delta$=+40:03:07.4),
``B''($\alpha$=20:31:10.30, $\delta$=+40:03:16.5) and
``C''($\alpha$=20:31:12.47, $\delta$=+40:03:19.9).
``A'' is resolved into multiple components.  The 2MASS
colours of ``A'' show it to be a reddened object without any IR excess
(Fig. \ref{JHKcol}).  However, it should be noted that the 2MASS
survey detected only one object here and it is probably dominated by 
the luminosity of the brightest source, which we label ``A''.
There are several fainter sources detected in the proximity of ``A''.
The 2MASS colours of ``B'' exhibit large reddening and IR excess and
place it in the region occupied by YSOs in the colour-colour diagram.
The MSX source is located 6.2\,arcsec SE of ``B'' and 3.7\,arcsec NE
of the IRAS position,  between ``A'' and ``B''.  The 3.6-cm radio
continuum source  of \citet{sridharan02}, \citet{beuther02d} and
\citet{beuther04a} (7.6\,mJy) is located closer to ``A'' than
``B'' (Fig. \ref{20293_KH2}) and is likely to be from one of the
sources within the cluster associted with ``A''.  ``C'' exhibits
reddening and excess (Fig. \ref{JHKcol}).  ``C'' is detected by
2MASS in $H$ and $K_s$ only.

Our continuum-subtracted H$_2$ image exhibits several interesting
features; the brighter ones are labelled on Fig. \ref{20293_KH2}.
The most prominent is the ring of H$_2$ emission encircling the
cluster containing ``A'' (also reported by \citealt{kumar02} and
\citealt{beuther04a}).  The ring appears elliptical and has a major
axis of $\sim$20\,arcsec inclined at $\sim$125$^{\circ}$ and a
minor axis of $\sim$15.5\,arcsec.
Embedded in this ring is the feature circled and labelled ``1''.
All of the features ``1--6'' appear to be split into multiple
components. ``2'', ``4'' and ``5'' appear to be bow shocks.  There
are also other fainter features seen in our image which are not
labelled here.  Obviously, this is a region where multiple star
formation is taking place.  Some of the possible drivers of the
outflows are ``B'', ``C'' and some of the fainter objects near
``A'' and ``C''.  The brightest of the 1.3-mm sources mapped in
this region by \citet{beuther04a} peaks near ``C''.  It should be
noted that,  in addition to ``A--C'', several other objects in
the field exhibit large reddening and excess, which warrants a 
detailed investigation of the field.  Deep infrared photometry 
and spectroscopy are needed to study the cluster, to identify the 
outflow driving sources and to understand the nature of the H$_2$ 
ring/shell.  The radio detection is not from the mm source or ``B''.  
With the 1\,arcec positional accuracy of the radio observations 
of \citet{beuther02d}, it is most probably from one of
the fainter objects located towards the north of ``A'' within
the cluster (shown in the inset on Fig. \ref{20293_KH2}).
The radio source is resolved (\citealt{beuther04a}, Figs. 3 and 7); 
it is much smaller than the H$_2$ ring.

\subsection{IRAS~20444+4629 -- {\it Mol 131 }\\ ({\small \it d = 2.42\,kpc, L = 3.34\,$\times$10$^3$\,L$_{\odot}$})}

The dense cloud core towards this region is detected in CS
\citep{bronfman96} and in NH$_3$ emission \citep{molinari96}.
\citet{molinari98} observed an unresolved radio source towards the
NH$_3$ core,  only one arcsec offset from the IRAS position.
\citet{wb89} detected asymmetric CO lines.
\citet{dobashi95} mapped the region in $^{12}$CO and $^{13}$CO lines.
The intensity of the $^{13}$CO emission detected by them peaks near
the IRAS source indicating a possible association.  They also detected
a bipolar outflow in $^{12}$CO (though this outflow was not evident
in the CO observations of \citealt{zhang05}).  Assuming a distance of
1.7\,kpc, they estimate the total mass of the cloud core to be
450\,M$_{\odot}$ and the YSO to be $\sim$8.5\,M$_{\odot}$.  A search
for 22.2-GHz water maser emission towards this source by
\citet{palla91} and \citet{wouterloot93} did not find any emission.
Nor was 6.7-GHz methanol maser emission detected \citep{macleod98b}.

\begin{figure}
\centering
\includegraphics[width=8.10cm,clip]{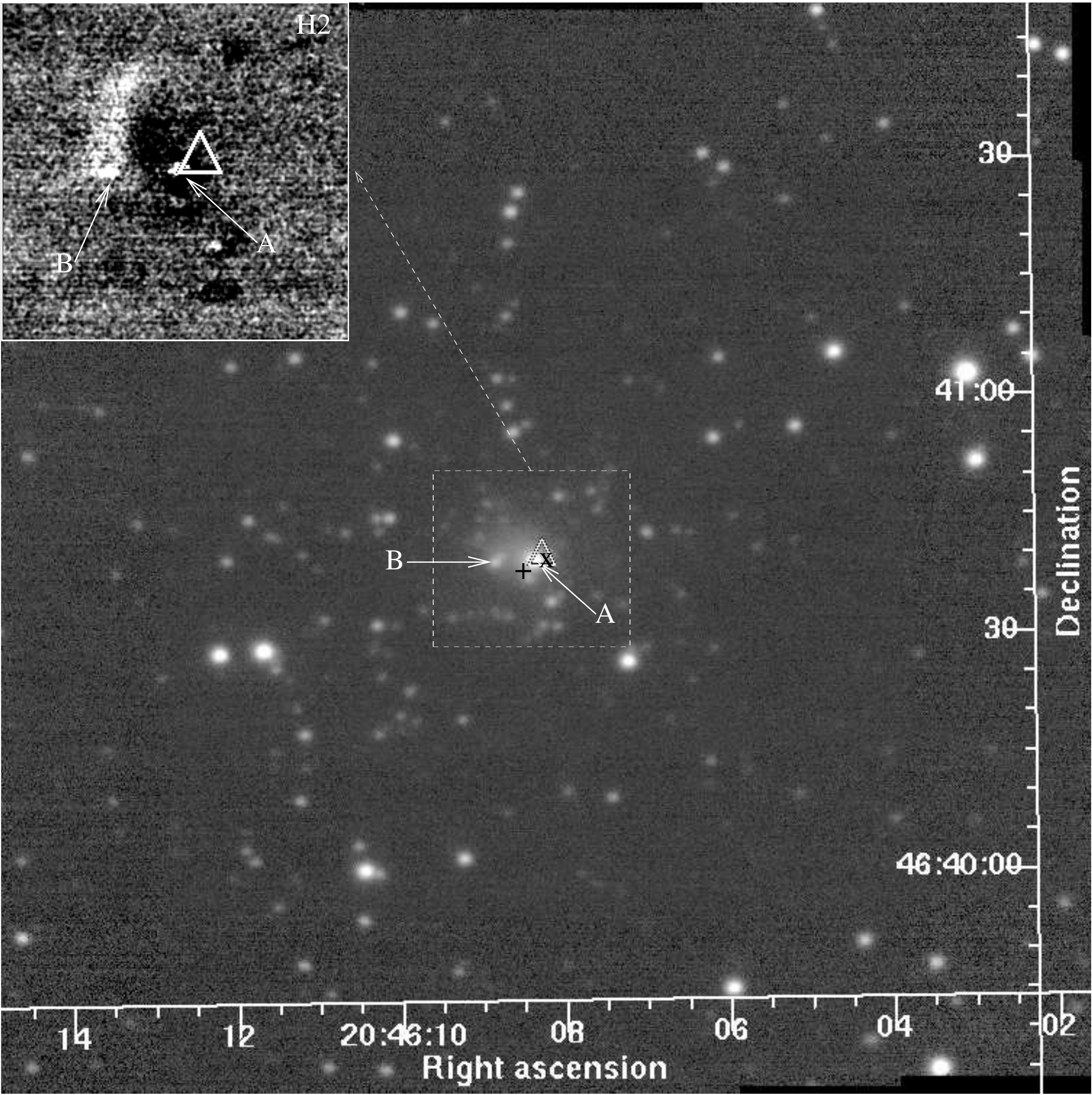}
\caption{The $K$-band image of IRAS~20444+4629.  A continuum-subtracted
H$_2$ image of the central region is shown in the inset.  ``x'' shows the
6-cm position of \citet{molinari98}.}
\label{20444_KH2}
\end{figure}

Fig. {\ref{20444_KH2}} shows our $K$-band image on which
the central region of the continuum-subtracted H$_2$ image is shown
in the inset.  The $K$-band image reveals a cluster of objects with 
a bright source ``A'' ($\alpha$=20:46:8.35, $\delta$=46:40:39.7) 
located close to the IRAS position with some fainter stars embedded 
in nebulosity.   The 2MASS colours of ``A'' do not appear to be highly 
reddened; it is at the boundary of the reddening band (Fig. \ref{JHKcol}).  
However, it should be noted that the source is poorly detected by 2MASS 
in $K_s$. ``B'' is an object nearby, which shows a comet-like nebulosity 
embedded in the nebula.  The cometary nebulosity associated with 
``B'' disappears in the continuum-subtracted H$_2$ image.  No bipolar 
emission is seen in this field in H$_2$ emission.  The H$_2$ image,
however, shows a faint arc-like line emission, originating from
the position of ``B'' and extending $\sim$7.5\,north and slightly
NW  (see the inset in Fig. Fig. {\ref{20444_KH2}}).  It is not clear 
if this is due to shock or fluorescence.  The IRAS position is 
1.45\,arcsec NW of ``A'';  the MSX source detected is offset 
3.7\,arcsec from the IRAS position and 2.25\,arcsec SE of ``A''.  
The unresolved 6-cm radio continuum source (flux density = 2.88\,mJy) 
detected by \citet{molinari98} is within 1.8\,arcsec of ``A''.  Given the 
agreement of the radio, IRAS, MSX and near-IR positions, ``A'' is 
likely to be the near-IR counterpart of the YSO; it has possibly 
evolved out of the very early phases of formation and may be in a 
UCH{\sc{ii}} phase.  The status of ``B'' needs to be investigated
more through multi-wavelength infrared observations with good 
angular resolution.

\subsection{IRAS~21078+5211 - {\it Mol 133}\\ ({\small \it d = 1.49\,kpc, L = 13.4\,$\times$10$^3$\,L$_{\odot}$})}

IRAS~21078+5211 is associated with a bright millimetre and
submillimetre source GH2O 092.67+03.07, located 30\,arcsec south
of a compact H{\sc ii} region (Bernard, Dobashi \& Momose 1999).
\citet{wb89} observed a red-asymmetry in the CO line implying a
possible outflow.  Through interferometric imaging in CS and line
observations of CS and CO,  \citet{bernard99} observed a circumstellar
disc (12\,M$_{\odot}$) and a very young outflow (0.6 M$_{\odot}$;
t$_{dyn} \simeq$ 3500\,yr) associated with GH2O 092.67+03.07.
The disc shows evidence for infall consistent with a high mass central
object and so GH2O 092.67+03.07 is believed to be protostellar.  They
derived a mass loss rate $\simeq$ 2$\times$10$^{-4}$ M$_{\odot}$ yr$^{-1}$
and a mass of the central star to be $\sim$6\,M$_{\odot}$.

IRAS~21078+5211 was selected as a candidate HMYSO by Molinari and
co-workers, who report the presence of a dense NH$_3$ core
\citep{molinari96} and a radio detection at 6\,cm \citep{molinari98}.
\citet{bronfman96} detected CS (2-1) with a velocity of -6.4\,kms$^{-1}$.
Neither \citet{szymczak00} nor \citet{slysh99} found any evidence for
CH$_3$OH maser emission. The source was included in the OH maser survey
of \citet{baudry97}, but again no detection was made.  H$_2$O maser
emission was detected by several investigators
\citep{palla91, brand94, miralles94, codella96, jenness95}.

Our $K$-band image is presented in Fig. \ref{21078_K}.  The
continuum-subtracted H$_2$ and Br$\gamma$ images do not show any
extended emission.  We do not see any significant IR sources, with
excess emission in 2MASS, near the IRAS or radio positions. The
three MSX sources detected in the field are towards the east of the
IRAS position and none of these appear to be associated with the
IRAS or the radio position of \citet{molinari98}.

There is a cometary UCH{\sc ii} region in this field detected by
\citet{miralles94} at 2 and 6\,cm with flux densities of 31$\pm$1\,mJy
and 40$\pm$1\,mJy, respectively.  This radio source is located towards
the SW of the IRAS position in our field (its location is shown in
Fig. \ref{21078_K}).  The 6-cm position of \citet{molinari98} is offset
from the position given by \citet{miralles94}, but agrees within the
angular extent of the core of the radio source at 6\,cm
(11.46\,arcsec$\times$16.16\,arcsec - Molinari et al., 19\,arcsec - Miralles et al).
The observations of \citet{molinari98} also revealed an extended halo at
6\,cm (60$\times$45\,arcsec) and an integrated flux density of 263.73\,mJy.
In Fig. \ref{21078_K},  we have labelled a bright 2MASS source 
``A'' ($\alpha$=21:09:21.77 $\delta$=52:23:10.8) detected near the radio
sources and surrounded by faint nebulosity.  ``A'' has multiple faint 
neighbours resolved in our images;  the coordinates given here
are those of the brightest object.  The 2MASS colours, which will be 
dominated by that of the brightest component ``A'' do not 
exhibit any IR excess (Fig. \ref{JHKcol}).  It is close to one of the 
MSX sources in this field, but the radio detections of both
\citet{miralles94}  and \citet{molinari98} are offset from it in 
a similar fashion suggesting that ``A'' is not the object responsible
for the radio emission.

The locations of the water maser (detected using the VLA) and 850-$\mu$m
continuum peak (observed using the JCMT) given by \citet{jenness95} agree
well; they fall outside the radio contours and are further SE from the
radio positions,  outside the field of view of our images.  The CS emission
from the disc detected by \citet{bernard99} also appears to be close to
the location of  the H$_2$O maser observed by \citet{jenness95}.   Hence,
it is possible that the youngest source in this region is located outside
our field, towards the SW. It is noteworthy that the outflow is associated
with the sub-mm source (located outside our field) and not from the UCH{\sc{ii}}
detected in the radio.  None of these fall within the error ellipse of
the IRAS position (20\,arcsec$\times$7\,arcsec).   It remains to be 
investigated whether the object implied by the IRAS detection is 
a deeply embedded and very young source different from those seen at 
longer wavelengths or if it is the combined emission from several objects.

\begin{figure}
\centering
\includegraphics[width=8.10cm,clip]{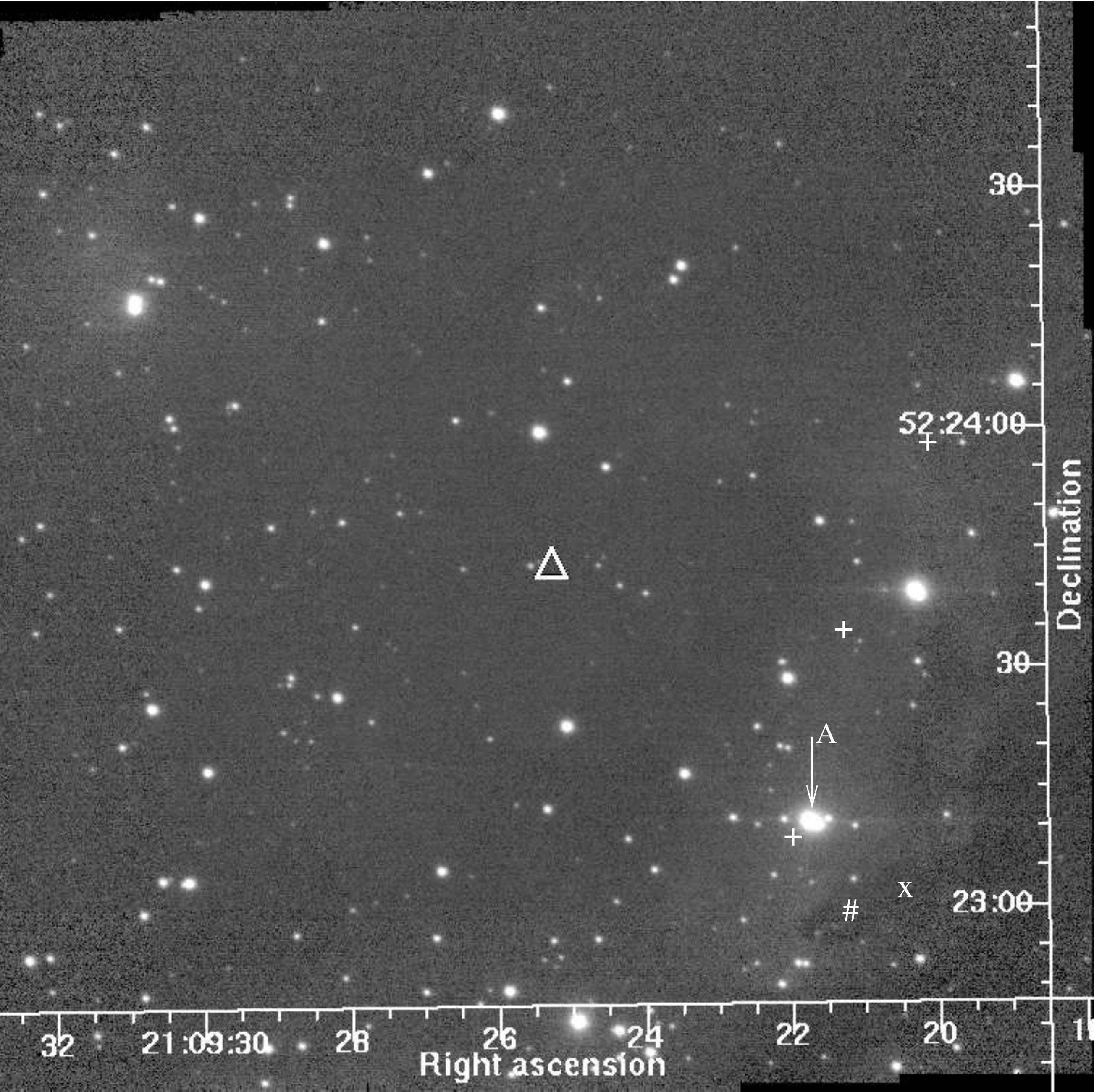}
\caption{$K$-band image of IRAS~21078+5211.  The 6-cm position of
\citet{molinari98} is shown by ``x'' and that of \citet{miralles94} is
shown by ``\#''}
\label{21078_K}
\end{figure}

\subsection{IRAS~21307+5049 - {\it Mol 136}\\ ({\small \it d = 3.6\,kpc, L = 4.0\,$\times$10$^3$\,L$_{\odot}$})}

A dense core was detected towards this source in NH$_3$ by
\citet{molinari96}.  \citet{molinari02} present high-spatial-resolution
observations of IRAS~21307+5049 in HCO$^+$ and 3.6-cm continuum
emission.  They find a compact molecular core with no free-free emission,
suggesting the presence of a very young, pre-UCH{\sc{ii}} object.  The
extended 3.6-cm emission detected by them is offset NE of their mm
continuum and HCO$^+$ emission peaks and the IRAS and MSX positions.
It is believed to be part of a supernova remnant and not connected
with the YSO.  IRAS~21307+5049 is also associated with water
maser \citep{palla91}.

Our near-IR images of this region reveal a cluster with a bright
object ``A'' ($\alpha$=21:32:30.60, $\delta$=+51:02:16.0) located
10.3\,arcsec SW of the IRAS position .  ``A'' is associated with
a cometary nebula (which opens towards the NW) in the $K$-band image,
and a few much fainter companions.  Fig. \ref{21307_KH2} shows our $K$-band
image on which source ``A'' is labelled. Our continuum-subtracted
Br$\gamma$ image does not show any extended emission.  The H$_2$ image
reveals a faint emission feature $\sim$3.7\,arcsec SE of ``A'',  which
is circled and labelled ``1'' on the figure. The region of the H$_2$
image close to ``A'' is shown in the inset in Fig. \ref{21307_KH2}.
The continuum-subtracted image exhibits a large negative residual on
the source ``A'', which implies a steep SED, probably arising from the
presence of heated dust.  This is consistent with the detection of
a single continuum emitting core at 3.4\,mm by \citet{molinari02}.
The MSX position is 10.2\,arcsec NE of ``A'' and is just 2.7\,arcsec
SE of the IRAS position, which casts doubt on ``A''  being the IR
counterpart of the YSO.  However, the location of the  3.4-mm core
imaged by \citet{molinari02} is only 0.5\,arcsec NE of ``A'' which
gives confidence in identifying ``A'' as the near-IR counterpart of
the HMYSO.  Moreover, the 2MASS $JHK_s$ colours of ``A'' exhibit 
large reddening and IR excess (Fig. \ref{JHKcol}), and place it in the
region occupied by YSOs in the colour-colour diagram (although the
source is poorly detected in $J$ and $H$ by 2MASS).  The faint
44-GHz CH$_3$OH maser is in fact detected just 2.2\,arcsec SE of
``A'' \citep{kurtz04}.  The location of ``A'' is offset by only
$\sim$1\,arcsec from the peak of the centroid of the 3-mm OVRO
source of \citet{fontani04}.  The peak position of their 850-$\mu$m
SCUBA source is also close to ``A''.  \citet{fontani04} also imaged
IRAS~21307+5049 in the near-IR $JHK_s$ bands and in H$_2$ and
Fe{\sc ii};  they did not detect line emission in H$_2$ and Fe{\sc ii}.

\citet{zhang05} detected a bipolar outflow in CO, the blue- and
red-shifted wings of which are oriented roughly in the NW-SE
direction.  This is in agreement with the location of the H$_2$ 
emission feature ``1'' SE of ``A''.  The centroid of the CO 
emission detected by them is 14\,arcsec west and north of the 
IRAS position, which is within their spatial resolution 
(Table \ref{resolutions}).  More recent $^{12}$CO maps 
\citep{fontani04}, made with better resolution and positional 
accuracy, also show a bipolar outflow, oriented NW-SE and centred 
on the 3-mm core, the location of which is close to that of 
``A'' (suggesting that ``A'' is the driving source of the 
outflow).  From its morphology, they suspect that the outflow 
is made up of two components.

\begin{figure}
\centering
\includegraphics[width=8.10cm,clip]{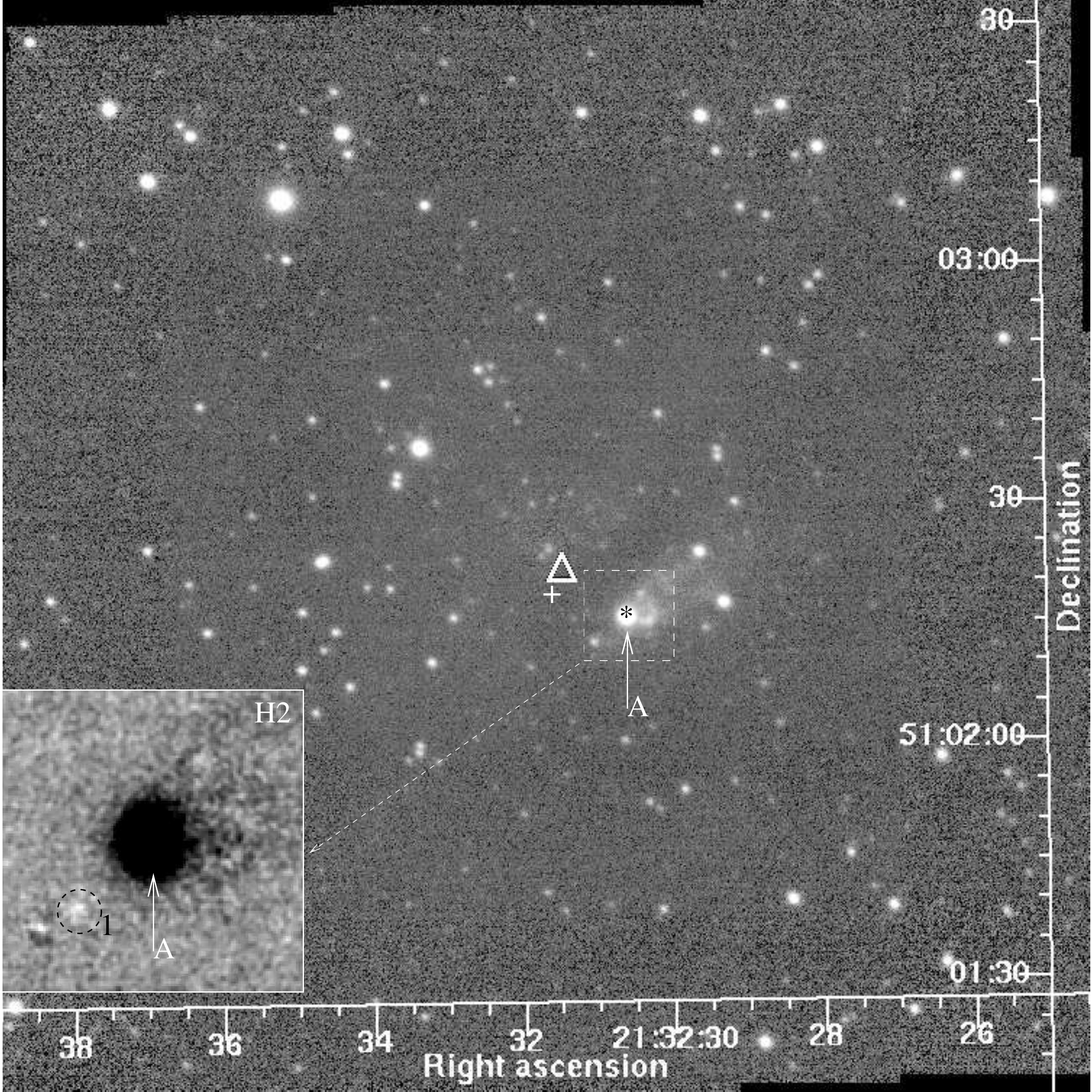}
\caption{Observed $K$-band image of IRAS~21307+5049.
Continuum-subtracted H$_2$ image of the region around the IR source
is shown in the inset. ``*'' shows the 3.4-mm continuum peak of
\citet{molinari02}.}
\label{21307_KH2}
\end{figure}

\subsection{IRAS~21391+5802 --{\it Mol 138}\\ ({\small \it d = 0.75\,kpc, L = (0.0939; 0.15)\,$\times$10$^3$\,L$_{\odot}$})}

\begin{figure*}
\centering
\includegraphics[width=16.5cm,clip]{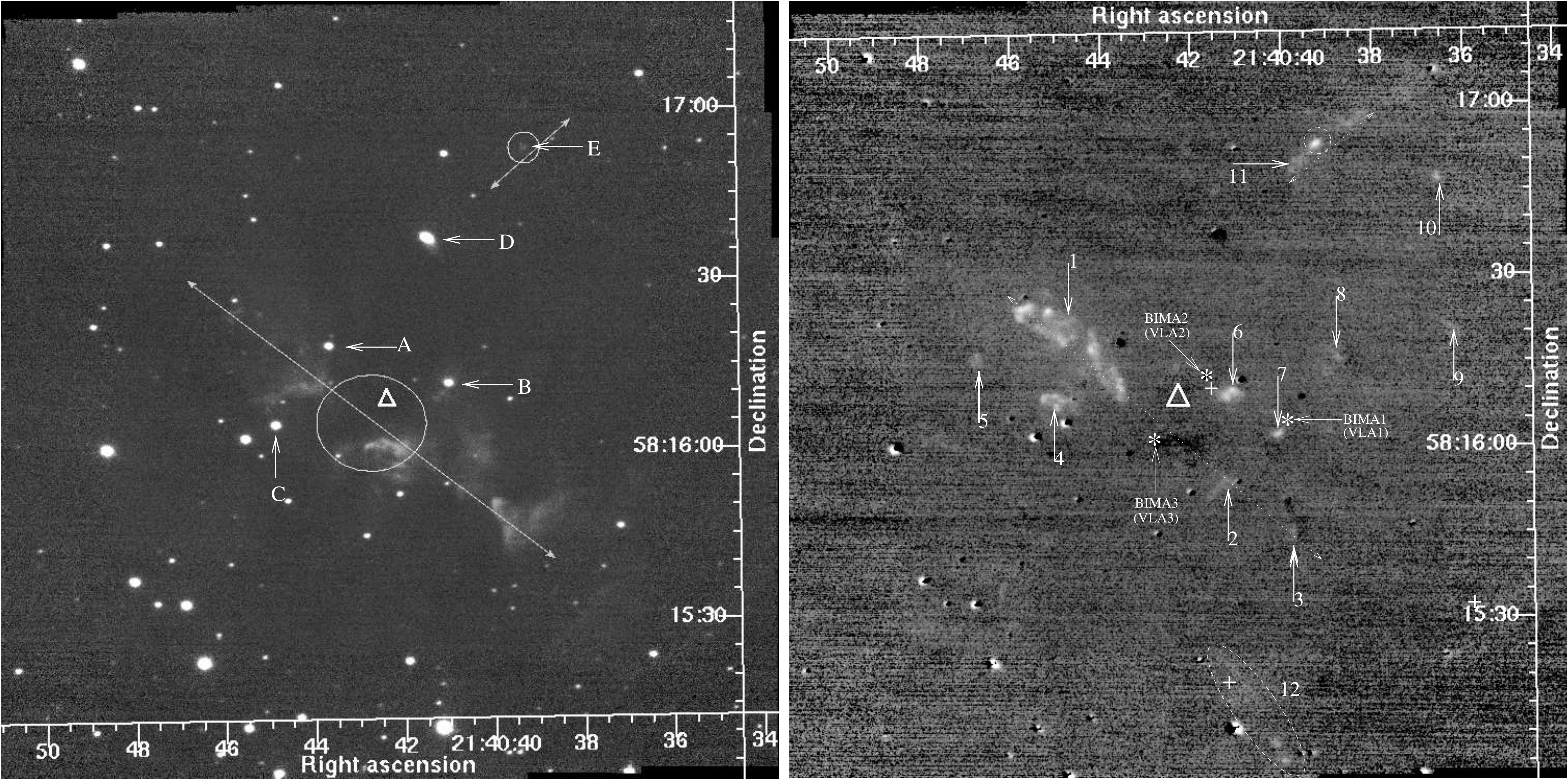}
\caption{The left panel shows the $K$-band image of IRAS~21391+5802. 
The right panel shows the continuum-subtracted H$_2$ image. The  
H$_2$ has been smoothed with a Gaussian of FWHM=2 pixels to enhance 
the faint emission features. The three mm and radio continuum sources 
detected by \citet{beltran02} are indicated by ``*'' and are labelled 
``BIMA1--3'' and ``VLA1--3''.   The locations of the brighter of the 
three (``BIMA2--3'') agree with those of the two mm continuum 
positions of \citet{codella01}}.
\label{21391_K}
\end{figure*}

IRAS~21391+5802 is embedded in the bright-rimmed globule IC1396(N)
in the H{\sc ii} region IC1396 in Cepheus. It was identified as
an intermediate-mass protostar by \citet{wilking93} and was
detected in NH$_3$ emission by \citet{molinari96}.  Methanol
masers from this region have been observed by \citet{kurtz04} and
water maser emission has been detected by Felli, Palagi \&
Tofani (1992), \citet{tofani95} and \citet{patel00}.  The
multi-epoch observations of \citet{patel00} allowed the proper
motions to be measured for ten water masers.  These were
consistent with the masers tracing the dominant bipolar outflow
in this source, even within $\sim$1AU of the source.

A number of outflows have been discovered in this region.
A bipolar outflow on arcminute scales was discovered by
\citet{sugitani89} in CO (J=1-0); it was subsequently mapped
by \citet{wilking90} (CO, J=2-1) and \citet{codella01}.
Shock-excited emission from [O{\sc i}] was seen in the
outflow in the ISO LWS spectra of \citet{saraceno96}; their 
spectra also revealed CO, OH and H$_2$O emission toward IRAS
21391+5802 showing the presence of warm, dense molecular gas.
They also obtained continuum observations from 350\,$\mu$m to 
1.3\,mm and derived a bolometric luminosity of 235\,L$_{\odot}$.

More recent studies of the region at millimetre wavelengths have since
revealed three embedded sources, the strongest of which (BIMA~2)
is associated with the IRAS source \citep{beltran02, beltran04}.
At higher angular resolution ($\sim$4\,arcsec), CO maps by
\citet{beltran02} show a collimated outflow oriented approximately
EW centred on ``BIMA~2'',  and a second, lower velocity
outflow running NS from ``BIMA~1''. The east-west flow is
easily identified as bipolar at high velocities (the
15.5-20.5kms$^{-1}$ channel in \citealt{beltran02}) whereas at low
velocity, the kinematics are complex and the outflow loses its
identity. It is speculated that other outflows may be confusing
the picture. The CS (2-1) emission echoes the characteristics of
the ``BIMA~2'' flow, being clearly bipolar at high velocity, though less
distinct at lower velocities. Observations in CH$_3$OH and CS (5-4)
reveal clumps of emission along the outflow axis. The total mass
of the EW outflow is estimated to be 13.6$\times
10^{-2}$M$_{\odot}$; the NS outflow mass is estimated to be
1.2$\times 10^{-2}$M$_{\odot}$ \citep{beltran02}.  Beltr\'{a}n et al.
also estimated a luminosity of $\sim$150\,L$_{\odot}$ for ``BIMA 2''
and suggest that ``BIMA~1'' and ``BIMA~3'' are more evolved lower-mass
objects.  No radio emission has been detected towards this region 
at 6\,cm by \citet{molinari98}. However, VLA observations of 
\citet{beltran02} at 3.6\,cm revealed three faint continuum 
sources ``VLA~1'' (0.21\,mJy), ``VLA~2'' (0.27\,mJy)
and ``VLA~3'' (0.43\,mJy), spatially coinciding with ``BIMA~1'', 
``BIMA~2'' and ``BIMA~3'' respectively.

Previous infrared observations by \citet{nisini01} show H$_2$
${\it{v}}=$1-0 S(1) emission extending along the ``BIMA~2'' outflow.
They also identify two additional outflows oriented EW, by
associating the H$_2$ emission with previous CO (J=2-1) maps from
\citet{codella01}. On larger scales, \citet{reipurth03}
discovered a new Herbig-Haro flow from IC1396N, with the
working surface located 0.6\,pc from the putative source newly
identified in their $JHK$ imaging and located on the HH 777 
flow axis.  They label this source as HH~777~IRS.
\citet{reipurth03} found no H$_2$ emission clearly associated
with the HH 777 flow, but do confirm most of the small scale flows
observed by \citet{nisini01}. 

Fig. {\ref{21391_K}} shows our $K$-band and continuum-subtracted
H$_2$ images.  No Br$\gamma$ emission is observed from this region.
Our observations cover a field much smaller than that obtained by
Nissini et al (2001), but have better spatial resolution.  There is
a spectacular outflow in the NE-SW direction, along the dotted arrow
and traced by the H$_2$ emission features labelled ``1--3''.
There is a second bipolar outflow emanating from the centre; the
features labelled ``4'' and ``5'' appear to be tracing one of the
the lobes of this bipolar outflow.  It is not clear if ``6'' and
``7'' trace the other lobe of this flow or represent another outflow.
The centre of the field around the IRAS position is devoid of
near-IR sources.  The region appears to be suffering from large
extinction and the driving sources of the jets revealed in H$_2$ are
not detected in $K$.  Four near-IR point sources closest to the IRAS
position are labelled ``A--D''.  All four sources exhibit
reddening and excess (Fig. \ref{JHKcol}).  However, from their
locations, these stars do not appear to be driving the two
outflows mentioned above.  
``D''($\alpha$=21:40:41.46, $\delta$=58:16:37.7) is identified
as HH~777~IRS detected by \citet{reipurth03} and our $K$-band 
image shows a cometrary nebula associated with it.  ``D'' is
the only source detected in the 11.6\,$\mu$m image obtained
by Reipurth et al.

The YSOs driving the two outflows mentioned above
are likely to be located within the circle shown near the centre
of the field, which includes the IRAS, MSX and mm continuum
positions.  Other H$_2$ emission features detected in this
field are labelled ``8--12''. It remains to be investigated
if ``8--9'' are associated with the outflow from ``D'', 
detected by \citet{reipurth03}.  The feature labelled ``11''
appears to be another bipolar outflow, probably produced by the
object labelled ``E'' ($\alpha$=21:40:39.26, $\delta$=58:16:53.2),
which is not detected by 2MASS.  Multiple stars of intermediate
and low mass are probably being formed in the central region.
This is confirmed by the multiple outflows suggested by the H$_2$
knots and the three millimetre sources detected by \citet{beltran02}.
The outflow defined by the direction of ``1'' is oriented at a
position angle of 50$^{\circ}$ and has a collimation factor of 6.5.
The blue-shifted lobe of the outflow from ``BIMA 2'' mapped in
CO by \citet{beltran02} roughly traces the jets ``1'' and/or ``2''.
\citet{wilking90} derived a position angle of 75\,degrees
for the CO outflow detected by them. The locations of the three
faint MSX sources in this field (detected only at 8.28\,$\mu$m)
are shown on Fig. {\ref{21391_K}}.  The first one is 5.9\,arcsec
NW of the IRAS position, close to ``B'' and ``BIMA2''.  The second
source is near the knotty H$_2$ emission feature ``12''.

\subsection{IRAS~21519+5613 - {\it Mol 139}\\ ({\small \it d = 7.3\,kpc, L = 19.1$\times$10$^3$\,L$_{\odot}$})}

\begin{figure*}
\centering
\includegraphics[width=16.5cm,clip]{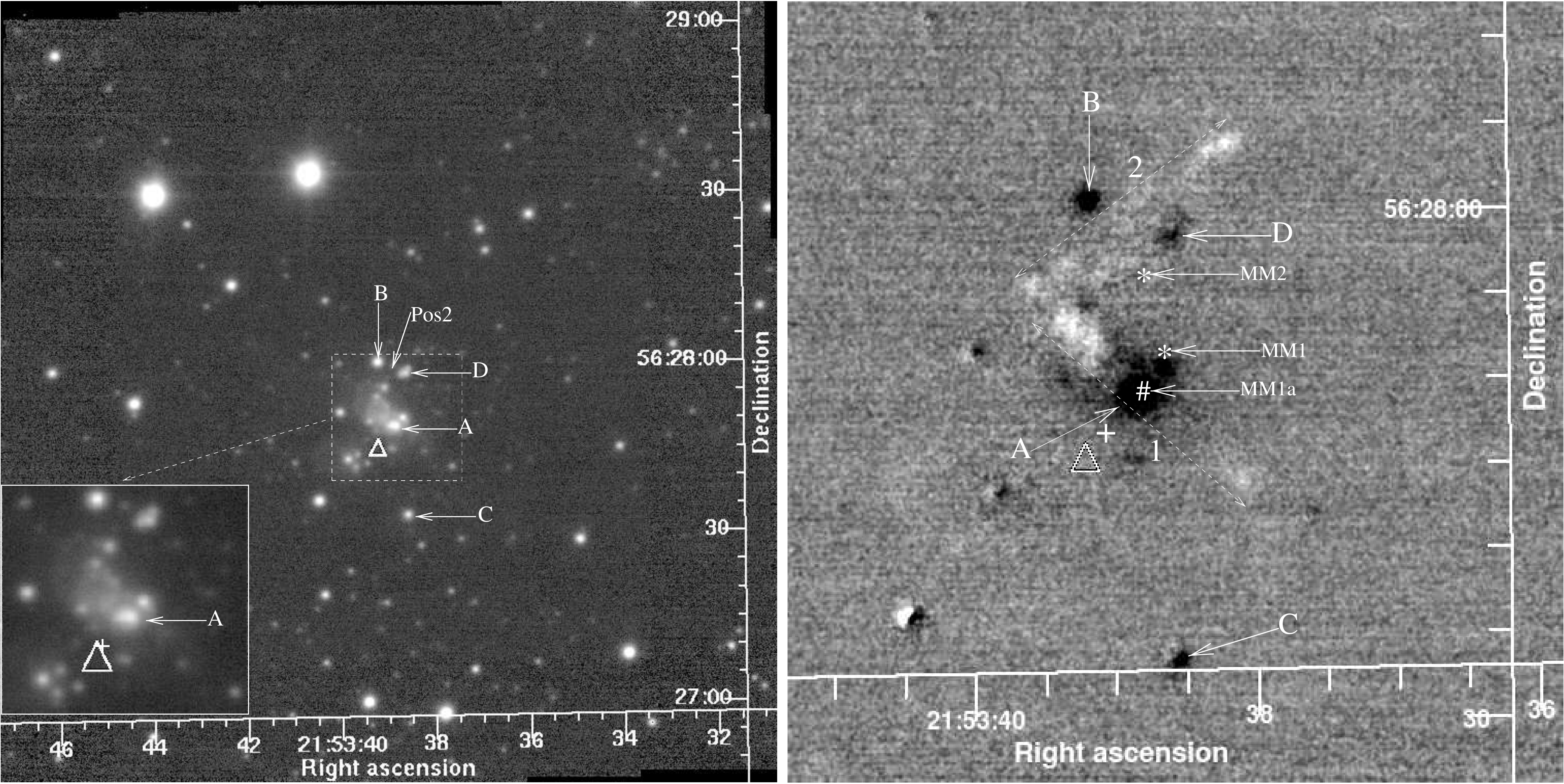}
\caption{Left: $K$-band image of IRAS~21519+5613. An enlarged
view of the central region is shown in the inset.  Right:
continuum-subtracted H$_2$ image of the central region.
``\#'' shows the 1.3-mm continuum peak and ``*'' shows the
two 3.4-mm continuum peaks detected by \citet{molinari02} .}
\label{21519_KH2}
\end{figure*}

Early observations by \citet{cesaroni88} found H$_2$O maser
emission from this source at 35\,Jy, though some of the later
attempts by \citet{comoretto90}, Gyulbudaghian, Rodriguez \&
Curiel (1990),  \citet{palagi93} and \citet{palla91} did not
result in a solid detection, yielding only upper limits.
However, \citet{wouterloot93} detected three-component H$_2$O
maser emission, $\sim$8.6\,arcsec NW of the IRAS position. No CH$_3$OH
maser emission was found \citep{macleod98b}.  A search
by \citet{wouterloot93} for OH masers also resulted in a null detection.
There was no 6-cm free-free continuum emission detected from this region
\citep{molinari98}.  \citet{molinari96} and \citet{bronfman96} detected a
core in NH$_3$ and CS(2-1) respectively.  CO lines were observed by
several investigators (\citealt{wb89}; \citealt{wouterloot93};
Su, Zhang and Lim 2004).  All of the detected line emissions were at
a v$_{LSR}$ close to -63 km s$^{-1}$.

No H$_2$ imaging of this region appears to have been done before.
Our $K$-band image reveals a cluster of IR sources close to the IRAS
position.  Most of the associated nebulosity disappears in our
continuum-subtracted H$_2$ image (right half of Fig. {\ref{21519_KH2}}).
However, the H$_2$ image reveals two faint well defined bipolar outflows,
the directions of which are shown by two arrows labelled  ``1'' and ``2''.
``1'' is oriented in the NE-SW direction at an angle $\sim$48$^{\circ}$;
it appears to be bipolar in nature about an IR source labelled
``A'' ($\alpha$=21:53:38.82, $\delta$=56:27:49.3).  This source appears
slightly extended EW and may be a binary. There appears to be
a companion $\sim$0.8\,arcsec east of ``A''; the seeing was
$\sim$1 arcsec when we observed this field,  so they are not very
well resolved.  There is another point source NW of ``A''
at a separation of $\sim$2.1 arcsec.  The 2MASS magnitude of ``A'',
which is a composite of all three sources, some other fainter components
and the faint nebulosity associated with the central region, shows IR
excess (Fig. \ref{JHKcol}). The outflow ``2'' is located north of
``1'' and is directed in the NW-SE direction with an angle of
$\sim$126.5$^{\circ}$.  The point source labelled
``B'' ($\alpha$=21:53:39.19 and $\delta$=56:28:0.7) is slightly offset
from the axis and centre of ``2''.  This source exhibits IR excess in
the 2MASS data.  ``D'' ($\alpha$=21:53:38.59, $\delta$=56:27:58.6)
is located near, but slightly offset from, the centre of ``2''.
``D'' is extended in NW-SE, implying the possible presence of
companions and the aggregate 2MASS colours do not exhibit any 
IR excess.  The coordinates given here are that of the brightest 
point source in ``D''.  There is a source labelled ``C''
($\alpha$=21:53:38.58 and $\delta$=56:27:33.6), which is detected
in $H$ and $K$ only by 2MASS.  The colours derived using the 2MASS
$H$ and $K$ magnitudes and the upper limit magnitude in $J$ place
it in the region of a reddened sources with IR excess in Fig. \ref{JHKcol};
deeper near-IR photometry is required to undertand its nature.  No
H$_2$ line emission is observed in its near vicinity.
Br$\gamma$ was not detected from either jet ``1'' or jet ``2''.

Our observations are consistent with the longer wavelength observations
of \citet{molinari02}. They discovered a very well defined core, extended
in the NS direction in HCO$^+$ emission.  They also detected two
continuum emitting cores at 3.4\,mm - one source named MM1 coinciding
with the brightest region of the HCO$^+$ core and the other, MM2,
located north of MM1.  A bright 1.3\,mm continuum source (MM1a) 
detected by them is nearly coincident with MM1 and is offset by only 
$\sim$2.2\,arcsec SE from MM1.  The mm continuum sources are close 
to the two jets discovered in our H$_2$ image and possibly harbour 
their driving sources. The millimetre positions are labelled on 
Fig. {\ref{21519_KH2}}.  The location of MM1a coincides  with that 
of ``A'', confirming that ``A'' is the near-IR counterpart of the 
YSO driving the outflow ``1''.  The location of MM2 is between 
``A'' and the ``BD'' axis.  MM2 is located close to the centre of 
``2'', but is slightly offset towards MM1.  The MSX mission also 
detected an IR source at wavelengths from and above 8.28\,$\mu$m 
(in C, D \& E bands).  The position of the detected MSX
source is 3.4\,arcsec SE of ``A'', close to the IRAS source which
is 5\,arcsec SE of ``A''.

\citet{su04} observed this region in 3-mm continuum and in
$^{12}$CO, $^{13}$CO and C$^{18}$O lines.
The 3-mm observations showed a peak, a few arcsec NW
of the IRAS position. The location of the 3-mm continuum peak
agrees well with the location of ``A'' to within 0.6 arcsec. Their
C$^{18}$O centroid also is very close to the 3-mm peak. The origin
of the outflow mapped in $^{12}$CO and $^{13}$CO appear to be
closely coincident with the centroid of the molecular gas and dust
continuum. Their velocity integrated $^{12}$CO line map shows blue-
and red-shifted lobes extending NE and SW respectively from the
3-mm peak, with a position angle of $\sim$75$^{\circ}$.  This is
consistent with the fact that, in our H$_2$ image, the NE lobe of
the jet labelled ``1'' is  brighter than the SW lobe.
It should be noted that the integrated $^{12}$CO map, especially
the blue-shifted lobe, also shows an elongation in the NW-SE
direction, which is consistent with the presence of the second
outflow (``2'') in our H$_2$ image.  Su et al. derived a
collimation factor of 1.6 for the observed outflow and 2.4 
after deconvolution. From our H$_2$ images, we estimate
collimation factors of 3.3 and 5.5, respectively, for the two
outflows.  \citet{zhang05} detected a bipolar outflow in CO, the
centroid of which is offset north and west of the IRAS location
by 14\,arcsec each, the direction of which is consistent with the
NW offset of ``A'' from the IRAS position, even though the offset
is within their angular resolution.

Overall, this region appears to host two collimated outflows.
``A'', and possibly ``B'' or its neighbours, appear to be the
near-IR counterpart/s of the IRAS and MSX object and are the
most likely candidate HMYSOs in this region.

\vskip 15mm
\subsection{IRAS~22172+5549 - {\it Mol 143}\\ ({\small \it d = 2.4\,kpc, L = 1.8$\times$10$^3$\,L$_{\odot}$})}

\begin{figure*}
\centering
\includegraphics[width=16.5cm,clip]{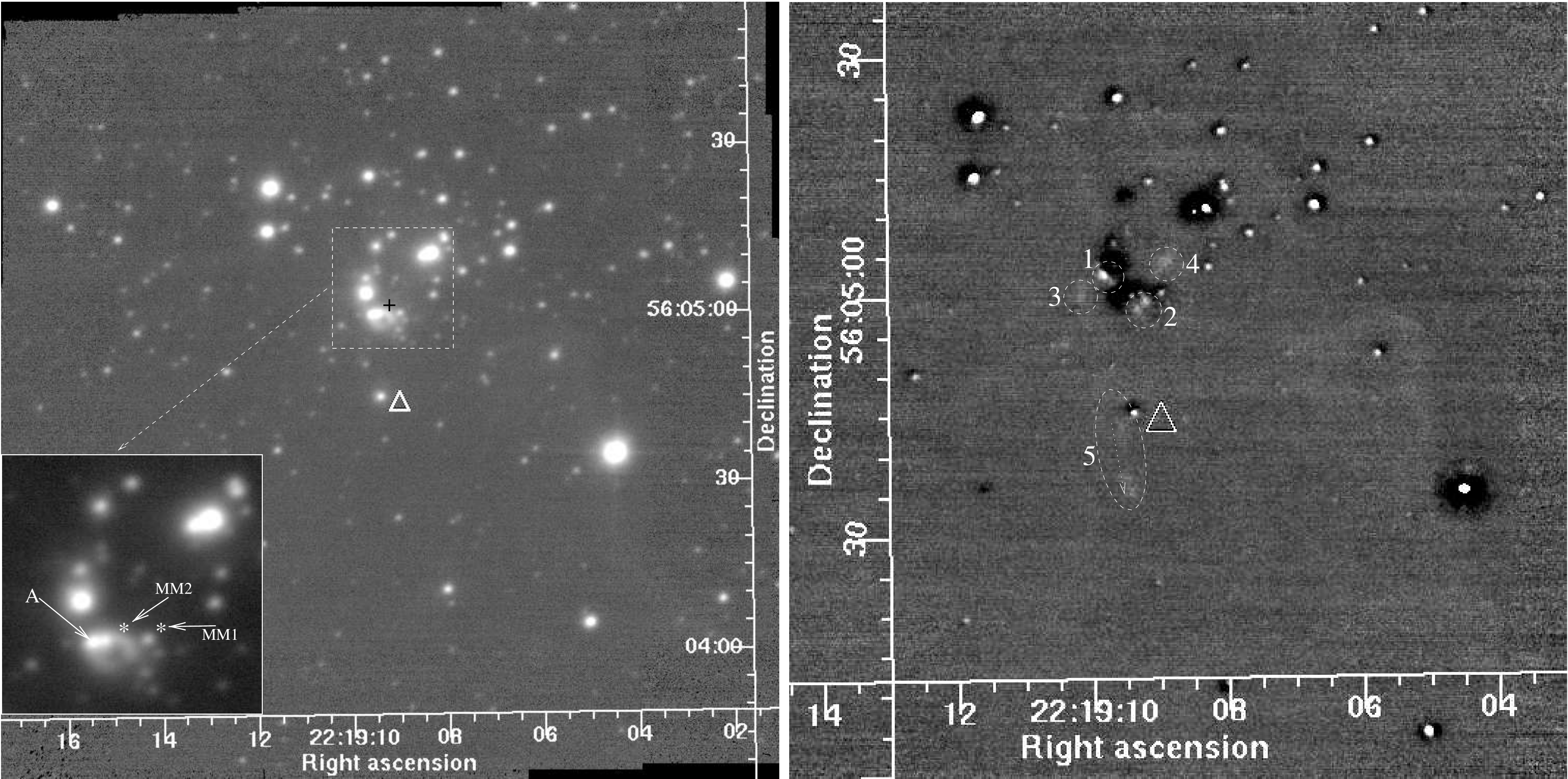}
\caption{Left: $K$-band image of IRAS~22172+5549.
An enlarged view of the central region is shown in the inset.  The
two 3.4-mm continuum sources of \citet{molinari02} are shown
by ``*'' in the inset.  Right: continuum-subtracted
H$_2$ image of the central region smoothed with a Gaussian of 2-pixel FWHM.}
\label{22172_KH2}
\end{figure*}

This is a known H{\sc ii} region. However, \citet{molinari98} did 
not find any radio emission from this source at 2\,cm or 6\,cm.
The dense core towards this region was detected in NH$_3$ emission 
by \citet{molinari96}.  The H$_2$O maser emission from this source 
is extremely faint.  \citet{palla91, palla93} did not detect H$_2$O 
maser emission.  Nor did \citet{comoretto90} or \citet{palagi93} 
who give only upper flux limits.  \citet{cesaroni88} and 
\citet{wouterloot93} detected faint H$_2$O maser emission.  A 
search by \citet{wouterloot93} for the 1665 MHz OH maser did not 
yield any detection.

Fich, Dahl \& Treffers (1990) detected H$_\alpha$ emission from this
source using Fabry-Perot observations.  \citet{molinari02} present
high-spatial-resolution HCO$^+$, 3.4-mm and 3.6-cm observations of
IRAS~22172+5549.  They find a compact molecular core with no free-free
emission, suggesting the presence of a very young pre-UCH{\sc{ii}}
protostar.  \citet{wb89} detected red-shifted  CO (1-0) emission.
\citet{zhang05} have also mapped a compact CO outflow toward this
source.  Recently, \citet{fontani04} have observed IRAS~22172+5549 
in a number of molecular tracers, and have obtained near-IR images 
in H$_2$ and [FeII].  They observed line emission features 
associated with a compact north-south CO outflow.

Our $K$ and H$_2$ images of this region show a cluster of objects
centred north of the IRAS position (Fig. {\ref{22172_KH2}}), with
some members of the cluster distributed in a ring-like pattern
where most of the activity appears to be happening.  The
continuum-subtracted H$_2$ image reveals two or more well collimated
jets.  \citet{molinari02} detected HCO$^+$ from the cluster, which
appears to be a single condensation in their integrated maps, but is
resolved into three separate condensations in their velocity resolved
maps.  Two of these cores show 3.4-mm continuum-emitting counterparts.
Their 3.4-mm position appears to be located within the cluster, NE of
the IRAS position.  The location of the centroid of HCO$^+$ emission is
also very close to the 3.4-mm peak.  The brighter of the two mm
sources (MM2) is only 2.7\,arcsec NW of the source labelled
``A'' ($\alpha$=22:19:9.48, $\delta$=56:5:0.3) in
Fig. {\ref{22172_KH2}}; the fainter mm peak (MM1) is 6.6\,arcsec NW of
``A''.  The 3-mm peak position of \citet{fontani04} also agrees
with the mm position of \citet{molinari02}.  The MSX mission detected
a mid-IR source at wavelengths 8.28\,$\mu$m and above.  The location of
the MSX source is $\sim$17.3 arcsec NE of the IRAS position and is
only 2.8\,arcsec NW of ``A'',  nearly coincident with MM2.  ``A'' itself
is double with the components separated by $\sim$1\,arcsec at an angle 
of 102$^{\circ}$. It is detected by 2MASS in $H$ and $K_s$ with an 
$H-K_s$ colour of 1.97.  It is not well detected by 2MASS in $J$, 
however, and the pair is not resolved.

There are several other fainter objects in the vicinity of ``A'', most
of which were not detected by 2MASS.   Our $K$-band image shows that
``A'' is associated with a broken ring-like nebulosity, with a diameter
of $\sim$2.8 arcsec, with ``A'' offset from the centre and embedded in
the ring.  The nebulosity disappears upon continuum-subtraction. Our
continuum-subtracted  H$_2$ image shows H$_2$ line emission on the NE
and SW sides of ``A'', which are labelled ``1'' and ``2''.  Also obvious
in the figure are the H$_2$ emission features circled and labelled
``3--5''.  ``5'' appears to contain the bow-shock of a jet;
there is some faint H$_2$ emission seen along the dotted arrow drawn
near ``5''.  The NS oriented bipolar outflow mapped in CO by
\citet{fontani04} has the centroid of the two lobes close to ``A''
and is in the direction of ``5''.  The near-IR images of
\citet{fontani04} also show most of the H$_2$ emission features
labelled on Fig. {\ref{22172_KH2}}.  Their H$_2$ and [Fe{\sc ii}]
images show the NE-SW oriented features in the direction of
``1--2'', with ``5'' located towards the outer region of the
red-shifted lobe of their CO map.  Hence it is possible that the
CO outflow mapped by them is composed of more than one.
Overall, this region appears to be active in star formation, hosting
at least two outflows.  With the positional accuracy quoted by
\citet{molinari02} and the proximity of the mm sources and the
centroid of the CO outflow to ``A'',  this source and/or some of
its very near neighbours are YSOs, which are responsible for the
outflows discussed here.  Deeper IR imaging at sub-arcsec resolution
is certainly warranted.

\subsection{IRAS~22305+5803 - {\it Mol 148}\\ ({\small \it d = 5.4\,kpc, L = 14.1$\times$10$^3$\,L$_{\odot}$})}

A dense core was detected towards IRAS~22305+5803 in CS
\citep{bronfman96} and NH$_3$ \citep{molinari96}.
The source is associated with H$_2$O maser emission \citep{palla91}.
\citet{molinari02} mapped the region at high spatial resolution
in HCO$^{+}$(1-0) and resolved a remarkable, clumpy ring of line
emission with the IRAS position located in the SW part of the ring.
However, their  3.4-mm maps did not reveal any continuum emitting
compact source within the ring.  3.6-cm radio continuum
observations obtained by them using the VLA detected a faint radio
source (0.12$\pm$0.03\,mJy) within the western part of the HCO$^{+}$
ring.  Their IRAM observations at 1.3\,mm revealed a compact core
associated with the source detected at 3.6\,cm.  From the radio
flux, they derive a B2 ZAMS star.  \citet{wb89} detected CO (1-0)
emission (with a blue shoulder).  \citet{zhang05} mapped a bipolar
outflow in CO line, close to the IRAS source.

Fig. \ref{22305_KH2} shows our $K$-band and continuum-subtracted
H$_2$ images. The $K$-band image reveals a compact cluster of IR sources
embedded in nebulosity at the centre of the field.  Most of the objects
in the cluster are located along a ridge extending roughly NS.  The
continuum-subtracted H$_2$ image shows large negative residuals associated
with the nebulosity towards the centre of the field.  Most of the diffuse
emission towards the northern part of the cluster disappears except for
three clumpy line emission features labelled ``1'' on the H$_2$ image.  
No point sources are seen at the location of these H$_2$ emission 
features; i.e, they are true H$_2$ emission features rather than 
residuals from continuum subtraction.  These features could be 
produced by one or more jets.

Three sources are labelled on Fig. \ref{22305_KH2} -
``A'' ($\alpha$=22:32:23.77, $\delta$=58:18:59.7),
``B'' ($\alpha$=22:32:23.77, $\delta$=58:19:07.7) and
``C'' ($\alpha$=22:32:24.09, $\delta$=58:19:11.4).
These bright sources are associated with the H{\sc ii} region
and the SW and western sides of the HCO$^{+}$ ring.
``A'' is not detected by 2MASS; the 2MASS detection is instead centred on
the nebulosity.  ``B'' exhibits slight excess in 2MASS data and ```C''
does not exhibit any excess (Fig. \ref{JHKcol}).  The location of the
IRAS object is 4.5\,arcsec SE of ``A''; the MSX source is 1.65\,arcsec
NE of ``A''.  The peak of the  3.6-cm emission detected by
\citet{molinari02} is only  0.98\,arcsec NW of ``A''.

\begin{figure*}
\centering
\includegraphics[width=16.5cm,clip]{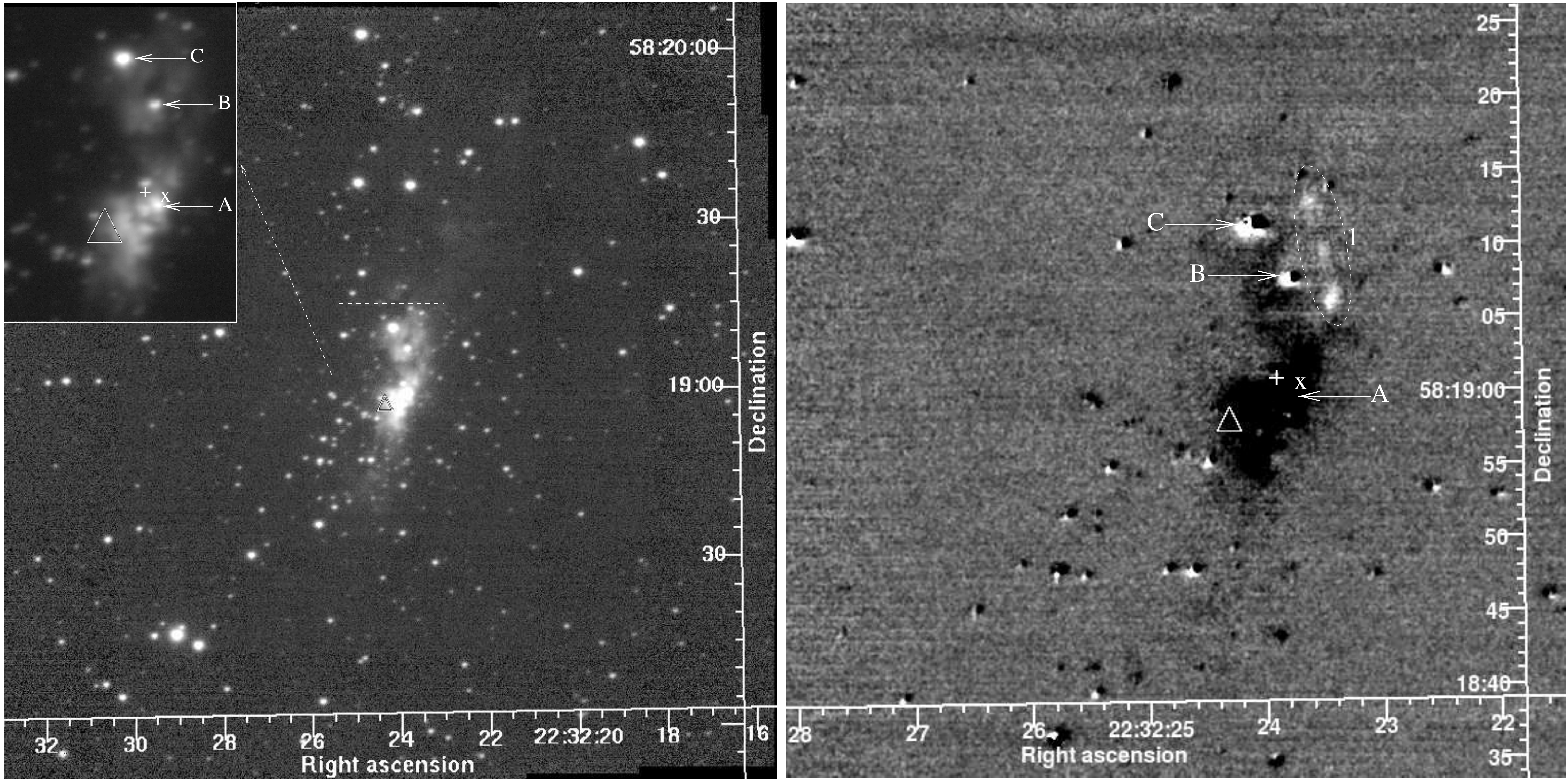}
\caption{The left panel shows the $K$-band image of IRAS~22305+5803.  An
enlarged view of the central region is shown in the inset. The 3.6-cm
radio continuum position of \citet{molinari02} is shown by ``x''. The
right panel shows the continuum-subtracted H$_2$ image of the central
region smoothed with a Gaussian of 2-pixel FWHM to enhance the faint
emission features.}
\label{22305_KH2}
\end{figure*}

The centre of the blue-shifted lobe of the CO outflow detected
by \citet{zhang05} appears close to the region of line emission
seen in our H$_2$ image.  The H$_2$ emission features  that we see
are therefore probably from the blue-shifted lobe of a highly
inclined jet.  We do not know if more than one outflow is
involved here.  However, the close proximity of the IRAS, MSX and
3.6-cm positions suggests that ``A'', rather than ``B'' or ``C'',
is the likely outflow source. The nature of ``A'' and its neighbours
needs to be probed through near- and mid-IR obsevations.

\subsection{IRAS~22570+5912 - {\it Mol 153}\\ ({\small \it d = 2.92; 5.1\,kpc, L = 20.1; 50.1$\times$10$^3$\,L$_{\odot}$})}

\begin{figure*}
\centering
\includegraphics[width=16.5cm,clip]{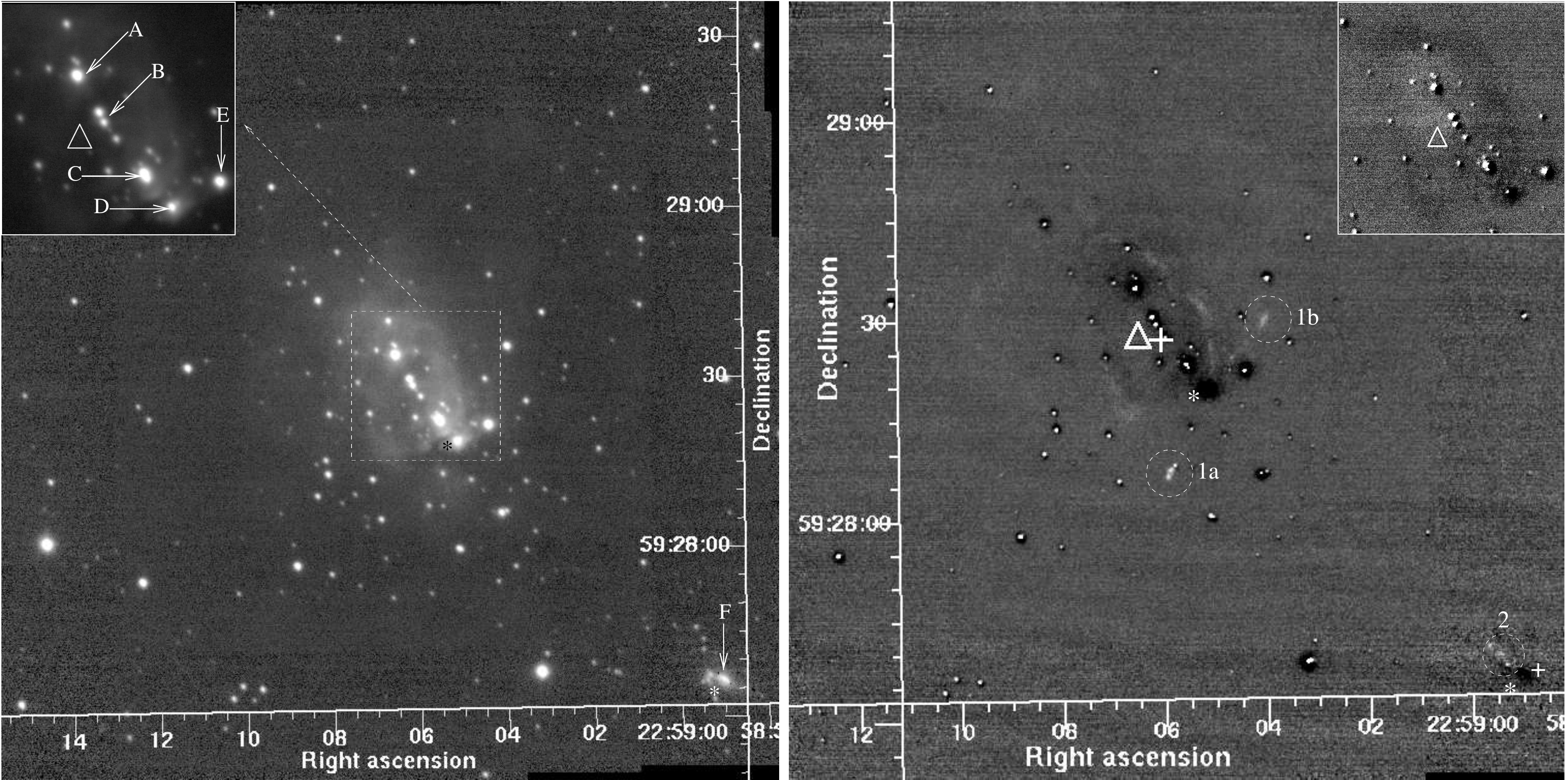}
\caption{The left panel shows the $K$-band image of IRAS~22570+5912.
The right panel shows the continuum-subtracted H$_2$ image of the
central region on which the continuum-subtracted Br$\gamma$ image of
the central cluster is shown in the inset. The H$_2$ image has been
smoothed with a Gaussian of 2-pixel FWHM.  The peak positions of the
two 1.2-mm sources of \citet{beuther02b} are shown by ``*''.}
\label{22570_KH2BrG}
\end{figure*}

High velocity CO was identified from IRAS~22570+5912 by \citet{shepherd96a}.
In their follow-up observations,  \citet{beuther02c} mapped the outflow
in CO (2--1) and found a complex morphology with the flow centred off the 
peak 1.2-mm position. This fact, coupled
with the broad extent of the flow at low angular resolution, leads them
to posit that there may be several outflows in the region.  The emission
extends over 1.5\,pc; the mass of the outflow is estimated at
73\,M$_{\odot}$. No emission from either H$_2$O \citep{palla91, sridharan02}
or CH$_3$OH \citep{macleod98b, slysh99, sridharan02} masers has
been found, nor emission from SiO or CH$_3$CN
\citep{sridharan02}. The dense core towards this region was
detected in CS emission \citep{bronfman96, beuther02b}. No ammonia
emission was detected by \citet{molinari96}.  Observations of the
dust continuum by \citet{beuther02b} at 1.2\,mm reveal four clumps,
with the brightest one located close to the centre of our field,
near the IRAS position.  The second brightest source is also within
our field, located towards the bottom right corner.  This structure
is partly repeated in their observations of dense gas tracers
(CS, C$^{34}$S), where three of the four 1.2-mm sources have
counterparts.

Fig. \ref{22570_KH2BrG} shows our $K$, H$_2$ and Br$\gamma$
images. The $K$-band image reveals a cluster, roughly centred on
the IRAS position.  A set of IR-bright objects is seen roughly
aligned in the NE-SW direction and embedded in nebulosity. The
nebula is roughly elliptical in morphology, the major axis of
which is closely aligned with the string of bright stars.
A few of the sources are labelled on Fig. \ref{22570_KH2BrG} -
``A'' ($\alpha$=22:59:06.52, $\delta$=+59:28:34.7),
``B'' ($\alpha$=22:59:06.18, $\delta$=+59:28:30.1),
``C'' ($\alpha$=22:59:05.54, $\delta$=+59:28:23.2),
``D'' ($\alpha$=22:59:05.12, $\delta$=+59:28:19.3),
``E'' ($\alpha$=22:59:04.40, $\delta$=+59:28:22.2) and
``F'' ($\alpha$=22:58:59.05, $\delta$=+59:27:36.4).
Most of these are resolved into multiple components in our
image and all exhibit reddening (Fig. \ref{JHKcol}).  ``A''
shows slight excess; it is resolved into a bright source with
a few fainter ones in the neighbourhood.  ``B'' is the object
closest to the IRAS position;  it is composed of a binary
with components of comparable brightness separated by
1.3\,arcsec.  The coordinates of ``B'' given here are of the
centroid of the pair. There is a third source $\sim$3\,arcsec
SW of the pair and a much fainter 4th object is located in
between the pair and the third source.  2MASS does not resolve
these and the 2MASS magnitudes, which are composites of these
four stars and dominated by the pair, do not exhibit any IR
excess.  ``C'' is again resolved into multiple components; the
coordinates given here are that of the brightest.  Its 2MASS
colours do not exhibit excess.  ``D'' is deeply embedded.
It is not detected by 2MASS in $J$ and $H$, where the
magnitudes given are only upper limits;  its 2MASS $K_s$
magnitude is of poor quality.  The 2MASS `magnitude limits'
locate ``D'' in the region of the giants (Fig. \ref{JHKcol}).
This is likely to be due to the large uncertainties in its
colours.  This object exhibits large reddening and is most
likely to be a YSO.  ``E'' also exhibits large reddening
and excess.

The MSX mission detected two objects in this field.   The brighter
one is only 2.9\,arcsec from the IRAS position and is located close
to ``B''.  However it should be noted that within the positional
uncertainties and spatial resolutions of IRAS and MSX, some or all 
of the objects labelled ``A--E'' may be contributing to the IRAS 
and MSX fluxes.

Several interesting H$_2$ emission features are seen in this region.
The most prominent of these are circled and labelled ``1a'' and
``1b'' on the continuum-subtracted image.  These appear to be the
two lobes of a bipolar jet.  Either ``D'' or ``E'' could be the
driving source of this outflow; considering the large reddening and 
the location close to the centroid of these two H$_2$ lobes, ``D'' is
the most likely candidate.  The brightest 1.2-mm peak of
\citet{beuther02b} is closest to ``D'' and is located merely
2.1\,arcsec from it implying that ``D'' is the YSO driving the
outflow traced by ``1a'' and ``1b''.   However, the 3.6-cm radio
continuum source (flux density = 29\,mJy) of \citet{sridharan02}
is located 7.3\,arcsec from the 1.2-mm peak, implying that the
radio emission is produced by a more evolved source in the cluster.
Our H$_2$ image also shows collimated emission features
SW of the cluster, in the bottom right corner of the figure, which
are circled and labelled ``2''.  ``2'' appears to emanate
from the object labelled ``F'' on the $K$-band image.  ``F'' is
detected by 2MASS in $K_s$ only.  This source is
nebulous, slightly elongated and is probably composed of one
or more objects embedded in nebulosity; the point sources within
``F'' are not resolved here.  The second MSX object is notably
within 1.2\,arcsec of ``F''.  The second brightest 1.2-mm peak
of \citet{beuther02b} is located only 2.7\,arcsec SE of ``F''.
We conclude that ``F'' hosts a YSO which produces the jet ``2''
detected in H$_2$.  In addition to these, there is also some faint
H$_2$ emission enveloping ``A--D'' on the NW and SE sides of
these diagonally aligned sources.  There is also some faint emission
extending east.  These are probably caused by fluorescent emission
due to the combined radiation from the more evolved stars within
this central cluster.

The CO map of \citet{beuther02c} could probably be the result of
at least two outflows from two different YSOs, one detected as
the near-IR source ``D'' and the second an embedded source
within ``F''.  From our H$_2$ image, we derive collimation factors
of 11.5 and 4.5 for the two outflows, ``1'' and ``2'', respectively.

There is very faint Br$\gamma$ emission detected towards the
central region of the cluster.  This may be due to emission from
the more evolved bright stars in the central region of the cluster.
The central portion of the Br$\gamma$ image is shown in the inset
on the H$_2$ image.  This region appears to host multiple YSOs
in different stages of evolution.  The presence of outflow/s,
multiple YSOs and the alignment of the bright objects nearly at
the centre makes this region interesting, warranting detailed study.

\subsection{IRAS~23139+5939\\ ({\small \it d = 4.8\,kpc, L = 25$\times$10$^3$\,L$_{\odot}$})}

Observations by \citet{beuther02b} reveal IRAS~23139+5939 to be single
peaked in both dust continuum emission and dense gas tracers. The CO (J=2-1)
emission from this source exhibits both red- and blue-shifted components
that are largely overlapping and centred on the millimetre continuum peak
\citep{beuther02c}.  The outflow is possibly nearly along the line of sight.
They estimate the mass of the gas in the outflow to be 57\,M$_{\odot}$.  The
multi-wavelength study of \citet{sridharan02} and \citet{beuther02d} reveal
a mid-infrared MSX source (positional accuracy of 5\,arcsec),  nearly coincident
with a 1.2-mm continuum source, H$_2$O masers and a faint 3.6-cm continum
source detected at two different epochs using the VLA (\citealt{tofani95},
0.92$\pm$0.15\,mJy; \citealt{sridharan02}, 1.4\,mJy).  The centres of all
these detections are offset to the north and east of the IRAS position by
8.7\,arcsec and 5.5\,arcsec respectively.

\citet{beuther02d} found three water masers associated with IRAS~23139+5939,
spatially coincident (within 1\,arcsec) with the radio,  millimetre and
mid-infrared positions.  Methanol maser emission is reported by
\citet{szymczak00}.   A multi-epoch observational study of the H$_2$O maser
emission has been carried out by \citet{goddi05} at very
high spatial resolution using the VLBA.  They resolved several components
and derived proper motions of the individual components of the maser.
Their observations support a conical outflow from the YSO and show that the
observed masers are preferentially arranged along the inner regions of the
blue-shifted lobe of the outflow, which is oriented nearly along the line
of sight.  A model for Keplerian motion in a disc does not fit the maser
data well.  They propose that the masers are preferentially associated
with collimated flows of gas found at the base of large-scale molecular
outflows.  The proper motions indicate a general expansion of the masers at
10's of km\,s$^{-1}$, as would be expected when the outflow along the
line-of-light is projected on to the sky plane.

IRAS~23139+5939 was included in the  $K$-band imaging survey of
\citet{hodapp94}, but no outflow drivers were detected. Observations by
Carnkner, Kozak \& Feigelson (1998) did not detect any X-ray emission 
from this source.

\begin{figure}
\centering
\includegraphics[width=8.10cm,clip]{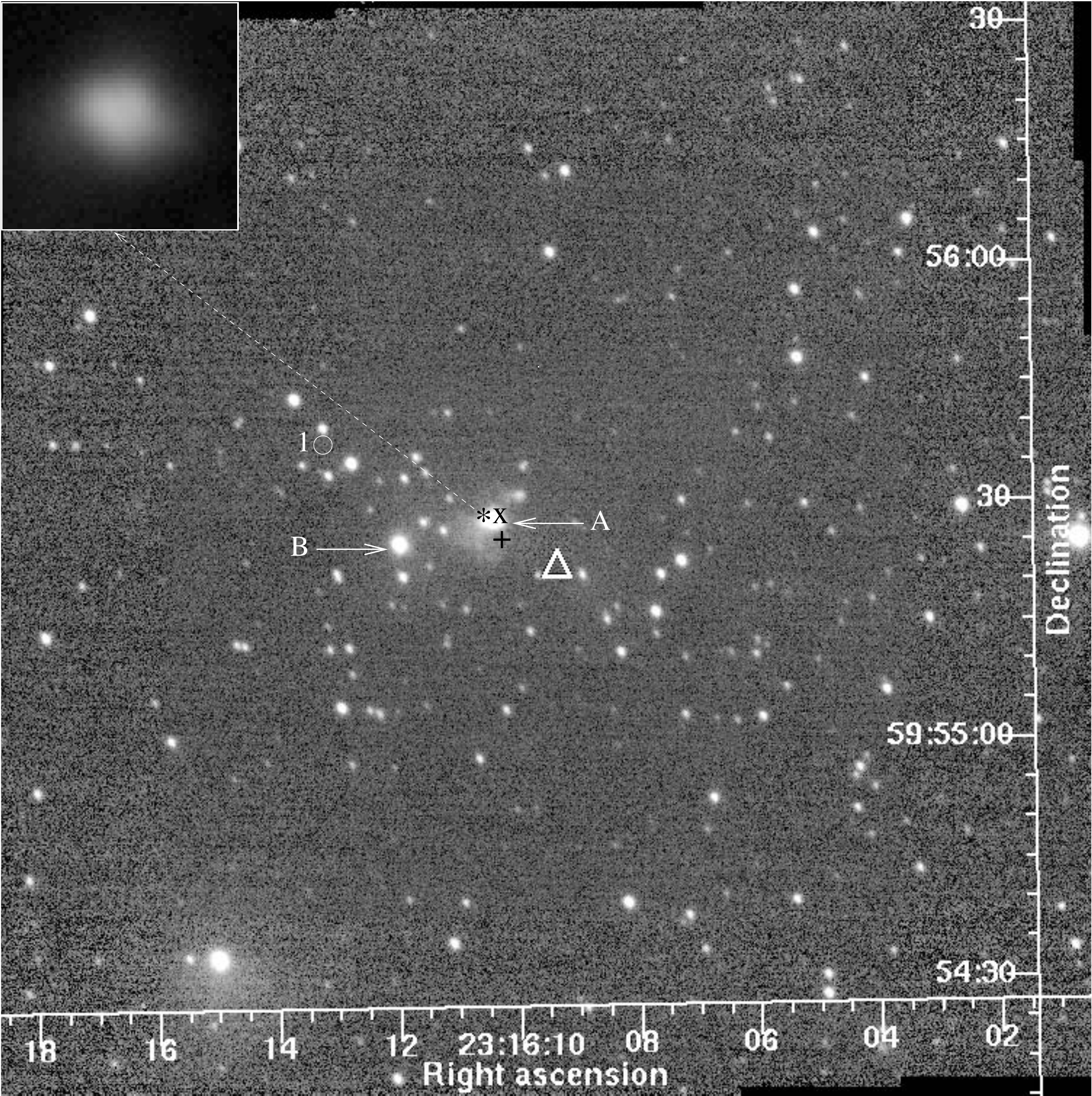}
\caption{$K$-band image of IRAS~23139+5939.  The inset shows an
enlarged view of the central object with the binary resolved . ``*'' shows the
peak of the 1.2-mm continuum emission of \citet{beuther02c} and ``x'' shows
the 3.6-cm peak position of \citet{tofani95}.}
\label{23139_K}
\end{figure}

Fig. \ref{23139_K} shows our $K$-band image of the field.  The bright
object ``A''($\alpha$=23:16:10.39, $\delta$=59:55:28.3), located
close to the IRAS position, is associated with a cometary nebula.
This object is just resolved as a binary in our images. The components of
the binary are at a separation of $\sim$0.4$\arcsec$ in the NW-SE direction,
the object in the SE being brighter than the one in the NW.  The inset in
Fig. \ref{23139_K} shows the source ``A'' resolved.  The continuum-subtracted
H$_2$ image (not shown here) revealed a faint H$_2$ emission features located
at the position enclosed by a circle and labelled
``1'' ($\alpha$=23:16:13.22, $\delta$=+59:55:38.1) on Fig. \ref{23139_K};
it is not clear whether ``1'' is produced by the outflow from ``A''.

There is a cluster associated with this star forming region towards the
centre of the observed field, with most of the cluster members located
south of ``A''. This object and a fainter object,
``B'' ($\alpha$=23:16:11.95, $\delta$=59:55:25.4), both exhibit IR excess
in their 2MASS colours and occupy the region of YSOs on the $JHK$
colour-colour diagram (Fig. \ref{JHKcol}).  ``A'' is located
9.8\,arcsec NE of the IRAS position (8.2\,arcsec east and 5.3\,arcsec
north).  The MSX position  is only 2.7\,arcsec SW of ``A'' (1.5\,arcsec
west, 2.24\,arcsec South).  It is important to note that the peak of
the 1.2-mm continuum located by \citet{beuther02b} is in very good
agreement (to within 0.5\,arcsec) with the location of ``A'', as well
as the centimetre source, the centroid of the CO outflow and the
location of  water maser spots.  We therefore identify ``A'' as the
YSO driving the main outflow in this field.  The faint radio emission
from ``A'' suggests that it is likely to be in a pre-UCH{\sc ii} stage.

\section[h]{MHO numbers}
\label{MHO-sect}
Only a few lines of Table \ref{MHO} are printed here. The complete
form of the table is available on-line only.
\begin{center}
\begin{table*}
\begin{centering}
\caption{MHO numbers of the H$_2$ line emission features identified
 in this paper (from http://www.jach.hawaii.edu/UKIRT/MHCat/).}
\label{MHO}
\begin{tabular}{@{}lllll}
\hline
No. &Object Name  &H$_2$ features       &MHO number             \\
    &(IRAS)       &in this paper$^{1}$&                         \\
\hline
1   &00420+5530   &1 - 1--4             &2900--2903     	\\
2   &04579+4703   &2 - 1,2; 3,4         &1000; 1001     	\\
3   &05137+3919   &3 - 1a,1b; 2         &1002,1003; 1004	\\
5   &05274+3345   &5 - 1                &1005           	\\[-0.5mm]
    &05274+3345   &5 - 2--7             &1006 -- 1011   	\\
6   &05345+3157   &6 - 1; 2--4          &1016; 1018     	\\[-0.5mm]
    &05345+3157   &6 - 5; 6             &1019; 1017     	\\[-0.5mm]
    &05345+3157   &6 - 7--11            &1012--1015     	\\
7   &05358+3543   &7 - 1; 2; 3          &1044; 1042; 1043       \\[-0.5mm]
    &05358+3543   &7 - 4; 5; 6          &1023; 1024; 1046       \\[-0.5mm]
    &05358+3543   &7 - 7; 8; 9          &1045; 1048; 1028       \\
8   &05373+2349   &8                    &734                    \\
9   &05490+2658   &9 - 1                &721                    \\
10  &05553+1631   &10 - 1--3            &1200                   \\[-0.5mm]
11  &06061+2151   &11 - 1--3            &1201--1203             \\[-0.5mm]
13  &18144-1723   &13 - 1               &2302                   \\[-0.5mm]
14  &18151-1208   &14 - 1; 3            &2201; 2212             \\[-0.5mm]
    &18151-1208   &14 - 2               &2202                   \\
18  &18264-1152   &18 - 1a; 1b,4        &2203; 2204             \\[-0.5mm]
    &18264-1152   &18 - 3               &2205                   \\
19  &18316-0602   &19 - 1--3            &2206--2208             \\[-0.5mm]
    &18316-0602   &19 - 4,5             &2209,2210              \\
20  &18345-0641   &20 - 1               &2211                   \\
30  &19374+2352   &30 - 1--3            &2600--2602             \\
31  &19388+2357   &31 - 1--3            &2613--2615             \\
32  &19410+2336   &32 - 1; 32 - 2--3    &2603; 2604             \\[-0.5mm]
    &19410+2336   &32 - 4--6, 11,12     &2606                   \\
    &19410+2336   &32 - 7,8             &2607                   \\
    &19410+2336   &32 - 9,10            &2620                   \\
33  &20050+2720   &33 - 1--3            &2608--2610             \\[-0.5mm]
    &20050+2720   &33 - 4; 5            &2611; 2612             \\
35  &20062+3550   &35 - 1a; 1b          &858; 859               \\
36  &20126+4104   &36 - 1; 2            &860; 861               \\[-0.5mm]
    &20126+4104   &36 - 3,4             &862                    \\
37  &20188+3928   &37 - 1--5            &863--867               \\
38  &20198+3716   &38 - I; II           &868; 869               \\
39  &20227+4154   &39 - 1               &870                    \\[-0.5mm]
    &20227+4154   &39 - 2--5            &871--874               \\
40  &20286+4105   &40                   &875                    \\
41  &20293+3952   &41 - 1--6            &876--881               \\
44  &21307+5049   &                     &882                    \\
45  &21391+5802   &45 - 1; 2,3          &2753; 2754             \\[-0.5mm]
    &21391+5802   &45 - 4,5             &2755                   \\[-0.5mm]
    &21391+5802   &45 - 6--10           &2756--2760             \\[-0.5mm]
    &21391+5802   &45 - 11; 12          &2761; 2762             \\
46  &21519+5613   &46 - 1; 2            &2764; 2763             \\
47  &22172+5549   &47 - 1--4            &2765                   \\[-0.5mm]
    &22172+5549   &47 - 5               &2766                   \\
48  &22305+5803   &48 - 1               &2700                   \\
49  &22570+5912   &49 - 1a,1b           &2701                   \\[-0.5mm]
    &22570+5912   &49 - 2               &2702                   \\
\hline
\multicolumn{4}{l}{$^{1}$Object numbers - H$_2$ line emission features
(labelled on Figs. A1--A50)}\\
\end{tabular}
\end{centering}
\end{table*}
\end{center}

\section[]{Source resolutions}
\label{resolutions-sect}
\begin{center}
\begin{table*}
\begin{minipage}{500mm}
\caption{Details of existing observations}
\label{resolutions}
\begin{tabular}{@{}llllll}
\hline
Reference               &Species        &frequency/     &Telescope used                         &HPBW/FWHM      &Pointing accuracy$^1$  \\
                        &               &wavelength     &                                       &(arcsec)       &(arcsec)      		\\
			\hline
Anglada et al. (1997)   &H$_2$O\,maser; NH$_3$&22.2; 23.7\,GHz  &Haystack 37-m, MA, USA   	&84       	&15                     \\
Bachiller et al. (1995) &CS(3-2); CO(2-1)&147; 230.5\,GHz&IRAM 30-m, Spain                      &16; 11         &3                      \\
Bernard et al. (1999)   &CS(2-1)        &98\,GHz        &NMA Interferometer, Japan              &3.4$\times$3\  &                       \\[-0.5mm]
                        &CS; CO         &98--115\,GHz   &NRO 45-m, Japan              		&$\sim$17       &$\sim$10               \\
Baudry et al. (1997)    &OH\,maser      &6.031; 6.035\,GHz&Effelsberg 100-m, Germany            &130            &                       \\
Beltr\'{a}n et al. (2002)&Continuum     &3.6\,cm        &VLA, New Mexico                        &$\leq$17.9$\times$8.8  &               \\[-0.5mm]
                        &Cont.;CS;CO;CH$_3$OH&96.25--244.936\,GHz&BIMA interferometer, CA &8.0$\times$6.7--1.1$\times$0.9 &       	\\
Beltr\'{a}n et al. (2004)&\multicolumn{3}{l}{see Beltr\'{a}n et al. (2002)}                     &               &                       \\
Beuther et al. (2002a)  &CO(1-0)        &115.27; 110.20\,GHz&IRAM 30-m                          &22             &                       \\[-0.5mm]
                        &CO(1-0)        &115.27\,GHz    &IRAM PdBI, Spain                       &4.1$\times$3.3 &                       \\[-0.5mm]
                        &SiO(2-1);H$^{13}$CO$^+$&86.85; 86.75\,GHz&IRAM 30-m                    &29             &                       \\[-0.5mm]
                        &SiO(2-1);H$^{13}$CO$^+$&86.85; 86.75\,GHz&IRAM PdBI                    &5.8$\times$5.36 &                      \\[-0.5mm]
                        &CO(2-1);CH$_3$OH&230.54; 241.79\,GHz&IRAM 30-m                         &11             &                       \\[-0.5mm]
                        &CO(6-5)        &691.47\,GHz    &CSO 10.4-m, Hawaii                     &11             &                       \\[-0.5mm]
                        &Continuum      &2.6\,mm        &IRAM PdBI                              &4$\times$3     &                       \\[-0.5mm]
Beuther et al. (2002b)  &Continuum      &1.2\,mm        &IRAM 30-m                              &11             &                       \\
                        &CS(2-1)        &97.98\,GHz     &IRAM 30-m                              &27             &                       \\[-0.5mm]
                        &CS(3-2)        &146.97\,GHz    &IRAM 30-m                              &17             &                       \\[-0.5mm]
                        &CS(5-4)        &244.936\,GHz   &IRAM 30-m                              &11             &                       \\[-0.5mm]
Beuther et al. (2002c)  &$^{12}$CO(2-1) &230.5\,GHz     &IRAM 30-m                              &11             &                       \\[-0.5mm]
Beuther et al. (2002d)  &H$_2$O\,Maser  &22.235\,GHz    &\multicolumn{3}{l} {Effelsberg 100-m; see Sridharan et al. (2002)}             \\[-0.5mm]
                        &CH$_3$OH\,Maser&6.7\,GHz       &\multicolumn{3}{l} {Effelsberg 100-m; see Sridharan et al. (2002)}             \\[-0.5mm]
                        &H$_2$O\,Maser  &22.235\,GHz    &VLA                                    &$\sim$0.4      &$<$1 (abs), $\sim$0.1 (rel.)\\[-0.5mm]
                        &CH$_3$OH\,Maser&6.7\,GHz       &ATCA, Australia                        &$\sim$1.9      &$\sim$1 (abs), $\sim$0.1 (rel.)\\[-0.5mm]
                        &Continuum      &1.2\,mm        &\multicolumn{2}{l} {IRAM 30-m; see Sridharan et al. (2002)}    &$\sim$5        \\[-0.5mm]
                        &Continuum      &2.6\,mm        &\multicolumn{2}{l} {IRAM PdBI + BIMA; see Beuther et al. (2002a)} &$\sim$1     \\[-0.5mm]
                        &Continuum      &3.6\,cm        &\multicolumn{2}{l} {VLA; see Sridharan et al. (2002)}  &1                      \\
Beuther et al. (2003)   &CO(1-0)        &115.27\,GHz    &IRAM 30-m                              &3.9$\times$3.6 &                       \\
Beuther et al. (2004a)  &Continuum      &1.3; 3\,mm     &IRAM PdBI                              &$\leq$1.91$\times$1.75;$\leq$5.1$\times$4.31   &       \\[-0.5mm]
                        &CO; SiO(2-1)	&1.3; 3\,mm     &IRAM PdBI + 30-m                       &$\leq$1.96$\times$1.79;$\leq$5.1$\times$4.36   &       \\
Beuther et al. (2004b)  &CN; CS         &113.385; 97.981\,GHz &IRAM PdBI                        &$\sim$1.46$\times$1.21--5.31$\times$3.42       &       \\[-0.5mm]
                        &CN; CS         &226.656; 244.936\,GHz &IRAM PdBI                       &$\sim$0.76$\times$0.61--1.52$\times$1.03       &       \\
Beuther et al. (2006)   &CO(2-1); Cont. &1.3\,mm        &SMA, Hawaii                            &3.6$\times$2.5--3.9$\times$2.5 &       \\[-0.5mm]
                        &CO(2-1)        &1.3\,mm        &IRAM 30-m+SMA data                     &4.2$\times$2.8 &                       \\
Brand et al. (1994)     &H$_2$O\,Maser  &22.235\,GHz	&Medicina 32-m, Italy                   &114            &$<$20                  \\
Brand et al. (2001)     &Different molecules &89.19--220.75\,GHz &IRAM 30-m                  	&27--17         &3                      \\[-0.5mm]
                        &HCO$^+$(4--3)  &356.73\,GHz    &KOSMA 3-m, Switzerland     		&70$\times$78   &                       \\
Bronfman et al. (1996)  &CS(2-1)        &97.981\,GHz    &15-m SEST, La Silla, Chile             &50             &$\sim$3                \\[-0.5mm]
                        &CS(2-1)        &97.981\,GHz    &OSO 20-m, Sweeden                      &39             &$\sim$3                \\
Carral et al. (1997)	&Continuum      &7\,mm		&VLA					&2.5$\times$0.5	&$\sim$0.2		\\
Carpenter et al. (1995) &CS(2-1)        &97.981\,GHz    &FCRAO 14-m, Massachusetts              &50$\times$100  &                       \\
Casoli et al. (1986)    &$^{12;13}$CO(1-0)&115; 110.2\,GHz&Bordeaux 2.5-m                       &264            &                       \\[-0.5mm]
                        &$^{12}$CO(1-0) &115\,GHz       &IRAM 30-m                              &22             &5-10                   \\[-0.5mm]
                        &$^{12}$CO(2-1) &230\,GHz       &McDonald 4.9-m, Texas                  &78             &                       \\
Cesaroni et al. (1988)  &H$_2$O\,Maser  &22.235\,GHz    &Medicina 32-m                          &114            &20                     \\
Cesaroni et al. (1997)  &Different molecules&           &IRAM 30-m                              &27--11         &4                      \\[-0.5mm]
                        &CH$_3$OH; HCO$^+$;             &88.93--92.25\,GHz&IRAM PdBI            &3.7--2.7       &                       \\[-0.5mm]
                        &CH$_3$CN, Cont.&               &                                       &               &                       \\
Cesaroni et al. (1999)  &Different molecules &88.6--241.7\,GHz &IRAM 30-m                    	&27--10         &4                      \\
Cesaroni et al. (2005)  &CS; CH$_3$OH   &96.356--241.144\,GHz&IRAM PdBI                         &$\sim$2.4--0.85 &                      \\
Chini et al. (1986)     &Continuum      &1.3\,mm        &3-m NASA IRTF, Hawaii                  &90             &                       \\
Chini et al. (2001)     &Continuum      &450; 850\,$\mu$m&15-m JCMT, Hawaii                     &8.3; 15        &                       \\[-0.5mm]
                        &Continuum      &1.3\,mm        &IRAM 30-m                              &10.7           &                       \\
Codella et al. (1996)   &H$_2$O\,Maser  &               &                                       &               &                       \\
Codella et al. (2001)   &SiO; CO; DCO$^+$;CS&86.8--230.5\,GHz&IRAM 30-m                         &29--10         &$\sim$3\               \\[-0.5mm]
                        &Continuum      &88; 92; 110\,GHz &OVRO interferometer, BP              &6.2$\times$3.4--5.2$\times$3 &         \\[-0.5mm]
                        &HCO$^+$(1-0)   &89.2\,GHz      &OVRO interferometer, BP                &6.2$\times$3.4 &                       \\
Comoretto et al. (1990) &H$_2$O\,Maser  &22.235\,GHz    &Medicina 32-m                          &114            &20                     \\
De Buizer et al. (2005) &Continuum      &11.7; 20.81\,$\mu$m&3-m NASA IRTF                      &1.3; 1.7       &$<$1                   \\
De Buizer et al. (2007) &Continuum      &12.5; 18.3\,$\mu$m&Gemini N. 8.1-m, Hawaii             &0.33; 0.48     &$<$0.1                 \\
\hline
\multicolumn{6}{l}{$^1$ a ``$<$'' implies that the pointing accuracy is better than the given value}\\
\end{tabular}
\end{minipage}
\end{table*}
\begin{table*}
\begin{minipage}{500mm}
\contcaption{}
\begin{tabular}{@{}llllll}
\hline
Dent et al. (1988)      &NH$_3$         &23.7\,GHz      &SERC 25-m, England                     &132            &                       \\[-0.5mm]
                        &NH$_3$         &23.7\,GHz      &Effelsberg 100-m                       &40             &                       \\[-0.5mm]
                        &HCO$^+$(1-0)   &89.188\,GHz    &OSO 20-m				&45             &                       \\[-0.5mm]
                        &HCO$^+$(3-2)   &267.557\,GHz   &3.8-m UKIRT, Hawaii                    &75             &                       \\
Dobashi et al. (1995)   &$^{12;13}$CO   &115\,GHz       &NRO 45-m				&16             &$<$10                  \\
Edris et al. (2005)     &OH\,Maser      &1.665;1.667\,GHz &MERLIN, England 			&0.174$\times$0.137     &0.025 (abs.), 0.010 (rel.) \\[-0.5mm]
                        &CH$_3$OH\,Maser&6.7\,GHz       &MERLIN interferometer                  &0.026$\times$0.026 &0.015 (abs.)       \\[-0.5mm]
                        &H$_2$O\,Maser  &22.2\,GHz      &MERLIN interferometer 			&0.04$\times$0.008  &0.012 (abs.)       \\
Estalella et al. (1993) &NH$_3$         &23.7\,GHz 	&Effelsberg 100-m			&40             &$<$8                   \\[-0.5mm]
                        &C$^{18}$O(2-1);CS(3-2)&219.56;146.97\,GHz&IRAM 30-m                    &$\sim$13;17    &$\sim$4                \\
Felli et al. (1992)     &H$_2$O\,Maser  &22.235\,GHz    &Medicina 32-m                          &114            &15                     \\
Fa\'{u}ndez et al., 2004&Continuum      &1.2\,mm        &15-m SEST                              &24             &                       \\
Fontani et al. (2004)   &Continuum      &850\,$\mu$m    &15-m JCMT                              &14             &                       \\[-0.5mm]
                        &$^{12}$CO(1-0) &115.271\,GHz   &OVRO interferometer                    &6.95$\times$6.12       &               \\[-0.5mm]
                        &$^{12}$CO(2-1) &230.538\,GHz   &NRAO 12-m, Kit Peak			&29             &                       \\[-0.5mm]
                        &$^{12}$CO; C$^{18}$O(2-1)&230.54; 219.56\,GHz&IRAM 30-m                &11; 12         &2                      \\[-0.5mm]
                        &H$^{13}$CO$^+$; SiO &217.11--86.75\,GHz&IRAM 30-m                      &12--29         &2                      \\[-0.5mm]
                        &CH$_3$C$_2$H   &222.17--102.55\,GHz&IRAM 30-m                          &11--29         &                       \\
Forster et al. (1978)   &Continuum      &1.3\,cm        &Hat Creek Interferometer               &6$\times$10.5  &               \\
Galt (2004)             &CH$_3$OH\,Maser&6.7\,GHz       &DRAO 25.6-m, Canada                    &420            &                       \\
Goddi et al. (2004)     &H$_2$O\,Maser  &22.235\,GHz    &European VLBI Network                  &(2.1$\times$1.1)$\times$10$^{-3}$      &(50--100)$\times$10$^{-6}$ \\
Goddi et al. (2005)     &H$_2$O\,Maser  &22.235\,GHz    &European VLBI Network                  &(1.1$\times$1--2.9$\times$1.4)$\times$10$^{-3}$ &      \\
Gyulbudaghian et al. (1990)&H$_2$O\,Maser&22.235\,GHz   &Haystack 37-m                          &90             &$\sim$20               \\
Hardebeck \& Wilson (1971)&OH\,Maser    &1.665;1.667\,GHz &OVRO Interferometer                  &               &5--15                  \\
Harju et al. (1998)     &SiO(2-1)       &86.847\,GHz    &15-m SEST				&57             &3                      \\[-0.5mm]
                        &SiO(3-2)       &130.2687\,GHz  &15-m SEST                              &40             &3                      \\[-0.5mm]
                        &SiO(2-1)       &86.847\,GHz    &OSO 20-m                            	&43             &4                      \\
Hofner \& Churchwell (1996)&H$_2$O\,Maser &22.235\,GHz  &VLA                                    &0.46$\times$0.4--0.75$\times$0.4 &0.1 (rel.)  \\
Hofner et al. (1999)    &Continuum      &7\,mm; 3.6\,cm &VLA                                    &1.7$\times$1; 0.5$\times$0.5 &         \\
Hughes \& MacLeod (1994)&Continuum      &6\,cm          &VLA                                    &$\sim$3        &                       \\
Hunter et al. (1995)    &CO(2-1)        &230\,GHz       &CSO 10.4-m                             &31             &4                      \\[-0.5mm]
                        &CO(3-2);CS(7-6)&$\sim$345\,GHz &CSO 10.4-m                             &20             &4                      \\[-0.5mm]
                        &$^{13}$CO(6-5) &661.068\,GHz   &CSO 10.4-m                             &11             &4                      \\[-0.5mm]
                        &H$_2$O\,Maser; Cont.&22.235; 8.4\,GHz&VLA                                &0.12$\times$0.11;0.28$\times$0.27 &	\\
Hunter et al. (1997)    &Continuum      &800\,$\mu$m    &CSO 10.4-m                             &               &5                      \\[-0.5mm]
                        &Continuum      &350; 450\,$\mu$m&CSO 10.4-m                            &12             &5                      \\[-0.5mm]
                        &$^{12;13}$CO;C$^{18}$O(2-1)&$\sim$230\,GHz     &CSO 10.4-m             &31             &5                      \\[-0.5mm]
                        &CO(3-2); CS(7-6)&$\sim$345\,GHz &CSO 10.4-m                            &20             &5                      \\[-0.5mm]
                        &CO(4-3); (6-5) &291; 436\,GHz  &CSO 10.4-m                             &15;11          &5                      \\[-0.5mm]
                        &$^{13}$CO; C$^{18}$O(1-0);     &110--115\,GHz  &OVRO Interferometer    &2.71$\times$2.11&                      \\[-0.5mm]
                        &CS(2-1); Cont. &               &                                       &               &                       \\
Hunter et al. (1999)    &H$_2$O Maser   &22.235\,GHz    &VLA                                    &0.0786$\times$0.0704   &               \\[-0.5mm]
                        &Continuum      &8.4\,GHz       &VLA                                    &1.4$\times$0.87 &                      \\[-0.5mm]
                        &HCO$^+$; H$^{13}$CO$^+$;       &$\sim$88\,GHz  &OVRO Interferometer    &2.7$\times$2.5 &                       \\[-0.5mm]
                        &SiO; Continuum      &               &                                  &               &                       \\[-0.5mm]
                        &CH$_3$CN       &$\sim$257\,GHz &CSO 10.4-m                             &30             &                       \\
Hunter et al. (2000)    &Continuum      &350\,$\mu$m    &CSO 10.4-m                             &11             &5                      \\
Jenness et al. (1995)   &Continuum      &800/850\,$\mu$m &15-m JCMT                             &17/18          &$\sim$4                \\[-0.5mm]
                        &Continuum      &450\,mm        &15-m JCMT                              &7/18           &$\sim$4                \\[-0.5mm]
                        &Continuum      &2\,cm          &RT (Ryle) Interferometer               &30$\times$30 cosec$\delta$ & \\[-0.5mm]
                        &CO(2-1); CS(5-4)&230.5--244.9\,GHz&15-m JCMT                           &21           &                       \\[-0.5mm]
                        &Continuum      &3.6\,cm        &VLA                                    &0.9            &                       \\[-0.5mm]
                        &H$_2$O\,Maser  &22.235\,GHz    &VLA                                    &0.4            &                       \\[-0.5mm]
Kraemer et al. (2003)   &Continuum      &12.5; 20.6\,$\mu$m&3-m NASA IRTF                       &1.1; 1.7       &                       \\
Kurtz et al. (1994)     &Continuum      &2; 3.6\,cm     &VLA                                    &$\sim$0.5; 0.9 &                       \\
Kurtz et al. (2004)     &CH$_3$OH\,Maser&44.069; 23.2\,GHz&VLA                                  &0.44$\times$0.38--4.18$\times$1.89 &0.5--0.6    \\
Kurtz \& Hofner (2005)  &H$_2$O Maser   &22.235\,GHz    &Effelsberg 100-m                       &40             &5                      \\[-0.5mm]
                        &H$_2$O Maser   &22.235\,GHz    &VLA                                    &0.52$\times$0.26--4.5$\times$2.3 &     \\
Lebr{\'{o}}n et al. (2006)&CO(2-1)      &230.5\,GHz     &IRAM 30-m                              &11             &                       \\
Little et al. (1988)	&HCO$^+$(1-0)	&89.19\,GHz 	&Mets\"{a}hovi 14-m, Finland		&50		&                       \\[-0.5mm]
			&HCO$^+$(3-2)   &267.56\,GHz    &3.8-m UKIRT            		&66             &                       \\[-0.5mm]
			&$^{12;13}$CO(1-0)&115.27; 110.2\,GHz&OSO 20-m                          &33; 35         &                       \\
McCutcheon et al. (1995)&Continuum      &450; 800; 1100\,$\mu$m&15-m JCMT                       &8; 14; 18      &                       \\
MacLeod et al. (1998a)  &OH; HCHO,CH$_3$OH&1.6--12.2\,GHz &HRAO 26-m, S. Africa                 &1800--210      &                       \\[-0.5mm]
                        &OH; CH$_3$OH\,Maser&1.6; 6.7\,GHz&DRAO 26-m                            &1800; 420      &                       \\[-0.5mm]
                        &H$_2$O\,Maser  &22.235\,GHz    &IRO 14-m, Brazil                       &222            &                       \\
\hline
\end{tabular}
\end{minipage}
\end{table*}
\begin{table*}
\begin{minipage}{500mm}
\contcaption{}
\begin{tabular}{@{}llllll}
\hline
MacLeod et al. (1998b)  &CH$_3$OH\,Maser&6.7; 12.2\,GHz &HRAO 26-m                              &420; 210       &                       \\[-0.5mm]
                        &CH$_3$OH\,Maser&6.7\,GHz       &DRAO 26-m                              &420            &                       \\
Matthews et al. (1986)  &$^{12;13}$CO(1-0)&115.27; 110.2\,GHz&OSO 20-m                          &33; 35         &                       \\[-0.5mm]
                        &$^{12;13}$CO(1-0)&115.27; 110.2\,GHz&3.8-m UKIRT                       &75             &                       \\
Menten (1991)           &CH$_3$OH\,Maser&6.7\,GHz       &NRAO 43-m, Green Bank, WV              &300            &                       \\
Migenes et al. (1999)   &H$_2$O Maser   &22.235\,GHz    &VLBA                                   &(0.21--0.31)$\times$10$^{-3}$ &        \\
Minier et al. (2001)    &CH$_3$OH\,Maser&6.7\,GHz       &EVN                                    &               &$\sim$0.03             \\[-0.5mm]
                        &CH$_3$OH\,Maser&12.2\,GHz      &VLBA                                   &               &$\sim$0.03             \\[-0.5mm]
                        &CH$_3$OH\,Maser&6.7; 12.2\,GHz &ATCA                                   &               &$\sim$0.3              \\
Minier et al. (2005)    &sub-mm Cont.   &450; 850\,$\mu$m&15-m JCMT                             &8; 15          &                       \\[-0.5mm]
                        &mm Cont.       &1.2\,mm        &15-m SEST                              &24             &                       \\
Miralles et al. (1994)  &radio Cont.    &2; 6\,cm       &VLA					&$\sim$5        &                       \\[-0.5mm]
                        &H$_2$O\,Maser,NH$_3$&22-23\,GHz&Haystack 36.6-m                        &$\sim$90       &$\sim$8                \\
Molinari et al. (1996)  &NH3            &23.7           &Effelsberg 100-m                       &40             &                       \\
Molinari et al. (1998)  &Continuum      &2\,cm          &VLA                                    &0.6$\times$0.47--1.3$\times$0.6  &     \\[-0.5mm]
                        &Continuum      &2\,cm          &VLA                                    &1.3$\times$0.57--2.2$\times$0.57  &    \\[-0.5mm]
                        &Continuum      &2\,cm          &VLA                                    &3.7$\times$1.7--9.3$\times$1.7  &      \\[-0.5mm]
                        &Continuum      &6\,cm          &VLA                                    &1.8$\times$1.4--4$\times$1.8  &        \\[-0.5mm]
                        &Continuum      &6\,cm          &VLA                                    &4$\times$1.7--6.5$\times$1.7  &        \\
Molinari et al. (2000)  &Continuum      &0.35--1.1\,mm  &15-m JCMT                              &18.5           &                       \\[-0.5mm]
                        &Continuum      &1.3; 2.0\,mm   &15-m JCMT                              &19.5; 27       &                       \\
Molinari et al. (2002)  &HCO$^+$,HCN,SiO&$\sim$88\,GHz  &OVRO interferometer                    &4$\times$5     &                       \\[-0.5mm]
                        &Continuum      &3.4; 1.3\,mm   &OVRO interferometer                    &4$\times$5     &                       \\[-0.5mm]
                        &Continuum      &3.6\,cm        &VLA                                    &9$\times$4     &                       \\
Moscadelli et al. (2000)&H$_2$O\,Maser  &22.235\,GHz    &NRAO VLBA                              &(1.2$\times$1.1)$\times$10$^{-3}$ & $<$0.2$\times$10$^{-3}$   \\
Moscadelli et al. (2005)&H$_2$O\,Maser  &22.235\,GHz    &EVN+VLBA                               &(0.6--0.8)$\times$10$^{-3}$ &$<$10$^{-3}$(abs.), \\[-0.5mm]
			&		&		&					&		&(0.1--100)$\times$10$^{-6}$(rel.)\\
Palagi et al. (1993)    &H$_2$O\,Maser  &22.235\,GHz    &Medicina 32-m                          &114            &25                     \\
Palla et al. (1991)     &H$_2$O Maser   &22.235\,GHz    &Medicina 32-m                          &114            &                       \\
Palla et al. (1993)     &H$_2$O Maser   &22.235\,GHz    &Medicina 32-m                          &114            &                       \\
Patel et al. (2000)     &H$_2$O Maser   &22.235\,GHz    &VLBA                                   &$\sim$0.8$\times$10$^{-3}$ &$\sim$10$^{-5}$(rel.)\\
Ramesh et al. (1997)    &HCO$^+$;CS     &86--99\,GHz    &NRO 45-m				&$\sim$16       &$<$5                   \\
Reipurth et al. (2003)  &Continuum      &11.6\,$\mu$m   &3.8-m UKIRT                            &               &$<$2                   \\
Ridge \& Moore (2001)   &$^{12}$CO(2-1);(1-0)&230\,GHz;115\,GHz&NRAO 12-m                       &27; 55         &                       \\[-0.5mm]
                        &$^{12}$CO(2-1) &230\,GHz       &15-m JCMT                              &21             &5                      \\
Ridge et al. (2003)     &$^{13}$CO;C$^{18}$O(1-0) &110.2; 109.78\,GHz &FCRAO 14-m               &               &5                      \\[-0.5mm]
                        &C$^{18}$O(2-1) &219.56\,GHz    &10-m HHT, Arizona                      &35             &2                      \\
Saraceno et al. (1996)  &Continuum      &350--1300\,$\mu$m &15-m JCMT                           &16--19         &                       \\
Schutte et al. (1993)   &CH$_3$OH Maser &6.7\,GHz       &HRAO 26-m                              &420            &                       \\
Shepherd et al. (1997)  &CO(1-0)        &115.271\,GHz   &BIMA interferometer                    &4.6$\times$5.7 &                       \\[-0.5mm]
                        &SiO;H$^{13}$CO$^+$;&86.847--86.639\,GHz&BIMA interferometer&6.8$\times$5.3;6.8$\times$5.4;&     		\\[-0.5mm]
			&SO$_2$;Continuum&		&					&7.2$\times$5.4	&			\\
Shepherd et al. (1998)  &CO(1-0)        &115.271\,GHz   &OVRO interferometer                    &5.55$\times$4.47 &0.2                  \\[-0.5mm]
                        &Continuum      &112.3\,GHz(2.7\,mm)&OVRO interferometer                &2.71$\times$2.15 &0.2                  \\
Shepherd \& Churchsell (1996a)&$^{12}$CO(1-0)&115.271\,GHz&NRAO 12-m                            &$\sim$60                               \\
Shepherd \& Churchsell (1996b)&$^{12;13}$CO(1-0)        &115.27; 110.2\,GHz&NRAO 12-m         	&$\sim$60 (at 115\,GHz) &		\\
Shepherd \& Kurtz (1999)&Continuum      &2.6\,mm        &OVRO interferometer                    &1.59$\times$1.48       &0.2            \\[-0.5mm]
                        &C$^{18}$O;$^{13}$CO (1-0) &$\sim$110\,GHz      &OVRO interferometer    &5.24$\times$4.59; 5.76$\times$5.24 &0.2 \\[-0.5mm]
                        &Continuum      &7\,mm; 3.6\,cm &VLA                                    &0.56$\times$0.45; 0.28$\times$0.25 &$<<$FWHM   \\[-0.5mm]
                        &Continuum      &3.6\,cm        &VLA                                    &0.28$\times$0.25       &$<<$FWHM       \\[-0.5mm]
                        &H$_2$O\,maser  &22.235\,GHz    &VLA                                    &1.01$\times$0.91--3.15$\times$2.84 &$<<$FWHM   \\
Shepherd et al. (2000)  &CO(1-0); Cont. &2.7; 1.3\,mm   &OVRO                                   &$<$7.5--3.5;$<$2.9--1.2 &              \\[-0.5mm]
                        &Continuum      &12.5; 17.9\,$\mu$m&10-m Keck, Hawaii                   &$\sim$0.4;0.5  &$<$1                   \\
Shepherd et al. (2001)  &Continuum      &3\,mm          &OVRO interferometer                    &2.05$\times$1.43, 2.96$\times$2.44 &   \\[-0.5mm]
                        &Continuum      &1\,mm          &OVRO interferometer                    &0.89$\times$0.79, 1.34$\times$0.97 &   \\
Shepherd et al. (2004a) &NH$_3$(1,1)    &23.695\,GHz    &VLA                                    &2.93$\times$2.55 &0.1                  \\[-0.5mm]
                        &Continuum      &1.3\,mm        &VLA                                    &2.93$\times$2.55 &0.1                  \\[-0.5mm]
                        &H$_2$O\,Maser  &22.235\,GHz    &VLA                                    &0.68$\times$0.27 &0.1                  \\[-0.5mm]
                        &H$_2$O\,Maser  &22.235\,GHz    &VLBA                                   &(0.86--1.1$\times$0.36)$\times$10$^{-3}$& \\[-0.5mm]
                        &Continuum      &850\,$\mu$m    &JCMT                                   &15             &                       \\
Shepherd et al. (2004b) &Continuum      &3.3\,mm        &OVRO                                   &4.9$\times$4.2 &                       \\[-0.5mm]
                        &H$^{13}$CO$^{+}$; SiO&86.75; 86.85\,GHz   &OVRO                        &5.3$\times$4.5 &                       \\[-0.5mm]
                        &Continuum      &6\,cm          &VLA (archival data)			&5.65$\times$2.98 &                     \\
Slysh et al. (1994)     &OH\,maser      &1.665; 1.667\,GHz&Nancay Radio Tel., France            &210$\times$1140 &                      \\
Slysh et al. (1999)     &CH$_3$OH\,Maser&6.7\,GHz       &Medicina 32-m                          &336            &                       \\
\hline
\end{tabular}
\end{minipage}
\end{table*}
\begin{table*}
\begin{minipage}{500mm}
\contcaption{}
\begin{tabular}{@{}llllll}
\hline
Snell et al. (1988; 1990)&$^{12}$CO(1-0) &115.271\,GHz  &FCRAO 14-m                             &45             &                       \\
Sridharan et al. (2002) &$^{12}$CO(2-1) &230.5\,GHz     &IRAM 30-m                              &11             &                       \\[-0.5mm]
                        &SiO(2-1),H$^{13}$CO$^{+}$(1-0)&86.9\,GHz       &IRAM 30-m              &29             &                       \\[-0.5mm]
                        &$^{13}$CO(1-0);CH$_3$CN(6-5) &110.2; 110.4\,GHz &IRAM 30-m             &22             &                       \\[-0.5mm]
                        &CH$_3$OH       &241.8\,GHz     &IRAM 30-m                              &11             &                       \\[-0.5mm]
                        &$^{12}$CO(2-1) &230.5\,GHz     &CSO 10.4-m                             &27             &                       \\[-0.5mm]
                        &NH3; H$_2$O\,Maser&$\sim$22.7; 22.2\,GHz&Effelsberg 100-m              &40             &                       \\[-0.5mm]
                        &CH$_3$OH Maser &6.7\,GHz       &Effelsberg 100-m                       &130            &                       \\[-0.5mm]
                        &Continuum      &3.6\,cm        &VLA                                    &0.7            &                       \\[-0.5mm]
                        &Continuum      &1.2\,mm        &IRAM 30-m                              &11             &                       \\
Sridharan et al. (2005) &Continuum      &2.2-4.7$\mu$m  &3.8-m UKIRT                            &$<$0.3         &0.15                   \\
Su et al. (2004)        &CO(1-0); Cont. &$\sim$3\,mm    &BIMA; NRAO 12-m                        &$\sim$6--10    &                       \\
Sugitani et al. (1989)  &$^{12;13}$CO(1-0) &115; 110\,GHz&NRO 4.5-m                             &17     	&                       \\
Szymczak et al. (2000)  &CH$_3$OH Maser &6.7 GHz        &32-m Toru\'{n} Radio Tel.              &330            &$<$25                  \\
Thompson et al. (2006)  &Continuum      &450; 850\,$\mu$m &15-m JCMT                            &8; 14          &$<$5                   \\
Tofani et al. (1995)    &Continuum      &3.6\,cm        &VLA                                    &0.3            &                       \\[-0.5mm]
                        &H$_2$O\,Maser  &22.235\,GHz    &VLA                                    &0.1            &$<$0.1(abs),$<$0.003(rel.)   \\[-0.5mm]
                        &H$_2$O\,Maser  &22.235\,GHz    &Medicina 32-m                          &114            &25                     \\
Torrelles et al. (1992b)&H$_2$O\,Maser  &22.235\,GHz    &VLA                                    &$\sim$1.1$\times$0.9 &10\% of beamsize \\[-0.5mm]
                        &Rad. Continuum &3.6\,cm        &VLA                                    &$\sim$4$\times$3.5\  &10\% of beamsize \\
van der Walt et al. (1995)&CH$_3$OH\,Maser&6.7\,GHz     &HRAO 26-m                              &420            &                       \\
Verdes-Montenegro       &NH3            &$\sim$23.7\,GHz&Haystack 37-m                          &84             &$\sim$15               \\[-0.5mm]
         et al. (1989)  &               &               &					&               &                       \\[-0.5mm]
                        &H$_2$O\,Maser  &22.235\,GHz    &Haystack 37-m                          &84             &$\sim$15               \\
Vig. et al. (2006)      &Continuum      &1.28\,GHz      &GMRT interferometer                    &7.5$\times$3.8 &                       \\[-0.5mm]
                        &Continuum      &610; 325\,MHz  &GMRT, India                            &8.4$\times$6.6; 14.4$\times$11.5 &     \\[-0.5mm]
                        &Continuum      &450; 850\,$\mu$m &15-m JCMT                            &$\sim$7$\times$11; $\sim$13$\times$17  &     \\[-0.5mm]
                        &Continuum      &130; 200\,$\mu$m &TIFR Balloon Facility 1-m            &$\sim$60       &$\sim$48               \\
Walsh et al. (1997)     &CH$_3$OH Maser &6.669\,GHz     &Parkes 64-m				&198            &                       \\
Walsh et al. (1998)     &CH$_3$OH Maser; Cont.&6.669; 8.64\,GHz &ATCA, Australia                &$<$2           &1 (abs.), 0.05(rel.)   \\
Walsh et al. (2003)     &Continuum      &450; 850\,$\mu$m&15-m JCMT                             &8; 15          &                       \\
Watson et al. (2003)    &H$_2$CO;H110$_{\alpha}$&4.83; 4.87\,GHz&Arecibo 305-m, Puerto Rico     &57             &                       \\
Wilking et al. (1989)   &CO(2-1)        &               &MWO 4.9-m, Texas, USA                  &72             &                       \\[-0.5mm]
                        &Continuum      &2.7\,mm        &OVRO interferometer                    &35             &                       \\[-0.5mm]
                        &Continuum      &2; 6\,cm       &VLA                                    &5.5; 16        &                       \\[-0.5mm]
                        &Continuum      &2; 6\,cm       &VLA (for IRAS~20227)                   &3.3$\times$1.2; 6.3$\times$3.8 &       \\
Wilking et al. (1990)   &$^{12;13}$CO(2-1)&230.5; 220.4\,GHz&NRAO 12-m                          &28             &5                      \\
Wilking et al. (1993)   &Continuum      &2.7\,mm        &OVRO interferometer                    &7.3$\times$7.5\&1.2                    \\[-0.5mm]
                        &Continuum      &1.25\,mm       &CSO 10.4-m                             &30             &5                      \\[-0.5mm]
                        &CS(2-1)        &97.97; 96.41\,GHz&NRAO 12-m                            &66             &8                      \\
Williams et al. (2004)  &Continuum      &450; 850\,$\mu$m &15-m JCMT                            &8; 14.4        &                       \\
Wu et al. (2001)        &$^{13}$CO (1-0)&110.201\,GHz   &13.7-mm QST                            &55             &                       \\
Wood \& Churchwell (1989b)&Rad. Cont.   &6\,cm          &VLA                                    &0.38$\times$0.38--1.6$\times$1.4 &     \\[-0.5mm]
                        &Rad. Cont.     &2\,cm          &VLA                                    &0.41$\times$0.41 &                     \\
Wouterloot et al. (1993)&H$_2$O\,Maser  &22.235\,GHz    &Effelsberg 100-m                       &40             &7--8                   \\[-0.5mm]
			&H$_2$O\,Maser  &22.235\,GHz    &Medicina 32-m                          &114            &15                     \\[-0.5mm]
			&OH\,Maser      &1.665\,GHz     &Effelsberg 100-m                       &468            &                       \\[-0.5mm]
			&CH$_3$OH\,Maser&12.179\,GHz    &Medicina 32-m                          &150            &                       \\[-0.5mm]
			&$^{12}$CO(2-1); (3-2)&230; 345\,GHz &KOSMA 3-m				&132; 66	&			\\[-0.5mm]
Wouterloot \&           &$^{12}$CO(1-0) &115\,GHz       &15-m SEST                              &43             &10                     \\
Brand (1989)            &$^{12}$CO(1-0) &115\,GHz       &IRAM 30-m                              &21             &5                      \\
Zapata et al. (2006)    &Continuum      &0.7; 1.3\,cm   &VLA                                    &$\sim$1.8$\times$1.6; $\sim$2.4$\times$1.6 &   \\[-0.5mm]
                        &Continuum      &3.6\,cm        &VLA                                    &4.7$\times$2.7 &                       \\
Zhang et al. (1998)     &NH$_3$         &$\sim$22.7\,GHz &VLA                                   &4.2$\times$2.8 &                       \\
Zhang et al. (1999)     &NH$_3$         &23.87; 24.14\,GHz  &VLA                                &1;0.31$\times$0.37;3.9$\times$3.1 &    \\
Zhang et al. (2001)     &$^{12}$CO(2-1) &230.5\,GHz     &NRAO 12-m;15-m JCMT                    &\multicolumn{2}{l}{see Zhang et al. (2005)} \\
Zhang et al. (2002)     &NH3            &$\sim$22.7; 24\,GHz    &VLA                            &3--5           &                       \\
Zhang et al. (2005)     &$^{12}$CO(2-1) &230.5\,GHz     &NRAO 12-m                              &$\sim$29       &                       \\[-0.5mm]
                        &$^{12}$CO(2-1); (3-2)&230.5; 345\,GHz  &CSO 10.4-m                     &$\sim$30       &                       \\
Zinchenko et al. (2000) &HNCO           &330--461\,GHz  &10-m HHT                               &25--18         &5                      \\[-0.5mm]
                        &HNCO; C$^{18}$O(2-1) &110--220\,GHz &SEST 15-m                         &47--24         &5                      \\[-0.5mm]
                        &HNCO; C$^{18}$O(1-0)&88--110\,GHz &OSO 20-m                         	&40--35         &5                      \\[-0.5mm]
                        &HNCO           &21.981\,GHz    &Effelsberg 100-m                       &40             &10                     \\
Zoonematkermani et al. (1990)&Continuum &20\,cm         &VLA                                    &$>$5.07        &$\sim$3                \\
\hline
\end{tabular}
\end{minipage}
\end{table*}
\end{center}

\bsp

\label{lastpage}

\end{document}